\documentclass[timesnewroman, 12pt]{report}
\usepackage[titletoc]{appendix}
\usepackage[a4paper]{geometry}
\usepackage[titletoc]{appendix}
\usepackage{authblk,lineno,color,lscape,url,sectsty}
\usepackage[autostyle]{csquotes}
\usepackage[dvipsnames]{xcolor}
\usepackage[labelfont=bf]{caption}
\usepackage{emptypage}
\usepackage{array,float,multirow,graphicx,setspace,cite,url,titlesec,tocloft}
\usepackage{amsmath,amsfonts,mathtools,notoccite}
\usepackage{chngcntr,rotating,pdflscape,enumitem,algpseudocode,algorithm,subcaption}
\usepackage[square,sort,comma,numbers,sort&compress]{natbib}
\usepackage[colorlinks=True,citecolor=blue,linkcolor=blue,anchorcolor=blue,filecolor=blue,urlcolor=blue]{hyperref}
\usepackage{tensor}
\usepackage{braket}
\usepackage{booktabs}
\usepackage[flushleft]{threeparttable}
\usepackage{lipsum}
\usepackage[Sonny]{fncychap}
\usepackage{tensor}
\DeclareUnicodeCharacter{0301}{\'{e}}
\def\blankpage{%
	\clearpage%
	\thispagestyle{empty}%
	\addtocounter{page}{-1}%
	\null%
	\clearpage}
\let\oldAA\AA
\renewcommand{\AA}{\text{\normalfont\oldAA}}
\begin{document}
\pagenumbering{roman}
\thispagestyle{empty}
\thispagestyle{empty}
\graphicspath{{Figures/PNG/}{Figures/}}
\begin{center}
{\bf {\Large Development of large scale CVD grown two dimensional materials for field-effect transistors, thermally-driven neuromorphic memory, and spintronics applications}}
\end{center}
\vspace{0.3 cm}
\begin{center}
    {\it \large By}
\end{center}
\begin{center}
    {\bf {\Large Sameer Kumar Mallik } \\ PHYS07201804016}
\end{center}
\begin{center}
\bf {{\large Institute of Physics, Bhubaneswar, India }}
\end{center}
\vskip 1.5 cm
\begin{center}
\large{
{ A thesis submitted to the }  \\
 {Board of studies in Physical Sciences }\\

In partial fulfillment of requirements \\
For the Degree of } \\
{\bf  DOCTOR OF PHILOSOPHY} \\
\emph{of} \\
{\bf HOMI BHABHA NATIONAL INSTITUTE}
\vskip 1.0 cm
\begin{figure}[H]
	\begin{center}
    \includegraphics[width=4.0cm, height= 4.0cm]{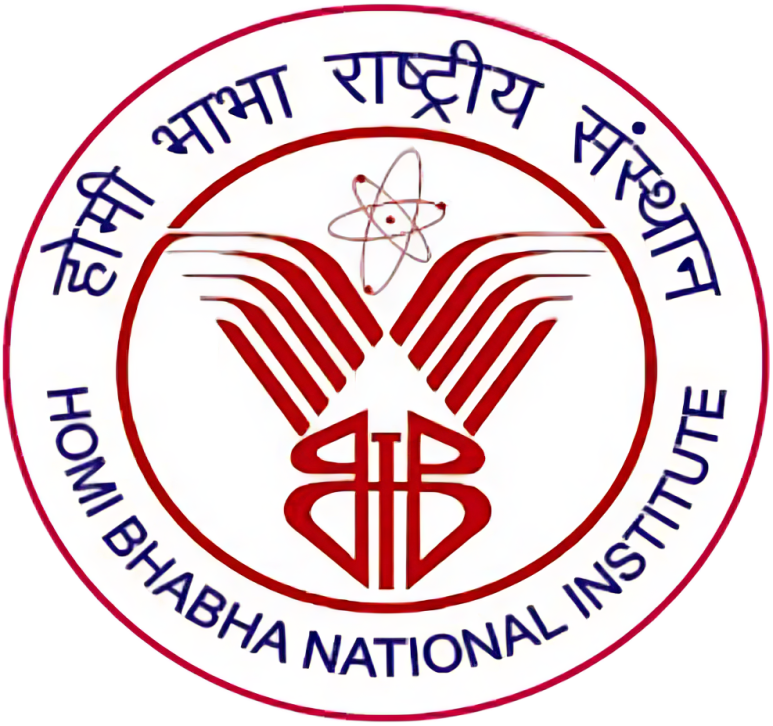}
	\end{center}
\end{figure}
\vskip 0.2 cm
\bf {May, 2024}
\end{center}
\thispagestyle{empty}
\blankpage     
\begin{center}
\textbf{\large \underline{List of Publications arising from the thesis}}
\end{center}
\begin{enumerate}

         \item \textbf{S. K. Mallik}, S. Sahoo, M. C. Sahu, S. K. Gupta, S. P. Dash, R. Ahuja, S. Sahoo
        \textit{Salt-assisted growth of monolayer $MoS_2$ for high-performance hysteresis-free field-effect transistor}.
        \href{https://doi.org/10.1063/5.0043884}{\textcolor{blue}{Journal of Applied Physics 129, 145106 (2021)}}.
        \\
         
         \item  \textbf{S. K. Mallik}, A. K. Jena, N. K. Sharma,  S. Sahoo, M. C. Sahu, S. K. Gupta, R. Ahuja, S. Sahoo
		\textit{Transition metal substituted $MoS_2/WS_2$ van der Waals heterostructure for realization of dilute magnetic semiconductors}.
		\href{https://doi.org/10.1016/j.jmmm.2022.169567}{\textcolor{blue}{Journal of Magnetism and Magnetic Materials 560 169567 (2022)}}. 
  \\

             \item A. K. Jena*, \textbf{S. K. Mallik*}, M. C. Sahu, S. Sahoo, A. K. Sahoo, N. K. Sharma, J. Mohanty, S. K. Gupta, R. Ahuja, S. Sahoo
		\textit{Strain‑mediated ferromagnetism and low‑feld magnetic reversal in Co doped monolayer $WS_2$}.
		\href{https://doi.org/10.1038/s41598-022-06346-w}{\textcolor{blue}{Sci Rep 12, 2593 (2022)}}. (* indicates authors of equal contributions.) 
  \\

        \item \textbf{S. K. Mallik}, R. Padhan, M. C. Sahu, S. Roy, G. K. Padhan, P. K. Sahoo, S. P. Dash, S. Sahoo.  
		\textit{Thermally-driven multilevel non-volatile memory with monolayer $MoS_2$ for Brain-inspired artificial learning}.
  \href{https://doi.org/10.1021/acsami.3c06336}{\textcolor{blue}{ACS Appl. Mater. Interfaces 2023, 15, 30, 36527–36538}}.\\

         \item \textbf{S. K. Mallik}, R. Padhan, M. C. Sahu, G. K. Padhan, P. K. Sahoo, S. P. Dash, S. Sahoo.  
		\textit{Ionotronic $WS_2$ Memtransistors for 6-bit Storage and Neuromorphic Adaptation at High Temperature}. 
  \href{https://doi.org/10.1038/s41699-023-00427-8}{\textcolor{blue}{npj 2D Mater Appl 7, 63 (2023)}}.\\

        \item \textbf{S. K. Mallik}, R. Padhan, S. Roy, M. C. Sahu, S. Sahoo, S. Sahoo.  
		\textit{Direct Transfer of Monolayer MoS$_2$ Device Arrays for Potential Applications in Flexible Electronics}. 
  \href{https://doi.org/10.1021/acsanm.3c05400}{\textcolor{blue}{ ACS Appl. Nano Mater. 2024, 7, 5, 4796–4804}}.\\

        \item \textbf{S. K. Mallik}, R. Padhan, A. K. Panigrahi, A. Kumar, S. Roy, M. C. Sahu, S. Sahoo, G. K. Padhan, S. Sahoo.
         \textit{Spectroscopic and Optoelectronic Investigations on Controlled Growth of MoS$_2$ using a Salt-assisted Modified CVD technique }. {\textcolor{blue}{(under review, 2024)}}.
\\

\end{enumerate}

\begin{center}
\textbf{\large \underline{List of Publications (Not included in the thesis)}}
\end{center}

\vspace{0.5 cm}

\begin{enumerate}

        \item M. C. Sahu, A. K. Jena, \textbf{S. K. Mallik}, S. Roy, S. Sahoo, R. S. Ajimsha, P. Misra, S. Sahoo 
        \textit{Reconfigurable Low Power $TiO_2$ Memristor for Integration of Artificial Synapse and Nociceptor}. 
        \href{https://doi.org/10.1021/acsami.3c02727}
        {\textcolor{blue}{ACS Appl. Mater. Interfaces 2023, 15, 21, 25713–25725}}.
        \\
     
        \item S. Sahoo, M. C. Sahu, \textbf{S. K. Mallik}, A. K. Jena, G. K. Pradhan, S. Sahoo 
        \textit{High Responsivity in Monolayer $MoS_2$ Photodetector via Controlled
Interfacial Carrier Trapping}. 
        \href{https://doi.org/10.1021/acsaelm.2c01567}{\textcolor{blue}{ACS Appl. Electron. Mater. 2023, 5, 2, 1077–1087}}.
        \\
     
        \item  A. K. Jena, M. C. Sahu, K. U. Mohanan, \textbf{S. K. Mallik}, S. Sahoo, G. K. Pradhan, S Sahoo
		\textit{Bipolar Resistive Switching in $TiO_2$ Artificial Synapse Mimicking Pavlov’s Associative Learning}.
		\href{https://doi.org/10.1021/acsami.2c17228}{\textcolor{blue}{ACS Appl. Mater. Interfaces 2023, 15, 2, 3574–3585}}.
  \\

        \item M. C. Sahu, S. Sahoo, \textbf{S. K. Mallik}, A. K. Jena, S. Sahoo
        \textit{Multifunctional 2D $MoS_2$ Optoelectronic Artificial Synapse with Integrated Arithmetic and Reconfigurable Logic Operations for In-Memory Neuromorphic Computing Applications}.
        \href{https://doi.org/10.1002/admt.202201125}{\textcolor{blue}{Adv. Mater. Technol. 2022, 2201125}}.
        \\

           \item \textbf{S. K. Mallik}, S. Sahoo, M. C. Sahu, A. K. Jena, G. K. Pradhan, S. Sahoo
        \textit{Polarized Moiré Phonon and Strain-Coupled Phonon Renormalization in Twisted Bilayer $MoS_2$}.
        \href{https://doi.org/10.1021/acs.jpcc.2c04201}{\textcolor{blue}{J. Phys. Chem. C 2022, 126, 37, 15788–15794}}.
        \\

       \item N. K. Sharma, S. Sahoo, M. C. Sahu, \textbf{S. K. Mallik}, A. K. Jena, H. Sharma, S. K. Gupta, R. Ahuja, S. Sahoo
        \textit{Electronic bandstructure modulation of MoX$_2$/ZnO (X:S,Se) heterostructure by applying external electric field}.
        \href{https://doi.org/10.1016/j.surfin.2022.101817}{\textcolor{blue}{Surfaces and Interfaces 29 (2022) 101817}}.
        \\
 
        \item A. K. Jena, M. C. Sahu, S Sahoo, \textbf{S. K. Mallik}, G. K. Pradhan, J. Mohanty, S. Sahoo
        \textit{Multilevel resistive switching in graphene oxide-multiferroic thin-film-based bilayer RRAM device by interfacial oxygen vacancy engineering}.
        \href{https://doi.org/10.1007/s00339-021-05243-9}{\textcolor{blue}{Appl. Phys. A 128, 213 (2022)}}.
        \\

        \item S. Sahoo, M. C. Sahu, \textbf{S. K. Mallik}, N. K. Sharma, A. K. Jena, S. K. Gupta, R. Ahuja, S. Sahoo
        \textit{Electric Field-Modulated Charge Transfer in Geometrically Tailored $MoX_2/WX_2$ (X = S, Se) Heterostructures}.
        \href{https://doi.org/10.1021/acs.jpcc.1c07218}{\textcolor{blue}{J. Phys. Chem. C 2021, 125, 40, 22360–22369}}.
        \\

        \item  M.C. Sahu, \textbf{S. K. Mallik}, S. Sahoo, S. K. Gupta, R. Ahuja, S. Sahoo
        \textit{Effect of Charge Injection on the Conducting Filament of Valence Change Anatase $TiO_2$ Resistive Random Access Memory Device}.
        \href{https://doi.org/10.1021/acs.jpclett.1c00121}{\textcolor{blue}{J. Phys. Chem. Lett. 2021, 12, 7, 1876–1884}}.
        \\

        \item S. Sahoo, {\bf S. K. Mallik}, M. C. Sahu, A. Joseph, S. Singh, S. K. Gupta, B. Rout, G. K. Pradhan, S. Sahoo
		\textit{Thermal conductivity of free-standing silicon nanowire using Raman spectroscopy}. \href{https://doi.org/10.1088/1361-6528/abb42c}{\textcolor{blue}{Nanotechnology 31 (2020) 505701}}. 
  \\




    \end{enumerate}

\blankpage
\vspace*{3.0in}
\hspace*{2.0in}
\begin{center}
        {\Large \emph{Dedicated to my beloved family members and friends}\\}
\hrule
\end{center}
    
\blankpage  
\begin{center}
\large \bf
ACKNOWLEDGMENTS
\end{center}
\vspace{0.5cm}
\par This thesis would not have been possible without the help, support, and guidance of many individuals, who have stood beside me in this beautiful journey. 
\par First, I would like to show my deep gratitude to my supervisor, Prof. Satyaprakash Sahoo for his guidance and belief in me. His scientific perspective, time-to-time insightful advice, and endless encouragement have shaped my research life to a great extent which made me grow academically and non-academically. His methodical approach has taught me to think positively and look at challenges as opportunities. I always believe he has a sixth sense and I will never forget his ample role in my PhD journey which led to fruitful publications. I am also thankful to his family members (Soumya Maam and Adi) for showing their love, care, and affection on multiple occasions.

\par Next, I would like to thank my labmates cum friends, Mousam C Sahu and Sandhyarani Sahoo who have shared the ups and downs during my research journey. I will never forget our laughter, gossip, sharing of food, collaborative discussions, and late-night experiments during the initial days of research. My gratitude also goes to the former postdoc member Dr. Anjan K Jena and former visiting scientist Dr. Neha K Sharma for providing necessary assistance during the COVID period and sharing their knowledge and experience in their respective domains. Roshan Padhan, I am grateful for being such a beautiful partner who always stood beside me like a younger brother and has shared enjoyable lab experiences with me. I also thank other popular figures of the lab Suman Roy, Ashis K Panigrahi, Alok Kumar, and Smruti R Senapati for their kind gestures and support in various experiments and teamwork. My special thanks to Dr. Harish C Das with whom I lived many cheerful moments, and who constantly guided me a way out through many difficulties. I want to thank my other scholar friends from IoP (Siddharth, Arpan, Chitrak, Pragyan, etc.) for being the gossip bodies and scholars from KIIT for extending their hands in carrying out Raman experiments.

\par I extend my appreciation to Prof. Gopal K Pradhan, KIIT for his unwavering support from the beginning of my PhD journey and for providing me the opportunity to use Raman and PL spectroscopy. I want to express my sincere gratitude to Prof. Sanjeev K. Gupta, St. Xavier's college; Ahmedabad, Prof. Prasana K Sahoo; IIT Kharagpur, and Prof. Saroj P Dash; Chalmers university of technology, Sweden for their invaluable insights, which enriched my understanding of the subject matter. I am also grateful to my doctoral committee members Prof. B. R. Sekhar, IoP: Committee Chairperson, Prof. Arijit Saha, IoP: Dean academic, Prof. Debakanta Samal, IoP, Prof. Kartik Senapati, NISER whose guidance contributed significantly to the smooth progression of my academic journey. Furthermore, I am thankful to the IoP administrative team for listening to my complaints and ensuring smooth functioning of my fellowships. Above all, I extend my thanks to IoP, HBNI, DAE, and SERB for their financial support, which facilitated the successful completion of my doctoral journey.

\par My special gratitude to my friends Dr. Saswat, Ayesha, and Sarthak, with whom I started my campaign to pursue for PhD after Batchlers. Together, we shared common challenges and provided each other with vital support during difficult times. Additionally, I am grateful to my seniors, Chinu bhai and Jaga bhai, for their enduring presence and guidance throughout this journey. Special appreciation also goes to Tanmaya, Rupashree, Vijaylaxmi, Sonali, Anjali, and Aakanshya for brightening my life in Bhubaneswar beyond the confines of work hours.

\par Finally, a heartfelt thank you to my parents (Bapa and Bou) and my little brother Bunty for the belief and support that have been the cornerstone of my success. Their endless sacrifices, whether it was providing emotional support or ensuring I had the resources to pursue my dreams, have been truly invaluable. I owe a debt of gratitude to my extended family members for their unconditional love and prayers,  without which none of my accomplishments would have been possible.     

\vspace{1.2cm}
\begin{flushright}
	\textbf{(Sameer Kumar Mallik)}
\end{flushright}

\setcounter{page}{7}
\hypersetup{linkcolor=blue}
\tableofcontents
\clearpage
\numberwithin{equation}{chapter}
\numberwithin{figure}{chapter}
\numberwithin{table}{chapter}
\addcontentsline{toc}{chapter}{Abstract}
\begin{center}
	{\large\textbf{Abstract}}
\end{center}
In recent years, the field of semiconductor research has witnessed a paradigm shift towards exploring two-dimensional (2D) materials as potential candidates for next-generation electronic devices. This shift is primarily driven by the limitations of state-of-the-art silicon technology, which is approaching its physical scaling limits. Among 2D materials, monolayer Transition Metal Dichalcogenides (TMDCs) have emerged as promising candidates due to their exceptional optoelectronic properties and potential for integration into advanced devices. This thesis explores the synthesis of 2D TMDCs like MoS$_2$ and WS$_2$ using modified Chemical Vapor Deposition (CVD) techniques and investigates their optical properties through Raman and photoluminescence (PL) spectroscopy. By optimizing the synthesis process using NaCl-assisted CVD method, various growth challenges such as premature growth, defects, and non-uniformity in monolayer MoS$_2$ samples are addressed. Additionally, device fabrication techniques for 2D material-based transistors are discussed, highlighting the superior performance of salt-assisted CVD-grown monolayer MoS$_2$ FETs, exhibiting hysteresis-free behavior and high field-effect mobility. Moreover, the research explores novel transfer techniques for 2D material devices, presenting a poly(methyl methacrylate) (PMMA)-assisted etching-free one-step approach for transferring device arrays onto diverse substrates, which demonstrates an improved transistor performance and applications in flexible optoelectronic devices.

Furthermore, this thesis delves into the multifunctionality of monolayer MoS$_2$ mem-transistors, which exhibit both high-performance transistor behavior at room temperature and synaptic multi-level memory operations at higher temperatures. By leveraging interfacial physics, ion dynamics, and hysteresis inversion, non-volatile memory with multi-level storage capabilities in monolayer MoS$_2$ mem-transistors is proposed, paving the way for efficient data processing and storage using 2D semiconductors. Additionally, the thesis explores the development of multi-bit ultra-high-density memory devices using monolayer WS$_2$, demonstrating over 64 storage states (6-bit memory operation). We also demonstrate key neuromorphic behaviors, such as biomimetic plasticity, near linear potentiation, and depression, making the 2D mem-transistors suitable for successful implementation in high-temperature neuromorphic computing. We further utilize ionotronic non-volatile conductance weights to establish excellent pattern recognition accuracy approaching software limit during artificial neural network (ANN) training of MNIST datasets. Moreover, with the implementation of extensive first-principles density functional theory (DFT) calculations, we demonstrate the MS$_2$ (M = Mo, W) monolayers, as well as their van der Waals (vdW) hetero-bilayers as promising candidates for the successful realization of 2D dilute magnetic semiconductors (DMS) with the incorporation of Mn and Co dopants. Our studies also reveal promising magnetic behaviors under strain engineering and doping configurations and the employment of micromagnetic simulations provide insights into the magnetic properties of Co-doped WS$_2$ monolayers, indicating potential applications in nanoscale spintronics and straintronics. The findings in this thesis offer significant contributions to the advancement of 2D materials and their applications in next-generation electronics and spintronic devices.

\clearpage 
\newpage
\setcounter{page}{1}
\pagenumbering{arabic}
\chapter{Introduction}
\label{C1} 
\section{CMOS scaling: Beyond Moore’s law}
The evolution of semiconductor technology has reached significant advancements over the last few decades, thanks to Moore’s law which has served as a remarkable guiding principle for miniaturization of future microelectronics.\cite{Schaller1997Jun} Transistors, known as the basic building blocks of a microelectronic chip have emerged as potential technology drivers in achieving high-performance computing platforms. To date, one microprocessor consists of more than 40 billion single transistors due to the continuous downscaling of Si-based process nodes which has been achieved by advancements in CMOS technology. Currently, the physical dimension of a transistor has also reached below 10 nm in Si-CMOS nodes following Moore’s Law.\cite{Thompson2006Jun} However, the parasitic effects arising from the physical limit of bulk materials such as Si, Ge possess many challenges in further downscaling of the process nodes. These challenges include area efficiency bottleneck due to the current transistor architecture that permits to design of only one top gate electrode to match the doping level of bottom and top channel surfaces. The scaling bottleneck severely affects the performance of the Si-transistors due to short channel effects such as carrier velocity saturation, unstable current ratios, and degradation of sub-threshold behaviour. The increase in the leakage current also affects the energy dissipation while miniaturizing the Si-based integrated circuits (ICs) beyond physical limits. As a result, there are still power and efficiency bottlenecks which makes difficult to overcome the Boltzmann tyranny and Landauer switching limit.\cite{Nadeem2021Apr,Liu}
\begin{figure}
	\centering
	\includegraphics[width=0.9\columnwidth]{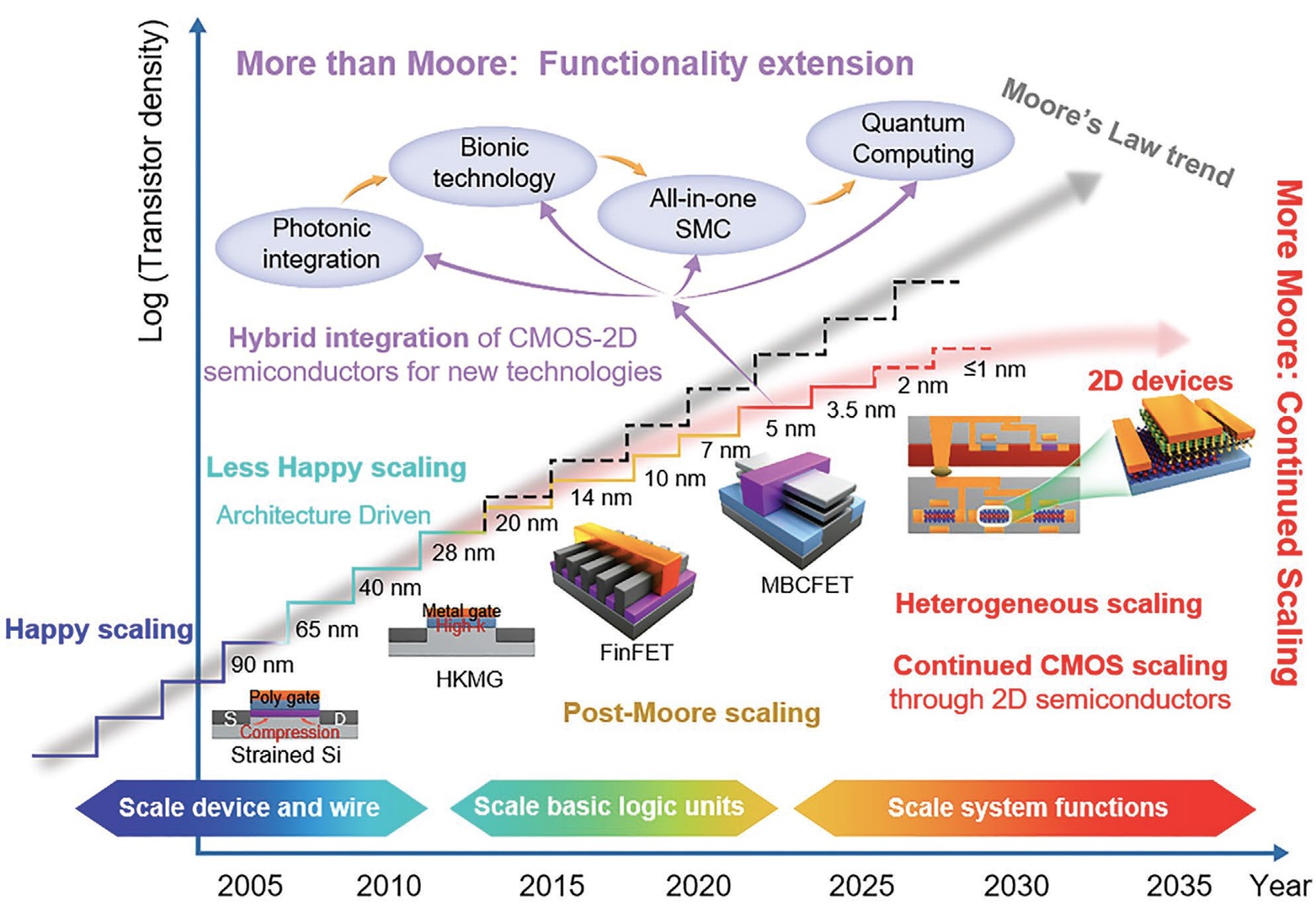}
	\caption{The evolution trend of the technology nodes for Si transistors. This figure is taken with permission from \cite{Peng_Zhou_2022}.}
	\label{Fig_1.1}
\end{figure}

Additionally, in this modern era of electronic devices, memory with ultrahigh data storage density is necessary for smarter computing platforms and vast information processing applications. The Von–Neumann computing architecture with physically separated memory and processing units generates a huge gap between high-speed writing and long retention times. In such architectures, the information is transmitted via data buses which results in energy loss and latency in Si-based microchips. The memory dilemma has created severe criticality (such as the Von-Neumann bottleneck) in the further development of ICs.\cite{Wali2023Oct} The journey to go beyond the limitations of Moore's Law begins with an understanding of the challenges that have arisen with traditional CMOS scaling. As transistor dimensions shrink to the nanoscale, quantum effects, power dissipation, and manufacturing complexities become formidable obstacles. Figure \ref{Fig_1.1} shows the trend of Moore’s law from the initial happy scaling to the architecture-driven less happy scaling represented by strained silicon and high-\textit{k} metal gate-based field-effect transistors (FETs).\cite{Peng_Zhou_2022} Aiming for maximizing gate control over the channel area to improve the short channel effects, the post-Moore scaling era has gone through different transistor configurations with state-of-the-art techniques such as fin FETs, gate-all-around FETs, and multibridge channel FETs.\cite{Peng_Zhou_2022} Further scaling beyond 5 nm intensifies the challenges possessed by silicon-based processors for which the golden era of Moore’s law is anticipated to flatten in the near future. To continue the scaling, the development of novel techniques with new emerging materials remains the only choice to realize the functionality extension beyond Von-Neumann architectures at ultra-scaled dimensions.  
\section{Emergence of atomically thin two dimensional materials}
\begin{figure}
	\centering
	\includegraphics[width=0.55\columnwidth]{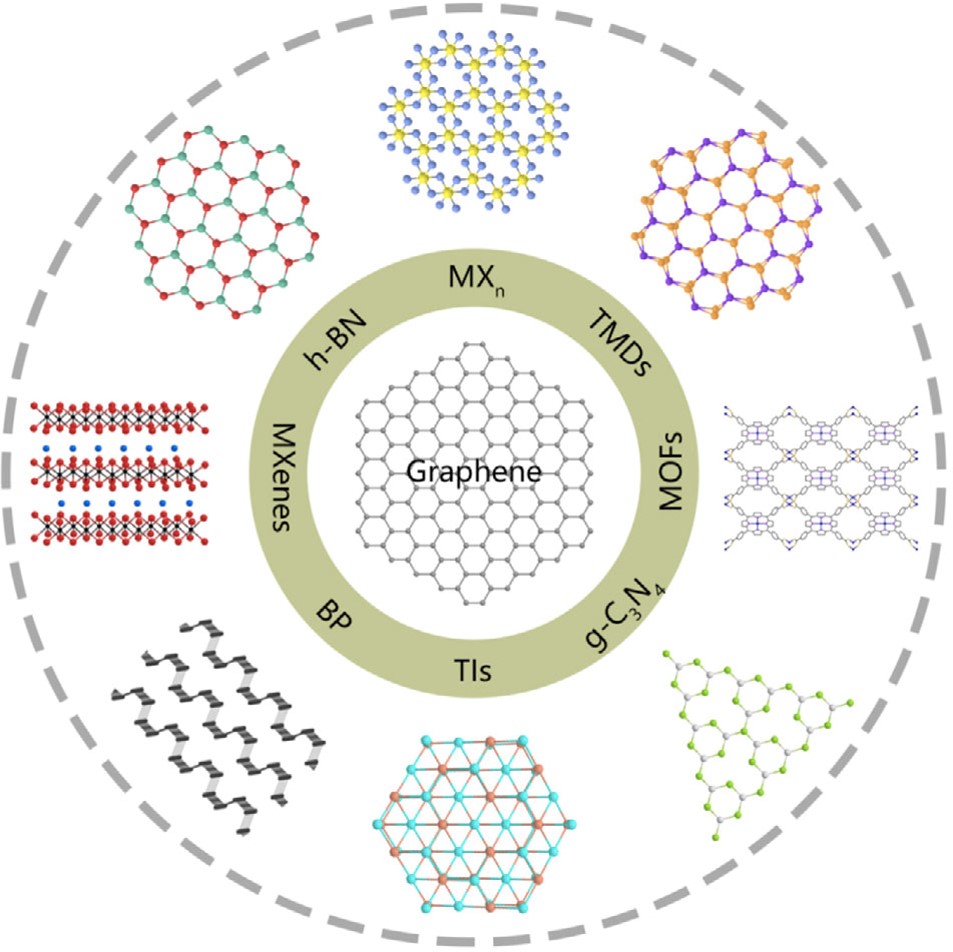}
	\caption{The family of atomically thin two dimensional materials. This figure is taken with permission from \cite{Guo_2019}.}
	\label{Fig_1.2}
\end{figure}

Two-dimensional (2D) materials represent a fascinating and rapidly evolving class of quantum materials that have captured the attention of scientists and engineers in recent years.\cite{Novoselov2016Jul} 2D materials (see Fig.\ref{Fig_1.2}) belong to the class of quantum materials that represent atomically thin layers with strong in-plane covalent bonds and weak out-of-plane van der Waal (vdW) bonds. The early research on 2D materials is primely focused on graphene (single layer of carbon atoms arranged in a hexagonal honeycomb lattice), a pioneering discovery in 2004 by A. K. Geim and K.S. Novoselov that has revealed extraordinary electronic, thermal, and mechanical properties at the limit of atomic thickness when compared to bulk graphite.\cite{Geim2007Mar} Though graphene possesses high carrier mobility and quantum mechanical confinement within $\sim$ 3.3 \AA, the semimetal nature with zero bandgap hinders its potential applications in switching devices such as transistors. Since then, other 2D materials have gained tremendous attention in both fundamental and applied research. Among them transition metal dichalcogenides (TMDCs), black phosphorus (BP), boron nitride, MXenes, and other 2D materials have since been unveiled, each presenting a distinctive set of characteristics and potential applications.\cite{Guo_2019} For example, graphene stands out for its exceptional electronic conductivity, while TMDCs provide semiconductor characteristics with a tunable band gap. BP, on the other hand, offers a versatile band gap, opening avenues for applications in optoelectronics.
The hexagonal boron nitride's (hBN's) insulating properties make it an ideal substrate for electronic devices. Hybrid structures like hBN/graphene bring synergistic benefits, capitalizing on the strengths of each constituent material. As researchers delve into the intricacies of these 2D materials, their combined use and novel configurations hold promise for advancing the CMOS platform. As the family of 2D materials continues to expand, researchers are uncovering novel ways to harness their properties for groundbreaking advancements in electronics, optoelectronics, and energy storage devices.\cite{Lemme2022Mar}  
\section{Transition metal dichalcogenides}
The 2D transition metal dichalcogenides (TMDCs) with a general formula of MX$_2$, where M is a transition metal (Mo, W, etc.) and X being a chalcogen (S, Se, Te), bear tremendous potential in miniaturizing electronic devices owing to their wide bandgap variability, high thermal and electrical conductivity, dangling-bond-free surfaces, and excellent mechanical flexibility.\cite{Manzeli2017Jun} In their monolayer form, TMDCs become direct bandgap semiconductors that possess large exciton binding energy, and strong spin-valley coupling that hold promising applications in the fields of optoelectronics, spintronics, valleytronics, and quantum optics.\cite{Shim2017Apr,Ahn2020Jun,Schaibley2016Aug,Turunen2022Apr} Monolayer MoS$_2$ being the most studied material in this family shows potential integration with silicon (ICs) that could further sustain Moore’s law scaling and improve device performance, as reported recently.\cite{Das2021Nov} Furthermore, the introduction of new members within the TMDCs family, including metallic NbSe$_2$ and VSe$_2$,\cite{Kezilebieke2020Jun} coupled with the ability to stack different 2D layers to form van der Waals (vdW) moiré superlattices,\cite{Zhang2017Jan} has unveiled a plethora of emergent phenomena. These phenomena encompass moiré excitons and various properties such as ferroelectricity, ferromagnetism, topological attributes, and superconductivity, all occurring at the 2D limit.\cite{Wilson2021Nov,Vincent2021Dec,Foutty2023Jun}
\subsection{Structural and electronic properties}
\begin{figure}
\centering
\includegraphics[width=0.8\columnwidth]{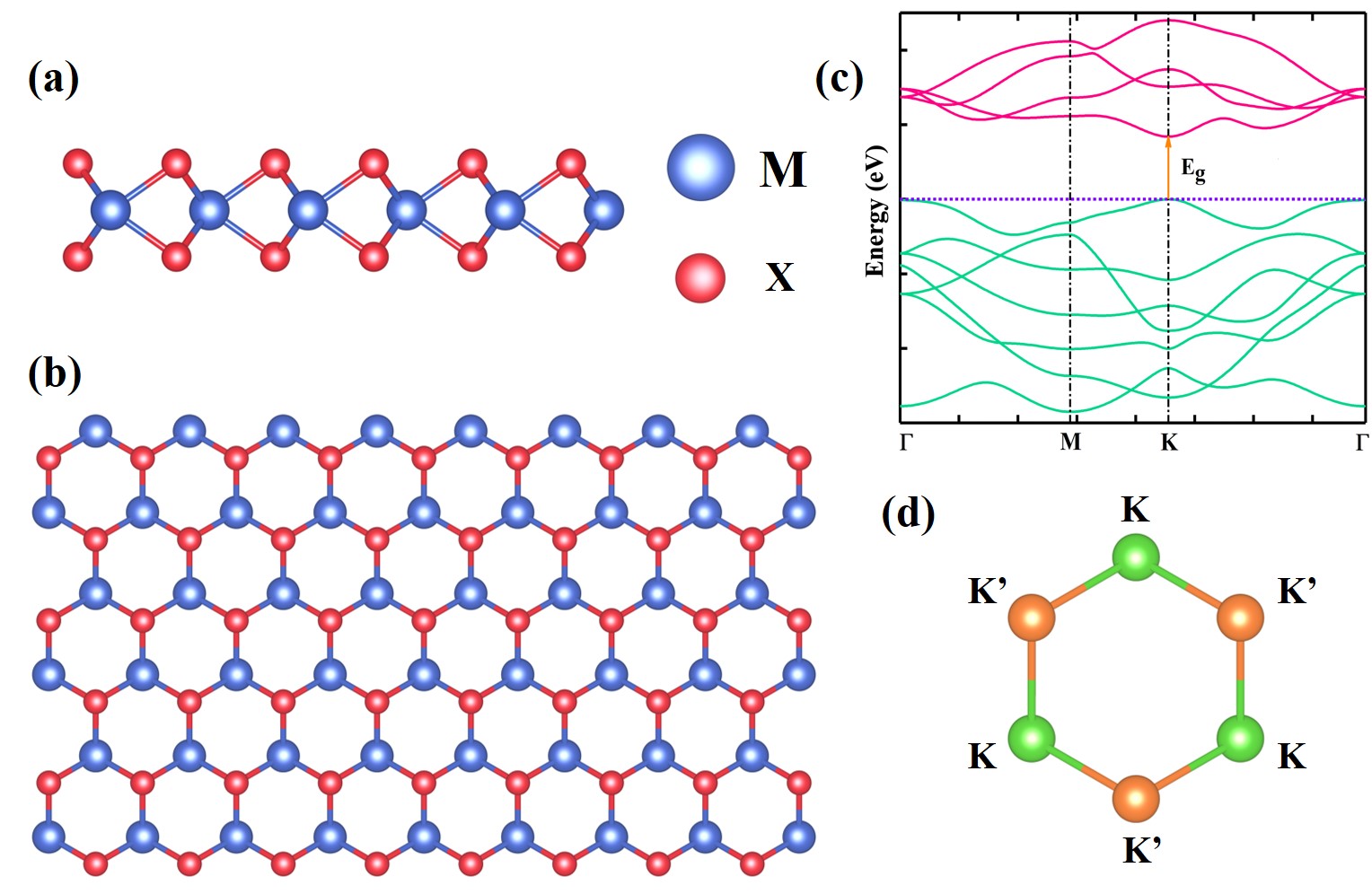}
\caption{(a) Side view of monolayer MoS$_2$ with sandwiched X-M-X structure (b) Top view showing the hexagonal crystal lattice (c) Calculated electronic bandstructure of monolayer MoS$_2$ showing direct bandgap at K point (d) Schematic representation of a Brillouin zone of MoS$_2$ with K, and K' valley points.}
\label{Fig_1.3}
\end{figure}

Unlike graphene, TMDCs are characterized by a unique crystal structure, where transition metal atoms are sandwiched between two layers of chalcogen atoms, as shown in Fig.\ref{Fig_1.3}(a). 
The thickness of a monolayer TMDC material is confined to 7 $\AA$. Depending on the atomic arrangements, TMDCs can form two different phases: a trigonal prismatic (2H) semiconducting phase and an octahedral (1T) metallic phase. For example, in semiconducting 2H MoS$_2$, each Mo atom is coordinated to six surrounding S atoms in a trigonal prismatic manner that forms a thermodynamically stable phase, as shown in the top view representation in Fig.\ref{Fig_1.3}(b). However, In 1T-MoS$_2$, six S atoms form a distorted octahedron around the Mo atom, resulting in a metastable phase. In bulk or few-layer form, TMDCs have indirect bandgaps in the range of 0.9-1.6 eV. However, when scaled down to a single monolayer, an indirect to direct bandgap transition takes place due to quantum confinement effects increasing the bandgap range to 1.6-2.0 eV. As the layer number thinned down to monolayer, the interaction between the d orbital of the transition metal and antibonding p$_z$ orbital of the chalcogen atom decreases, raising the conduction band minimum at the $\Gamma$ point of the Brillouin zone and hence increasing the bandgap. A typical calculated electronic band structure of the monolayer MoS$_2$ results in a direct bandgap at the K-points, which is depicted in Fig.\ref{Fig_1.3}(c). The six K-points in the reciprocal space Brillouin zone also form a hexagonal arrangement, as illustrated in Fig. \ref{Fig_1.3} (d). These band extrema (also called valleys) give rise to optical interband transitions from the visible to near-infrared range. Owing to their controllable optoelectronic responses, the TMDCs could serve as potential replacements beyond silicon in the front-end-of-line (FEOL) technologies in the alternative back-end-of-line (BEOL) integration, and Internet of Things (IoT) based embedded sensors.

\section{Synthesis of transition metal di-chalcogenides}
\subsection{Top down approach}
In the top down synthesis approach, monolayer TMDCs can be obtained using a micromechanical exfoliation technique commonly known as the “scotch tape method”. This remains the most familiar synthesis technique for the past two decades since the exfoliation of graphene by using this technique. Due to the weak vdW force between adjacent layers and strong intralayer covalent bonding, high-quality single or few-layer 2D flakes can be mechanically peeled out from the bulk crystal and transferred to the target substrate by applying pressure. While this method yields flakes with the highest purity and cleanliness, making it ideal for studying the fundamental properties and demonstrating high-performance devices, however, it is not scalable, and there is limited control over the size and thickness of the flakes. Other exfoliation methods such as liquid phase ultrasonication, and electrochemical intercalation can increase the production of exfoliated nanosheets. However, due to major limitations, such as low yield of monolayer samples during ultrasonication, and loss of semiconducting properties in case of intercalation, these techniques limit the large-scale integration and can only be used for specific applications.
\subsection{Bottom up approach}
\begin{figure}
	\centering
	\includegraphics[width=0.8\columnwidth]{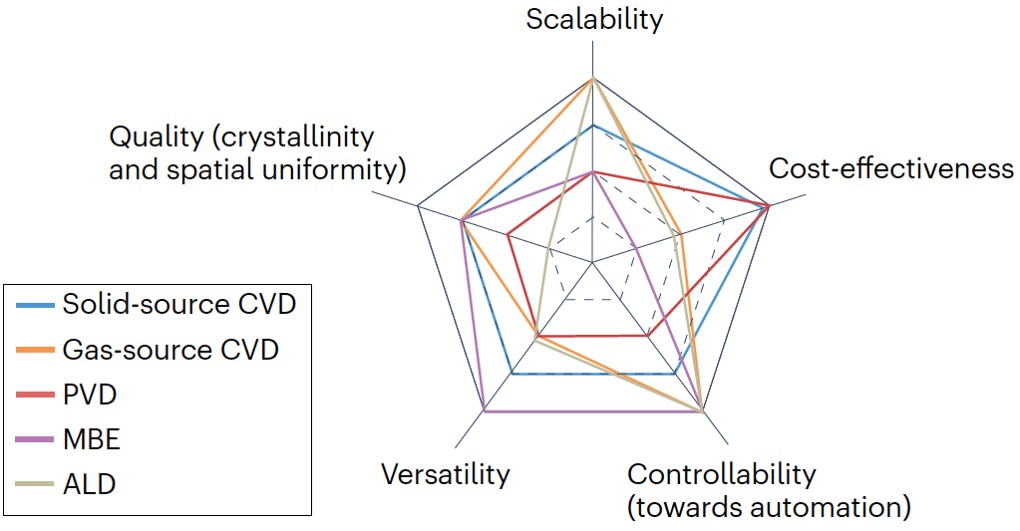}
	\caption{Comparison of various bottom up synthesis approaches. This figure is taken with permission from \cite{Zhang2023Dec}}.
	\label{Fig_1.4}
\end{figure}

The implementation of large-scale 2D crystals with consistent thickness for future CMOS circuits encounters limitations using the top-down approach. In order to achieve high-quality 2D films with layer-controllable thickness, several bottom up synthesis methods have been suggested. These techniques include physical vapour deposition (PVD), atomic layer deposition (ALD), molecular beam epitaxy (MBE), and various chemical vapour depositions.\cite{Zhang2023Dec} A comparative analysis of these approaches, considering scalability, quality (spatial uniformity, crystallinity, and defect density), versatility (applicable 2D layered chalcogenide materials), controllability for large-scale growth, and cost-effectiveness, is illustrated in Fig.\ref{Fig_1.4}. Although no perfect approach has emerged, gas-source CVD holds transformative potential with further tailoring for a broader range of TMDs at reduced operational costs. Conversely, solid-source CVD and PVD, being versatile and cost-effective, suit lab-scale explorations with unexplored materials; however, controllability issues must be addressed for automated, reproducible, and uniform large-scale fabrication. MBE can produce high-quality crystals, but it is expensive. ALD is still in its early stages of development, but it has the potential to produce high-quality, uniform crystals. 
Recent years have witnessed rapid progress in the exploration of diverse synthesis techniques, revealing numerous exciting developments over the period of time, which is illustrated in Fig.\ref{Fig_1.5}. Significant strides have been taken to enhance the versatility, scalability, quality, and manufacturability of vapor-phase-deposited 2D TMDCs, marking crucial breakthroughs for their practical applications.

\subsection{Chemical Vapour Deposition}
\begin{figure}
	\centering
	\includegraphics[width=1\columnwidth]{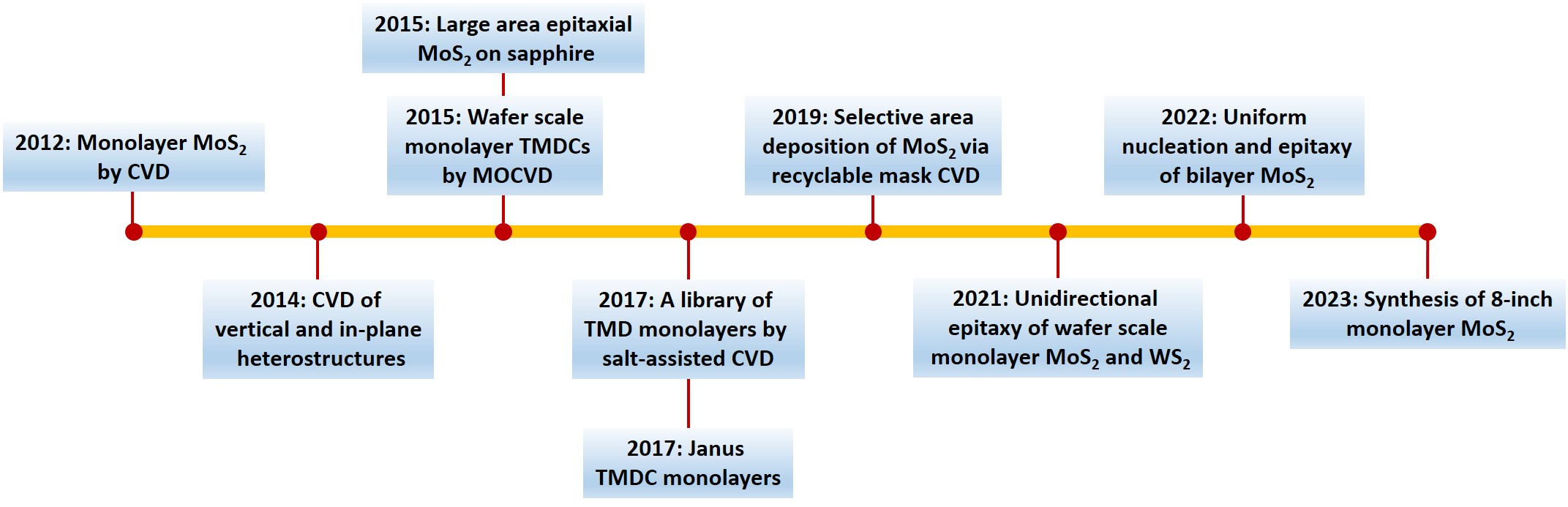}
	\caption{Recent advancements in synthesis approaches of TMDCs.}
	\label{Fig_1.5}
\end{figure}

Chemical vapour deposition (CVD) is one of the most explored synthesis routes to produce large-scale high-quality monolayer 2D semiconductors as channel materials for transistors. This technique involves vapour phase reaction of metal and chalcogen precursors under some specific growth parameters and finally, the product is deposited on the growth substrate. The precursors used for this technique can be solid powder of metal or gas containing metal-organic precursors such as carbonyls. The latter technique is referred to as metal-organic CVD (MOCVD) method where the metal ((Mo(CO)$_6$ or (W(CO)$_6$) and chalcogen gas precursors are introduced through carrier gas inside the CVD wall for the growth of 2D materials. The conventional CVD approach uses powdered metal oxides (i.e. MoO$_3$ and WO$_3$ for MoS$_2$ and WS$_2$ growth, respectively) as the metal source precursors, which are kept at a distance from the chalcogen (S, Se) precursors. Both precursors evaporate at high temperatures and react in vapour phase inside the CVD chamber. The growth conditions can be tuned by regulating certain growth parameters which determine the size, quality, and thickness of the samples. So far, various methods and mechanisms have been reported to understand the growth process to improve the crystal quality and uniformity for practical device applications,\cite{Zhang2023Dec} as shown in Fig.\ref{Fig_1.5}. The recently developed salt-assisted CVD synthesis in 2017 addresses such challenges by increasing reaction rate and degree of reproducibility. The most studied TMDCs material, MoS$_2$ grown by CVD method provides potential advantages in the progressive scaling of ongoing silicon technology. Recently, CVD-grown monolayer MoS$_2$ samples have predominantly been explored as an active material in the fabrication of intrinsic hysteresis-free transistors,\cite{Si2018Jan} high-responsive photodetectors,\cite{Wang2023Oct} high-performance CMOS arrays,\cite{Xu2018Nov} and neuromorphic computing devices.\cite{Wali2023Oct} Such broad applications require the supply of high-quality materials that necessitate in-depth studies of controlled synthesis and reproducibility. 
\section{Applications of transition metal di-chalcogenides}
\subsection{Field-effect transistors}
\begin{figure}
	\centering
	\includegraphics[width=0.9\columnwidth]{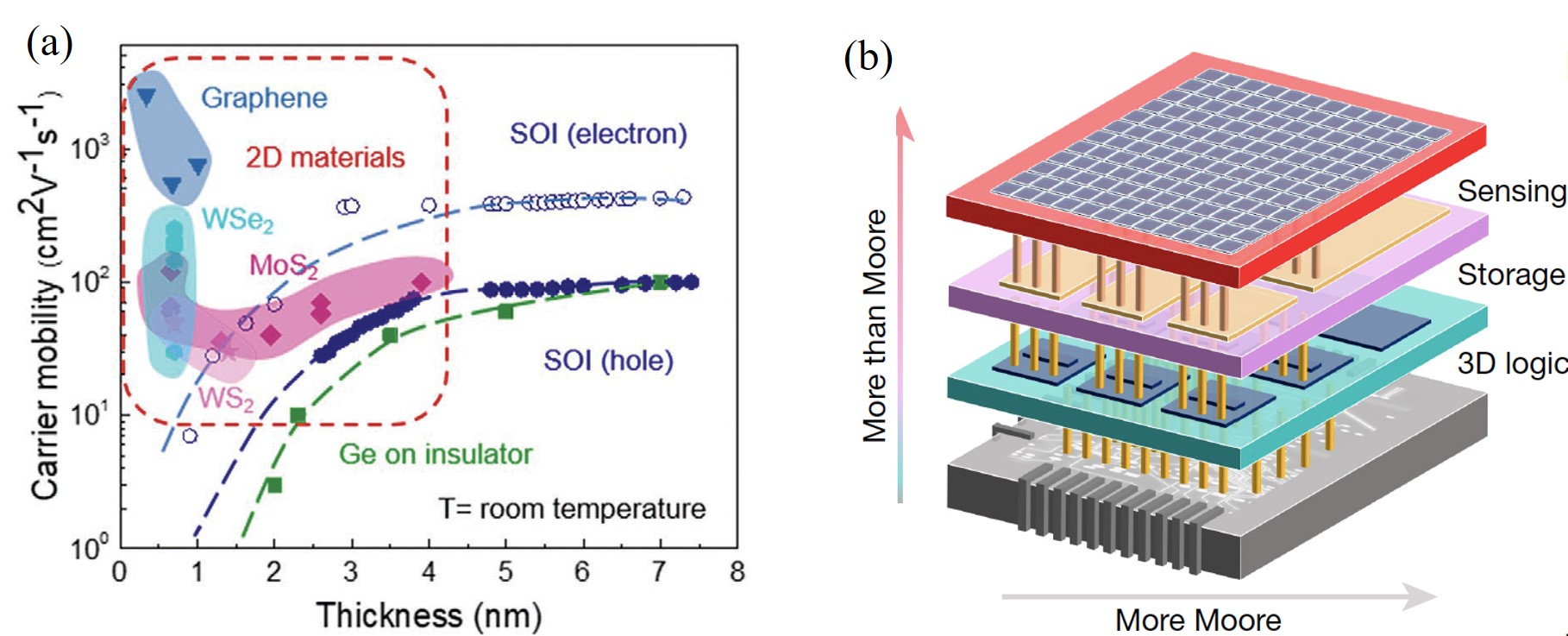}
	\caption{(a) Comparison of carrier mobilities as a function of channel thickness scaling in case of FETs based on Silicon, Germanium, graphene, and 2D TMDCs. This figure is taken with permission from \cite{Peng_Zhou_2022}. (b) Schematic diagram of the 3D monolithic integration of 2D FETs in ‘More Moore’ and ‘More than Moore’ architecture technologies. This figure is taken with permission from \cite{Jayachandran2024Jan}}.
	\label{Fig_1.6}
\end{figure}
A field-effect transistor (FET) functions as an electrical device with three terminals, wherein the modulation of electrical current flowing through the semiconductor channel occurs through the application of an external electric field. In essence, the FET device can transition from a state of complete conductivity when in its ``on-state" to being highly resistant to current flow when in its ``off-state." This transition between these two states is governed by the electric field applied at the gate terminal. The primary utilization of FETs lies in their capacity as switches, particularly within the domain of digital electronics, although they also serve as signal amplifiers for analog electronic applications. Presently, FETs serve as foundational components in the vast spectrum of modern electronic technologies.\cite{Das2021Nov}

Traditionally, silicon-based FETs have dominated the semiconductor industry, but recent advancements have sparked interest in exploring 2D materials, particularly TMDCs for FET applications. TMDC-based semiconductors such as molybdenum disulfide (MoS$_2$) and tungsten diselenide (WSe$_2$) possess atomically thin layers that allow for precise control over key transistor parameters such as carrier mobility, threshold voltage, and hysteresis. Carrier mobility, representing the ability of charge carriers to move through the material under an electric field, is a crucial parameter for transistor performance. TMDC-based FETs often demonstrate high carrier mobilities, which contribute to their excellent switching speeds and overall device efficiency.\cite{Sebastian2021Jan} The room temperature carrier mobilities corresponding to the channel thickness are plotted for various FETs, as depicted in Fig. \ref{Fig_1.6}(a).\cite{Peng_Zhou_2022} As indicated from the graph, the electron and hole mobilities of bulk silicon and germanium experience significant degradation, sharply decreasing below 4 nm technology nodes. In contrast, the mobilities of 2D materials like graphene, MoS$_2$, and WSe$_2$ remain unaffected by thickness reduction and can maintain high mobility even at the sub-nanometer scale. Moreover, TMDCs exhibit a distinct threshold voltage, the minimum electric field required to induce conductivity in the channel, which can be finely tuned by controlling the material thickness or applying external gate bias. This tunability enables the optimization of device characteristics for specific applications, such as low-power electronics or high-speed computing. Another important parameter in FETs is hysteresis, which refers to the difference in threshold voltage during the forward and reverse sweep of gate voltage. TMDC-based FETs typically exhibit minimal hysteresis, attributed to their high crystalline quality and reduced trap states, resulting in stable and reliable device operation. The sizable bandgap possessed by different TMDCs plays a crucial role in determining the device's functionality, allowing them to operate as either semiconductors or insulators. Furthermore, Jayachandran et. al. has successfully demonstrated a wafer scale 3D integration of 2D materials-based FETs with increased interconnect density and reduced electrostatic coupling.\cite{Jayachandran2024Jan} As shown in Fig. \ref{Fig_1.6}(b), the monolithic integration of FET arrays provides advanced packaging solutions for future miniaturized devices. These properties all together enable the design of TMDC-based FETs with tailored electronic properties, suitable for diverse applications ranging from digital electronics to optoelectronics. 

\subsection{Non-volatile memory}
Nonvolatile memory retains data for longer periods, exceeding 10 years, even in the absence of a power source, making it essential in our information-driven society.\cite{Meena2014Dec} The rise of  Big data and cloud computing creates a growing need for storing larger amounts of data in smaller spaces. Owing to their atomic thickness, high mobility, and sustainable miniaturization properties, 2D materials are considered one of the most promising alternatives to existing silicon-based integrated memory devices.\cite{Lemme2022Mar} As silicon-based flash memories have reached their physical architecture-driven limits, emerging 2D materials such as graphene, TMDCs, and h-BN show potential advancements in improving existing memory technologies and also pave the way for future high-density reliable storage devices. The key factors of a non-volatile memory include high stability and endurance which can produce repeatable memory operations. Another figure of merit is high extinction ratios which increase the multi-level memory capacity in a single device.
\begin{figure}[hbt!]
	\centering
	\includegraphics[width=0.9\textwidth]{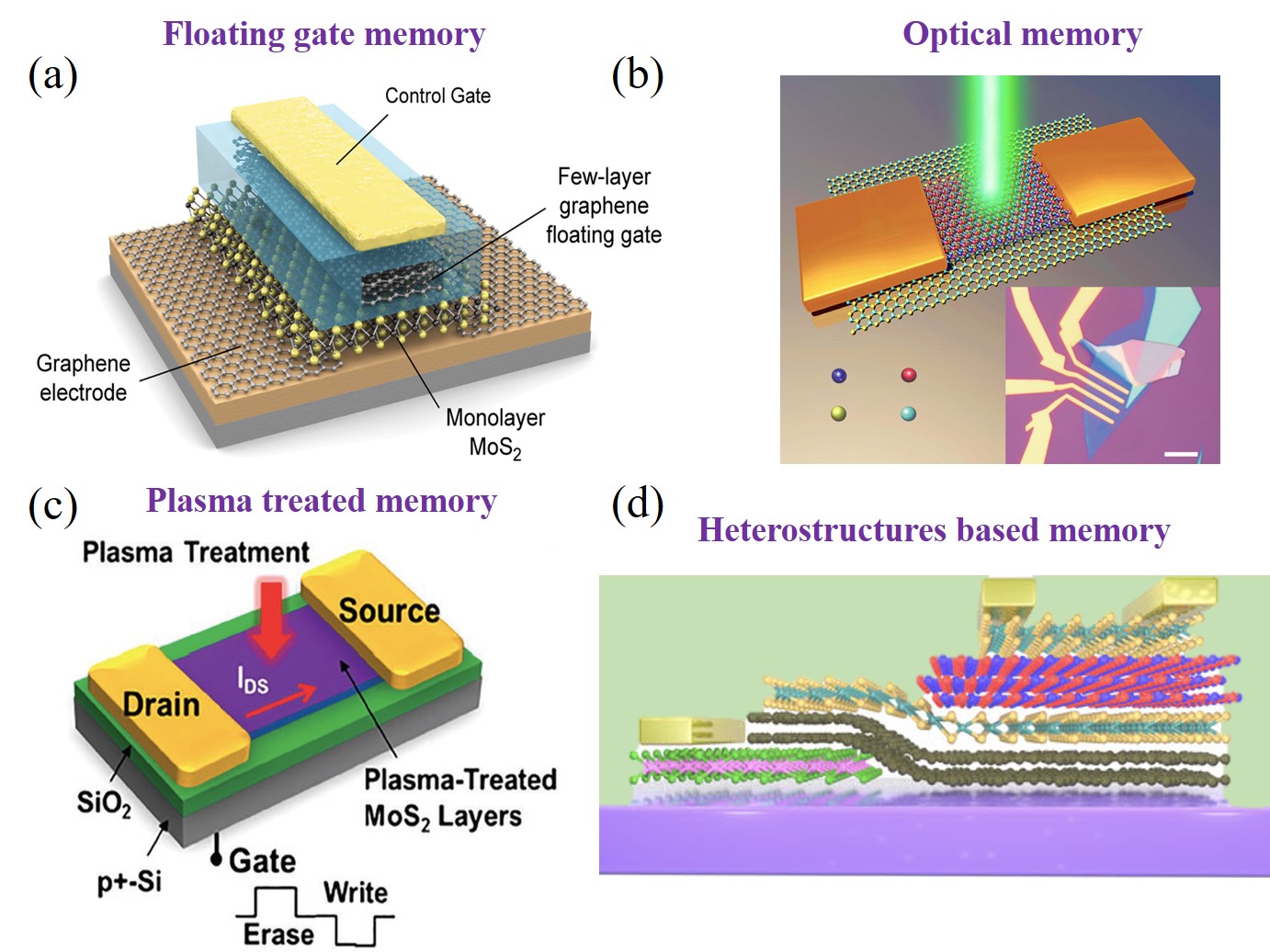}
	\caption{Various types of 2D materials based memory device architectures. Schematic illustration of (a) Floating gate memory (b) Light-driven optical memory,(c) Plasma treated memory and (d) 2D heterostructures-based memory devices. These figures are taken with permissions from \cite{Bertolazzi2013Apr,Xiang2018Jul,Chen2014Apr}}.
	\label{Fig_1.7}
\end{figure}

Considering the superior charge trap capacity of conventional $Al_2O_3/HfO_2/Al_2O_3$ gate stack and few-layer MoS$_2$ as the channel, Zhang et al. establish a large memory window ($\sim$20 V) with efficient modulation of the program and erase operation.\cite{Zhang2015Jan} A stable two-terminal floating gate memory cell is realized by Vu et al. on a vertically stacked MoS$_2$/hBN/graphene heterostructure providing program/erase operations with retention $>$ 10$^4$ s.\cite{Vu2016Sep} Opto-electronically controlled floating gate (FG) and charge trap-based devices exhibit fast operation and consume minimal power, resulting in efficient memory performance in recent years.\cite{Gao2021Nov, Liu2021Jun} The schematics of various popular memory device architectures are represented in Fig.\ref{Fig_1.7}. Moreover, making these memory devices involves complex methods using the latest technology which includes precisely adding different semiconductor layers and using many lithography processes. This results in complicated mem-transistors with lots of floating metals and oxide layers. With the growing desire to save energy, making circuits smaller than Moore's law requires looking into alternative architectures using 2D materials.

\subsection{Neuromorphic computing}
\begin{figure}[hb!]
	\centering
	\includegraphics[width=0.9\columnwidth]{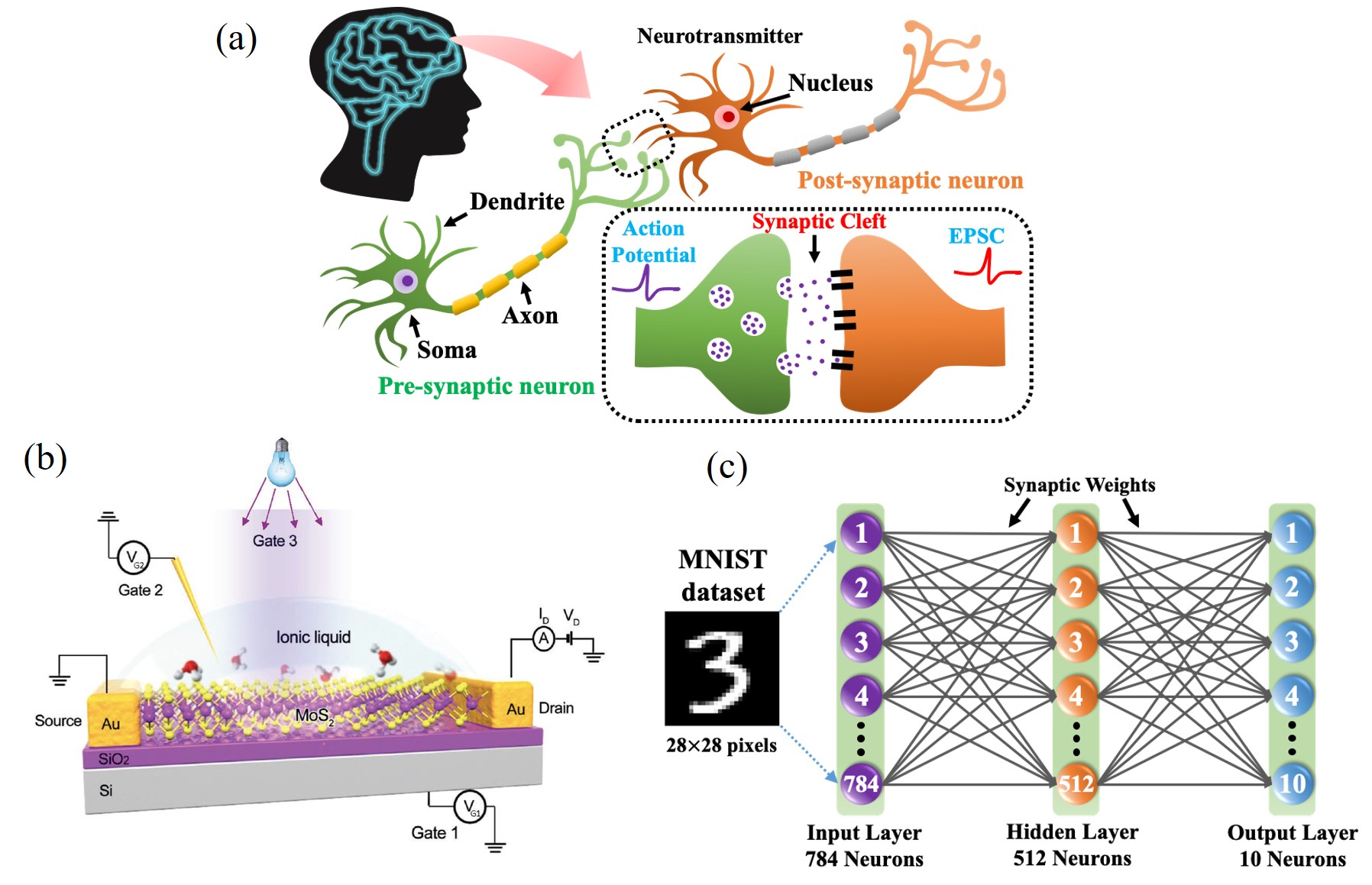}
	\caption{ (a) Demonstration of biological synapse operation between pre and post-synaptic neurons generating EPSC. (b) An architecture of the various artificial synapses using multiple gating mechanisms to the MoS$_2$ channel. (c) Schematic illustration of an artificial neural network performing AI tasks such as hand-written image recognition of the MNIST datasets. The figures (a) and (c) are taken with permission from \cite{Mallik2023Aug}. The figure (b) is taken with permission from \cite{John2018Jun}}.
	\label{Fig_1.8}
\end{figure}

In recent years, there has been a surge of interest and significant progress in the field of neuromorphic computing, a cutting-edge approach that aims to mimic the structure and functionality of the human brain.\cite{Bian2021Dec,Cao2021Jan,SubbulakshmiRadhakrishnan2022Dec} A biological synapse in the human cerebrum transmits neurological impulses through electrical and chemical signals between pre and post-synaptic neurons. Neurotransmitters released from the axon of the presynaptic neuron attach to neuro-receptors at the dendrite of the postsynaptic neuron, generating an excitatory postsynaptic current (EPSC), as illustrated in Fig.\ref{Fig_1.8}(a). The magnitude of EPSC depends on the synaptic weight which corresponds to the dynamic memory states in an artificial synapse. 

By leveraging the inherent properties of 2D materials, researchers are exploring novel architectures and devices for neuromorphic computing, with the potential to revolutionize artificial intelligence, machine learning, and cognitive computing.\cite{SubbulakshmiRadhakrishnan2022Dec} There are different types of synaptic devices with various structures, including two-terminal, three-terminal, and multi-terminal, each having unique signal transmission paths and methods for weight modulation. Two-terminal devices like memristors,\cite{Hu2014Jan} phase-change memories,\cite{Xu2020Dec} and atomic switches\cite{Bose2017Nov} are often small in size and easy to integrate due to their simple structure. However, signal transmission and learning processes happen independently, leading to signal inhibition during learning operations with the output signal providing feedback to the synaptic device. On the other hand, three-terminal synaptic transistors can manage both signal transmission and self-learning processes simultaneously, offering high stability and repeatability. Figure \ref{Fig_1.8}(b) represents a three-terminal architecture-driven artificial synapse based on monolayer MoS$_2$ that mimics the key biological synaptic functionalities.\cite{John2018Jun} The multiple gating techniques such as electronic, photoactive, and ionotronic offer a versatile platform to study these emerging materials for synaptic operations. As shown in Fig. \ref{Fig_1.8}(b), a synergistic gating provides multiple degrees of freedom to initiate the pre-synapse, and a post-synaptic current is recorded from the drain/source terminals. These materials also offer both analog and digital switching characteristics that have potential applications in brain-inspired computing and artificial intelligence. Specifically, in a crossbar architecture, the dynamic states of the synaptic devices can also be employed to perform online learning tasks such as image, vocal, and pattern recognition. In Fig. \ref{Fig_1.8}(c), a three-layer artificial neural network (ANN) is shown consisting of 784 input neurons and 10 output neurons. The ANN structure is designed to train and recognize MNIST handwritten datasets of 28$\times$28 pixels, where the synaptic weights can be derived from the plasticity of the synaptic devices such as potentiation, depression, STDP, etc. Therefore, understanding and leveraging the properties of 2D materials-based devices are crucial for their potential applications in developing novel neuromorphic devices.        

\subsection{Spintronics}

Spintronics signifies a revolutionary shift in computing device architecture, utilizing magnetic materials to harness the spin of electrons alongside their charge for computation.\cite{Barla2021Apr} This approach enables the development of energy-efficient memory and switching devices, offering advantages such as nonvolatility for embedded memory applications and lifetime endurance. Moreover, spintronics ensures the scalable fabrication of devices with minimal lithographic steps, rendering it a cost-effective and highly competitive technology.\cite{Makarov2016Oct}

\begin{figure}
	\centering
	\includegraphics[width=0.9\columnwidth]{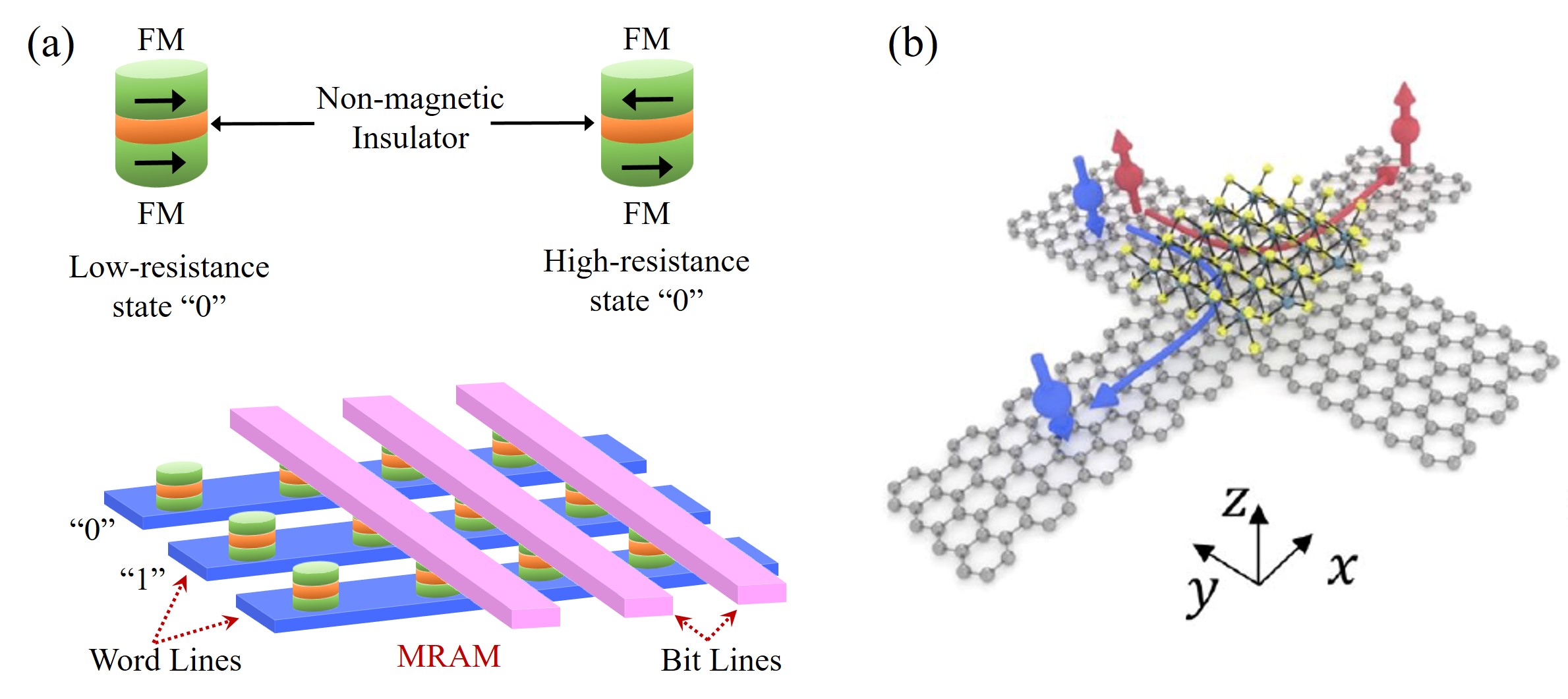}
	\caption{ (a) Schematic illustration of a magnetic tunnel junction (MTJ) spin valve device and its high-density integration into MRAM crossbar architecture. (b) The spin Hall effect at room temperature is observed in graphene when it is in proximity to a high spin-orbit coupled 2D material MoS$_2$. The figure (b) is taken with permission from \cite{Safeer2019Feb}.}
	\label{Fig_1.9}
\end{figure}

In spintronics research, the fundamental component is the magnetic tunnel junction (MTJ),\cite{Ikeda2007Apr} which comprises two ferromagnetic layers separated by a thin insulating non-magnetic layer. The MTJ stores information based on the relative orientation of the magnetization of these layers, as depicted in Fig.\ref{Fig_1.9}(a). When the ferromagnetic layers align antiparallel, the MTJ exhibits a high-resistance state, functioning as a memory-storing bit ``0" or an OFF switch. Conversely, when the layers align parallel, the MTJ stores ``1" as memory or operates as ON switch. Consequently, computation is simplified to toggling the MTJ between parallel and antiparallel states. Currently, MTJs are predominantly constructed using bulk magnetic materials. However, as these devices are scaled down to meet advanced technology node requirements, they encounter challenges. Interface roughness becomes comparable to individual layer thickness, leading to significant interlayer diffusion of atomic species. This results in degraded device performance and device-to-device variability. A potential solution to these challenges lies in employing a combination of 2D van der Waals (vdW) magnetic and insulating nonmagnetic materials to construct MTJs. By their inherent nature, vdW materials maintain atomically smooth interfaces and sharp boundaries even when reduced to single-layer thickness, offering an ideal pathway for creating scalable and energy-efficient spintronic devices.\cite{Li2019Jul} A proposed schematic illustration of an MRAM crossbar architecture utilizing MTJs is presented in Fig.\ref{Fig_1.9}(a). As the first experimental demonstration, Mohiuddin et al. reported exfoliated monolayer graphene in a Permalloy (NiFe)-based MTJ structure where the Tunneling Magnetoresistance (TMR) ratio was only 0.4 $\%$.\cite{Mohiuddin2008Nov} Subsequent studies utilizing similar device configurations, including monolayer, bilayer, or few-layer graphene, as well as other 2D materials such as MoS$_2$, WS$_2$, and h-BN, also yielded TMR ratios of less than a few percent.\cite{Ahn2020Jun} These findings underscore the need for further advanced studies aimed at optimizing the MTJ integration process with TMDC materials. 

Other spin modulator devices encompass spin transistors,\cite{Semenov2007Oct} a subject of extensive research aimed at electrically controlling spin current within a three-terminal architecture. In a spin transistor, facilitated by the Rashba spin-orbit interaction, spin injected from the source ferromagnetic (FM) contact propagates towards the drain FM contact via Larmor precession.\cite{Koo2020Dec} Typically, a spin transistor operates in the ON state (i.e., depletion mode), as zero gate voltage results in no precession, aligning the spin direction parallel to the drain FM. By varying the gate voltage, the spin precession angle can be adjusted, leading to periodic modulation of the source-to-drain current. To further advance electrical control of magnetism, the principle of current-based spin-orbit torque switching offers promising solutions for low-power operation. This principle exploits materials with high spin-orbit coupling, where electrons scatter based on their spin states, giving rise to the spin Hall effect.\cite{Wunderlich2010Dec} This phenomenon allows for the conversion of electric charge current flow into an orthogonal spin current, as illustrated in Fig.\ref{Fig_1.9}(b). These emerging phenomena enrich the properties of two-dimensional (2D) materials and present potential pathways for their widespread and diverse applications in modern technology. 

Furthermore, dilute magnetic semiconductors (DMS) represent a class of materials with intriguing potential for applications in spintronics, a field focusing on the manipulation of electron spin for information processing and storage. By introducing small concentrations of magnetic ions into semiconductor matrices, DMS offers the intriguing possibility of combining semiconductor functionality with magnetic ordering. This combination enables precise control over electron spin, making DMS promising candidates for various spintronic applications. For instance, DMS have been explored for spin injection and detection in spin valves and spin transistors, as well as for spin-polarized light-emitting diodes (spin-LEDs) and spin-based memory devices.\cite{Hirohata2020Sep} Additionally, their tunable magnetic and electronic properties make them suitable for applications in magnetic sensors and quantum computing. The reports by Dietl et al.\cite{Dietl2010Dec} provide foundational insights into the properties and potential applications of DMS. Moreover, many reports have been dedicated recently to investigate the underlying mechanism behind ferromagnetic orderings in DMS by using first-principle calculations coupled to Monte Carlo simulations, which underscores their importance in advancing spintronics research.\cite{Sato2010May, Zhang2017Mar, Xing2023Feb, Ramasubramaniam2013May} Nonetheless, the 2D vdW non-magnetic materials have profound applications in electronics and optoelectronics, the magnetic behavior in doped 2D materials and their heterostructures remains an open question that needs to be addressed to expand their fundamental and technical applications. Moreover, their compatibility with existing semiconductor fabrication techniques makes DMS an attractive candidate for integrating spintronic functionalities into conventional electronic circuits, paving the way for next-generation computing and data storage technologies.
\section{Organization of the thesis}
This thesis consists of seven chapters which are mainly focused on synthesizing 2D transition metal dichalcogenides using chemical vapour deposition techniques and studying their potential applications in transistors, memory, and neuromorphic devices. Also, by employing first principle density functional theory calculations, room temperature dilute magnetic semiconductors using monolayer and heterostructures of 2D TMDCs are investigated under strain and various structural modifications. 

Chapter-\ref{C1} provides a brief introduction to the device scaling beyond Moore's law and presents a road map for 2D materials in the silicon age. This chapter also reviews various types of 2D materials such as TMDCs and their properties, promising synthesis routes, and vast applications in electronics and computing platforms. In Chapter-\ref{C2}, we outline the synthesis of 2D TMDCs such as MoS$_2$ and WS$_2$ using a modified CVD setup and their various optical characterization using Raman and PL spectroscopic technique. In this chapter, we demonstrate an optimization and manipulation of NaCl-assisted CVD synthesis of layered MoS$_2$ under diverse growth conditions. The growth kinetics are systematically analyzed under the Kinetic Wulff Construction (KWC) theory by varying the ratio of the precursor's flux during the growth. Specifically, newly developed key parameters such as geometry confinements, and precursor/promotor ratio are discussed to achieve controlled edge terminations, planar and self-seeding nucleation. Furthermore, this study unravels several growth challenges such as premature growth, defects, sulfur vacancies, and non-uniformity in salt-assisted CVD-grown monolayer MoS$_2$ samples. The low-frequency and polarization-dependent Raman studies are also discussed on the CVD-grown MoS$_2$ samples. In the end, we also provide the salt-assisted synthesis route to obtain large-scale WS$_2$ monolayers.

In Chapter-\ref{C3}, the device fabrication of 2D materials-based transistors using the photolithography technique is initially demonstrated. Then we discuss some important device parameters such as contact electrodes, contact resistances, and Schottky barrier heights (SBH) and various device geometries such as the two-probe method, and the transfer length method (TLM). The optimally-grown monolayer MoS$_2$ manifests excellent device performances and intrinsic photo response behavior under wavelength selective illumination. Furthermore, we demonstrate the realization of high-performance hysteresis-free field-effect transistors (FETs) based on salt-assisted CVD-grown monolayer MoS$_2$ samples. First principles DFT calculations are implemented to inspect the effects of the barrier height and metal-induced gap states between the Ag electrodes and MoS$_2$ for the superior carrier transport. The role of absorbed molecules and oxide traps on the hysteresis are studied in detail. A hysteresis-free intrinsic transistor behavior is obtained by an amplitude sweep pulse I$\sim$V measurement with varying pulse widths. Under this condition, a significant enhancement of the field-effect mobility up to 30 cm$^2V^{-1}s^{-1}$ is achieved. Moreover, to correlate these results, a single pulse time-domain drain current analysis is carried out to unleash the fast and slow transient charge trapping phenomena. Our findings on the hysteresis-free transfer characteristic and high intrinsic field-effect mobility in salt-assisted monolayer MoS$_2$ FETs could be beneficial for future device applications in complex memory, logic, and sensor systems. Transfer techniques of two dimensional (2D) materials and devices offer nanoscale integration with the existing silicon-based technology and flexible electronics. To date, the chemical etching technique is being widely used for the transfer of 2D materials and there has been constant effort in improving the method to achieve an etching-free, and clean transfer of 2D materials without affecting the device performances. In the following part, we demonstrate a poly(methyl methacrylate) (PMMA)-assisted etching-free one-step approach to transfer device arrays consisting of monolayer MoS$_2$ and metal electrodes to different substrates (i.e. SiO$_2$/Si$^{++}$, flexible). The crystalline quality, strain relaxation, and interfacial coupling effects of the transferred devices are analysed using Raman and photoluminescence spectroscopy. The room temperature gate-tunable drain current measurements of the transferred devices on SiO$_2$/Si$^{++}$ substrate show a reduced Fermi-level pinning effect. Furthermore, the observation from temperature-dependent threshold voltage shifts, mobilities, and hysteresis evolution indicates an improved transistor performance in the transferred device. The proposed one-step transfer method could be useful to transfer large array of 2D material devices on arbitrary substrates for future flexible opto-electronic applications. 

The demands of modern electronic components require advanced computing platforms for efficient information processing to realize in-memory operations with a high density of data storage capabilities towards developing alternatives to von Neumann architectures. In Chapter-\ref{C4}, we demonstrate the multifunctionality of monolayer MoS$_2$ mem-transistors which can be used as a high-geared intrinsic transistor at room temperature; however, at a high temperature ($>$ 350 K), they exhibit synaptic multi-level memory operations. The temperature-dependent memory mechanism is governed by interfacial physics, which solely depends on the gate field modulated ion dynamics and charge transfer at the MoS$_2$/dielectric interface. We have proposed a non-volatile memory application using a single FET device where thermal energy can be ventured to aid the memory functions with multi-level (3-bit) storage capabilities. This work paves the way towards reliable data processing and storage using 2D semiconductors with high-packing density arrays.

To effectively manage the massive amount of data and global memory requirements, the upcoming nonvolatile memory (NVM) technologies must furnish multi-bit ultra-high-density storage capacity with a significant extinction ratio, endurance, retention, and low energy consumption. In Chapter-\ref{C5}, we have demonstrated a mem-transistor device based on monolayer WS$_2$ that addresses more than 64 storage states (6-bit) and synaptic operations, emulating biomimetic plasticity. The direct salt-assisted synthesis of monolayer WS$_2$ on SiO$_2$/Si$^{++}$ substrate using a cost-effective CVD method has the potential for scalability. In addition, it is advantageous to directly fabricate three-terminal 2D memory devices with robust ion-gating mechanisms and device functionalities. The Na$^+$ diffused SiO$_2$ can be treated as an ionic gating medium instead of dielectric gating at high temperatures due to the thermal activation of mobile Na$^+$ ions.  Using band-bending characteristics, we employ a fingerprint mechanism to address the carrier dynamics under electrostatic doping fluctuations induced by local ion movements. The WS$_2$-based ionotronic memory is equipped with endurance tests ($>$ 400 pulse cycles), $10^5$ extinction ratio with stable retention, low device-to-device variations, and 64 distinct nonvolatile storage levels with reliable signal-to-noise ratios using combinational pulses. We also demonstrate the excitatory postsynaptic current with varying time scales, near linear potentiation, and depression for successful implementation in artificial neural network (ANN). The simple and cost-effective synthesis and fabrication methods presented in this work could substantially enhance the distinctive attributes of WS$_2$ and other emerging layered 2D materials, thus representing potential for their advancement as scalable memory solutions. The associated bio-inspired synaptic capability delineates a viable path to constructing next-generation 2D mem-transistors towards advanced computing platforms at high temperatures and leverages potential memory applications in harsh electronics as demanded in the aerospace, military, and automotive industries.

Renewed interests in dilute magnetic semiconductors (DMS) based on magnetic/non-magnetic ion doping in two-dimensional (2D) materials such as graphene, phosphorene, h-BN, and transition metal dichalcogenides (TMDCs) have been the focus of extensive research for the past few years to meet several applications in nanoelectronics and spintronics. The criteria for strong ferromagnetism and large magnetic anisotropy in 2D DMS demands investigations of magnetic impurities along with various doping strategies. In Chapter-\ref{C6}, with the implementation of extensive first-principles calculations, we demonstrate the MS$_2$ (M = Mo, W) monolayers, as well as their van der Waals (vdW) hetero-bilayers as promising candidates for the successful realization of 2D DMS with the incorporation of Mn and Co dopants. Under various pairwise doping configurations at different host atom sites, we report the electronic properties modifications induced change in magnetic exchange interactions. The magnetic coupling among the dopant pairs can be tuned between ferromagnetic (FM) and anti-ferromagnetic (AFM) orderings via suitable doping adjustments. The developed interlayer exchange coupling between the vdW layers leads to strong and long-ranged FM interactions which unleash robust magnetic moments with stable doping configurations. Our findings address the novel magnetic behavior of the layered vdW heterostructures and may further guide the future experimental efforts for the possible applications in modern electronics and nanoscale magnetic storage devices. Strain-mediated magnetism in 2D materials and dilute magnetic semiconductors hold multi-functional applications for future nano-electronics. We studied a possible emergence of FM in Co-doped WS$_2$ ML under strain engineering by using first-principles DFT calculations and micromagnetic simulation, which have tremendous applications in TMDC based straintronics. Co-doping marks a significant change in magnetic properties with an impressive magnetic moment of 2.58 $\mu _B$. The magnetic exchange interaction is found to be double exchange coupling between Co and W and strong p-d hybridization between Co and nearest S, which is further verified from spin density distribution. We find that the resultant impurity bands of the Co-doped WS$_2$ plays a role of seed to drive novel electronic and magnetic properties under applied strain. Among several biaxial strains, the magnetic moment is found to be a maximum of 3.25 $\mu _B$ at 2\% tensile and 2.69 $\mu _B$ at -2\% compressive strain due to strong double exchange coupling and p-d hybridization among foreign Co and W/S. Further magnetic moments at higher applied strain is decreased due to reduced spin polarization. In addition, the competition between exchange splitting and crystal field splitting of Co d-orbital plays a significant role to determine these values of magnetic moments under the application of strain. From the micromagnetic simulation, it is confirmed that the Co-doped WS$_2$ monolayer shows slanted ferromagnetic hysteresis with a low coercive field. The effect of higher strain suppresses the ferromagnetic nature, which has a good agreement with the results obtained from DFT calculations. Our findings indicate that induced magnetism in WS$_2$ monolayer under Co-doping promotes the application of 2D TMDCs for the nano-scale spintronics, and especially, the strain-mediated magnetism can be a promising candidate for future straintronics applications. At the end, doped monolayers of MoS$_2$ and WS$_2$ are successfully synthesized using the CVD method. Magnetization measurements conducted with a SQUID magnetometer revealed hysteresis loops indicative of weak ferromagnetic order at both 5 K and 300 K, confirming their status as dilute magnetic semiconductors. Nevertheless, additional characterizations are necessary to fully understand and harness the potential of these doped materials for spintronic applications. Finally, a comprehensive overview and the future prospects of the current thesis work are outlined in the last chapter (Chapter-\ref{C7}). 
\blankpage
\chapter{Salt-assisted CVD synthesis of TMDCs: structural and optical characterization}
\label{C2} 

\section{Introduction}
Atomically thin two-dimensional (2D) materials, particularly monolayer Transition metal dichalcogenides (TMDCs),\cite{Manzeli2017Jun} represent a cutting-edge frontier in semiconductor research due to their remarkable optoelectronic properties and potential for integration into advanced devices.\cite{Tedstone2016Apr} While exfoliation from bulk crystals has been the go-to method for proof-of-concept studies and spectroscopic experiments, the scalability limitations necessitate the exploration of alternative approaches.\cite{Zhao2021Nov} In this context, direct bottom-up synthetic techniques like chemical vapor deposition (CVD) exhibit potential advantages towards scalable and high-quality 2D layers production,\cite{Zhou2018Apr} paving the way for practical applications of TMDCs in low-power electronics. The most studied TMDC material, Molybdenum disulfide (MoS$_2$) grown by the CVD method provides potential advantages in the progressive scaling of ongoing silicon technology. Recently, CVD-grown monolayer MoS$_2$ samples have predominantly been explored as an active material in the fabrication of intrinsic hysteresis-free transistors,\cite{Mallik2021Apr} highly responsive photodetectors,\cite{Sahoo2023Feb} high-performance CMOS arrays,\cite{Xia2023Nov} and neuromorphic computing devices.\cite{Sahu2023Jan} Such broad applications require the supply of high-quality materials that necessitate in-depth studies of controlled synthesis and reproducibility.       
\par The conventional CVD approach uses powdered metal oxides (i.e. MoO$_3$ and WO$_3$ for MoS$_2$ and WS$_2$ growth, respectively) as the metal source precursors. However, relatively low vapour pressure, high melting and boiling points of the transition metals often result in low chemical reactivity and affect reproducibility.\cite{Cai2018Jul} Moreover, the formation of nucleation and growth modes are correlated with several key growth parameters, including growth temperature, growth time, precursors ratio, growth pressure, and promoters. So far, various methods and mechanisms have been reported to understand the growth process to improve the crystal quality and uniformity for practical device applications.\cite{Zhang2023Dec} The recently developed salt-assisted CVD synthesis addresses such challenges by increasing reaction rate and degree of reproducibility.\cite{Zhou2018Apr} In such growth techniques, alkali halides such as NaCl, KI reduce the sublimation temperature of metal precursors by forming an intermediate product known as metal oxy-halides which enhances the nucleation density and fastens the growth rate.\cite{Zhou2018Apr} However, the growth kinetics and thermodynamics during CVD synthesis involve various non-equilibrium processes such as decomposition of precursors, attachment towards substrates, and diffusion of adatoms etc.\cite{Xu2024Jan}   

In this chapter, we discuss various CVD components and the growth parameters that play an important role in the growth dynamics of 2D materials. An optimization and manipulation of NaCl-assisted CVD synthesis of layered MoS$_2$ under diverse growth conditions is demonstrated. The growth kinetics are systematically analyzed under the Kinetic Wulff Construction (KWC) theory by utilizing Raman and photoluminescence spectroscopic studies and spatial mapping analysis. Specifically, some of the crucial parameters such as geometry confinements, and precursor/promotor ratio are discussed to achieve controlled edge terminations, planar and self-seeding nucleation of MoS$_2$ layers. Furthermore, this study unravels several growth challenges such as premature growth, defects, sulfur vacancies, and non-uniformity in salt-assisted CVD-grown monolayer MoS$_2$ samples. We also discuss polarization Raman spectroscopy to address the presence of external perturbation such as strain in as-grown samples. In the end, a method to grow large-scale triangular domain monolayer WS$_2$ is provided that represents potential applications in multi-level memory devices. 
\section{Modified chemical vapour deposition setup}
\begin{figure}
	\centering
	\includegraphics[width=0.9\columnwidth]{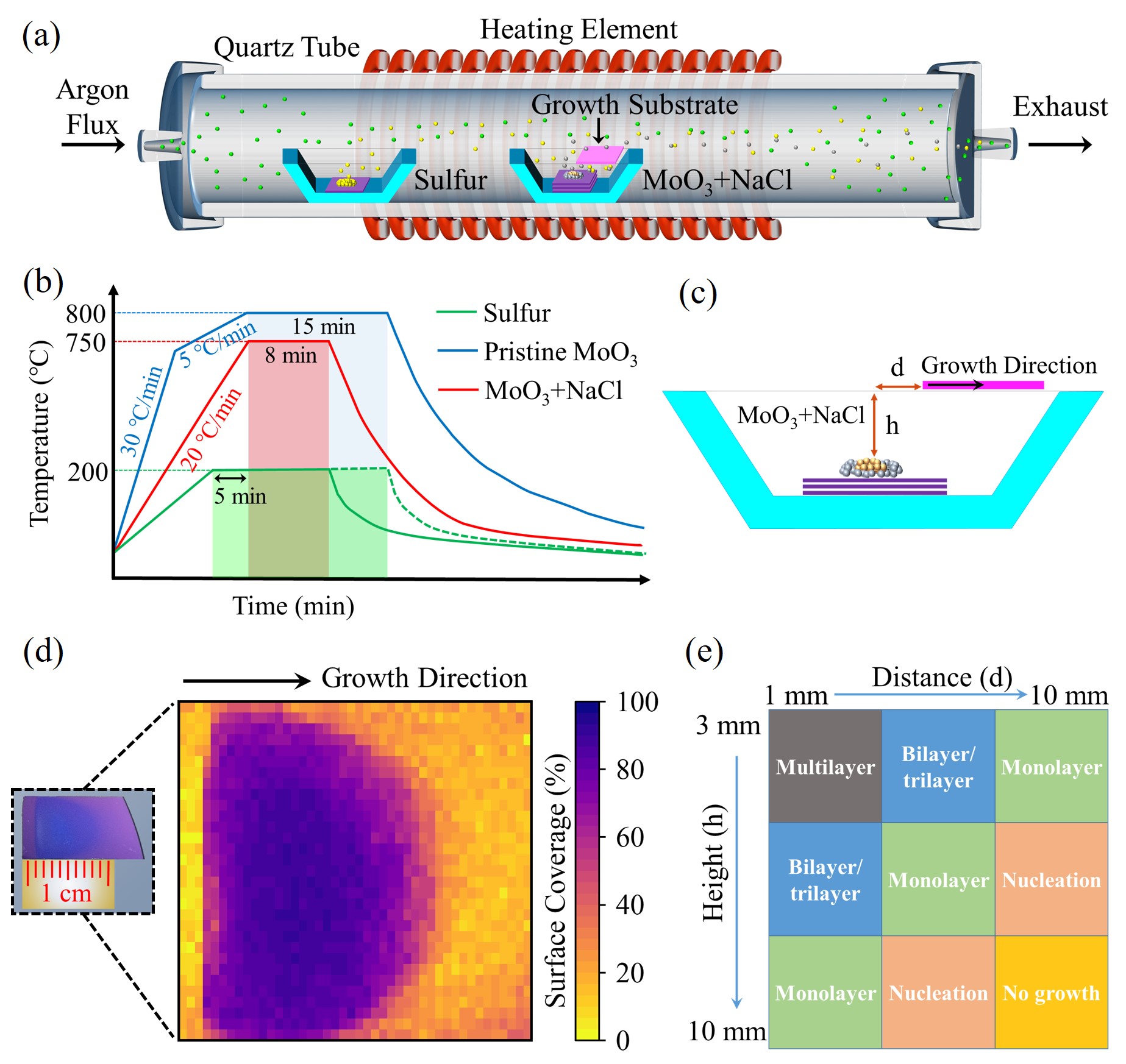}
	\caption{(An illustration of NaCl-assisted CVD method for synthesis of MoS$_2$ crystals. (a) The schematic diagram represents cross-sectional view of a CVD system and the relative positioning of precursors (S, MoO$_3$+NaCl) and growth substrate. (b) Temperature profile of S, pristine MoO$_3$, and MoO$_3$+NaCl during two different CVD processes. (c) Cross sectional schematic view of the boat containing MoO$_3$+NaCl and face down SiO$_2$/Si growth substrate with varied geometry confinements i.e. height “h” and distance “d” between precursor and growth substrate. (d) Surface coverage mapping of monolayer MoS$_2$ grown on SiO$_2$/Si substrate. The growth direction inside the CVD system is indicated with arrow mark. The left inset shows the optical image of the whole growth substrate post-CVD synthesis. (e) A brief sketch illustrating MoS$_2$ growth dynamics based on the competition between the confinement parameters i.e. height (h), and distance (d) where h varies from 3 to 10 mm and d varies from 1 to 10 mm.}
	\label{Fig_2.1}
\end{figure}

The MoS$_2$ crystals examined in this study are synthesized using a home-built CVD system. The various components of the CVD systems include a single zone tube furnace with temperature gauge and controller, Mass-flow controller, Rotary pump, pressure gauge etc. which are assembled for the vapour phase deposition of 2D materials. A modified salt-assisted synthesis strategy is utilized to grow controllable MoS$_2$ domains on SiO$_2$/Si substrates. Figure \ref{Fig_2.1}(a) demonstrates the positioning of metal and chalcogen precursors, growth substrates etc. inside a two-inch quartz tube using a single-zone CVD furnace. The boat containing metal precursor and promoter (MoO$_3$+NaCl powder) is placed at the center of the furnace. A SiO$_2$ (285 nm)/Si (1$\times$1 cm$^2$) growth substrate is cleaned using the ultrasonication method with acetone, isopropyl alcohol, and de-ionized (DI) water, followed by drying with pressurized dry N$_2$ (99.999$\%$) to eliminate any moisture. The growth substrate is carefully positioned on the boat (containing metal precursors) with face down to the precursors. Another boat containing sulfur powder (S) as chalcogen precursor is placed in the upstream region towards the end of the heating element where the temperature reaches exactly the melting conditions of S during the CVD process. The tube is initially pumped down to 0.2 mbar, and then purged three times with 500 sccm of high-purity Ar (99.999$\%$) to eliminate any oxygen contaminants inside the tube. The system is heated to target temperature at a certain ramp speed using Ar as the carrier gas. The pressure inside the tube is maintained by a pressure controller knob during the growth. Finally, the system is cooled down rapidly to ambient temperature by partially opening the furnace. The optimized temperature profile of two different growth processes (i.e. conventional synthesis without using salt promoter, and modified salt-assisted approach) are demonstrated in Fig.\ref{Fig_2.1}(b). In the conventional growth process, the central heating zone containing pristine MoO$_3$ undergoes a gradual temperature increase to 700 $^{\circ}$C at a rate of 30 $^{\circ}$C/min, subsequently reaching the target temperature of 800 $^{\circ}$C at a rate of 5 $^{\circ}$C/min. It is crucial to highlight that transition metal oxides like MoO$_3$ and WO$_3$ exhibit low vapor pressure due to their relatively high melting points, resulting in a sluggish evaporation rate.\cite{Zhou2018Apr} Hence, the deliberate slow temperature ramp, particularly beyond 700 $^{\circ}$C in our synthesis method, becomes essential to compensate for the gradual evaporation of MoO$_3$. Additionally, the evaporation of S is carefully initiated 5 minutes before the commencement of the growth process. The growth duration is maintained at approximately 15 minutes at 800 $^{\circ}$C, followed by a rapid cooling step achieved by opening the furnace. In case of salt-assisted synthesis, the temperature of the MoO$_3$+NaCl precursor is ramped at a rate of 20 $^{\circ}$C/min to achieve a lower target temperature of 750 $^{\circ}$C. The growth time is also reduced to 8 mins due to the accelerated growth kinetics from the fusion of MoO$_3$ and NaCl promoter. According to Shinde et. al,\cite{Shinde2018Feb} the Na-containing intermediate solid (Na$_2MoO_4$) reduces the energy of the reaction during the sulfurization process and the formation of the final product (MoS$_2$) can be realized at a lower temperature. As shown in Fig.\ref{Fig_2.1}(b), unlike the case of conventional CVD, the NaCl-assisted CVD synthesis involves only three steps and provides a much faster route to produce controllable MoS$_2$ crystals.
The geometry confinements during MoS$_2$ growth play a crucial role in regulating chalcogen flux and reactant concentrations at the vapor phase near the growth substrate. Hsieh et. al.\cite{Hsieh2016Jan} reported the production of high-quality graphene using close-packed stacks of graphite and growth substrate due to a self-limited nucleation mechanism. As depicted in Fig.\ref{Fig_2.1}(c), we have varied the confined growth space by using height “h” and distance “d” between the precursors and growth substrate to achieve stable growth conditions and lowering the flowrate at the inner surface of the growth substrate facing down to the precursors. The surface coverage and various growth zones of a typical SiO$_2$/Si growth substrate after MoS$_2$ synthesis are shown in Fig.\ref{Fig_2.1}(d). The coverage mapping is extracted using ImageJ software from an optimally grown monolayer MoS$_2$ as displayed in the left inset. It is important to find that the central zone of the substrate has 100 $\%$ coverage which represents a continuous layer of monolayer MoS$_2$. The edges surrounding the central zone covered up to 40 to 80 $\%$ are dominated by individual MoS$_2$ domains. The average domain size of the samples varies from 5 to 100 $\mu$m depending on other growth parameters that will be discussed in the subsequent sections. Figure \ref{Fig_2.1}(e) illustrates the dependence of growth area confinement towards obtaining the MoS$_2$ crystals. A balanced ratio between height and distance leads to various types of sample growth on SiO$_2$/Si substrate. For example, in case of positioning the growth substrate with maximum distance and height from the precursors results in no growth of MoS$_2$, while positioning the growth substrate with minimum distance lead to overgrowth that results in multilayer MoS$_2$ flakes. However, the optimal growth conditions for mono- and bilayer crystals can be achieved by maintaining the growth conditions between these two extreme situations. 

\subsection{Growth parameters}

The characteristics of 2D materials, such as their dimensions, structure, and interfaces, play a crucial role in determining their properties. These attributes can be deliberately engineered and fine-tuned through precise adjustments in the processes of chemical vapor deposition (CVD). Understanding the underlying mechanisms of CVD growth is essential for controlling the synthesis of these materials effectively. This involves comprehending how various parameters like precursor composition, flow rate, pressure, and temperature influence processes such as mass and heat transfer, interfacial reactions, and ultimately, material growth. 

For instance, minor fluctuations in temperature within the source zone can lead to substantial variations in the saturation pressure of gaseous precursors, thus impacting the growth of transition metal dichalcogenides (TMDCs). Similarly, temperature variations in the reaction zone affect the transport of species and their reactions at the vapor-solid interface, providing a means to modulate reaction rates. Higher temperatures in the source zone result in elevated gas concentrations, promoting controlled growth through chemical reactions, while lower temperatures lead to growth mechanisms limited by mass transport. Additionally, investigations into materials like MoSe$_2$ and WSe$_2$ have highlighted the systematic influence of temperature on parameters such as layer size, number of layers, and morphology, emphasizing its critical role in controlling the growth of TMDCs.\cite{Liu2015Jun, He2016Jul} Similarly, the flow rate of the carrier gas was observed to impact the morphology, size, and density of flakes. Decreasing the gas carrier flow during the heating stage leads to an increase in the density and size of the flakes. Additionally, the shape of MoS$_2$ flakes transitioned from zigzag to triangular with decreasing gas carrier flow, which is directly linked to the amount of sulfur in the gas phase. With increased carrier flow, sulfur transport became more efficient, suppressing the evaporation of MoO$_3$, resulting in a decrease in MoO$_3$ partial pressure and favoring zigzag edge termination. The pressure within a CVD chamber can fluctuate significantly, ranging from a few atmospheres to several millitorrs or even lower, and this greatly impacts gas flow dynamics. Lower pressures result in increased volume flow and gas velocity for the same molar flow, while also causing a decrease in precursor concentration according to the ideal gas equation PV = nRT. Consequently, the combination of low concentration and high velocity in precursor mass feed enhances reaction controllability. Therefore, for achieving the growth of large-scale TMDC film, low-pressure CVD methods are typically employed. At low pressure, the nucleation of the second layer occurs solely at grain boundaries, whereas at high pressure, nucleation happens randomly atop the first layer, resulting in a mixture of monolayer and multilayer products.
\section{Characterization using optical spectroscopy}
\subsection{Raman spectroscopy}
Raman spectroscopy is a highly effective method for characterizing materials by analyzing their unique lattice vibrational frequencies. It involves inelastic scattering of light, occurring when incident light interacts with material phonons, which results in shifts in photon energy. These shifts reveal Raman-active phonon modes, which are observed as peaks in Raman spectra. Each peak corresponds to a specific Raman-active phonon mode, and the frequency (wavenumber) and full width at half maximum (FWHM) of the peaks provide information about material properties like purity, thickness, and doping level.
\begin{figure}
\centering
\includegraphics[width=0.9\textwidth]{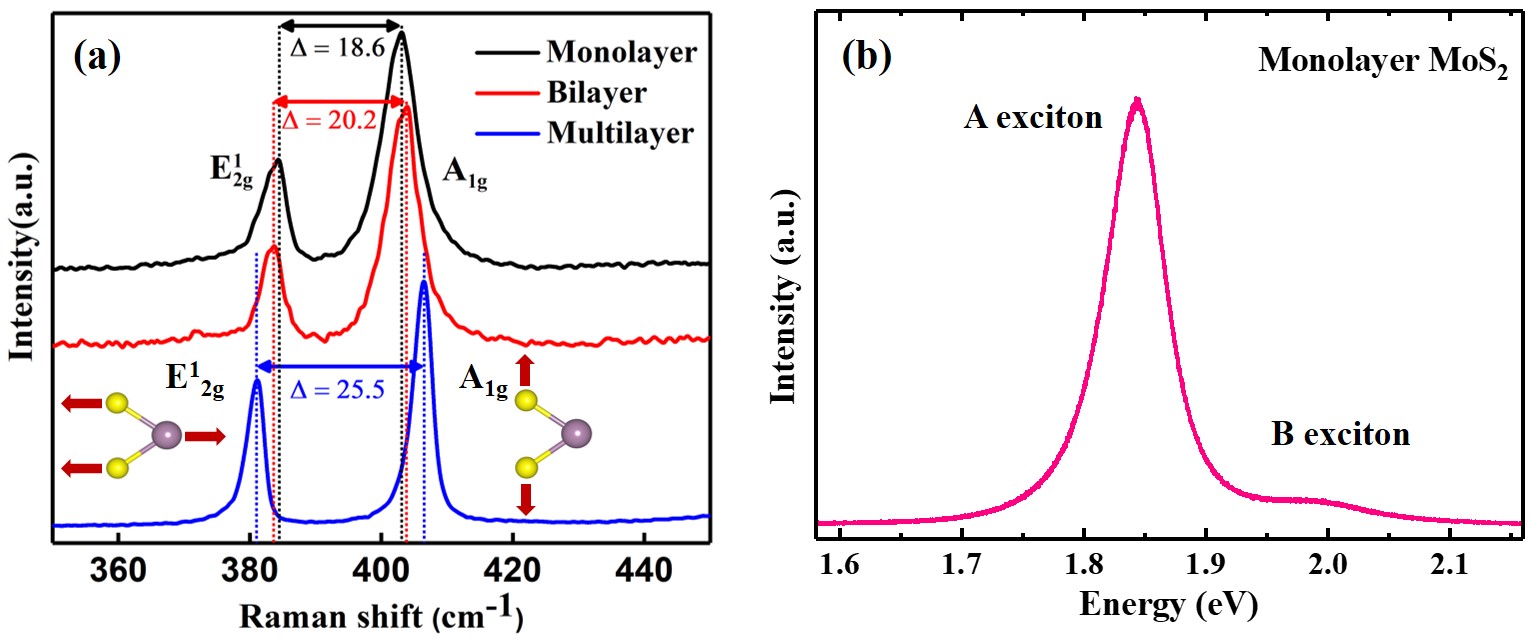}
\caption{(a) Raman spectra with a 532 nm excitation wavelength showing two predominant Raman modes (E$^1_{2g}$ and A$_{1g}$) of monolayer and multilayer MoS$_2$. The frequency difference ($\Delta$) between the two modes confirms the monolayer nature of MoS$_2$. (b) Room temperature photoluminescence of monolayer MoS$_2$ showing direct bandgap at 1.87 eV.}
\label{Fig_2.2}
\end{figure}

In our CVD-grown samples, Raman spectra were collected using a confocal micro-Raman spectrometer (Renishaw Invia) with a laser excitation wavelength of 532 nm, employing a backscattering configuration with a 100x (NA = 0.8) objective. To prevent local heating, the laser power on the sample was maintained at a low level. Figure \ref{Fig_2.2}(a) displays the in-plane and out-of-plane Raman modes (E$^1_{2g}$ and A$_{1g}$, respectively) for monolayer, bilayer and multilayer CVD-grown MoS$_2$. The frequency difference ($\Delta$) between these modes serves as a reliable identifier for the number of layers present in the sample MoS$_2$. Further detailed studies utilizing Raman spectroscopy are discussed for various CVD samples in subsequent sections.

\subsection{Photoluminescence spectroscopy}
Semiconductor materials are broadly categorized as direct and indirect bandgap semiconductors based on their bandgap structure. In direct bandgap semiconductors, electron-hole recombination occurs efficiently, resulting in photon emission. Conversely, in indirect bandgap semiconductors, electron-hole recombination requires additional phonon emission or absorption, resulting in low photon yield. Direct bandgap semiconductors exhibit luminescence, making them suitable for optoelectronic devices. Photoluminescence (PL) spectroscopy, is used to excite electrons across the bandgap and detect emitted photons, utilizing higher energy radiation, which provides insights into bandgap size and carrier lifetime. TMDCs demonstrate layer-dependent bandgap structures, transitioning from indirect to direct bandgap with decreasing layer thickness. This tunability makes PL spectroscopy valuable for both electronic property analysis and thickness estimation. In this work, PL spectroscopy was conducted using the above Raman spectrometer with a 532 nm laser excitation wavelength. Figure \ref{Fig_2.2}(b) shows the excitonic PL peaks for monolayer MoS$_2$. The strong PL at 1.87 eV indicates the direct bandgap nature of the monolayer MoS$_2$ which lies in the visible range.  

\begin{figure}
\centering
\includegraphics[width=0.9\textwidth]{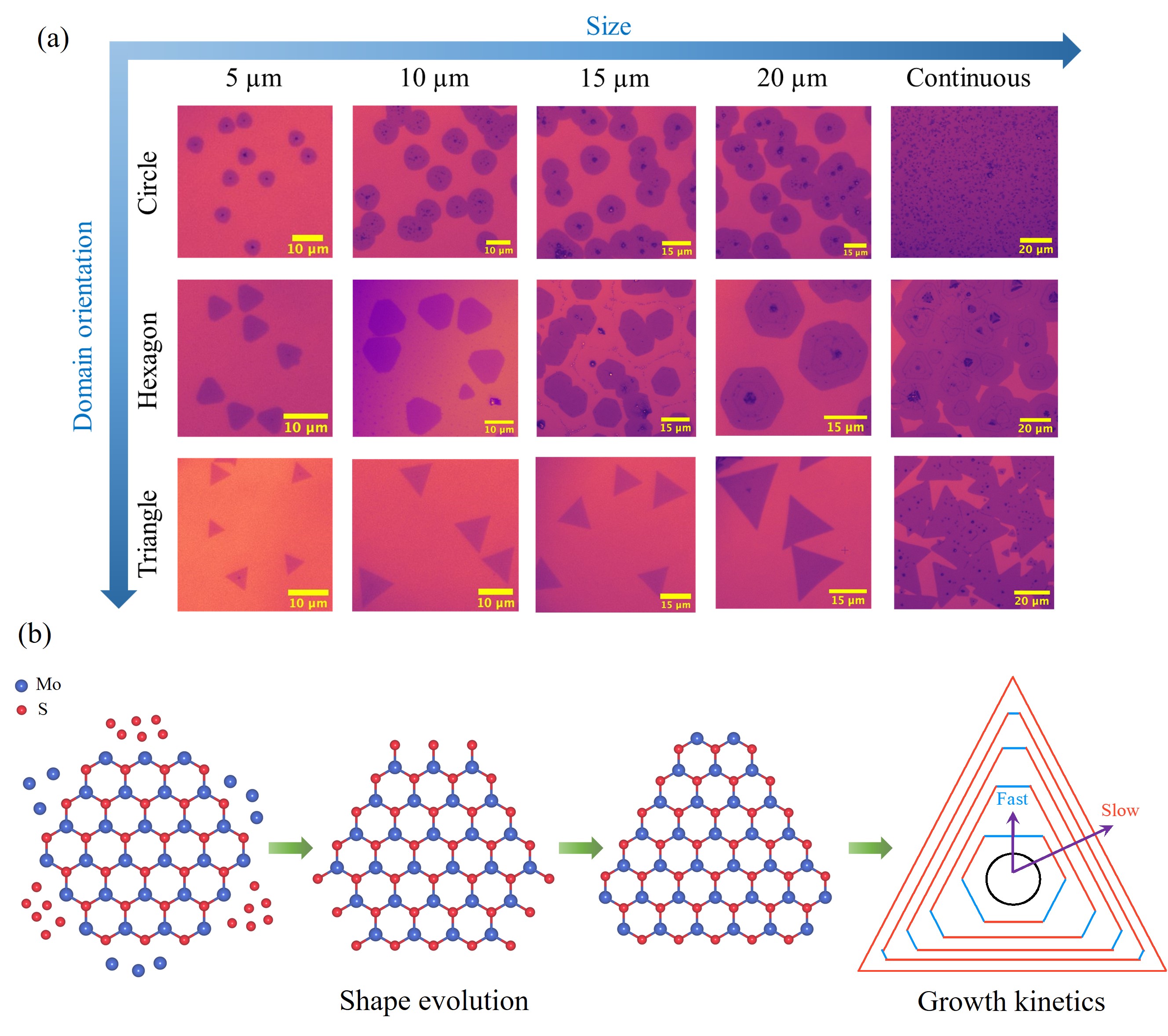}
\caption{(a) Optical images of the CVD grown monolayer MoS$_2$ under different size and domain configurations and the evolution of the 2D islands with adjusting the growth parameters. (b) Schematic illustration of shape evolution and growth kinetics of 2D islands under KWC theory.}
\label{Fig_2.3}
\end{figure}

\subsection{Evolution of MoS$_2$ domains based on KWC theory}
Precisely controlling the edge geometry and crystal shape of monolayer MoS$_2$ is crucial for their effective use in various applications. In the present study, the MoS$_2$ samples are obtained in various sizes and domain orientations using the modified synthesis of the NaCl-assisted CVD technique, as indicated in Fig.\ref{Fig_2.3}(a). Morphology (domain orientation) control is achieved by varying only the NaCl concentration while keeping other parameters constant. It is important to note that the varying NaCl concentration, i.e., varying NaCl:MoO$_3$ ratios, has a significant impact on the chemical potentials of individual reactants during the reaction. The variations of chemical potentials indeed result in Mo-rich or S-rich conditions, which facilitate the edge reconstruction in 2D domain growth. It is important to note that in thermodynamic equilibrium, a crystal’s shape can be determined by the Wulff construction geometry which states that the shape and morphology of a 2D material domain can be predicted in such a way that the edge energy is minimized. In such cases, the ratio of the distance between its center and each edge is proportional to the corresponding edge energy. Additionally, the kinetic Wulff construction (KWC) theory explains that under very fast growth conditions, the shape of a 2D material is defined by the slowest-growing edge. The growth of MoS$_2$ typically begins with planar nucleation of nanoparticles, which then expand in all directions to form triangular monolayers. Despite it's hexagonal crystal structure, the MoS$_2$ flakes often exhibit a triangular shape due to the KWC mechanism.\cite{Kong2023Jun} In our case, maintaining the NaCl:MoO$_3$ weight ratio at 1:10 results in a low NaCl concentration, leading to a slow vaporization rate of MoO$_3$. Consequently, the growth of Mo-termination edges proceeds slowly, resulting in MoS$_2$ growth with a dodecagon structure typically observed as circular domains in optical microscopy (see Fig.\ref{Fig_2.3}(a)). Within these structures, both zigzag (ZZ) and armchair (AC) edges are present, giving rise to the phenomenon of edge magnetism due to differences in local edge magnetic moments. Increasing the NaCl concentration enhances the Mo-concentration, expediting Mo-ZZ growth along the edges. When the growth rates of Mo-ZZ and S-ZZ reach equilibrium, hexagonal terminations dominate among other islands, representing an equilibrium state where Mo-ZZ and S-ZZ edges appear alternatively. Further increasing the NaCl:MoO$_3$ ratio to 1:3 accelerates Mo-ZZ edge growth, leading to early termination and the predominance of S-ZZ edges. This results in the formation of triangular domains in the final MoS$_2$ flake, as depicted in Fig.\ref{Fig_2.3}(b). The increase in the domain sizes appear in various local regions in individual samples, the smallest of 5 $\mu$ ms from the substrate edge area to the continuous films at the centre of the substrate.
\begin{figure}[H]
\centering
\includegraphics[width=0.9\textwidth]{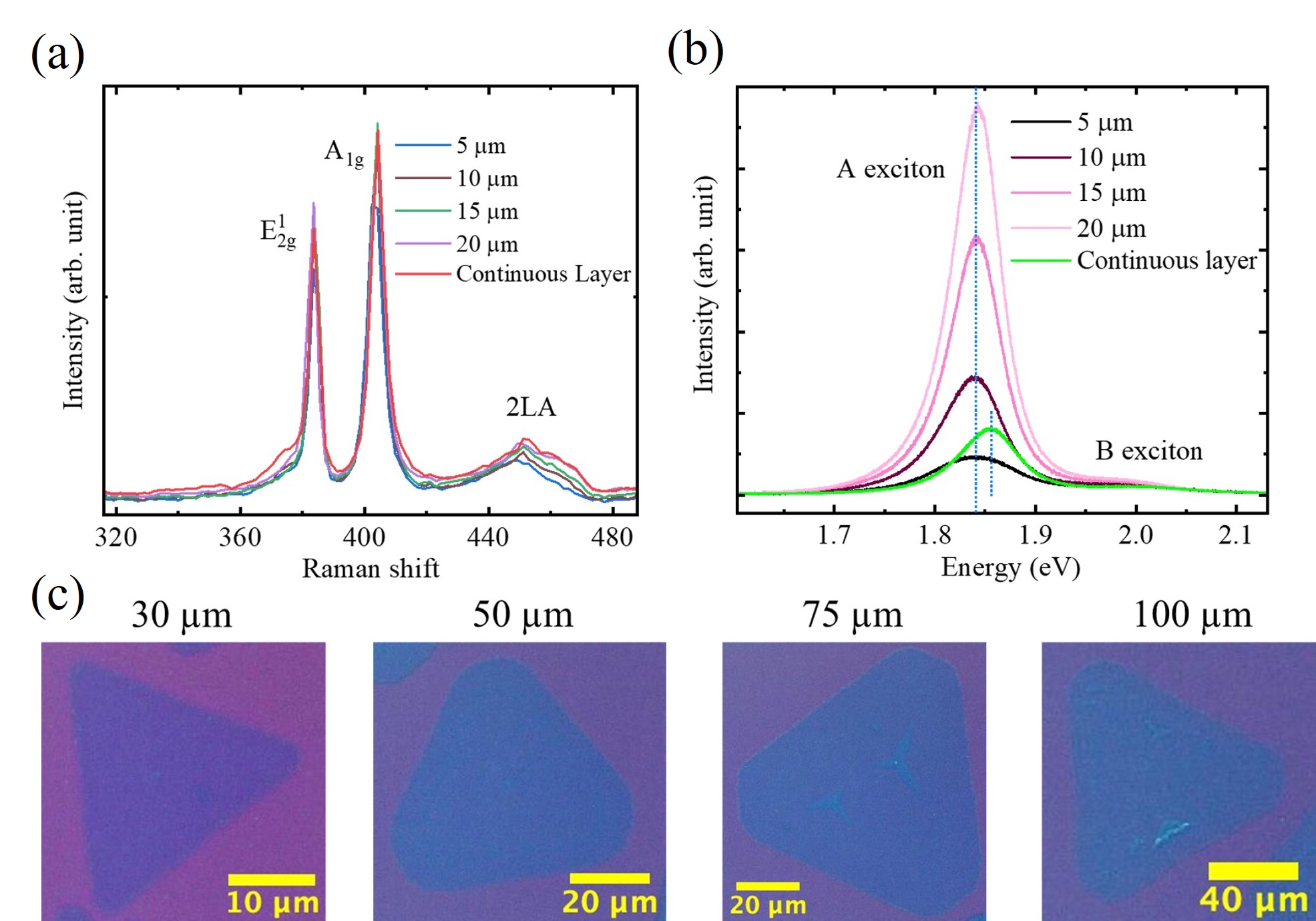}
\caption{(a) The vibrational studies using Raman spectroscopy showing three predominant Raman active modes (E$^1_{2g}$, A$_{1g}$, and 2LA) of monolayer MoS$_2$ with increasing domain sizes from 5 to 20 $\mu$m along with the continuous layer. (b) Room temperature photoluminescence studies show the evolution of A excitonic transition for increasing MoS$_2$ domain sizes. (c) The optical image of the large-scale monolayer MoS$_2$ domains produced from our modified synthesis technique with minor adjustments to the growth parameters.}
\label{Fig_2.4}
\end{figure}

The Raman and PL spectroscopic techniques are employed to gain additional insights into the optical properties of various MoS$_2$ samples. In Fig.\ref{Fig_2.4}(a), shows the vibrational Raman modes of monolayer MoS$_2$ with different size of triangular domains. It is essential to note that notably, there is no significant change in the characteristic Raman modes (E$^1_{2g}$, A$_{1g}$, and 2LA) as the domain size increases, except for a small peak preceding the E$^1_{2g}$  mode in the continuous layer. This peak is attributed to the substrate and becomes more prominent due to the large surface area of the continuous film but diminishes in the case of isolated flakes. Figure \ref{Fig_2.4}(b) represents the evolution of PL excitonic peaks with increasing flake size. It is interesting to observe the sharp increase in the A exciton peaks with the triangle size increases from 5 to 20 $\mu$m. However, in case of continuous layers, due to the presence of grain boundary and bilayer nucleation centers the PL peak drastically decreases. Additionally, a slight blueshift of the A exciton transition is observed, potentially due to trion state quenching. Further discussion, in-depth analysis and related experiments at low temperatures could offer more insights into excitonic states in continuous layers. As shown in Fig.\ref{Fig_2.4}(c), the size of the MoS$_2$ domains can be increased up to 100 $\mu$m or more by adjusting few other parameters. These parameters include increasing the growth pressure from 500 to 600 mbar which is found to be beneficial in controlling precursor evaporation and slowing down the growth process. This adjustment also extends the total growth time of the vapor phase reaction from 8 to 15 minutes, depending on other conditions. It is worth mentioning that some nucleated bilayers are observed on large monolayers due to the additional growth time and loss of control in some parameters. However, consistent yields of MoS$_2$ samples are obtained through repeated growth conditions, indicating the ease of controlling growth parameters using the NaCl-assisted CVD method.
\begin{figure}[hbt!]
	\centering
	\includegraphics[width=0.9\columnwidth]{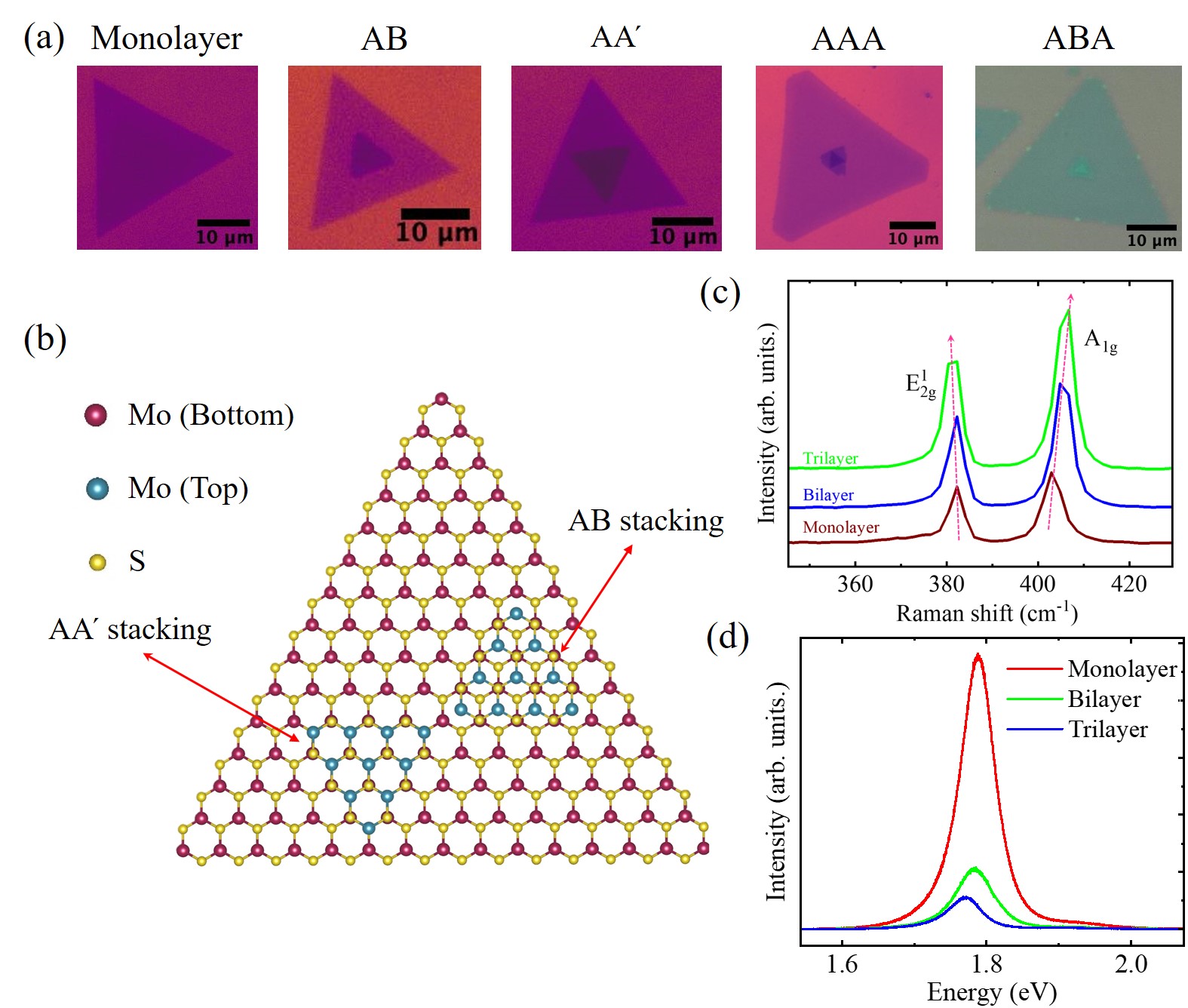}
	\caption{(a) The optical images of various stacking configurations of mono, bi (AB, AA'), and tri-layer(AAA, ABA) MoS$_2$ samples grown using the NaCl-assisted CVD method. (b) Atomistic arrangements (top view) showing the two different stable stacks in the case of bilayer MoS$_2$ generally occurred from the CVD synthesis. (c) Raman and (d) PL spectroscopic studies in the case of mono, bi, and tri-layer MoS$_2$. The frequency variations and PL intensities are consistent with the number of layers.}
	\label{Fig_2.5}
\end{figure}
\subsection{Self-seeding nucleation and layer stacking}
With shape and size controlled growth of monolayer MoS$_2$, our NaCl-assisted CVD technique can also produce layer dependent samples. Figure \ref{Fig_2.5}(a) shows the optical image of a monolayer,  AB and AA’ stacked bilayers, as well as AAA and ABA stacked tri-layers of MoS$_2$. The stacking of the layers can only be realized at relatively high temperatures around 800 $^{\circ}$C. With the rise of the temperature, the influence of NaCl becomes more pronounced. With the reaction rate directly linked to temperature, there is a sharp increase in reactant concentration. Consequently, the nucleation mechanism shifts from 2D planar to self-seeding nucleation. In such cases, the seeding of MoO$_{3-x}$S$_y$ occurs at the center of the monolayer domains, effectively facilitating the multilayer growth of MoS$_2$. An atomistic model is also shown in Fig.\ref{Fig_2.5}(b) to distinguish between the two stable stacking of bilayer MoS$_2$ i.e. AB and AA’ stacking, generally occurred from the CVD synthesis. Figure \ref{Fig_2.5}(c) represents the results from the Raman measurements on the monolayer, bilayer and trilayer MoS$_2$. These results reveal layer-dependent variations in the E$^1_{2g}$  and A$_{1g}$  modes, consistent with previous findings. As the layer numbers increase, a red shift in the E$^1_{2g}$  modes is observed, attributed to long-range Coulombic interlayer interactions. Similarly, the blue shift in the A$_{1g}$  modes suggests an increase in restoring forces due to multiple layer stacking. The frequency difference between these modes often corresponds to the number of layers present in the MoS$_2$ samples. The layer-dependent photoluminescence (PL) analysis, depicted in Figure \ref{Fig_2.5}(d), demonstrates a sharp decrease in the intensity of the PL excitonic peak with an increase in the number of layers. Additionally, red shifts in the A exciton peaks suggest the emergence of an indirect band gap in bilayer and trilayer samples.
\begin{figure}[hbt!]
	\centering
	\includegraphics[width=0.85\columnwidth]{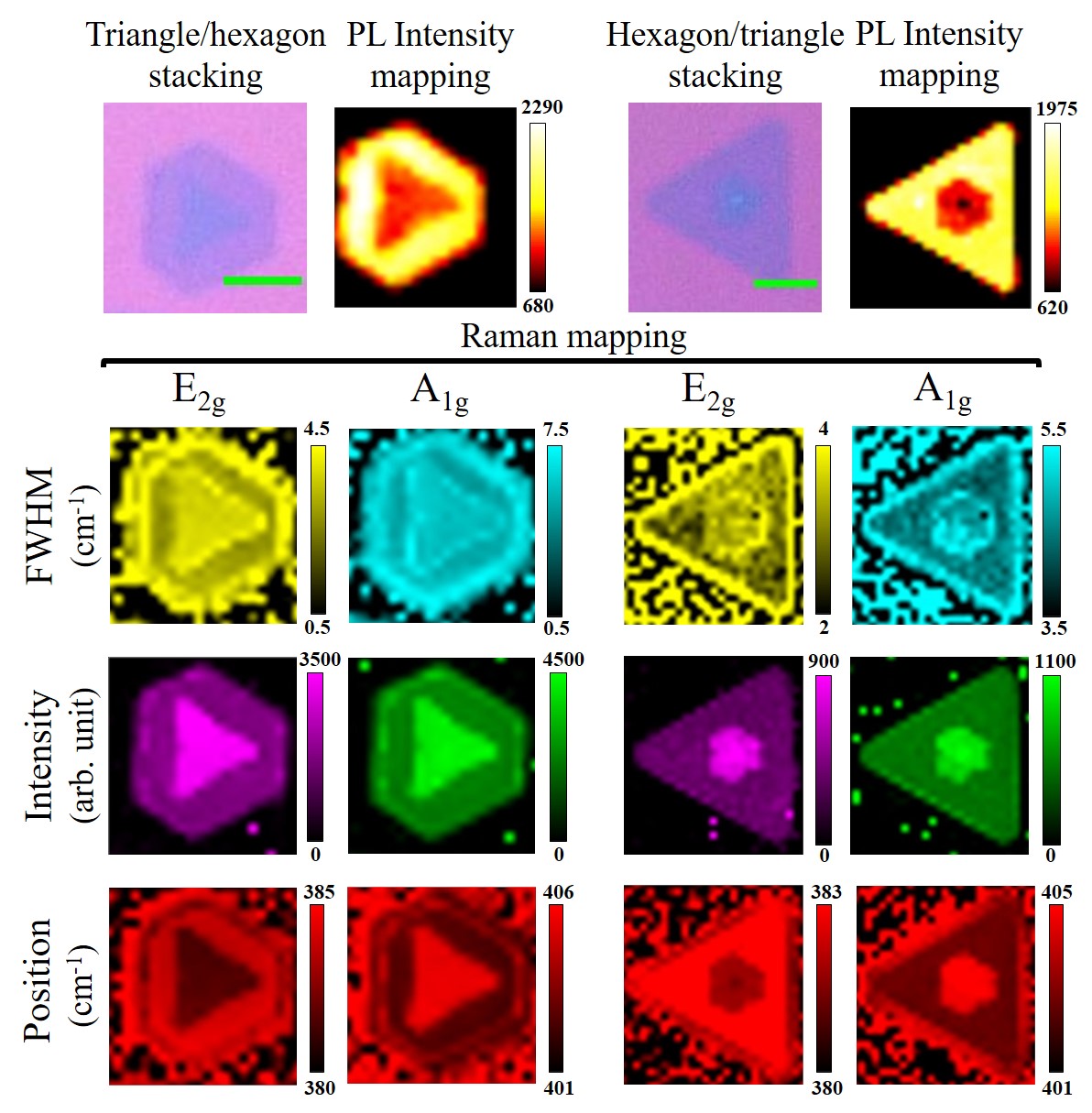}
	\caption{The optical images of two special stacking configurations along with PL and Raman mapping results are plotted for bilayer MoS$_2$. The scale bar is 5 $\mu$m.}
	\label{Fig_2.6}
\end{figure}

We also study the Raman and PL mapping analysis on few special type of stacking geometries of the CVD grown MoS$_2$ domains. These include triangle/hexagon stacking and hexagon/triangle stacking as represented in Fig.\ref{Fig_2.6}. The PL mapping and Raman mapping results demonstrate the growth of high-crystalline quality samples under optimal conditions using modified CVD parameters. We also observe a slight reduction in PL intensity at the edges of hexagon domains compared to triangular domains in monolayer MoS$_2$.

\section{Challenges of CVD growth of MoS$_2$}
Defects in 2D materials play a significant role in determining their properties and functionalities. These defects can arise from various sources including intrinsic lattice imperfections, extrinsic impurities, and environmental factors. Understanding and controlling defects in 2D materials are crucial for their practical applications in electronics, and photonics. During the optimization of the NaCl-assisted synthesis of MoS$_2$, we often resulted in various growth challenges such as premature growth, non-uniformity, etc. which may arise due to the adjustments of the growth parameters causing imbalance growth reactions. Such factors must be monitored while synthesizing high crystalline-quality samples. In this section, we have discussed some of the commonly noticed challenges of CVD growth that may hinder optical and electrical performances.
\begin{figure}[hbt!]
	\centering
	\includegraphics[width=0.9\columnwidth]{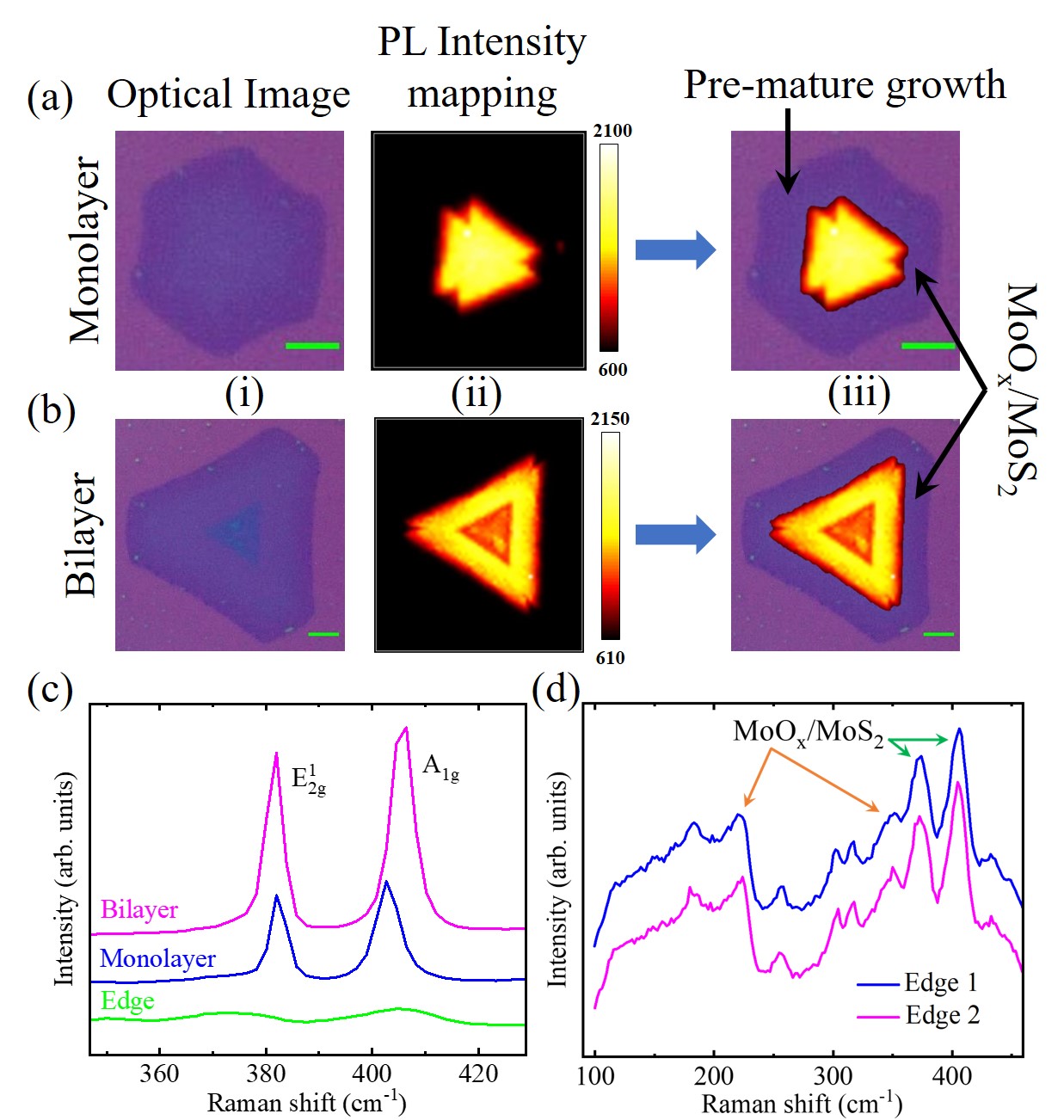}
	\caption{(i) The optical image (ii) PL intensity mapping (iii) Stacked mapping/optical image of a (a) monolayer and (b) bilayer MoS$_2$ domain showing the premature growth region near edge area. (c) The comparison of Raman modes between monolayer(blue), bilayer(magenta) and prematurly grown edge(green) region. (d) The magnified view of the Raman shifts at the low-frequency region showing the complex peaks arising from MoO$_x$ and very weak peaks from MoS$_2$. The scale bar is 5 $\mu$m.}
	\label{Fig_2.7}
\end{figure}
\subsection{Premature growth}
The optical image of a monolayer MoS$_2$ domain is shown in Fig.\ref{Fig_2.7}(a) (i). A uniform hexagonal domain is observed in the high-resolution optical microscope. However, while studying the PL mapping results of the same domain, we encounter a bright luminescence arising only from the center of the hexagonal flake, as represented in Fig.\ref{Fig_2.7}(a) (ii). The bright luminescence is surrounded by a boundary-like low luminescence region which abruptly vanishes. In Fig.\ref{Fig_2.7}(a) (iii), we overlay the PL mapping results with the optical image on the same scale to identify the area of the sample from which no luminescence is emitted. We designate these areas as premature growth regions, a classification verified by Raman spectroscopy.

In Fig.\ref{Fig_2.7}(b) we notice similar observations in the PL mapping results in the case of bilayer MoS$_2$. A reduced PL intensity is identified in the bilayer region which is located at the center. However, akin to the monolayer case, there are corresponding areas near the edge where no luminescence is detected. The Raman modes for the three different regions (monolayer, bilayer, edge/premature area) are plotted in Fig.\ref{Fig_2.7}(c). It is interesting to note that similar to PL intensity, the Raman modes are nearly absent at the edges compared to the highly intense Raman modes observed in monolayers and bilayers. We can only see a few humps in the Raman spectra at the edges. Fig.\ref{Fig_2.7}(d) represents the magnified region near the humps of the Raman spectra at two different edges for the sake of repeatability. Interestingly, we observe mixed modes of MoO$_3$ and MoS$_2$ which is referred to the premature growth. Such growth conditions may arise due to sudden cooling during the domain growth or inadequate flow rate or sulfur concentration. As the intensity of the Raman modes is found to be very weak, further studies could provide valuable insights into these premature growth conditions.
\begin{figure}[hbt!]
	\centering
	\includegraphics[width=0.9\columnwidth]{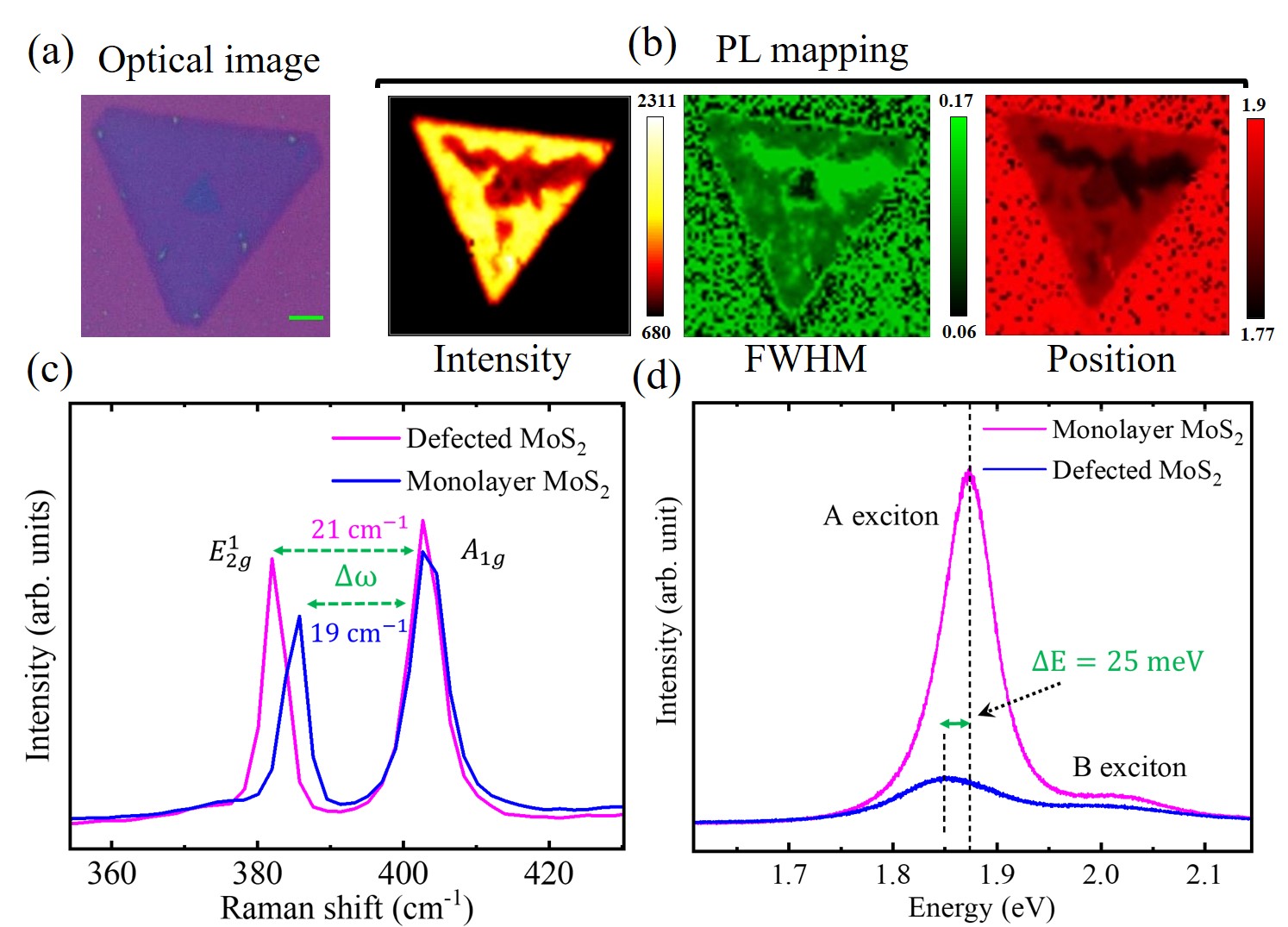}
	\caption{(a) The optical image of a monolayer MoS$_2$ having a small bilayer triangle at the center. (b) The mapping results from the PL peak intensity, FWHM, and position show the degree of non-uniformity present in the sample. (c) The Raman signatures are plotted between well-grown and defected MoS$_2$ showing significant shifts in the E modes. The frequency between the two modes is increased from 19 to 21 cm$^{-1}$ in case of defected MoS$_2$. (d) The corresponding PL spectroscopic results also depicts the PL quenching in case of defected MoS$_2$.The scale bar is 5 $\mu$m.}
	\label{Fig_2.8}
\end{figure}
\subsection{Non uniformity}
In addition to the premature growth as discussed in the previous section, we also observe other types of defects such as non-uniform luminescence throughout the sample. Figure \ref{Fig_2.8}(a) represents an optical image of a fully grown monolayer MoS$_2$ flake with a small bilayer triangle at the center. However, upon characterizing with various PL mappings, we observe non-uniform optical responses, particularly from the monolayer region. The PL intensity is reduced and varies locally throughout the region. The FWHM and position mapping results, as shown in Fig.\ref{Fig_2.8}(b), also illustrate similar non-uniformity present in the sample. We further employ Raman spectroscopy to observe variations in the Raman modes, if any. Interestingly, as depicted in Fig.\ref{Fig_2.8}(c), the E$^1_{2g}$  modes at the defected MoS$_2$ have shifted by about 2 cm$^{-1}$, whereas the A$_{1g}$  modes remain unchanged in the pure monolayer region. This shift in the E mode could be attributed to the strain associated with the defects or the in-plane strain developed during bilayer nucleation. Furthermore, the PL spectra shown in Fig.\ref{Fig_2.8}(d) provide additional insights into the defected region. As illustrated, the excitonic transition is severely affected in the defected region, exhibiting intensity quenching and a clear red shift of 25 meV. Such redshifts also confirm the presence of strain near the defect region, consistent with findings reported in previous articles\cite{Li2020Mar, He2016Octstrain}

\begin{figure}
	\centering
	\includegraphics[width=0.9\columnwidth]{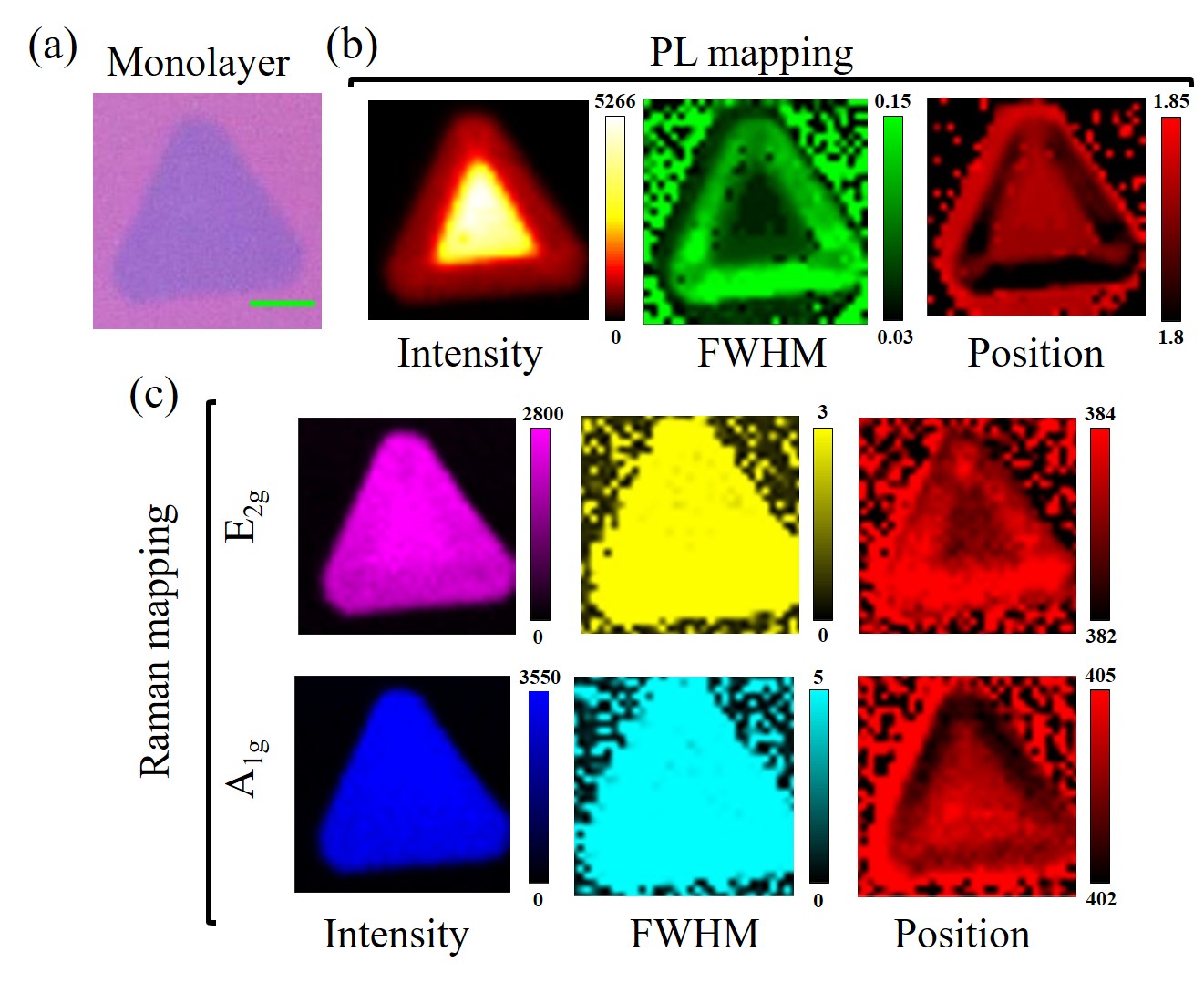}
	\caption{(a) The optical image of a CVD-grown monolayer MoS$_2$. (b) Intensity, FWHM, and position PL mapping profiles of the monolayer MoS$_2$ as shown in (a). (c) The similar Raman mapping images are shown for E$^1_{2g}$ and A$_{1g}$  modes. The scale bar is 5 $\mu$m.}
	\label{Fig_2.9}
\end{figure}

\subsection{Sulfur vacancies and Defect healing}
The prevalent defects commonly observed in CVD-grown samples are sulfur vacancies, which also play a role in facilitating n-type doping in MoS$_2$-based transistors. Figure \ref{Fig_2.9}(a) presents an optical image of a monolayer MoS$_2$ grown using the CVD technique. However, despite the optically uniform appearance of the MoS$_2$ domain, sharp variations are evident in the PL mapping results, as depicted in Figure \ref{Fig_2.9}(b). The bright triangular PL shape in the intensity mapping signifies complete MoS$_2$ growth, while the low PL intensity at the edge area (red in colour) indicates the presence of sulfur vacancies within the same flake The FWHM has also broadened at the edges which is also a signature of the vacancy related defects. The position of the PL at the edge area has also red shifted which is an indication of presence of defects. However, the Raman mapping results, as shown in Figure \ref{Fig_2.9}(c), indicate minimal effects on the vibrational modes due to sulfur defects. The intensity and FWHM modes remain relatively consistent between well-grown MoS$_2$ and MoS$_2$ facilitated by sulfur vacancies. Although the position mapping displays some indications of defects, the intensities are too weak for further analysis.
\begin{figure} [hbt!]
	\centering
	\includegraphics[width=0.9\columnwidth]{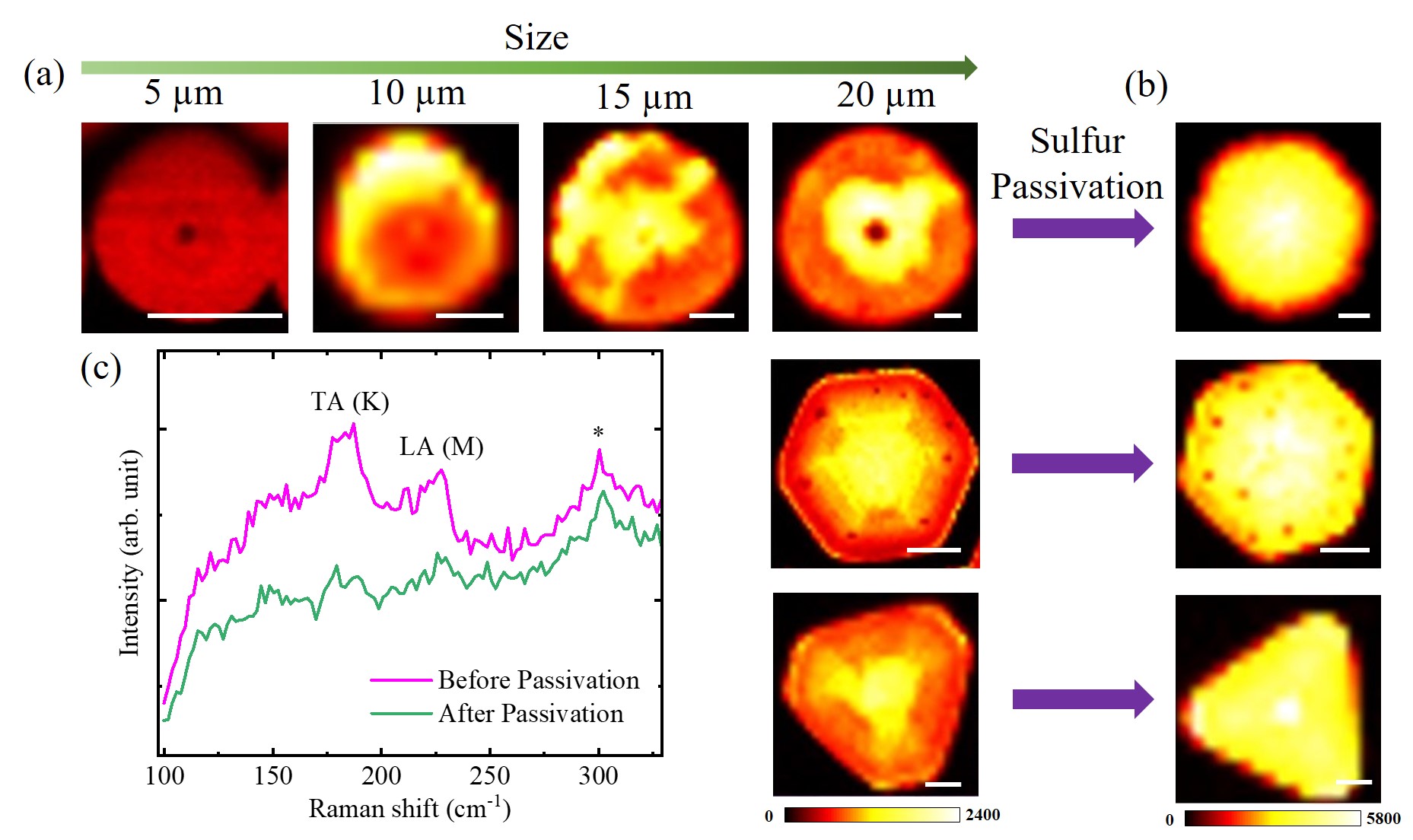}
	\caption{(a) The PL intensity mapping images of MoS$_2$ domains with increasing size of the islands. The bright spot can be identified from the yellow regions. (b) The PL mapping images after the sulfur passivation process. (c) The Raman spectroscopic signatures show the defect-activated acoustic modes, TA and LA in the case of before passivated MoS$_2$. It also plots the Raman signatures for passivated MoS$_2$ samples No Raman shifts near the acoustic region is observed. The scale bar is 5 $\mu$m.}
	\label{Fig_2.10}
\end{figure}

The dynamics of the sulfur vacancies have been employed with respect to the increase of the domain sizes. As shown in Fig.\ref{Fig_2.10}(a), The MoS$_2$ flake measuring 5 $\mu$m exhibits sulfur vacancies throughout. However, as the dimensions increase to 20 $\mu$m, the evolution of defect healing through sulfur passivation becomes apparent, indicated by the emergence of bright luminescence. Initially, the passivation process occurs non-uniformly across the flakes, eventually stabilizing at the center when the flake size reaches around 20 $\mu$m. High concentrations of sulfur vacancies degrade the electrical performance of MoS$_2$-based devices, leading to increased contact resistance, reduced carrier mobility, and decreased device lifetime. Addressing defects like sulfur vacancies in MoS$_2$ films often involves defect healing strategies. One approach is post-growth annealing in a sulfur-rich environment to facilitate the migration of sulfur atoms to vacancy sites, thereby reducing vacancy concentrations, enhancing sulfur incorporation, and minimizing vacancy formation. The sulfur vacancies shown in the 20 $\mu$m flakes are further employed for the sulfur passivation process in Ar atmosphere for 15 mins. The resulting strong and bright luminescence from the passivated MoS$_2$ flakes is shown in Fig.\ref{Fig_2.10}(b) indicating the complete growth of MoS$_2$ with the absence of sulfur vacancies. Furthermore, we examine the Raman vibrational modes at the low-frequency sides to detect the presence of any defect-activated peaks in the sulfur vacancy region. Figure \ref{Fig_2.10}(c) illustrates the Raman modes before and after passivation in the sulfur vacancy region. Interestingly, two defect-activated modes, namely Transverse (TA) and Longitudinal Acoustic (LA) modes, are observed near 200 cm$^{-1}$, consistent with findings in previous literature.\cite{Mignuzzi2015May} However, the Raman signature of the sample after passivation indicates the absence of these defect-activated peaks, consistent with the PL mapping results. Thus, defect healing techniques aim to improve the structural and electronic quality of MoS$_2$ films, leading to enhanced device performance and reliability.

\section{Synthesis of salt-assisted WS$_2$}
\label{S 2.7}
In this specific experiment, high purity NaCl (99 $\%$, SRL) and WO$_3$ (99.998 $\%$, Alfa Aesar) powders are used as metal precursors and sulfur (99.995 $\%$, Alfa Aesar) powder is used as a chalcogen source. These precursors are placed in separate alumina boats. The SiO$_2$ (285 nm)/Si (1$\times$1 cm$^2$) substrate is cleaned using the ultrasonication method with acetone, isopropyl alcohol, and de-ionized (DI) water, followed by drying with pressurized dry N$_2$ (99.999$\%$) to
eliminate any moisture. Before placing the cleaned substrate face down on the boat with the precursors, the precursor (WO$_3$+NaCl) height is adjusted to minimize the distance between the precursor and growth substrate to 3mm, to channelize the sulfur feeding during the growth period. The boat containing the growth substrate and precursors is then moved to the center of the furnace, while the boat containing sulfur is placed upstream. The tube is initially pumped down to 0.2 mbar, and then purged three times with 500 sccm of high-purity Ar (99.999$\%$) to eliminate any oxygen contaminants inside the tube. The system is heated to 900 $^{\circ}$C at a ramp speed of 30 $^{\circ}$C/min using Ar (75 sccm) as the carrier gas. The pressure inside the tube is maintained at 500 mbar throughout the 10 min growth time. Finally, the system is cooled down rapidly to ambient temperature
by partially opening the furnace.
\section{Chapter summary}

In this Chapter, we discuss various important aspects of the synthesis of 2D materials using the modified CVD technique. The growth parameters are observed to play a critical role in achieving stable growth conditions. We demonstrate a salt-assisted CVD synthesis technique with few modified geometries and parameters. The evolution of various sizes and domain orientations of the MoS$_2$ samples are studied using Raman and photoluminescence spectroscopic studies and spatial mapping analysis. We also discuss the dynamics of domain growth under the KWC theory. Furthermore, this study delves into critical parameters like geometry constraints and precursor/promoter ratios aimed at achieving controlled edge terminations, planar, and self-seeding nucleation of MoS$_2$ layers. Additionally, it unveils various growth challenges encountered in salt-assisted CVD-grown monolayer MoS$_2$ samples, including premature growth, defects, sulfur vacancies, and non-uniformity. The sulfur vacancies are then removed by a defect-healing process in which the MoS$_2$ samples are further treated in a sulfur passivation process. Finally, a method for growing large-scale triangular domain monolayer WS$_2$ is presented, offering potential applications in multi-level memory devices.

\blankpage
\chapter{Device fabrication, transfer, and electrical characterization of monolayer MoS$_2$ field-effect transistors}
\label{C3} 

\section{Introduction}
MoS$_2$ based field-effect transistors (FETs) have paramount importance in future electronic applications such as communication, data storage, memory, optoelectronics, and quantum optics owing to their high electron-hole mobilities, high ON/OFF current ratios, low operation power, good photoresponsivity, high sensitivity for molecular detection, etc.\cite{Radisavljevic2011Mar, Wu2013Apr, Yin2012Jan, Late2013Jun, Sarkar2014Apr} Indeed, a typical room temperature mobility range of about 0.8-18 $cm^{2}V^{-1}s^{-1}$ is very common in an unencapsulated monolayer MoS$_2$ with current ON/OFF ratios in order of 10$^5$ - 10$^7$.\cite{Jeon2015Jan, Najmaei2013Aug, vanderZande2013Jun, Lee2012May} However, other studies have demonstrated high carrier mobilities up to 30 $cm^{2}V^{-1}s^{-1}$ in monolayer MoS$_2$ FETs prepared by MOCVD method.\cite{Sebastian2021Jan, Arnold2017Mar} An unusual hysteresis in the transfer curves of MoS$_2$ based FETs plays a critical factor in regulating the electrical performance of the devices such as threshold voltage instability, low carrier mobility, drain current reduction, etc.\cite{Shu2016Jan, Jayachandran2020Oct} In literature, several remarks on the origin of hysteresis have been discussed to avoid or minimize such unwanted behaviors in MoS$_2$ FETs.\cite{Li2014Sep, DiBartolomeo2017Oct} For example, using a high workfunction metal such as Pt results in the reduction of hysteresis in transfer curves with minimal charge trapping effects.\cite{Jayachandran2020Oct} The carrier trapping mechanism, which is considered to be a major factor for hysteretic behavior, is being studied widely to control the broadening of the hysteresis.\cite{Li2017Dec, Guo2015Mar} However, the majority of the previous works on hysteresis studies cover a few layers of exfoliated MoS$_2$, and a very limited study on CVD-grown monolayer samples is recently reported.\cite{Datye2018Oct} Moreover, a comprehensible understanding of the localized interface trap states using the pulsed I$\sim$V technique is still in its infancy. Although the newly developed salt-assisted CVD-growth has emerged as a prevalent method for producing high-quality monolayer MoS$_2$, the transient charge trapping mechanisms and hysteresis-free transfer characteristics have not been reported so far. Furthermore, the estimation of intrinsic field-effect mobility from the hysteresis-free transfer characteristic in salt-driven CVD-grown MoS$_2$ FETs is still lacking.

 In this chapter, we demonstrate the fabrication of devices using salt-assisted CVD-grown monolayer MoS$_2$ towards the realization of high-performance hysteresis-free field-effect transistor (FET). Density functional theory calculations are implemented to monitor the effects of the Schottky barrier and metal-induced gap states between our metal electrodes and MoS$_2$ for achieving high carrier transport. The role of absorbed molecules and oxide traps on the hysteresis are studied in detail. For the first time, a hysteresis-free intrinsic transistor behavior is obtained by an amplitude sweep pulse I$\sim$V measurement with varying pulse widths. Under this condition, a significant enhancement of the field-effect mobility up to 30 $cm^2 V^{-1}s^{-1}$ is achieved. Moreover, to correlate these results, a single-pulse time-domain drain current analysis is carried out to unleash the fast and slow transient charge trapping phenomena. Our findings on the hysteresis-free transfer characteristics and high intrinsic field-effect mobility in salt-assisted monolayer MoS$_2$ FETs will be beneficial for future device applications in complex memory, logic, and sensor systems. 

\section{Fabrication of transistors}



\begin{figure}
	\centering
	\includegraphics[width=0.95\textwidth]{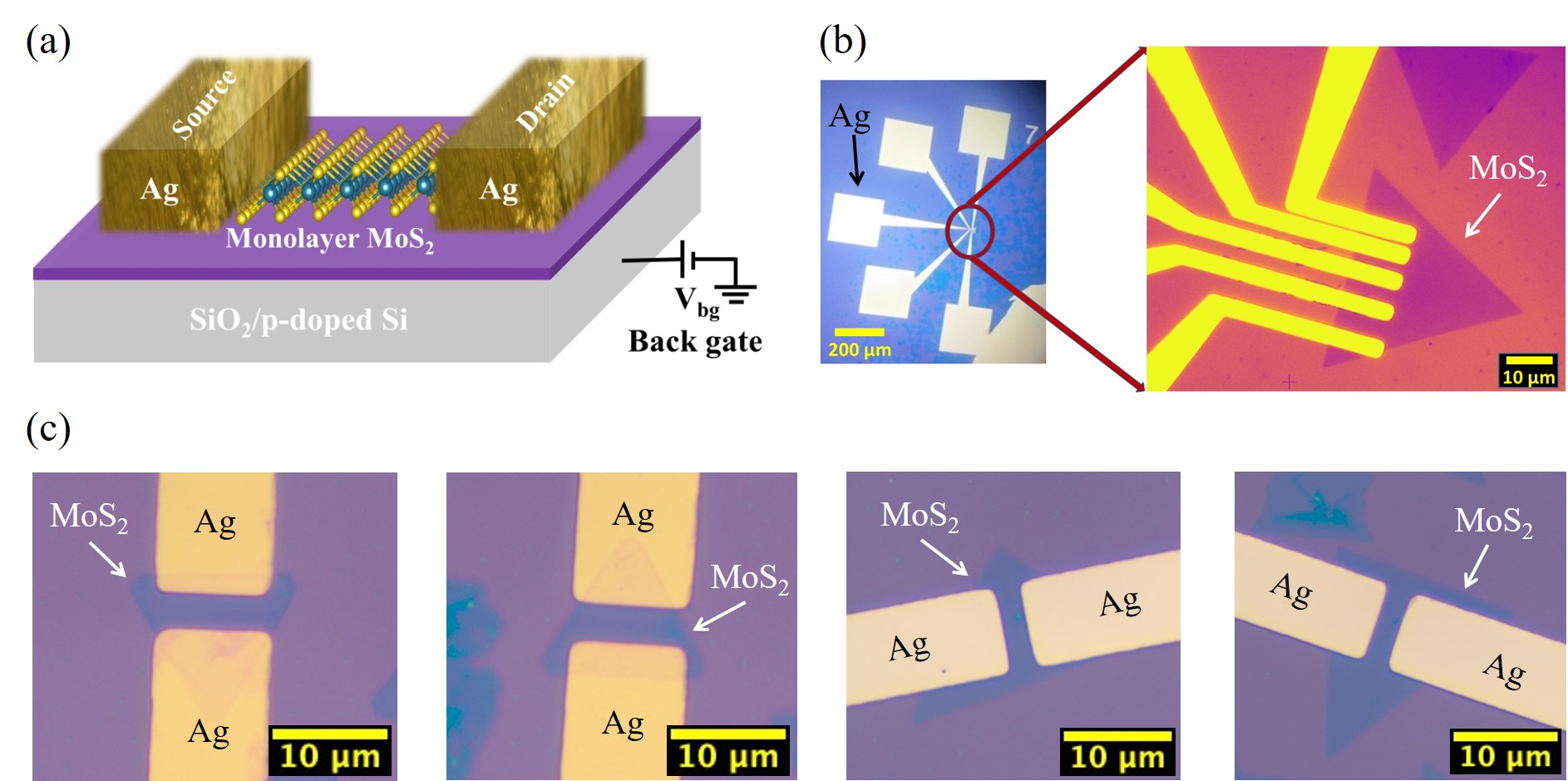}
	\caption{(a) Schematic of monolayer MoS$_2$ FET with SiO$_2$/Si back gate, Ag source/drain contacts. (b) Microscopic image of fabricated devices with different channel lengths on a single MoS$_2$ domain. A low-magnification optical image is shown on the left of (b), where the whole pattern of Ag contact pads is visible. The right side of (b) shows a high-magnification image near the MoS$_2$ channel area. (c) High-resolution optical images of fabricated devices on different monolayer MoS$_2$ flakes.}
	\label{Fig_3.1}
\end{figure}

 The monolayer MoS$_2$ based FETs are fabricated by using a photolithography (Heidelberg $\mu$PG 101) system. First, the MoS$_2$ grown SiO$_2$/Si substrate is coated with a positive photoresist (ma-p-1205) by a spin coater (SUSS Microtech) and then baked at 80 $^{\circ}$C for 1 min. Under the inspection of a high-resolution microscope, the required contact patterns are exposed on the single-layer MoS$_2$ flakes with a 405 nm laser in photolithography. The exposed patterns are developed with an alkaline solution (1:4, NaOH:DI water) and the sample is mounted on a thermal evaporation chamber for the deposition of the contact metal (Ag) followed by a lift-off process. The back gate of the FETs is facilitated by the heavily doped Si substrate. Finally, the devices are mounted on the micromanipulating four-probe stage (Lakeshore cryogenic probe station) for probing the contacts, and the DC and pulsed electrical measurements are recorded by using a Keithley 4200-A SCS parameter analyzer both in air and in high vacuum ($\sim$10$^{-5}$ mbar) conditions. We limit our measurements to a V$_{bg}$ (back gate voltage) of $\sim$ 38 V (with compliance) to keep the device safe for more extended measurements. To prevent photo-excitation of charge carriers, all experiments are performed in dark conditions. A back-gated monolayer MoS$_2$ FET is schematically shown in Fig.\ref{Fig_3.1}(a) with necessary electrical connections so that a carrier injection from metal to semiconductor channel material can be realized pictorially. The optical microscopic images of some fabricated devices on CVD-grown Monolayer MoS$_2$ using the photolithography process are also depicted in Fig.\ref{Fig_3.1}(b)\&(c).   

\section{Contact electrode}
Carrier transport in a FET channel is generally affected by various factors such as the van der Waals (vdW) gap between metal and semiconductor, Schottky barrier, metal-induced gap states, etc. Several metals and their combinations with varying work function values have been studied extensively as an electrode to MoS$_2$ based devices. However, very little research enlightens on getting low Schottky barrier contacts with MoS$_2$ by using low work function metals such as Sc and Ti. However, in that context, Ag is the least explored material with a work function value of 4.46 eV (a little higher than for monolayer MoS$_2$) even though having excellent electrical and thermal conductivity. Furthermore, Ag (111) surface has a hexagonal geometry with a lattice constant (2.93 $\AA$) that nearly matches with monolayer MoS$_2$ (3.18 $\AA$) making it a suitable candidate as a metal electrode to MoS$_2$.
\subsection{Projected band alignment of MoS$_2$/Ag (111) heterostructure}
\label{SS 3.1}
To ensure our proposition and experimental validation, a comprehensive DFT study is implemented on the MoS$_2$/Ag interface. All electronic calculations are performed by using the density functional theory (DFT) and by employing the Quantum Espresso software package. To include the interaction between the core and the valence electrons, the ultrasoft pseudopotential (USPP) along with a plane wave basis set is considered for the calculations, and the exchange-correlation interaction of the system is included by using the Perdew–Burke–Ernzerhof (PBE) functional. keeping the computational cost low, the plane wave cut-off energy is taken to be 80 Ry and the cut-off energy for the charge density is chosen as 800 Ry as obtained during the optimization process. All structures are optimized by using the conjugate gradient minimization scheme.
\begin{figure}[hbt!]
	\centering
	\includegraphics[width=0.9\textwidth]{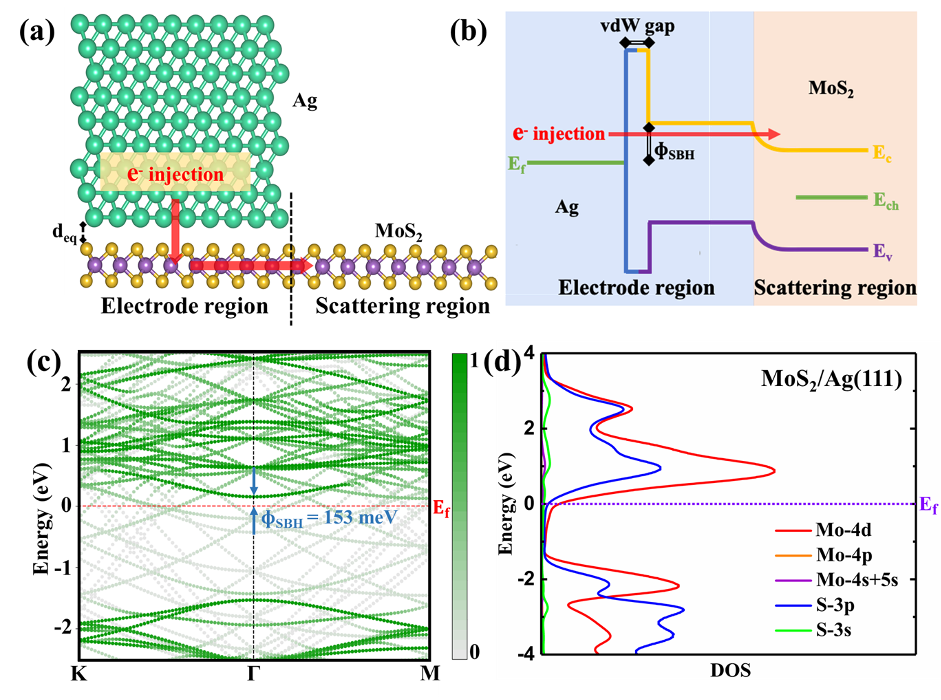}
	\caption{(a) Schematic of a cross-sectional view of Ag top contact to monolayer MoS$_2$ forming electrode and scattering region. Electron injection from Ag to MoS$_2$ is shown in red arrow. (b) Schematic illustration of band alignments near the Ag-MoS$_2$ interface. The red arrow shows the electron injection through the vdW gap and Schottky barrier. Ef and Ech correspond to metal Fermi level and channel potential, respectively. (c) Projected band profile of MoS$_2$/Ag (111) heterostructure showing the projection of MoS$_2$ bands (dotted green lines) with scalebar, The n-type Schottky barrier height ($\phi_{\rm SBH}$) is shown in blue color. (d) Corresponding PDOS of MoS$_2$ showing small gap states for hybridized Mo-d orbital.}
	\label{Fig_3.2}
\end{figure}
The Brillouin zone integrations are carried out within un-shifted Monkhorst-pack k point mesh sizes of 10$\times$10$\times$1 and 6$\times$6$\times$1 in the case of the unit cell and supercell, respectively. During the structural relaxation, the ion dynamics are regulated under BFGS algorithms. The lattice constant for the relaxed monolayer unit cell is optimized to be 3.188 $\AA$. A sufficient vacuum in the z direction is provided in all of the calculated structures to avoid interaction between the two layers. The self-consistent calculations are performed until a total energy threshold convergence of 10$^-6$ Ry is reached and the force on the atoms is converged up to a difference of 10$^-3$ Ry/$\AA$. The heterostructures of Ag (111) and MoS$_2$ are constructed with a 3$\times$3$\times$1 supercell, ensuring a small lattice mismatch and an equilibrium separation is fixed at 2.5 $\AA$ as per the energy minimization. The orbital projected band structure and DOS calculations are performed with the above-mentioned parameters in the case of the unit cell and supercell.

The band structure and corresponding partial density of states (PDOS) of monolayer MoS$_2$ 3$\times$3$\times$1 supercell are obtained to compare the band alignment of MoS$_2$/Ag (111) heterostructure.\cite{Mallik2021Apr} The direct bandgap of 1.7 eV in our case explains a small underestimation as compared to the experimental value (1.8 eV) in the case of monolayer MoS$_2$. This is well known that DFT-based calculations underestimate the bandgap compared to the experiment. The Brillouin zone folding in the case of supercell results in a direct bandgap at gamma point instead of K point. The carrier injection (in this case electron) from metal to MoS$_2$ channel can be comprised of two steps as shown in Fig.\ref{Fig_3.2}(a), the tunneling of the electron from metal to the MoS$_2$ underneath Ag (considered as electrode region) followed by the transport of electrons from electrode region to MoS$_2$ channel (known as scattering region). The corresponding band alignments at both regions with the vdW gap and Schottky barrier height ($\phi_{SBH}$) are shown schematically in Fig.\ref{Fig_3.2}(b). Further band structure calculation of the electrode region is performed to measure $\phi_{SBH}$ which is nothing but the energy difference between $CBM(E_c)/VBM(E_v)$ of MoS$_2$ for n/p-type Schottky barrier respectively and Fermi level of MoS$_2$/Ag (111) heterostructure. The equilibrium separation distance $(d_{eq})$ between the Ag and MoS$_2$ is optimized for 2.5 $\AA$ by the energy minimization. Figure \ref{Fig_3.2}(c) shows the projected band profile of MoS$_2$/Ag (111) heterostructure (electrode region) where the color bar represents the percentage of orbital contributions of the MoS$_2$ underneath Ag. The $\phi_{SBH}$ is calculated to be 153 meV by the usual definition (forming an n-type Schottky contact) which is a reasonably smaller value as compared to the other high work function metals. Moreover, the projected band structure of MoS$_2$ is seen to retain its original form compared to the band profile of pristine monolayer MoS$_2$. This indicates a negligible strain effect of Ag (111) layers on MoS$_2$ due to the similar crystal structures and nearly equal lattice constants. A PDOS calculation is further carried out to gain insight into the effect of hybridization between Ag and MoS$_2$ on the orbital projections of MoS$_2$ at the bandgap region. As seen from Fig.\ref{Fig_3.2}(d), a small amount of overlapped Mo-4d state is noticed at the Fermi level and band-gap region when comparing with PDOS of pristine MoS$_2$. This spreading of band edge Mo states may arise due to the weakening of Mo-S covalent bonds by metal adsorption.\cite{Gong2014Apr} However, considering low work function metals for getting metalized states with low electrical conductivity and high work function metals having relatively high electrical conductivity but with higher Schottky barriers, Ag plays a perfect balancing role having high electrical conductivity and forming low Schottky barrier with MoS$_2$. 


\section{Towards high performance hysteresis-free FETs}
In literature, it has been reported that Ag forms a smoother interface with monolayer MoS$_2$ and by using it as a metal contact, high carrier transport is achieved in a MoS$_2$-based transistor compared to Ti.\cite{Yuan2015Jan, Abraham2017Sep} This result is consistent with our projected band structure calculations as discussed in the section \ref{SS 3.1}. Moreover, the lowest contact resistance is also realized by using Ag or Au as a drain/source metal contact of monolayer MoS$_2$ FET.\cite{Schauble2020Nov, Smithe2018Jul} This motivated us to fabricate single-layer MoS$_2$ FET on the as-grown CVD sample with Ag as contact metal for drain and source electrodes by using the photolithography process. The back gate contact of the FET is facilitated by heavily doped Si substrate and 300 nm SiO$_2$ beneath the MoS$_2$ layer provides the required gate capacitance. The high-magnification optical image of the as-fabricated device is shown in Fig.\ref{Fig_3.3}(a). The channel length (L) and width (W) are calculated to be $\approx$ 2.7 and 8.1 $\mu$m, respectively. The inset shows a low-magnification optical image of the complete FET with a scalebar of 200 $\mu$m.
\subsection{Output and transfer characteristics}

\begin{figure} [hbt!]
	\centering
	\includegraphics[width=0.9\textwidth]{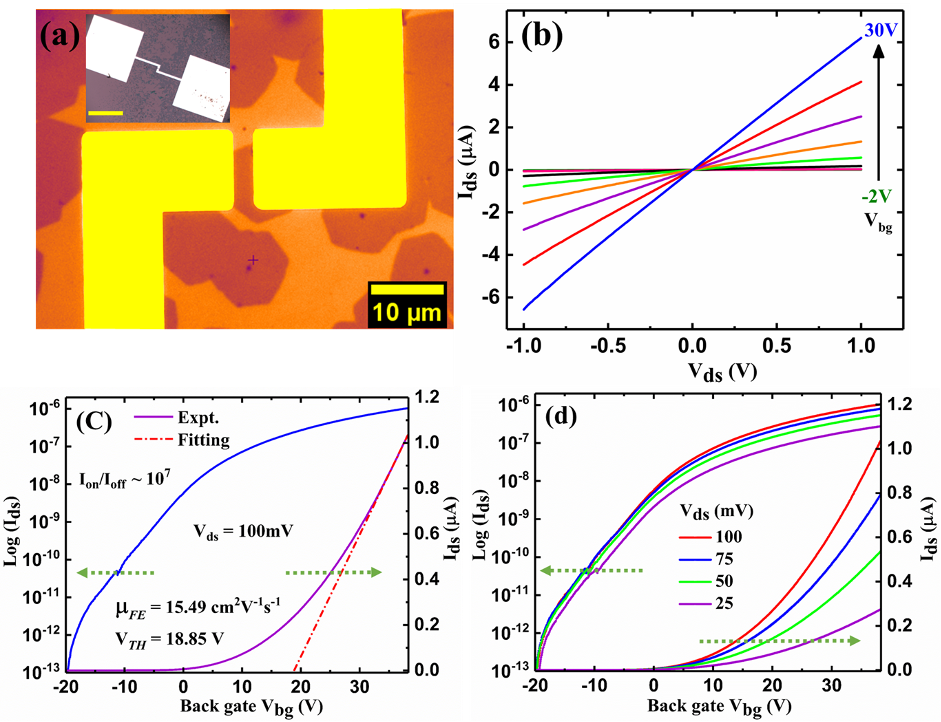}
	\caption{(a) An optical microscopic image showing as-fabricated monolayer MoS$_2$ FET (Ag electrode is shown in false yellow color), Inset shows the low magnification image of the device with Ag contact pads (scale bar is 200 $\mu$m). (b) Output characteristics i.e. drain current $(I_{ds})$ vs drain-source voltage (V$_{ds}$) with different fixed gate bias (V$_{gs}$) from -2 V to 30 V. (c) mobility and threshold voltage extraction from linear I$_{ds}$ transfer characteristics (right axis), corresponding semi-logarithmic drain current (left axis) showing high current on/off ratio at V$_{ds}$ = 100 mV. (d) Room temperature transfer characteristics at different V$_{ds}$ in both linear (right axis) and semi-logarithmic (left axis) scales showing the stability of the FET.}
	\label{Fig_3.3}
\end{figure}

Nearly linear output characteristics i.e. variation of the drain current (I$_{ds}$) with drain-source voltage (V$_{ds}$) can be noticed in Fig.\ref{Fig_3.3}(b), which indicates exemplary Ohmic contact behavior for the fabricated single-layer MoS$_2$ FET. It is interesting to note that the output currents are in the range of microampere, which elucidates the high pristine quality of NaCl-assisted CVD-grown MoS$_2$ FET. The drain current is recorded for different gate voltages starting from small negative voltage (-2 V) to large positive voltage (30 V) showing the excellent electrostatic gate doping effect.

The electrical measurements are performed in both ambient and high vacuum ($\sim10^{-5}$ mbar) conditions to understand the effect of the surrounding environment on channel mobility. Figure \ref{Fig_3.3}(c) shows transfer characteristics i.e. curve between drain-source current (I$_{ds}$) vs back gate voltage (V$_{gs}$) at a constant drain-source voltage (V$_{ds}$ = 100 mV) in vacuum for a single-layer MoS$_2$ back gated FET at room temperature. The field-effect mobility of the charge carriers is extracted by using the expression,\cite{Radisavljevic2011Mar}
\begin{equation}
    \mu_{FE} = \frac{dI_{ds}}{dV_{bg}}\times\frac{L}{WC_{ox} V_{ds}} 
\label{eq:mobility_eq}
\end{equation}
where L and W are the channel length and width respectively (in our case, L = 2.7 $\mu$m and W = 8.1 $\mu$m), C$_{ox}$ is the capacitance value of the back gate dielectric per unit area ($C_{ox} = \epsilon_0 \epsilon_r/d$; in our case, relative permittivity of SiO$_2$, $\epsilon_r$ = 3.9, thickness of SiO$_2$ layer d = 300 nm ). At a bias voltage V$_{ds}$ = 100 mV, from the slope of the transfer curve, the room temperature mobility is calculated to be $\sim$15.49 $cm^{2}V^{-1}s^{-1}$, which is higher or comparable to previously reported results on monolayer MoS$_2$ without surface passivation by high-k dielectrics.\cite{Liu2015Sep, Ahn2017Jun, Kang2014Mar}

It should be mentioned here that this mobility value could be an underestimated/overestimated channel mobility as it includes the contact resistances at the two electrodes. Excluding such effects by the four-probe measurement method might provide the actual field-effect mobility.\cite{Nazir2016Jun} However, sticking to our main aim to focus on the hysteresis at large positive gate bias, where the effect of contact resistances on the hysteresis would be not significant anymore, all the measurements are carried out using two probe geometry to get effective mobility. The threshold voltage ($V_{TH}$) is found to be 18.85 V and is obtained from the linear curve fitting of drain current values by using the expression,
\begin{align}
I_{ds}= \frac{W}{L}\mu_{FE}C_{ox}V_{ds}(V_{bg}-V_{TH})
\label{eq:VTH_eq}
\end{align}
In literature, it is seen that the presence of the grain boundary influences the carrier transport of the FETs and hence the mobility to a great extent due to bandgap broadening, and strain effects.\cite{vanderZande2013Jun} As we are getting good mobility, we expect our MoS$_2$ sample is free from such grain boundaries and the obtained mobility is only subjected to substrate effects. The room temperature transfer characteristics for different drain-source voltages with a forward sweep of back-gate voltage in both linear (right axis) and semi-logarithmic (left axis) scale are shown in Fig.\ref{Fig_3.3}(d). The observed OFF currents are in the order of $10^{-13}$ A, while the ON current for V$_{ds}$ = 100 mV saturates at $\sim10^{-6}$ A, illustrating a magnitude of $\sim10^7$ order current on/off ratio for our device. Our results establish an excellent agreement with previously reported I$_{on}$/I$_{off}$ ratios values for back-gated monolayer MoS$_2$ transistors. Similar measurement for transfer characteristics is also carried out in ambient pressure condition (air medium) with mobility and $V_{TH}$ extraction where a high degradation in the saturation drain current value ($\sim10$ times decrease in magnitude) is noticed with a decrease in mobility of 3.86 cm$^{2}V^{-1}s^{-1}$.\cite{Mallik2021Apr} The current on/off ratio is reduced to $\sim10^4$ leaving a minor change in threshold voltage ($V_{TH}$ = 16.86 V).  This is expected due to the adsorption of water/oxygen molecules on the surface of the MoS$_2$ channel acting as charge trapping and scattering centers under the application of high positive gate voltage.

\subsection{Effect of environment on hysteresis}

\begin{figure}
	\centering
	\includegraphics[width=0.9\textwidth]{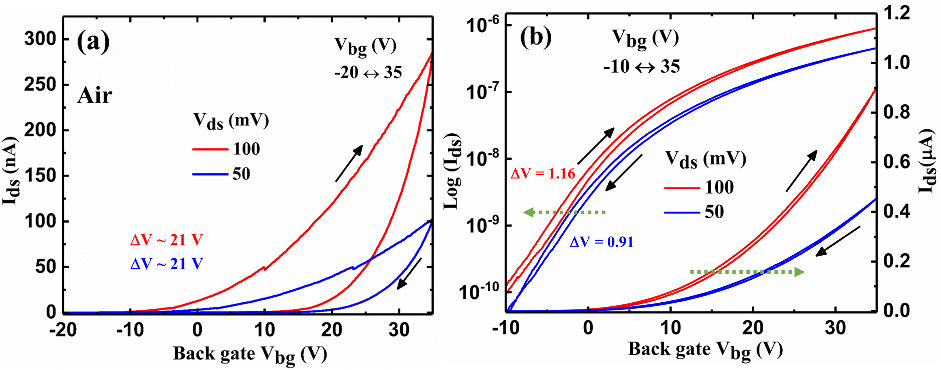}
	\caption{Hysteresis loops in the transfer characteristics of MoS$_2$ FET in (a) air (b) high-vacuum environment. The curves are shown in both linear (right axis) and semi-logarithmic scale (left axis) for two different V$_{ds}$ = 100 mV (red), 50 mV (blue). For Air condition, only linear scale is shown due to broad hysteresis. $\Delta V$ represents the maximum hysteresis width during forward and backward sweeps. Black arrows indicate the directions of drain currents with back gate sweeping voltages.}
	\label{Fig_3.4}
\end{figure}

To explore the intrinsic/extrinsic contributions to the hysteretic behavior in transfer characteristics of the MoS$_2$ FET, the DC gate voltage is further applied in a dual sweep mode. The device is first exposed to ambient pressure to observe the hysteretic nature under gate bias stress. The sweep range is varied from negative gate voltage (-20 V) to high positive gate voltage (35 V) and again sweep back to the negative gate voltage (-20 V) in a cyclic manner as shown in Fig.\ref{Fig_3.4}(a). As in the literature, it is observed that voltage sweep rates can significantly influence the hysteresis of the device.\cite{Cho2013Sep} To avoid such additional effects a fixed sweep rate of  0.1 V/s is maintained for each measurement. The observed hysteresis is measured by the maximum voltage shift in the transfer characteristics between forward and backward sweeps known as hysteresis width ($\Delta$V). The dual-sweep is recorded at two fixed drain-source voltages (V$_{ds}$) i.e. 100 mV and 50 mV to examine any possible variations on the hysteresis width. The sweeping directions induce a broad clockwise hysteresis width of $\Delta$V $\sim$21 V, as shown in Fig.\ref{Fig_3.4}(a). This behavior is quite similar to the previous reports, which dominantly attributes to a large amount of charge trapping at the MoS$_2$ surface due to the adsorption/desorption of ambient gases and water molecules. At the beginning of the forward sweep, when a particular negative bias is applied to the gate terminal, an initial desorption process of the trapped molecules takes place, releasing additional charge carriers to the conduction band of the MoS$_2$ channel, which causes a left shift of the transfer characteristics. As we reach a high positive gate bias, the adsorption of ambient molecules is increased, which captures more electrons filling the trapping sites. As a consequence, during the backward sweep starting from the high positive gate bias, the channel of MoS$_2$ experiences fewer charge carriers due to trapped charges causing degradation in drain current values, which basically leads to a right shift of the transfer characteristics. This total offset in transfer curves, also known as hysteresis width, is mainly related to two plausible factors at room temperature, i.e. (i) transfer of electrons between MoS$_2$ channel and external adsorbates\cite{Late2012Jun} (ii) oxide traps at MoS$_2$/SiO$_2$ interface,\cite{Guo2015Mar} which is regarded as an intrinsic factor for trapping of charges. However, to study this intrinsic trapping mechanism for better control of MoS$_2$ and other 2D material-based upcoming devices, such a large amount of hysteresis caused by extrinsic molecules/gases is certainly avoidable. To rule out these external factors, we proceed with our hysteresis study in a high-vacuum ($\sim 10^{-5}$ mbar) environment keeping the same above gate bias sweeping conditions as shown in Fig.\ref{Fig_3.4}(b). Interestingly, a substantial decrease in the hysteresis width is noticed upon switching the environment from ambient conditions to high-vacuum. The considerable reduction of $\Delta$V in high-vacuum provides the testimony of elimination of adsorbate effects. However, from Fig.\ref{Fig_3.4}(b) it is evident that the effect of the environment could not reduce the hysteresis to 100 $\%$.  As discussed, the oxide traps at the interface also contribute to the hysteresis and are not affected by external factors like the one adopted here (high vacuum condition). This residual hysteresis can be merely attributed to the oxide trapping at the MoS$_2$/SiO$_2$ interface. These charge traps essentially come from unpreventable dangling bonds at the SiO$_2$ surface.\cite{Guo2015Mar} Another plausible factor is that presence of trapped H-bonded water molecules between MoS$_2$ and SiO$_2$ interface may additionally contribute to the charge traps as described in earlier reports.\cite{Late2012Jun, Guo2015Mar} However, in our case, we don’t expect trapped water molecules between the MoS$_2$/SiO$_2$ interface as the samples are prepared at high-temperature CVD method.

\subsection{Pulsed I$\sim$V characterization technique }
In order to obtain the intrinsic transfer characteristics of the MoS$_2$ FET, a further study on the remnant hysteresis induced by oxide-based traps is certainly required. As per the conventional DC measurement technique is concerned, the trapping/de-trapping process at the MoS$_2$/SiO$_2$ interface is inevitable for the complete removal of hysteresis in the transfer characteristics due to the more extended measurement periods. However, in this situation, the pulsed I$\sim$V technique is demanded due to its shorter measurement time, which might truncate the effect of trapping hence minimizing the hysteresis further. Recently, numerous approaches have been addressed to quantify the charge trapping densities in bulk gate oxide as well as at the interface of MoS$_2$ and gate-dielectric by using pulsed I$\sim$V methodology.\cite{Datye2018Oct, Park2018Mar} A high-temperature hysteresis study by pulsed measurements with the formation of memory step due to slow relaxation of trapped charges indicates elevated temperature applications of MoS$_2$ thin-film devices.\cite{Jiang2015Feb} Furthermore, controllable hysteresis with different gas pressure conditions corroborates MoS$_2$ as a versatile candidate for broad-range memory-based applications.\cite{Urban2019Sep} Therefore, room temperature pulse I$\sim$V measurements are carried out to inspect the intrinsic charge trapping mechanism causing hysteresis. The pulsed gate voltages with different pulse widths (P$_w$) are deployed to study the change in hysteresis at a fixed drain-source DC bias (V$_{ds}$ = 100 mV). 
\begin{figure}
	\centering
	\includegraphics[width=0.9\textwidth]{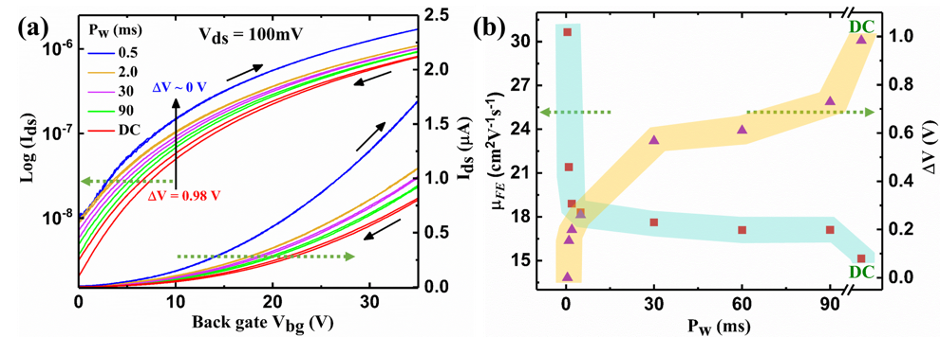}
	\caption{(a) Pulsed I$\sim$V characterization with short-pulse amplitude dual-sweeping technique for base voltage fixed at 0 V. Pulse amplitude is swept monotonically at a step of 0.1 V to the highest amplitude of 35 V at a fixed DC bias of $V_{ds}$ = 100 mV. The transfer curves are obtained for gate voltage pulse widths, $P_w$ = 90, 30, 2, 0.5 ms, and corresponding drain currents are shown in both semi-logarithmic (left axis) and linear (right axis) scales. A DC transfer characteristic is shown with the same sweeping rate and dual-sweeping range for the comparison. The vertical black arrow indicates a decrease in hysteresis widths ($\Delta V$) from DC to pulse currents with decreasing pulse widths. (b) Variation of mobility (left axis) and hysteresis width (right axis) with different pulse widths, $P_w$, and comparison with DC mobility.}
	\label{Fig_3.5}
\end{figure}

Figure \ref{Fig_3.5}(a) shows the dual sweep transfer characteristics by using the pulse gate voltages with pulse widths varying from 0.5 ms to 90 ms. However, in the pulse gate voltage amplitude sweep technique, the corresponding drain currents are recorded using a spot mean measurement method in the plateau region of each pulse, as demonstrated elsewhere.\cite{Mallik2021Apr} In order to circumvent the hysteresis in DC measurement and obtain intrinsic trap-free drain currents, the voltage pulse widths are optimized in the millisecond (ms) time-domain range. The base voltage is set to 0 V in each measurement and the pulse amplitude is monotonically increased up to 35 V at the step of 0.1 V followed by a reverse sweep, in the same manner, generating a complete set of I$_{ds}\sim$ V$_{gs}$ curve. The DC hysteresis is also shown to collate with pulsed hysteresis measurements having the highest hysteresis width of $\Delta$V = 0.98 V. To obtain the hysteresis as a function of pulse width only, the rise and fall time of the pulses are maintained as short as possible at 1 $\mu$s. From Fig.\ref{Fig_3.5}(a) it can be seen that the hysteresis width ($\Delta$V) is kept on decreasing as we decrease the gate voltage pulse width from 90 ms to 2 ms and at a particular pulse width (P$_w$ = 0.5 ms) it practically vanishes. At the pulse width of 0.5 ms, the backward sweep drain currents overlap with the forward sweep drain current values eliminating hysteresis at the same time providing near intrinsic transfer characteristics for the MoS$_2$ FET. On comparing to the previous report by Datye et al.,\cite{Datye2018Oct} where a maximum of 89 $\%$ reduction of hysteresis is observed by pulsed I$\sim$V technique with similar device configuration. In our case, a $\sim$100 $\%$ elimination of hysteresis is observed for the first time in NaCl aided CVD-grown monolayer MoS$_2$ FET. The presence of NaCl suppresses the nucleation and promotes the lateral domain growth resulting in chemically pure MoS$_2$ with high crystallinity and electronic quality compared to the conventional CVD method. This results in the reduction of active charge trap centers in the channel region which may lead to the elimination of hysteresis as observed in our case. In case of conventional CVD grown monolayer MoS$_2$ as report earlier,\cite{Datye2018Oct} no significant increase in the drain current values is observed with the decrease in pulse width. Whereas, in our salt-assisted case, a systematic increase in current values is noticed with the decrease in pulse width results in higher mobility. The right Y-axis of Fig.\ref{Fig_3.5}(b) illustrates the variation of hysteresis widths of the transfer curves obtained by pulse I$\sim$V measurements with different pulse widths and DC measurements. The population of trapped charges (N$_t$ in cm$^{-2}$) can be calculated by using the equation,
\begin{align}
N_t = \Delta V\times \frac{C_{ox}}{q}
\label{eq:trap density_eq}
\end{align}
where, C$_{ox}$ is the oxide capacitance in $F/m^2$, q being the elementary charge in C. In the case of DC I$\sim$V, the hysteresis width of $\Delta$V = 0.98 V corresponds to $\sim7\times10^{10}$ $cm^{-2}$ number of trapped charges present at the interface of MoS$_2$ and gate dielectric. However, in the case of pulsed I$\sim$V, the number of trapped charges are reduced to $5.2\times10^{10}$, $4\times10^{10}$, and $1.3\times10^{10}$ $cm^{-2}$ for P$_w$ = 90, 30, and 2 ms, respectively. For P$_w$ = 0.5 ms a vanishing hysteresis suggests the absence of trapped of charges during forward and backward sweeps. The physical apprehension of hysteresis-free drain currents interprets higher mobility values due to the trap-free flow of charge carriers through the MoS$_2$ channel as compared to the stressed DC measurements. The left Y-axis of Fig. 5 (b) describes the pulse width-dependent field-effect carrier mobility values and a fair comparison with DC mobility. The DC mobility is calculated to be 15.13 $cm^{2}V^{-1}s^{-1}$, whereas the pulsed mobility with pulsed width, P$_w$ = 90 ms is extracted to be 17.11 $cm^{2}V^{-1}s^{-1}$, and is keep on increasing until we get a trap-free near intrinsic mobility of around 30.65 $cm^{2}V^{-1}s^{-1}$ at a pulse width, P$_w$ = 0.5 ms. All the mobility values are calculated using forward sweep transfer curves. For this transfer curve having P$_w$ = 0.5 ms, we have also carried out Y-function mobility as reported by previous article\cite{Sebastian2021Jan} and the calculated value is found to be around 32 $cm^{2}V^{-1}s^{-1}$. It is worth noticing that the mobility extracted from the hysteresis-free drain currents (corresponds to P$_w$ = 0.5 ms) is increased by $\sim$100 $\%$ than the DC mobility value elucidating considerable current enhancement on the application of pulsed gate bias sweep as compared to the previous report.\cite{Datye2018Oct} 
Our findings on hysteresis-free transistor characteristics are comparable with recently reported hysteresis-free FET characterization on
monolayer MoS$_2$ grown by the MOCVD method \cite{Dodda2020Sep}.
\subsection{Waveform measurements}
During a pulse I$\sim$V measurement, the hysteresis minimization is obtained by considering two possible approaches. Here, we discuss them one by one in detail. Case-1: A pulse voltage in a waveform has three components i.e., rise time, pulse width, and fall time consisting of a pulse period. The time difference between the two pulses (i.e., the falling edge of one pulse and the rising edge of the next pulse) is denoted as delay time. Charge trapping occurs during the pulse width, whereas throughout the delay time, de-trapping process is expected.\cite{Guo2015Mar} As it is a well-known fact that de-trapping is a much slower process than trapping. However, if one keeps the delay time the same as the pulse width (which is in our case), the number of trapped charges will not get enough time to recover. Now during the pulse amplitude sweep of the gate voltage, as discussed in Fig.\ref{Fig_3.5}(a), we apply a pulse train with monotonically increasing its amplitude. During the forward sweep, the accumulation of charges will occur at the trapping sites because, in our pulse time parameters, the pulse width and delay time are the same. This accumulation of trapped charges causes hysteresis in the transfer curve by pulse I$\sim$V measurements. Now, one way to avoid this accumulation is by giving more time for the charges to be de-trapped i.e., increasing the delay time. In the previous reports, various research groups have opted for this method to reduce the hysteresis in transfer characteristics using pulsed I$\sim$V measurements. Although using a much longer delay time minimizes the hysteresis a little bit further, however, none of the consensuses have ever been reported to completely rule out the hysteresis by this method \cite{Lu2016Apr, Datye2018Oct}.
\begin{figure}[hbt!]
	\centering
	\includegraphics[width=0.9\textwidth]{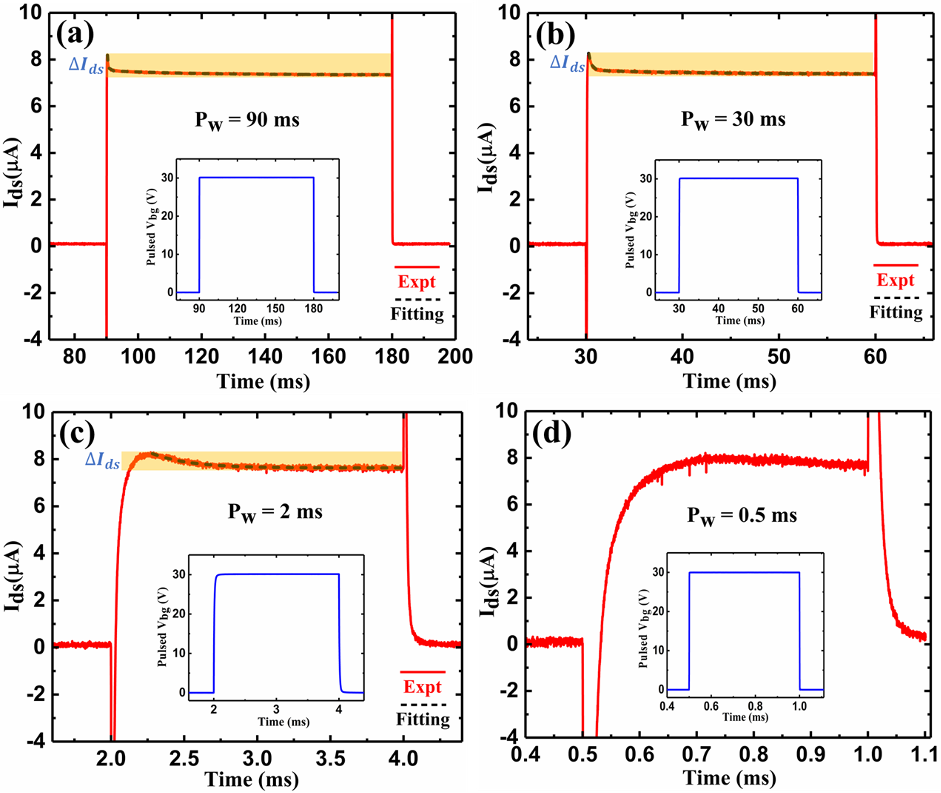}
	\caption{Transient drain current characteristics (red curve) by using single pulse waveform measurement for different pulse widths, Pw = (a) 90 ms (b) 30 ms (c) 2 ms (d) 0.5 ms. Insets show the pulse gate voltage (blue curve) from 0 V to 30 V as a function of time. The drain current reduction $(\Delta I_{ds})$ in each case is highlighted in yellow color. The fittings to the drain currents are shown in the black dotted line by using single and double exponential decay equations given in the main text.}
	\label{Fig_3.6}
\end{figure}

Case-2: In this context, we propose another approach that not only eliminates the hysteresis completely but also provides a faster measurement method for getting intrinsic transistor properties. As shown in Fig.\ref{Fig_3.5}(a), decreasing the pulse width of the gate voltage pulse train and simultaneously measuring the drain current values produce transfer curves with reduced hysteresis and as the pulse width is sufficiently low i.e. at P$_w$ = 0.5 ms the hysteresis is eliminated giving rise to high ON-current values as compared to other transfer curves. To have a clear understanding, we switch to a waveform transient drain current analysis by employing a single voltage pulse to the gate terminal, as shown in Fig.\ref{Fig_3.6}. The drain-source voltage is fixed at 1 V for each measurement. The pulse width is varied from 90 to 0.5 ms to inspect the minimum voltage stress time required to induce charge trapping significantly at the MoS$_2$/SiO$_2$ interface. The humps in the current values at the pulse transitions stem from capacitive charging effects from the higher cable capacitance. At the rising edge of the pulse from 0 V to 30 V, the drain current increases suddenly to $\sim$8 $\mu$A due to the accumulation of charge carriers into the MoS$_2$ channel. At the beginning of the plateau region of the pulse voltage, the drain current drops quickly indicating an initial fast charge trapping followed by a slower, continuous decrease in current values with the voltage pulse maintained at 30 V. The reduction of drain current ($\Delta I_{ds}$), shown in the highlighted region (yellow), basically attributes to some of the channel electrons getting captured at the trapping sites due to the voltage stress during the pulse width. As seen in Fig.\ref{Fig_3.6}, $\Delta I_{ds}$ is more in case of wider pulse width (P$_w$ = 90 ms), stipulating more trapping of charges, which gradually decreases with the decrease in pulse widths. However, the transient curve associated with P$_w$ = 0.5 ms experiences no degradation in the current values elucidating inadequate response time for electrons to be trapped. These results explain the reason behind the shifting of the ON-current values for different pulse width conditions corresponding to the spot-mean degraded current values except for P$_w$ = 0.5 ms, where a high, hysteresis-free drain current values are recorded. This is the required fastest pulse, which might provide actual drain current values, free from any significant charge trapping effects and hence hysteresis, representing intrinsic transistor behavior.

The fleeting drain current values at the plateau region of the pulse voltage are fitted with two trap models for P$_w$ = 90 and 60 ms whereas for P$_w$ = 2 ms, the drain current values are best fitted with one trap model as given by the following equations.
\begin{align}
Two\ trap\ model: I=I_0+A(e^\frac{-(t-t_0)}{\tau_1})+B(e^\frac{-(t-t_0)}{\tau_2})
\label{eq:twotrap_eq}
\end{align}
\begin{align}
One\ trap\ model: I=I_0+A(e^\frac{-(t-t_0)}{\tau_1})  
\label{eq:onetrap_eq}
\end{align}
Where, $\tau_1$ and $\tau_2$ are the time constants for fast and slow trapping processes, respectively (see Table \ref{table 3.1}). The parameters A and B are initial amplitudes (currents) for the two trapping contributions, $I_0$ and $t_0$ are adjustable offsets in current and time frames, respectively. The values of these fitting parameters and constants for different pulse widths are shown in Table \ref{table 3.1}.
\begin{table}
	\centering
	\caption{Fitting parameters for different pulse widths extracted from trap models}
	\label{table 3.1}
	\begin{tabular}{|l|l|l|l|l|}
		\hline 
		P$_w$ (ms) & A (Amp) & B (Amp) & $\tau_1$ (s)& $\tau_2$ (s) \\ \hline
		90        & 6.6$\times10^{-7}$ & 1.8$\times10^{-7}$ & 3.2$\times10^{-4}$  & 2.6$\times10^{-2}$ \\ \hline
		30        & 7.1$\times10^{-7}$ & 1.6$\times10^{-7}$ & 2.8$\times10^{-4}$  & 1.0$\times10^{-2}$ \\ \hline
		2         & 6.0$\times10^{-7}$ & - & 2.6$\times10^{-4}$  & - \\ \hline 
	\end{tabular}
\end{table}

The two aforementioned models appear to fit suitably the entire region of transient currents when the transistor is turned ON by the pulse voltage. The trapping time constants indicate the occurrence of multiple trapping processes for different pulse width conditions. The faster trapping process may be attributed to the tunneling of electrons at the MoS$_2$/SiO$_2$ interface sub-gap states, whereas the relatively longer trapping process may arise due to the jumping of channel electrons into deeper defect states of SiO$_2$.\cite{Park2018Mar, Urban2019Sep} However, by decreasing the pulse width to 2 ms, the longer trapping process is eliminated due to insufficient time for the tunneling of electrons into deeper states. Our findings on NaCl assisted CVD grown monolayer MoS$_2$ FET not only unfold the fundamentals of the charge trapping mechanism but also offer a meticulous approach towards intrinsic, hysteresis-free transistor performances by using pulse I$\sim$V and fast transient measurement techniques for practical device applications.

\section*{Conclusions}
In summary, a progressive study on the time-domain drain current responses is demonstrated to minimize the fast transient charge trapping effects at MoS$_2$/SiO$_2$ interface in a NaCl-aided CVD-grown monolayer MoS$_2$ FET. The role of Ag as a contact metal for drain/source electrodes is critically analyzed by using a comprehensive DFT study to explore the underlying interfacial properties. The Schottky barrier at the Ag/MoS$_2$ interface is calculated to be 153 meV from the projected band profile of the heterostructure. The room temperature hysteresis analysis using DC and pulse I$\sim$V techniques is performed in both ambient and vacuum conditions to realize different trapping mechanisms. An increase in the field-effect mobility by 100 $\%$ is perceived along with hysteresis-free, intrinsic transfer characteristics in the case of short pulse amplitude sweep measurement than the corresponding DC measurement. Finally, single-pulse transient measurements are carried out to get insight into the fast and slow trapping phenomena at the interface between MoS$_2$ and gate-dielectric. Our experimental results are helpful in achieving the fundamentals of trapping mechanisms to obtain the enhanced trap-free channel mobility value in the case of monolayer MoS$_2$. This proposed study indicates the importance of controlling the interfacial charge traps for further advancement of the device performances, not only in monolayer MoS$_2$ but also in other two-dimensional materials and their heterostructures.
\section{2D transfer techniques}
Chemical vapor deposition (CVD) is one of the most explored synthesis routes to produce large-scale, and high-quality monolayer 2D semiconductors such as MoS$_2$. The growth of such crystals at high temperatures ($>$ 650 $^{\circ}$C) often modifies the clean interface of the growth substrate due to charge/ion doping which could hinder their opto-electronic response for transistor applications.\cite{Cai2018Jul} For example, recently, we have reported Na$^+$ ion doping in the gate dielectric via a salt-assisted CVD method that resulted in large hysteresis in the transfer characteristics of monolayer MoS$_2$ and WS$_2$-based field-effect transistors (FETs).\cite{Mallik2023Aug, Mallik2023Sep} Though controlled interfacial engineering in such cases has their own advantages in multi-bit memory and neuromorphic computing applications, these CVD-grown samples could be further explored to realize intrinsic transistor properties and flexible electronic applications by transferring them to application-specific target substrates.\cite{Kim2023Feb}

So far, considerable efforts have been made to transfer atomically thin 2D materials from their growth substrates to preferred target substrates. These approaches include recently developed techniques such as etchant-based poly(methyl methacrylate) (PMMA)-assisted wet transfer method, water-soluble layer transfer method, polymer-free transfer method, metal-assisted transfer method, poly(dimethylsiloxane) (PDMS)-based dry transfer method, and more.\cite{Watson2021May, Schranghamer2021Oct} The conventional chemical etchants such as NaOH, and KOH degrade the film quality and hence significantly affect the performance of the devices. Recently, Zhang et. al. demonstrated a capillary force-driven transfer method with several different TMDCs/substrate combinations that are free from etchant-induced damage.\cite{Zhang2017Dec} However, lack of mechanical support resulted in unwanted bubbles and wrinkles creating unfavorable conditions to achieve desired electrical characteristics. In another approach, a double support layer of PMMA/rosin was utilized for facile transfer of large area 2D materials to fabricate high-performance electronic devices.\cite{Zhang2019May} In another study, Desai et. al. demonstrated an improved exfoliation technique to obtain selective monolayer MoS$_2$ by using a thin layer of gold on bulk MoS$_2$.\cite{Desai2016Jun} Additionally, dry deterministic transfer methods using viscoelastic stamping such as PDMS layer has recently been explored to fabricate van der Waals (vdW) heterostructure devices.\cite{Frisenda2018Jan, Iwasaki2020Feb} 

Transfer methods involving wet-chemical etching may susceptible to process-related damages, trapped bubbles at interfaces, residual polymers, wrinkles, and undesired doping.\cite{Watson2021May} Presence of such imperfections can be revealed by the spectroscopic investigations and electrical measurements. For example, the impact of doping and strain in transferred 2D materials has been investigated by using Raman and photoluminescence (PL) spectroscopy.\cite{Mlack2017Feb, Hong2018Nov} In other studies, it was reported that wrinkles, cracks, and residual polymers in transferred films can affect the device conductivity and carrier mobility.\cite{Xu2014Apr, Liu2019Nov} The quality of the interface between semiconductor and metal electrodes, and charge injection mechanisms are crucial for nano-electronics and flexible devices. There are few recent reports on transfer of metal electrodes directly on 2D materials which could reduce interfacial roughness across the contact region and improve the carrier injection.\cite{Liu2022Sep, Liu2022May, Makita2020Mar} The state-of-art transfer of printed electrodes using graphene, hexagonal-BN, and polyvinyl alcohol (PVA) to the desired 2D films allow the formation of vdW contacts with low contact resistances.\cite{Liu2022Sep} It is important to note that the transfer techniques that have been discussed so far are limited to only 2D material or metal electrodes transfer. However, a method involving whole device (both channel materials and electrodes) transfer is still lacking. Furthermore, the efficient scaling of complementary metal-oxide-semiconductor (CMOS) compatible devices by stacking engineering and monolithic integration of 2D materials demand development of clean, damage-free, and cost-effective one-step transfer of whole device arrays. 
In this regard, a facile one-step PMMA-assisted etchant-free technique is demonstrated to transfer monolayer MoS$_2$ with metal (Ag) contacts from one substrate to preferred target substrates. The delamination of device arrays from their growth substrates is facilitated by several optimizations of spin-coating parameters, baking time, temperature, etc. Furthermore, Raman and PL studies reveal the release of strain and interfacial coupling effects before and after transfer. Our direct device transfer (DDT) method yields improvement in the electrical transport outcomes of the transferred devices. In particular, the transistor performances such as on/off ratios, carrier mobilities, Fermi-level pinning, nature of the hysteresis, and contact between Ag/MoS$_2$ are investigated. The temperature-dependent transport measurements are carried out in MoS$_2$ FETs to study the carrier dynamics and trapping mechanisms before and after the DDT method. Our transfer technique could be beneficial to efficiently transfer large array of 2D materials-based devices and sensors for flexible opto-electronic applications.
\section{Direct transfer of monolayer MoS$_2$ device arrays}
\begin{figure}
	\centering
	\includegraphics[width=0.9\textwidth]{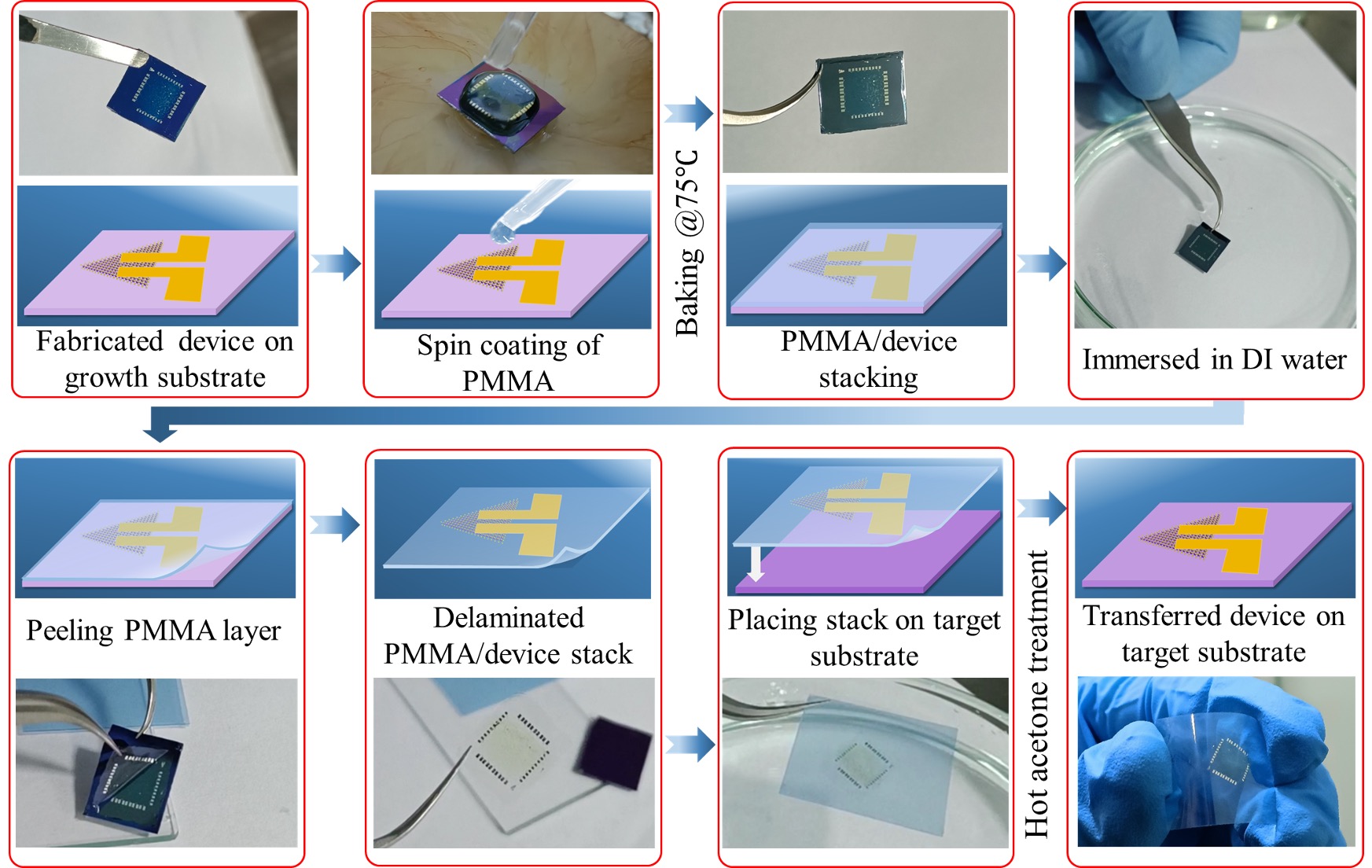}
	\caption{Realtime optical images and corresponding schematic representation of various steps during PMMA-assisted transfer of monolayer MoS$_2$-based device arrays from growth substrate (SiO$_2$/Si) to target substrate (flexible PET).}
	\label{Fig_3a}
\end{figure}

The monolayer MoS$_2$ isolated triangular domains are obtained using a NaCl-assisted CVD synthesis technique. The number of layers and crystalline qualities are investigated by using Raman and PL spectroscopic studies.\cite{Mallik2022Sep} Two terminal geometry-based device arrays (as shown in the top-left of Fig.\ref{Fig_3a})  are fabricated on the CVD-grown monolayer MoS$_2$ using the photolithography process. Our proposed DDT method involving various key steps is shown with schematic representation in Fig.\ref{Fig_3a}. In the first step, a comparatively thick layer of PMMA is spin-coated on a 1$\times$1 cm$^2$ growth substrate containing MoS$_2$ devices with a speed of 500 rpm for 120 s. The PMMA-coated substrate is baked at 75 $^{\circ}$C for 300 s and then, the edges of the substrate are scratched by using a blade. The PMMA/Device stack is then immersed in DI water for 180 s. This allows the water penetration through the scratched region between the PMMA film and the growth substrate facilitating initial process of delamination. As shown in Fig.\ref{Fig_3a}, the PMMA/device stack is gently peeled out from one of the corners of the substrate. Extreme precautions are taken care during the peeling of stacked layers to get rid of any strain-induced damage. Nonetheless, PMMA serves as a robust supporting layer with good flexibility and adhesive contact to atomically thin monolayers as compared to other polymers.\cite{Watson2021May} To further realize the wide applicability of our transfer method, we find that printed metal electrodes deposited on a fresh substrate can easily be transferred. Also, we have successfully transferred monolayer MoS$_2$ grown by conventional CVD synthesis technique (without NaCl).
The digital camera image of the delaminated array of devices consisting of monolayer MoS$_2$ with PMMA adhesive layer is shown in Fig.\ref{Fig_3a}. Subsequently, the PMMA/device stack is transferred onto a clean flexible substrate. It is important to note that the PMMA-assisted transfer process generally induces a large amount of polymer residues on the 2D surfaces.41 In previous reports, these residues have been removed by various post-transfer methods such as annealing,42 plasma treatment,43 mechanical cleaning via conducting mode atomic force microscope,44 and other cleaning methods.45 However, we demonstrate a simple process to achieve near residue-free surfaces after PMMA-assisted DDT transfer by cleaning in hot acetone. The post-transferred substrate is then dipped in hot acetone (45 $^{\circ}$C) for 15 minutes to not only dissolve the sacrificial PMMA layer but also successfully remove the polymer residues over monolayer MoS$_2$ channels. Hot acetone is also beneficial to avoid any deposition of the moisture on the target substrate after successful transfer of the devices. An image of the transferred device to a flexible poly(ethylene terephthalate) (PET) substrate is shown in the bottom-right of Fig.\ref{Fig_3a}. It is worth mentioning that, to address the stability and reproducibility of our transfer method, we have transferred different set of whole device arrays by several times on SiO$_2$/Si$^{++}$, and flexible substrates. It is important to highlight that the electrodes and channel materials are found to be intact in all transfer attempts, except in one case where only one damaged metal electrode is noticed. The optical properties of the transferred films associated with strain and coupling are further examined by Raman and PL analysis before and after the DDT process. The electrical characterizations of as-fabricated and transferred devices on SiO$_2$/Si$^{++}$ substrate are performed using temperature-dependent transport measurements.

\section{Structural and optical characterization of transferred MoS$_2$}
\begin{figure}[hbt!]
	\centering
	\includegraphics[width=0.9\textwidth]{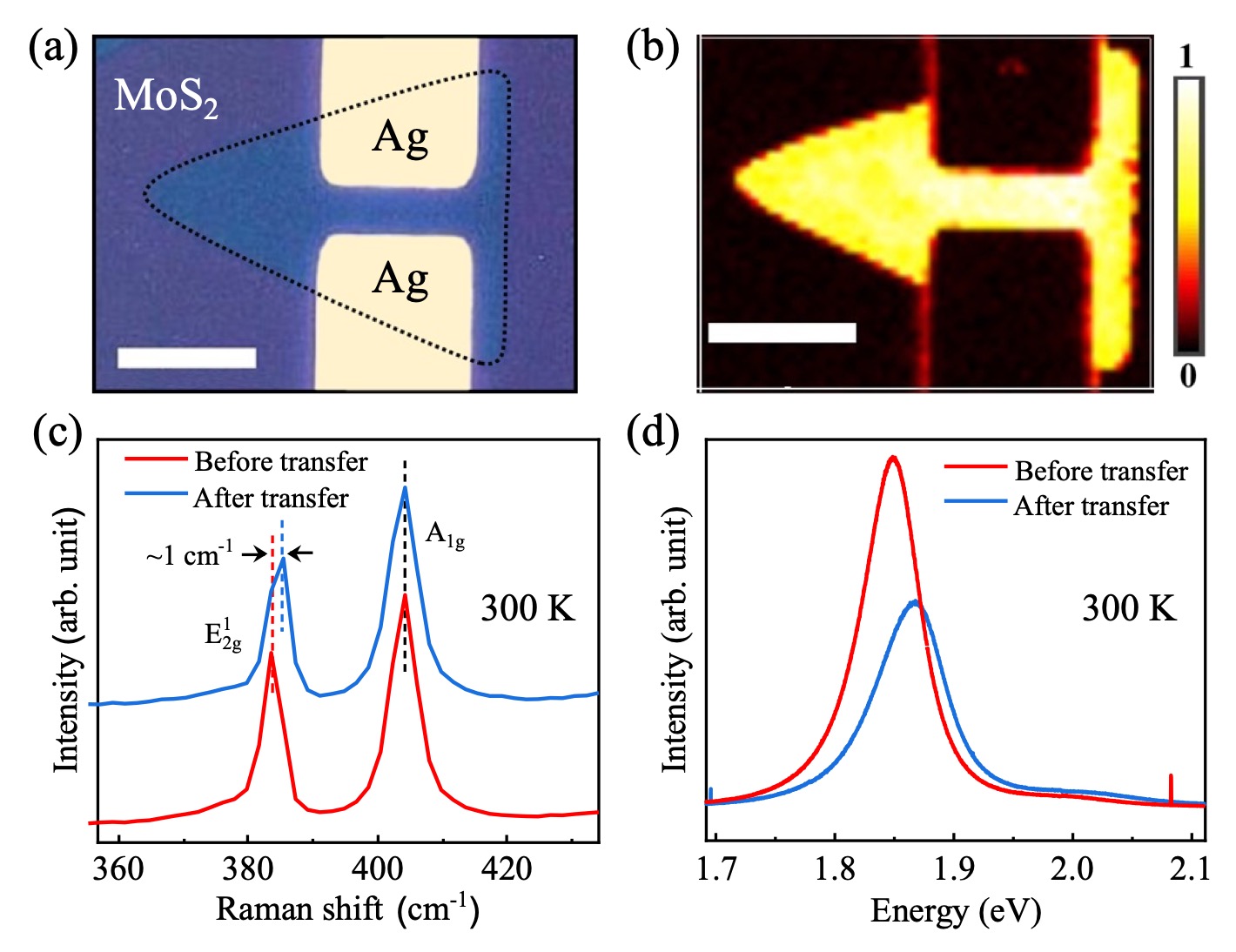}
	\caption{(a) Optical microscopy image of one of the transferred devices (Ag/MoS$_2$) on a target SiO$_2$/Si$^{++}$ substrate. (b) corresponding PL mapping. (c) Raman and (d) PL spectra collected from the MoS$_2$ channel before and after the device transfer. The scale bar is 10 $\mu$m.}
	\label{Fig_3b}
\end{figure}

The optical characterizations such as Raman and PL have been used to understand the crystalline quality, strain, doping, etc. in the transferred TMDCs.\cite{Rao2019May} Figure \ref{Fig_3b}(a) represents an optical image of a typical device consisting of monolayer MoS$_2$ channel with Ag electrodes, which is transferred to a pre-cleaned SiO$_2$/Si$^{++}$ substrate. A spatial mapping of spectrally integrated PL counts is initially performed on the transferred devices as shown in Fig.\ref{Fig_3b}(b). The bright and uniform luminescence from the monolayer MoS$_2$ indicates that the crystal quality of the MoS$_2$ is preserved even after the transfer process. In a previous article, Jain et. al.\cite{Jain2018May} demonstrated dark spots and lines in PL mapping of PDMS-based transferred MoS$_2$ flakes which suggests the presence of interfacial bubbles and wrinkles. In our case, absence of such detectable spots/lines in the PL mapping throughout the sample indicates a nearly clean and damage-free surface after hot-acetone treatment during the proposed DDT technique.
Furthermore, we employ micro-Raman spectroscopy (equipped with 532 nm laser excitation) to investigate whether the transferred device (the MoS$_2$ channel area) is associated with strain and doping. As shown in Fig.\ref{Fig_3b}(c), the two prominent Raman active modes i.e. $E_{2g}^1$ and, $A_{1g}$ are obtained in the high-frequency region that corresponds to the in-plane and out-of-plane vibrations and the frequency difference ($\Delta\omega$) between these two modes is the signature of the number of layers/thickness of MoS$_2$.\cite{Mallik2022Sep} By carefully analyzing the peak positions, it is found that the $E_{2g}^1$ mode of the transferred sample is blue shifted by 1 cm$^{-1}$, however no change or frequency shift is observed for $A_{1g}$ mode. This observation indicates strain relaxation ($E_{2g}^1$ is sensitive to strain) and the absence of any charged interfacial impurities ($A_{1g}$ is sensitive to doping) in monolayer MoS$_2$ during the transfer process. A similar shift in the Raman frequency has been reported in PDMS-based transfer technique.48 The origin of the strain in CVD grown MoS$_2$ could arise due to the lattice mismatch and presence of dangling bonds on the SiO$_2$ substrate during the high temperature synthesis.40 The growth-induced strain is then released after the transfer process causing a blue shift of $E_{2g}^1$ mode. We also compare the full-width half maxima (FWHM) of the observed Raman modes before and after transfer. The FWHM of $A_{1g}$ mode ($\sim$ 4.8 cm$^{-1}$) remains constant which indicates our transfer process doesn’t affect the crystalline quality of the channel materials. This finding is in consistent with the PL mapping results. Moreover, the FWHM of $E_{2g}^1$ mode is decreased from 3.5 to 2.7 cm$^{-1}$ which is also a signature of strain relaxation after transfer, as reported earlier.\cite{Mallik2022Sep} Furthermore, room temperature PL measurements are performed on the MoS$_2$ channel and the results are shown in Fig.\ref{Fig_3b}(d). It is observed that before the transfer, the PL spectra of MoS$_2$ are characterized by a strong excitonic transition at 1.84 eV (A exciton) and a relatively weak excitonic peak at 1.98 eV (B exciton), arising from the spin-orbit coupling induced valence band splitting. However, after transfer, a blue shift of 0.02 eV is observed for A excitons. The release of additional electron doping and strain from the growth substrate could cause the observed blue shift and the decrease in the intensity can be attributed to the quenching of charged excitons after the transfer. The results from Raman and PL spectra indicates our DDT technique could reduce the mechanical damage during the transfer process while preserving the crystal quality.             
\section{Transistor performance of transferred devices}
To investigate the electrical properties of the transferred device, we perform output characteristics, i.e. drain current (I$_{ds}$) vs drain-source voltage (V$_{ds}$) for gate voltages ranging from 0 to 40 V which are shown in Fig.\ref{Fig_3c}(a). The heavily doped $Si^{++}$ is treated as a back gate and the 300 nm SiO$_2$ serves as gate capacitance in our MoS$_2$ FET. The channel length and width of the devices are calculated to be 3 and 10 $\mu$m, respectively.
\subsection{Fermi-level pinning and current switching ratio}
\begin{figure}[hbt!]
	\centering
	\includegraphics[width=0.9\textwidth]{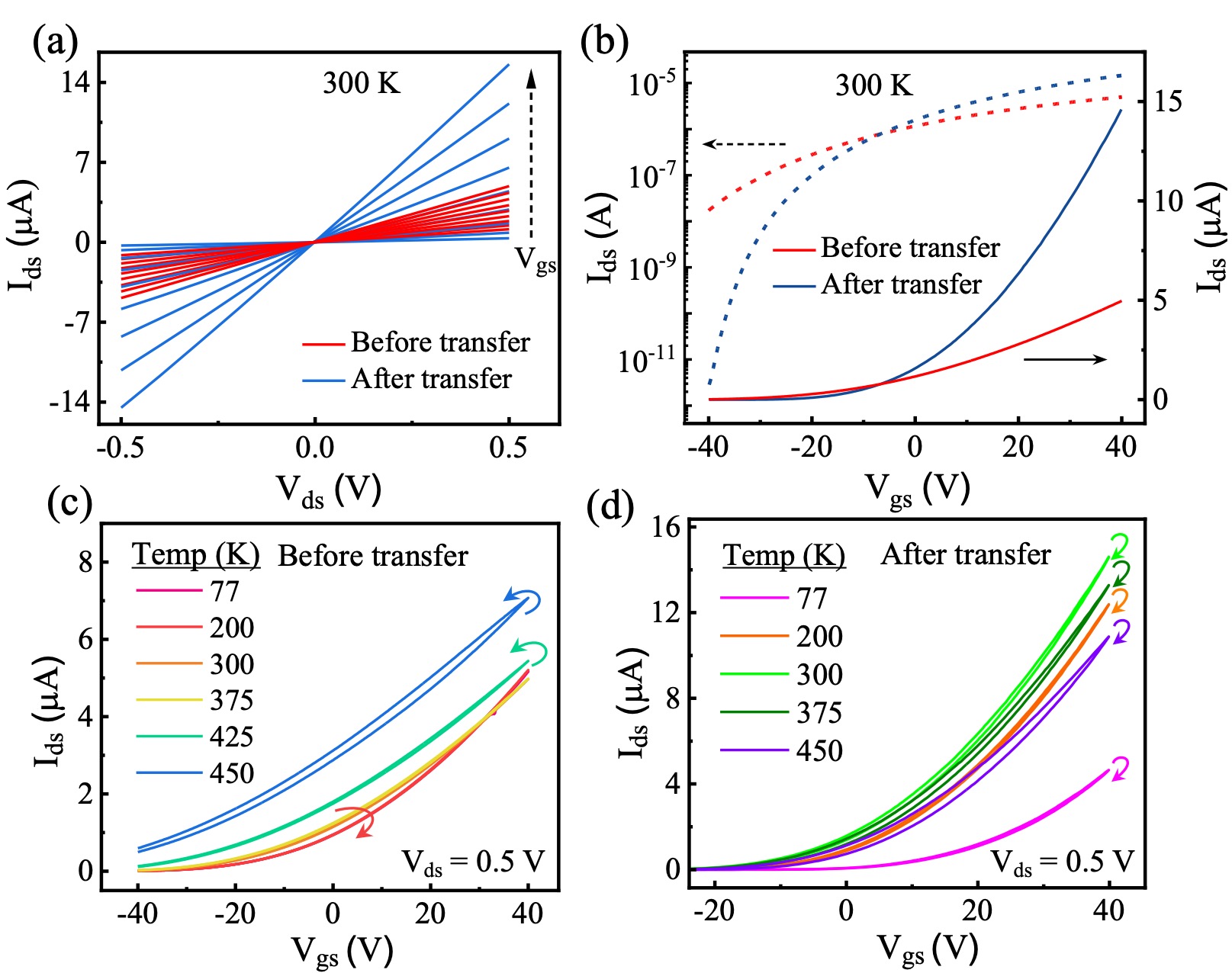}
	\caption{Room temperature (a) output and (b) forward sweep linear (left Y-axis) and semi-logarithmic (right Y-axis) transfer characteristics of a monolayer MoS$_2$ FET before and after the device transfer. The red and blue lines in (a) represent the output curves of before and after transferred device. The V$_{gs}$ is varied from 0 to 40 V with the step of 5 V. Temperature dependence (ranging from 77 to 450 K) of dual sweep transfer curves of the corresponding device (c) before and (d) after the DDT method. The curved arrows represent the clockwise or anti-clockwise nature of the hysteresis loop.}
	\label{Fig_3c}
\end{figure}

The linear output characteristics are observed indicating the Ohmic contact between Ag and MoS$_2$ for both before and after transfer cases. Such an Ohmic nature of MoS$_2$-based FETs using Ag electrodes has also been reported in previous articles.\cite{Mallik2021Apr} Interestingly, we notice an enhanced gate tuning of drain currents in transferred devices. As shown in Fig.\ref{Fig_3c}(a), the large field-effect modulation from the gate voltage variations in the transferred FET indicates the absence of Fermi-level pinning effects. The pinning effect can be seen for as-fabricated devices due to additional charge doping on the substrate during the CVD growth. The transfer characteristics (Ids vs gate-source voltage (V$_{gs}$)), shown in Fig.\ref{Fig_3c}(b) indicate similar drain current improvement after the transfer. With the gate voltage sweep from -40 to 40 V, and a fixed V$_{ds}$ at 0.5 V, we obtain a low transistor on/off ratio of about 10$^2$ (in case of before transfer) which increases to $>$ 10$^6$  (in case of after transfer). Recently, similar on/off ratios ($\sim$ 10$^6$) has been reported in MoS$_2$ back-gated FET array with transferred Ag contacts.\cite{Liu2022May} 

To understand the carrier dynamics and trapping of the transferred device, the dual sweep transfer measurements are performed at various temperatures ranging from 77 to 450 K. In general, hysteresis is often noticed in the transfer characteristics of CVD grown MoS$_2$-based FETs owing to their large surface-to-volume ratio and charge trapping/de-trapping at the channel/dielectric interfaces.\cite{Mallik2021Apr} As shown in Fig.\ref{Fig_3c}(c), before the transfer, we observe a narrow clockwise hysteresis and minimal variation of transfer curves with increasing temperature up to 400 K. However, above 400 K we notice a hysteresis inversion from clockwise to anti-clockwise direction and with increase in the hysteresis window. This inversion in the hysteresis path is also observed in our previous studies which could provide interesting applications in memory and neuromorphic computing.\cite{Mallik2023Aug, Mallik2023Sep} The hysteresis inversion arises due to additional doping of charge carriers at high temperatures and the detail of such studies can be found elsewhere.\cite{Mallik2023Aug} Interestingly, as shown in Fig.\ref{Fig_3c}(d), the transistor characteristics of the transferred device don’t show temperature-dependent hysteresis inversion indicating our transfer technique is capable of removing the additional charge doping effects arising from the growth substrate.  
\subsection{Schottky barrier height}
\begin{figure}
	\centering
	\includegraphics[width=0.9\textwidth]{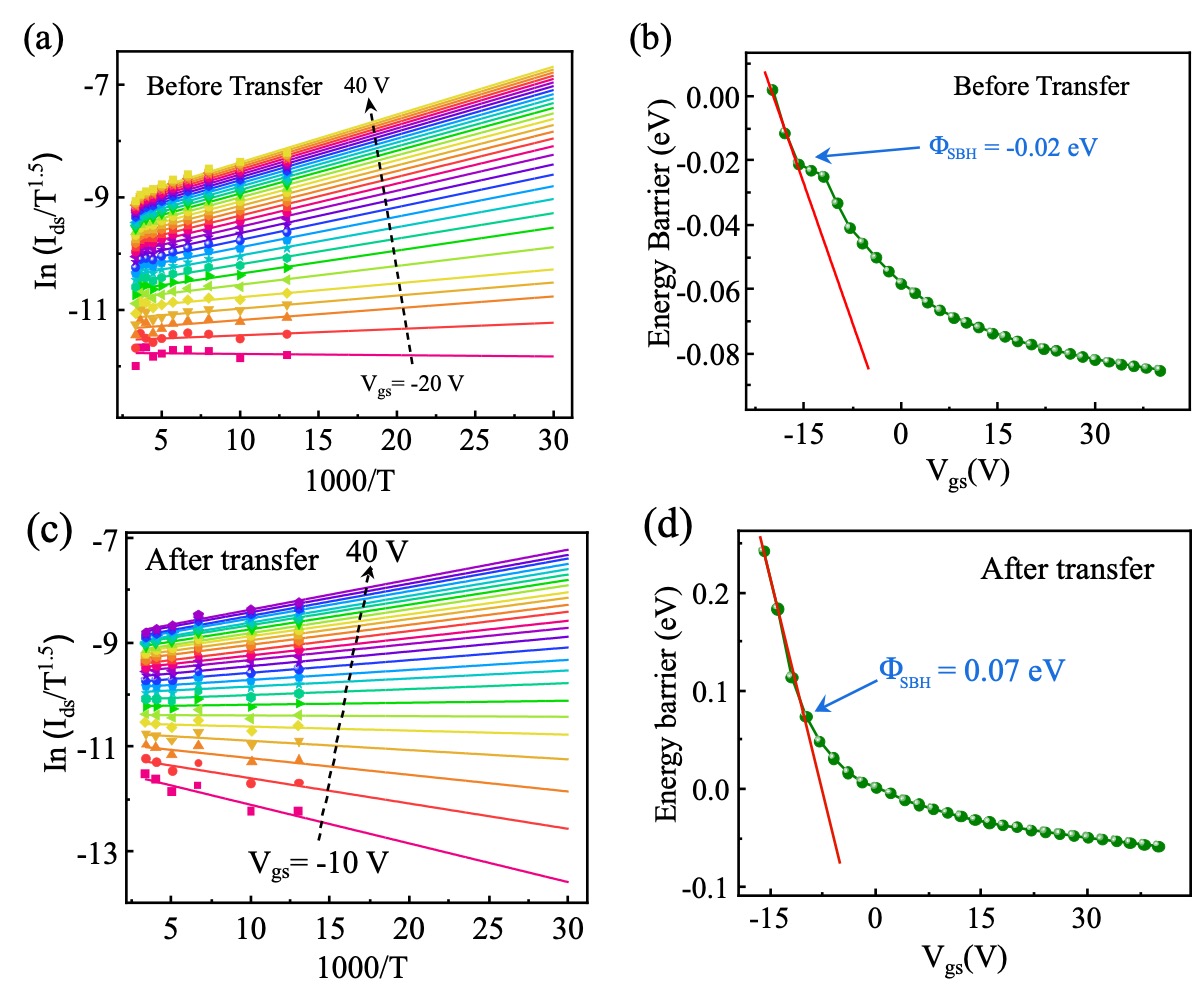}
	\caption{(a) Arrhenius plots for extraction of barrier heights (b) Barrier heights vs V$_{gs}$ and extraction of SBH from the linear fitting (red line) for Ag/MoS$_2$ contacts after device transfer. }
	\label{Fig_3d}
\end{figure}
So far, from the discussion of optical and electrical results, the proposed transfer method appears to be a convenient technique for whole-device transfer. In order to further evaluate the effectiveness of the transfer process some of the key device parameters need to be examined. As our transfer process involves the delamination of whole devices from the substrate, analysis of the metal/channel interface is crucial for smooth carrier injection from metal to semiconductor channel. The Schottky barrier height (SBH) between Ag and MoS$_2$ could address the influence of transfer on the carrier injection in the transferred devices. From the low-temperature transfer characteristics, the SBHs in both before and after transferred devices are calculated using the following 2D thermionic emission equation\cite{Xie2022Mar}

\begin{align}
    I = A_{\rm 2D}\ W \ T^{\frac{3}{2}} \ e^{\left(\frac{-q \Phi_B}{\rm K_B T}\right)} \ \left [1-e^{\left(\frac{-q V_{\rm ds}}{\rm K_B T}\right)}\right]
\label{eq:SBH_eq}
\end{align}
where, $A_{2D}$ is the Richardson’s constant for 2D systems51, $W$ is the channel width, $q$ is the elementary charge, $K_B$ is the Boltzmann’s constant, and $\Phi_B$ is the barrier height. Using this expression, we construct the Arrhenius plot as shown in Fig.\ref{Fig_3d}(a) and the barrier heights as a function of V$_{gs}$ are extracted from their slopes. The SBH corresponds to the barrier height at flat band condition that can be identified at the point where the extracted barrier height deviates from its linear dependence.\cite{Xie2022Mar} As shown in Fig.\ref{Fig_3d}(b), the extracted SBH of the transferred MoS$_2$ device is about 0.07 eV. These results are comparable to the SBH value of -0.02 eV in the case of before-transferred devices as shown Fig.\ref{Fig_3d}(a). The near-equal SBH values further supports robust and damage-free transfer of the devices.
\begin{figure}[hbt!]
	\centering
	\includegraphics[width=0.9\textwidth]{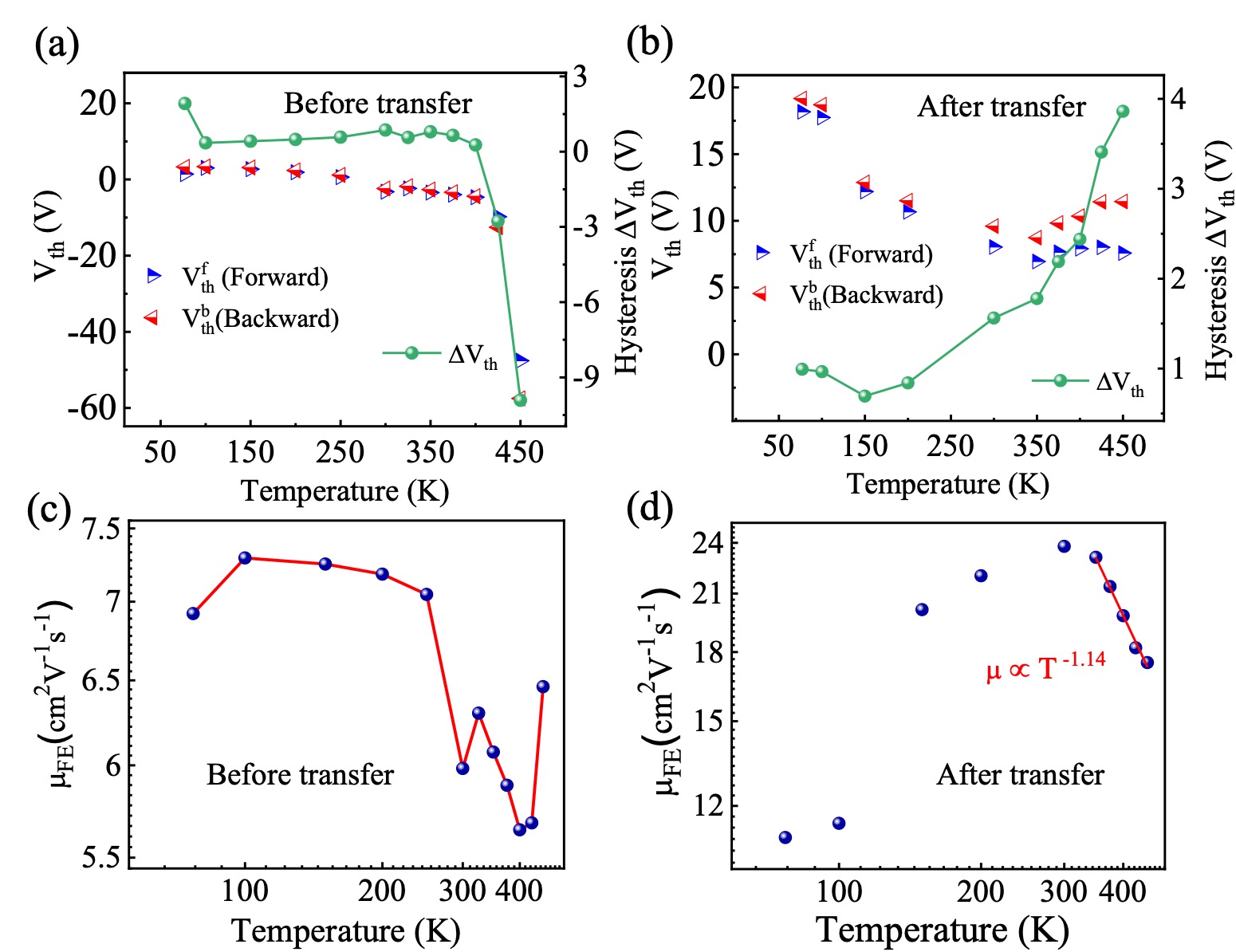}
	\caption{Temperature dependence of threshold voltage shifts, $V_{th}^f$, $V_{th}^b$, (left Y-axis) and hysteresis window, $\Delta V_{th}$ (right Y-axis) of (a) before and after (b) device transfer. Temperature dependence of Field-effect mobility ($\mu_FE$) variations (c) before and (d) after the transfer. A linear fitting (red line) of the decrease of mobilities is shown at high temperatures ranging from 350 to 450 K in (d). The solid lines in (a, b, and c) are guides to the eye.}
	\label{Fig_3e}
\end{figure}
\subsection{Threshold voltage variation with temperature}
The dependence of the threshold voltage shifts, evolution of hysteresis window ($\Delta V_{th}$), and calculated field-effect mobilities are plotted vs the temperature, as shown in Fig.\ref{Fig_3e}(a-d). The parameter $\Delta V_{th}$ is defined by the threshold voltage difference between forward ($V_{th}^f$) and backward ($V_{th}^b$) sweeps. Before transfer, as shown in Fig.\ref{Fig_3e}(a), the $\Delta V_{th}$ changes it’s sign with the temperature, which is due to the hysteresis inversion 400 K, as discussion earlier. Whereas, in case of the transferred device, the value of $\Delta V_{th}$ increases with temperature as shown in Fig.\ref{Fig_3e}(b). This could be due to the thermally activated deep trapping states with enhanced carrier trapping and de-trapping during forward and backward sweeps at high gate voltages. 
\subsection{Phonon limited carrier mobility}

We also compare the temperature dependence of the mobilities ($\mu_{FE}$) before and after the device transfer, which are calculated by using the equation \ref{eq:mobility_eq}. The extracted mobilities don’t follow a significant variation with temperature in case of as-fabricated device shown in Fig.\ref{Fig_3e}(c). This could be due to the low carrier mobility of the as-grown samples. However, in case of the transferred device, the mobility values are increased by three times and the temperature dependence of the mobility curve shows a sharp decrease above room temperature due to increased thermal screening of the phonons. Similar temperature dependence variation is also observed in the MoS$_2$ FET by Radisavljevic et. al.\cite{Radisavljevic2013Sep} As shown in Fig.\ref{Fig_3e}(d), the phonon limited carrier mobility is found to have $T^{-1.14}$ dependency above room temperature, which is comparable to previous report.\cite{Radisavljevic2013Sep} For the sake of completeness, an overview of various support layers used in transferring 2D materials and printed metal contacts with their structural and electrical implications along with our results are listed in Table \ref{table 3.2}.

\begin{table}
	\centering
	\caption{An overview of various support layers used in transferring 2D materials, and printed metal contacts with their structural and electrical implications.}
	\label{table 3.2}
	\renewcommand{\tabcolsep}{0.05cm}
	\renewcommand{\arraystretch}{0.4}
         \setlength{\leftmargini}{0.4cm}
         \scalebox{0.9}{
	\begin{tabular}{|m{2.5cm}|m{2.5cm}|m{5cm}|m{4.1cm}|m{2cm}|}
		\hline 
		\centering Support Layer & \centering Transferring Material &  \centering Structural \par Implications & Electrical Implications &  \hspace{8.0 cm} Ref. \\ \hline
		\centering PMMA (Etching-free) & 1L MoS$_2$+ Metal (Ag) electrodes & 
                \begin{itemize}
                    \item Residue free transfer 
                    \item Removal of strain and \par additional charge doping 
                \end{itemize} &
                \begin{itemize}
                     \item Increase in electron mobility transistor ON/OFF ratio
                     \item Hysteresis \par modulation
                \end{itemize}      
                & This work \\ \hline
                
		\centering PMMA (Etching-free) & \centering  MoS$_2$ &
        \begin{itemize}
		    \item Wrinkle and Crack free
		\end{itemize} & \begin{itemize}
		    \item High mobility 
            \item High ON/OFF ratio
		\end{itemize}  & \hspace{15.0 cm} \cite{Liu2019Nov} \\ \hline
  
		\centering PMMA/PVA (Cuetching) PMMA/PVA (KOH \par etching) & \centering Graphene MoS$_2$ & 
            \begin{itemize}
		    \item Presence of ionic/metal residue
		    \end{itemize} &
            \begin{itemize}
		    \item Wide Dirac voltage distribution
            \item Enhanced mobility
            \item Low subthreshold swing
		    \end{itemize}  & \hspace{15.0 cm} \cite{VanNgoc2016Sep} \\ \hline
  
        \centering PDMS & \centering Graphene MoS$_2$ & \begin{itemize}
            \item Adhesion instability
            \item Tensile strain
        \end{itemize} & \centering --- & \hspace{15.0 cm} \cite{Kim2021May} \\ \hline

        \centering TRT \par (Cu etching) & \centering MoS$_2$ & \begin{itemize}
            \item Residue and wrinkle-free
        \end{itemize} & \centering --- & \hspace{15.0 cm} \cite{Lin2015Dec} \\ \hline
        
        \centering Graphene & \centering Metal (Au, Ag, Cu, Pt, Ti, Ni) & 
        \begin{itemize}
            \item Formation of Vandel Waals contact
        \end{itemize} &
        \begin{itemize}
            \item Tuned Schottky \par contacts
            \item Batch production with uniform \par electrical \par characteristics
        \end{itemize} & \hspace{15.0 cm} \cite{Liu2022May} \\ \hline

        \centering PVA+PMMA \par (Etching-free) & \centering Metal (Au) &
        \begin{itemize}
            \item Damage free transfer 
        \end{itemize} & 
        \begin{itemize}
            \item High Mobility 
            \item Low subthreshold swing
            \item Low contact \par resistance
        \end{itemize} & \hspace{15.0 cm } \cite{Makita2020Mar} \\ \hline
	\end{tabular}}
\end{table}

\section{Applications in flexible electronics}
\begin{figure}
	\centering
	\includegraphics[width=0.6\textwidth]{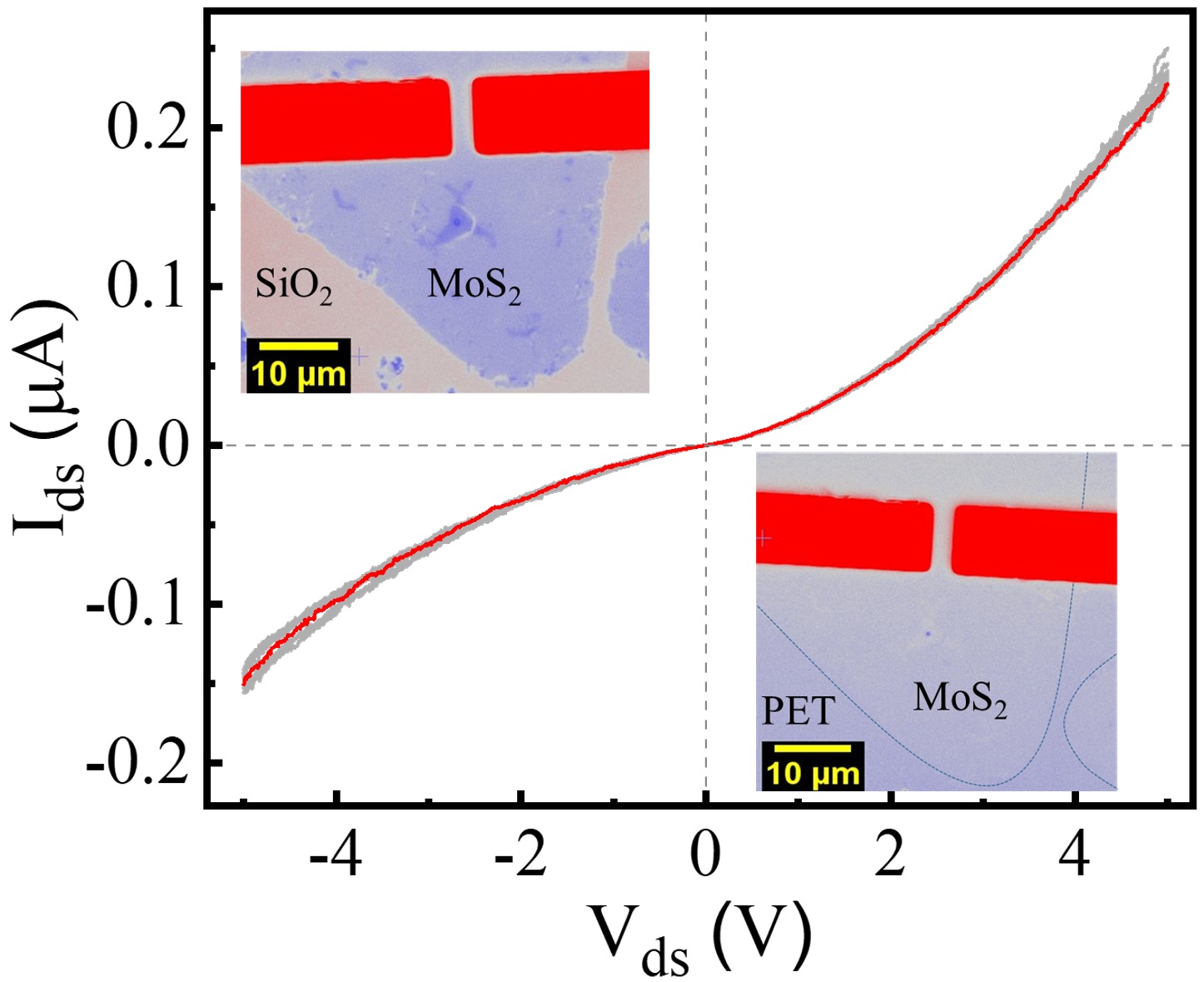}
	\caption{The repeatability of the device performances for 10 number of I$\sim$V cycles. The left and right inset indicates the optical image of a typical device (Ag electrode with monolayer MoS$_2$ as channel material) before and after DDT transfer, respectively.}
	\label{Fig_3f}
\end{figure}
In the present study, we want to emphasize that transferring whole devices (including metal electrodes and channel materials) provides unique opportunity to study and compare the same set of devices on various substrates. Our transfer method could be beneficial for direct integration of heterostructure based devices for flexible electronics.  Furthermore, we have carried out several I$\sim$V measurements on the devices transferred to the flexible substrate (PET). These results are discussed in Fig.\ref{Fig_3f}. As shown in Fig.\ref{Fig_3f}, the negligible fluctuations in the I$\sim$V results indicate stable electrical characteristics of the monolayer MoS$_2$ on flexible substrate. Our DDT method can be utilized to directly transfer the array of devices of 2D materials and heterostructures skipping various cumbersome steps in between. Furthermore, The proposed technique could be beneficial towards efficient FEOL and BEOL integrations of 2D materials with existing CMOS technology and flexible nanoelectronics.
\section{Chapter summary}
In summary, a progressive study on the time-domain drain current responses is demonstrated to minimize the fast transient charge trapping effects at MoS$_2$/SiO$_2$ interface in a NaCl-aided CVD-grown monolayer MoS$_2$ FET. The role of Ag as a contact metal for drain/source electrodes is critically analyzed by using a comprehensive DFT study to explore the underlying interfacial properties. The Schottky barrier at the Ag/MoS$_2$ interface is calculated to be 153 meV from the projected band profile of the heterostructure. The room temperature hysteresis analysis using DC and pulse I$\sim$V techniques is performed in both ambient and vacuum conditions to realize different trapping mechanisms. An increase in the field-effect mobility by 100 $\%$ is perceived along with hysteresis-free, intrinsic transfer characteristics in the case of short pulse amplitude sweep measurement than the corresponding DC measurement. Finally, single-pulse transient measurements are carried out to get insight into the fast and slow trapping phenomena at the interface between MoS$_2$ and gate-dielectric. Our experimental results are helpful in achieving the fundamentals of trapping mechanisms to obtain the enhanced trap-free channel mobility value in the case of monolayer MoS$_2$. This proposed study indicates the importance of controlling the interfacial charge traps for further advancement of the device performances, not only in monolayer MoS$_2$ but also in other two-dimensional materials and their heterostructures. 

We have also demonstrated one-step PMMA-assisted etchant-free technique to directly transfer the device arrays consisting of monolayer MoS$_2$ as channel materials and Ag as source/drain electrodes on arbitrary substrates (i.e. SiO$_2$/Si$^{++}$, flexible). The Raman and PL spectroscopic investigations reveal the release of strain and removal of additional electron doping originating from the growth substate. Furthermore, the room temperature electrical characterizations of the transferred MoS$_2$ devices on SiO$_2$/Si$^{++}$ substrate show a reduced Fermi level pinning effect, mobility improvement, and increase in the transistor on/off ratio by a factor of 10$^4$. We also find that our transfer technique could completely arrest the temperature-dependent hysteresis inversion in the transferred devices by removing the effect of additional charge doping arising from the growth substrate during high-temperature CVD synthesis. Our work on the proposed transfer technique provides a facile approach toward efficient stacking and complex engineering of 2D materials based future flexible nanodevices.

\blankpage 
\chapter{Thermally-driven multi-level non-volatile memory in monolayer MoS$_2$ mem-transistors}
\label{C4} 

\section{Introduction}
In this modern era of electronic devices, ultrahigh data storage density is imperative for smarter computing platforms and vigorous information processing applications. Memtransistors (integration of transistors with memristive functionalities) based on novel materials have been
proposed with various architectures showing controlled transport with digital switching ratios,\cite{Lee2020Nov, Farronato2022Aug} multibit optoelectronic memory,\cite{Lee2016Nov} neuromorphic computing applications, etc.\cite{Sahu2023Jan, Wang2019Jun} Especially, the electrolyte-gated transistors show better conductance modulation benefited from its ion-gating mechanism over electrostatic charge trap phenomena, making them viable candidates for brain-inspired computation and logic-in-memory applications.\cite{Jiang2017Aug, Yao2021Jun} Despite the recent development of synaptic ion-gated transistors, multilevel memory
based on ion-gating mechanisms is still lacking. However, liquid electrolytes practically limit the large-density integration of devices and high-temperature applications. Consequently, it is desirable to instigate new multilevel memories with eccentric functionalities that confer the possibility of robust device architecture with higher throughput. Furthermore, with the growing packing density of FET arrays on a single wafer, high-performance integrated circuits can reach an operating temperature as high as 500 K,\cite{Schroder2003Jul} making it essential to understand and exploit novel properties of 2D material-based devices at high temperatures. Thermally driven memory is one of the applications where thermal energy can aid memory functions with multilevel storage capabilities.

In this chapter, we demonstrate the multifunctionality of monolayer MoS$_2$ mem-transistors which can be used as a high-geared intrinsic transistor at room temperature; however, at a high temperature ($>$ 350 K), they exhibit synaptic multi-level memory operations. The temperature-dependent memory mechanism is governed by interfacial physics, which solely depends on the gate field modulated ion dynamics and charge transfer at the MoS$_2$/dielectric interface. We have proposed a non-volatile memory application using a single FET device where thermal energy can be ventured to aid the memory functions with multi-level (3-bit) storage capabilities. Furthermore, our devices exhibit linear and symmetry in conductance weight updates when subjected to electrical potentiation and depression. This feature has enabled us to attain a high classification accuracy while training and testing the Modified National Institute of Standards and Technology datasets through artificial neural network simulation. This work paves the way towards reliable data processing and storage using 2D semiconductors with high-packing density arrays for brain-inspired artificial learning. 
\section{Fabrication of monolayer MoS$_2$ mem-transistors}
\begin{figure}
	\centering
	\includegraphics[width=0.9\textwidth]{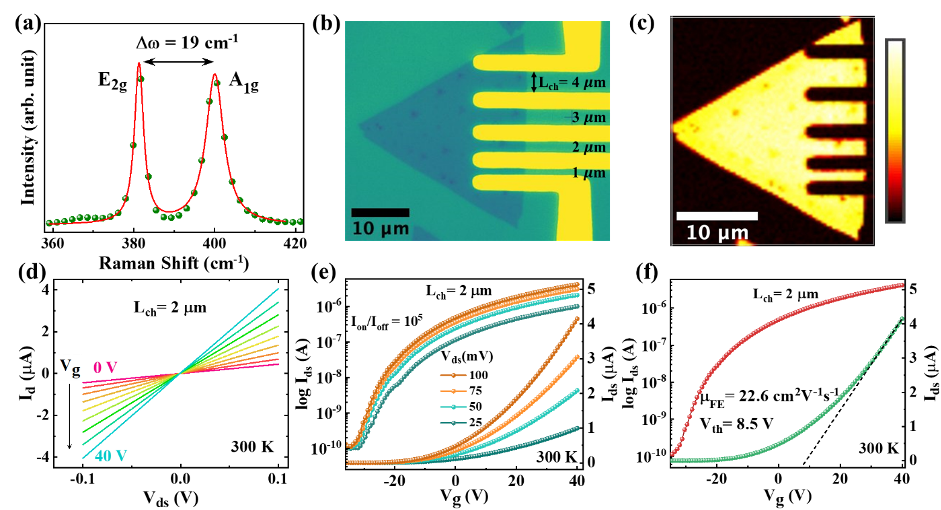}
	\caption{Structural characterization, design, and electrical properties of MoS$_2$ transistor (a) The Raman spectroscopic signature of ML MoS$_2$ show E$_{2g}$ and A$_{1g}$ modes. The Lorentz fit results (solid red lines) provide the high crystallinity of CVD grown ML MoS$_2$ sample. (b) Optical image of the MoS$_2$ device having equal channel widths (10 $\mu$m) and length varies from 1 to 4 $\mu$m, Ag is being used as source/drain electrodes (shown in false color). (c) Photoluminescence mapping confirms the uniform optical grade quality of the MoS$_2$ sample with robust device fabrication. (d) The output characteristics (I$_d$ vs V$_{ds}$) for fixed gate voltages (V$_g$ ranging from 0 to 40 V) show the Ohmic contact nature of the as fabricated device. (e) The transfer characteristics (I$_d$ vs V$_{g}$) is displayed in both linear and semi-logarithmic scale for fixed drain-source voltages (25 to 100 mV) showing stable electrostatic gate control and the current on/off ratio $\sim 10^5$. (f) The field-effect mobility ($\mu_{FE}$) and threshold voltage (V$_{th}$) is extracted from the slope of the linear region of the transfer curve.}
	\label{Fig_4.1}
\end{figure}

Figure \ref{Fig_4.1}(a) displays the prominent Raman active modes i.e. E$_{2g}$ and A$_{1g}$ of as-grown MoS$_2$. The frequency difference ($\Delta\omega$ = 19 cm$^{-1}$) between these phonon modes are characteristic signature of ML nature of our sample. The Lorentz fit results provide the FWHM of E$_{2g}$  ($\sim$ 2.5 cm$^{-1}$) and A$_{1g}$ ($\sim$ 5.2 cm$^{-1}$) modes comparable to that of exfoliated MoS$_2$ which depict high crystallinity of our CVD grown samples. Figure \ref{Fig_4.1}(b) displays the optical micrograph of as-fabricated MoS$_2$ devices with silver (Ag) as drain/source electrodes. The near equal work function of Ag and electron affinity of ML MoS$_2$ allows the formation of a low Schottky barrier with negligible strain effect and superior carrier transport when Ag is used as metal contact to ML MoS$_2$. The channel lengths (L$_{ch}$) of proposed devices are varied from 1 to 4 $\mu$m with channel width fixed at 10 $\mu$m. Photoluminescence (PL) mapping of the MoS$_2$ device shown in Fig.\ref{Fig_4.1}(c) provides intense and uniform luminescence, which illustrates exemplary optical properties of the ML MoS$_2$ throughout the channel regions.

Initially, the room temperature basic transistor properties are obtained for L$_{ch}$ = 2 $\mu$m in a high vacuum environment ($10^{-6}$ mbar) under dark conditions. Figure \ref{Fig_4.1}(d) demonstrates the linear output characteristics (i.e., drain current I$_d$ vs. drain-source voltage V$_{ds}$) for the V$_{ds}$ sweep range from -100 to 100 mV at fixed back gate voltages (V$_g$ = 0 to 40 V in steps of 5 V). Linear $I_d\sim V_{ds}$ curves and I$_d$ values in few $\mu$A ensure the Ohmic contact has been formed between Ag and MoS$_2$,\cite{Batool2023Mar} which drives excellent carrier transport with low power consumption as per the previous literature.\cite{Mallik2021Apr} The stable electrostatic gate control with n-type transistor behavior can be observed in Fig.\ref{Fig_4.1}(e) during the single sweep transfer characteristics (I$_d$ vs. V$_g$) for V$_g$ sweep range from -40 to 40 V with fixed V$_{ds}$ values (25, 50, 75, and 100 mV). The semi-logarithmic plot shows a high on/off current ratio (I$_{on}$/I$_{off}$) of about 10$^5$ providing excellent transistor switching operations. The field effect mobility ($\mu_FE$) and threshold voltage (V$_{th}$) is extracted from the slope of the linear region of the transfer curve for V$_{ds}$ = 100 mV, which is shown in Fig.\ref{Fig_4.1}(f). In our case, such high carrier mobility of $\sim$22 $cm^2V^{-1}s^{-1}$ is free from grain boundary and strain effects and therefore meets excellent agreement with previously reported results on unpassivated ML MoS$_2$.\cite{Nasr2019Jan} The as-prepared back-gated MoS$_2$ devices with Ag as metal electrodes show superior transistor performance at room temperature and are further investigated for temperature-dependent behavior.
\section{Clockwise and anticlockwise hysteresis}
The temperature-dependent dual sweep transfer characteristics are conducted to perceive the evolution of hysteresis in as-fabricated devices under high vacuum ($\sim$ 10$^{-6}$ mbar). The hysteresis behavior over the temperature range from 100 to 450 K with gate voltage forward sweep from -60 to 40 V and backward sweep from 40 to -60 V for different channel lengths are obtained. The indiscernible dual sweep curves from 100 to 400 K stipulate weak hysteretic behavior manifesting lessened gate stress effects,\cite{Cho2013Sep} unvaried threshold voltages during gate operations, and hence elicit excellent transistor performance. Nevertheless, noticeable hysteresis is observed above 400 K, which inflates further at elevated temperatures. According to previous reports, the origin of hysteresis in MoS$_2$ based FETs can have various source factors at room temperature, such as adsorption of oxygen/water molecules onto MoS$_2$ channel surface at high bias in ambient pressure,\cite{Late2012Jun} charge trapping/detrapping at MoS$_2$-dielectric interface,\cite{Guo2015Mar} etc. However, at high temperatures, the charge trap density increases due to the thermal activation of deep trap states in the dielectric, allowing a faster trapping process which essentially causes outspread hysteresis. A recent report by He et. al.\cite{He2016Oct} explains that a charge injection from the back gate to dielectric that drives unique hysteretic crossover characteristics at relatively higher temperatures. Kaushik et. al.\cite{Kaushik2017Oct} describe independent mechanisms occurring at room and high temperatures in the case of FETs based on multilayered exfoliated MoS$_2$, that switches the hysteresis loop from clockwise to anticlockwise.
\begin{figure} [hbt!]
	\centering
	\includegraphics[width=0.9\textwidth]{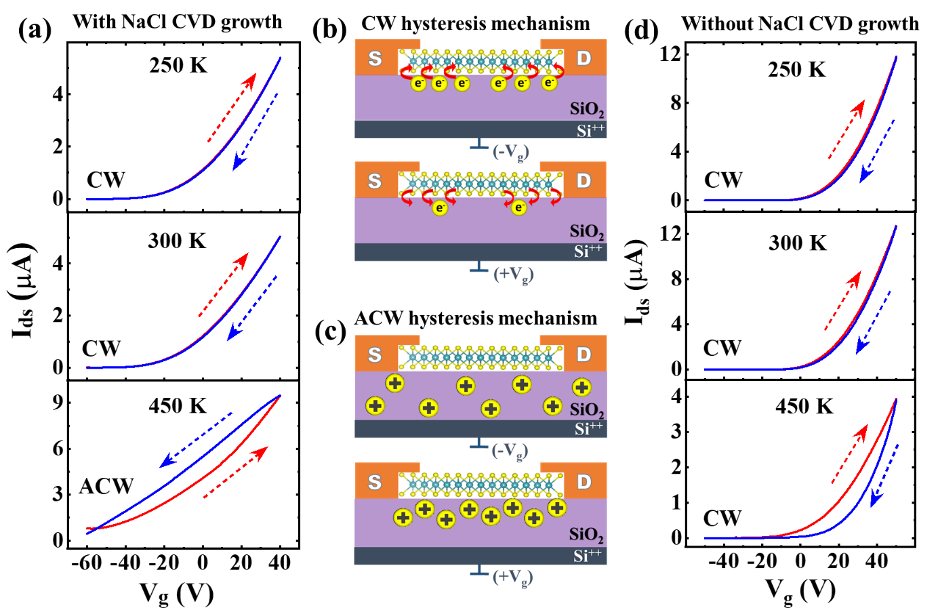}
	\caption{Temperature-dependent hysteresis evolution and proposed mechanism. (a) Dual sweep transfer curves of the MoS$_2$ device (L$_{ch}$ = 2 $\mu$m) are plotted against temperatures at 250 K, 300 K, and 450 K. The gate voltage sweep range is taken from -60 to 40 V and again back to -60 V with a sweep rate of 0.5 V/s; V$_{ds}$ is fixed at 100 mV. The forward (backward) sweep is represented by solid red (blue) lines indicating a clockwise hysteresis loop marked by arrows. Cross-sectional schematic representations of the device portray the possible interfacial mechanism that involves (b) charge-(de)trapping processes in clockwise hysteresis at high negative (V$_g$ $<$ 0) and positive gate voltages (V$_g$ $>$ 0). (c) the movement of mobile ions in gate oxide under V$_g$ $<$ 0 and V$_g$ $>$ 0. (d) Dual sweep hysteresis behaviors on CVD grown MoS$_2$ synthesized using conventional techniques (without NaCl). The transfer curves show the emergence of clockwise hysteresis with respect to temperature. As shown, the degree of hysteresis increases maintaining the same direction (clockwise) as we increase the temperature from 250 to 450 K.}
	\label{Fig_4.2}
\end{figure}

The room temperature transfer characteristics in our case show robust transistor performance due to its minimal hysteresis behavior. A transfer curve at temperature 250 K and 300 K for L$_{ch}$ = 2 $\mu$m  is shown in Fig.\ref{Fig_4.2}(a), along with possible mechanisms illustrated in Fig.\ref{Fig_4.2}(b) that justify the nature of the hysteresis loop. Low backward sweep current (BSC) indicates clockwise hysteresis in a dual sweep transfer curve, whereas high BSC reflects anti-clockwise hysteretic nature. The red (blue) curve signifies a forward (backward) sweep to realize the clockwise hysteresis nature at low and room temperature having hysteresis width less than 1 V. As shown in Fig.\ref{Fig_4.2}(b), two cross-sectional schematic representations of the device portray the possible interfacial mechanism that involves charge-trapping processes at high negative and positive gate voltages. As illustrated in the schematic, at the beginning of the forward sweep, when a particular negative bias is applied to the gate terminal, an initial de-trapping process in the SiO$_2$ releases additional electrons to the MoS$_2$ channel, which causes a left shift of the transfer characteristics. As we reach a high positive gate bias, the capturing of electrons at the trapping sites of SiO$_2$ dominates. As a consequence, during the backward sweep, the channel of MoS$_2$ contains fewer electrons, which leads to low BSC.\cite{Mallik2021Apr} 

The lower panel of Fig.\ref{Fig_4.2}(a) shows the transfer characteristics at a high temperature ($\sim$450 K), depicting a high BSC and hence anti-clockwise hysteresis behavior with a large voltage window. The earlier results on the anti-clockwise hysteresis in the case of MoS$_2$ FETs are reported at high temperature where a completely different form of hysteresis has been observed, associated with a sudden current jump near zero gate bias. He et. al.\cite{He2016Oct} proposed a distinct mechanism that entails the charge injection from the gate (n-type Si) to the thermally activated deep trap states in SiO$_2$, creating a repulsive and attractive gate field depending on the bias polarization. However, the nature of the anti-clockwise hysteresis in our case is distinctly different from that of previous reports and requires an independent mechanism that can corroborate our forthcoming results. The fundamental idea that propels the origin of anti-clockwise hysteresis in our case is as follows; in NaCl-assisted CVD growth of MoS$_2$, the minute amount of mobile charges such as sodium ions (Na$^+$) exist in the bulk of SiO$_2$ gate dielectric is inevitable during high temperature synthesis process. It is apparent that the mobile ions in gate oxide can be a plausible factor responsible for threshold voltage shift and could induce hysteretic transfer characteristics in FETs.\cite{Egginger2009Jul} As shown in the schematic of Fig.\ref{Fig_4.2}(c), under the application of high negative bias at the gate terminal, the Na$^+$ ions move away from the channel-interface region, and a random distribution of ions can be realized in the gate dielectric during the forward sweep. However, under high positive bias, these ions drift towards the MoS$_2$ channel, which eventually increases the drain current values during the backward sweep.

We reproduce the above results on evolution of hysteresis by fabricating devices with different channel lengths (L$_{ch}$ = 1, 3, and, 4 $\mu$m). Even with different channel lengths the repeatability of our results is highly reproducible over the temperature range from 100 to 450 K indicating the high-temperature anti-clockwise hysteresis behavior is not limited to a confined channel length. To support the Na$^+$ ion-driven mechanism, it is extremely important to compare these results with conventional (without NaCl) CVD grown MoS$_2$ based transistors. For this, we have conducted the dual sweep transfer characteristics on such CVD grown pristine MoS$_2$ samples to examine the role of Na$^+$ ions on device performances. It is worth mentioning that the hysteresis behavior of without NaCl CVD grown MoS$_2$ based transistors show a consistent clockwise behavior throughout the temperature range from 250 to 450 K, as shown in Fig.\ref{Fig_4.2}(d). This behavior is attributed to the increased thermally activated trap centers at MoS$_2$/SiO$_2$ interface which trap/de-trap charge carriers depending on the voltage polarization. This comparison confirms our proposed mechanism of the Na$^+$ ion-driven high-temperature anti-clockwise hysteresis in MoS$_2$ transistors. In the subsequent sections, we shall expound the mechanism more lucidly and observe how the ions migrate inside SiO$_2$, generating a diffusion-like drain current behavior in the conduction channel with temperature.
\section{Transient current analysis}
\begin{figure}
	\centering
	\includegraphics[width=1\textwidth]{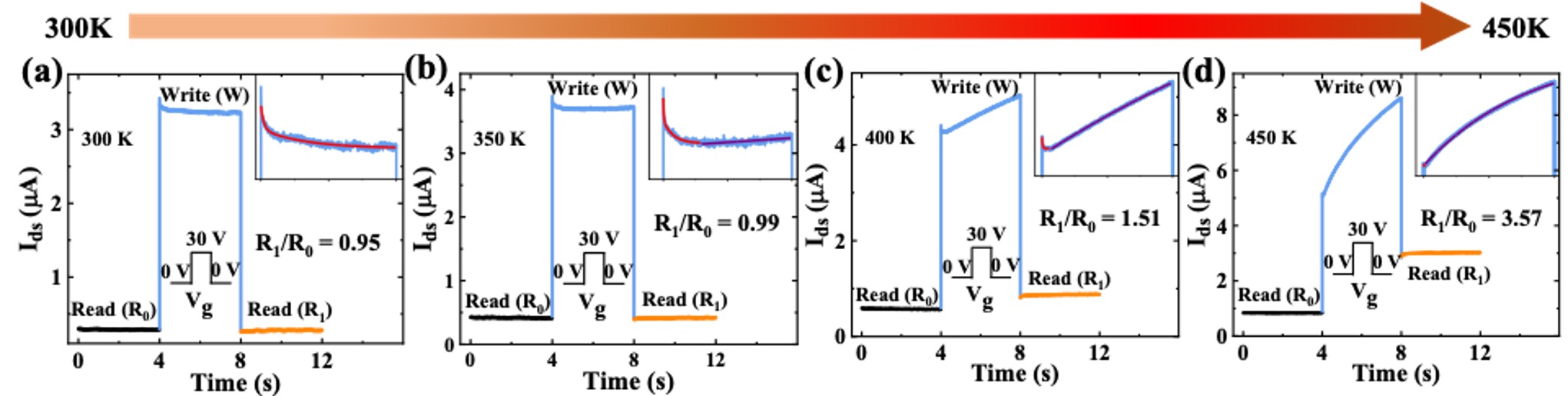}
	\caption{Transient drain current analysis and proposed energy band diagrams (a-d) represent the waveform nature of the drain current by employing a single pulse gate voltage at four different temperature regions starting from 300 to 450K. The WRITE and READ operations and also indicated. The drain-source voltage is kept fixed at 300 mV. The insets of each figure represent the magnified view of the WRITE process and its progression with temperature. The fittings to the fleeting drain currents during the WRITE operations are represented by red and magenta colors for the two trap models and the current monitoring model, respectively. The READ ratios ($R_1/R_0$) are calculated for each temperature case.}
	\label{Fig_4.3}
\end{figure}

The transient drain current analysis provides an elementary route to understand subsequent intermediate mechanisms at different time scales that alter with temperature under positive bias at the gate terminal. Figure \ref{Fig_4.3}(a-d) represents the waveform nature of the drain current by employing a single pulse gate voltage at four different temperature regions starting from 300 to 450 K. Each pulse waveform has a period of 12 s which consists of three components of equal intervals; an initial READ (R$_0$) at V$_g$ = 0 V, WRITE (W) at V$_g$ = 30 V followed by a final READ (R$_1$). A constant V$_{ds}$ of 300 mV is used for all the above operations. The insets of each figure represent the magnified view of the WRITE process and its progression with temperature. At 300 K, READ provides initial drain current values, which increase suddenly as the pulse amplitude increases from 0 to 30 V due to the accumulation of charge carriers into the MoS$_2$ channel, as shown in Fig.\ref{Fig_4.3}(a). However, during the WRITE process, while the voltage pulse is maintained at 30 V, the drain current drops quickly, followed by a slight decrease attributed to two types of charge trapping, i.e., fast and slow at the MoS$_2$/SiO$_2$ interface. This degradation of drain current due to positive gate stress is well consistent with previously reported results.\cite{Mallik2021Apr} The final READ (R$_1$) process provides nearly identical drain current values as that of initial READ (R$_0$) such that the ratio between them, i.e., R$_1$/R$_0$ remains near unity.
\subsection{Charge trapping}
The drain current reduction during WRITE operation is best fitted with the two-trap model equation \ref{eq:twotrap_eq} as described in the previous section. At 350 K, the fleeting drain current during the WRITE operation shows a near similar behavior, except a slow increase of current values can be observed after an initial degradation of current, as shown in Fig. 3 (b). In this case, the R$_1$/R$_0$ ratio still provides near unity value. However, when the temperature exceeds 350 K, i.e., at 400 K, the drain current during the WRITE process reforms its behavior and is observed to increase sharply till the gate stress is applied. Interestingly, as shown in Fig. 3 (c), the R$_1$/R$_0$ ratio also increases to 1.5, providing an initial signature for possible memory applications of our MoS$_2$ FETs. When the temperature is further increased to 450 K, a more pronounced increment in the drain current values is observed during the WRITE operation, as shown in Fig. 3 (d). In this case, the enhancement in R$_1$/R$_0$ ratio to 3.57 indicates thermal modulation of drain currents in the presence of gate bias. 
\begin{figure}
	\centering
	\includegraphics[width=0.6\textwidth]{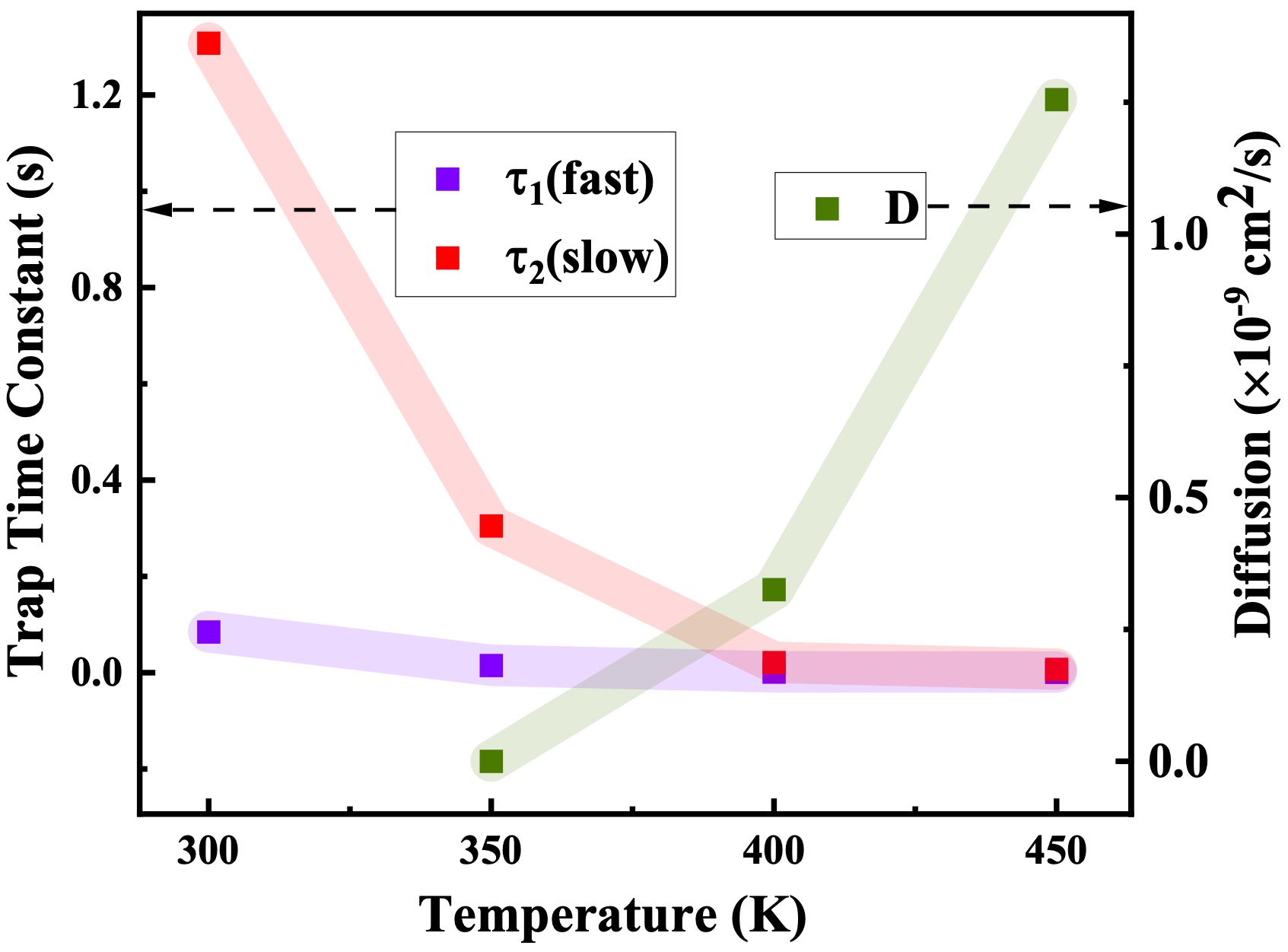}
	\caption{Temperature-dependent progression of interfacial mechanisms where the left Y axis represents the trap time constants (outcome of trap model) and the right Y axis represents the diffusion coefficient (outcome of current monitoring model).}
	\label{Fig_4ex}
\end{figure}
\subsection{Diffusion of ions}
It is worth mentioning that the sharp increment of current during the WRITE process always follows a very short-lived degradation of drain currents (as indicated by an arrow in the inset), which uncovers two possible mechanisms that are taking place at high temperatures. The origin of the initial degradation of drain currents is previously discussed, which corresponds to the trapping of charge carriers at the MoS$_2$/SiO$_2$ interface. However, the trend is carefully analyzed to shed light on the mechanism of the increased drain current. We notice that the increasing trend of drain current closely resembles a diffusion-like current and is accordingly best fitted with the following current monitoring equation,\cite{Wang2014Mar}
\begin{align}
I=I_0-A(e^\frac{-(D \pi^2)}{L^2t})  
\label{eq:diffusion_eq}
\end{align}
where, I$_0$, and A are standard derived parameters, L is the channel length, and D is the diffusion coefficient. The temperature-dependent progression of underlying mechanisms is shown in Fig.\ref{Fig_4ex}, where the left Y axis represents the trap time constants (outcome of trap model), and the right Y axis represents the diffusion coefficient (outcome of current monitoring model). As the temperature rises, the two trap constants (fast and slow) initially provide finite values and then become negligible around 400 K. However, the diffusion coefficient has a finite value above 400 K, stipulating its dominant nature at high temperatures. The charge trapping process is more favourable at room temperature, quenched by the migration of ions near the MoS$_2$/SiO$_2$ interface at high temperature. Here, it is worth mentioning that the accumulation of ions enroute the memory capabilities in ML MoS$_2$ FETs, which will be further discussed in the following sections.
\section{Energy band diagram and operation mechanism}
\begin{figure}
	\centering
	\includegraphics[width=0.9\textwidth]{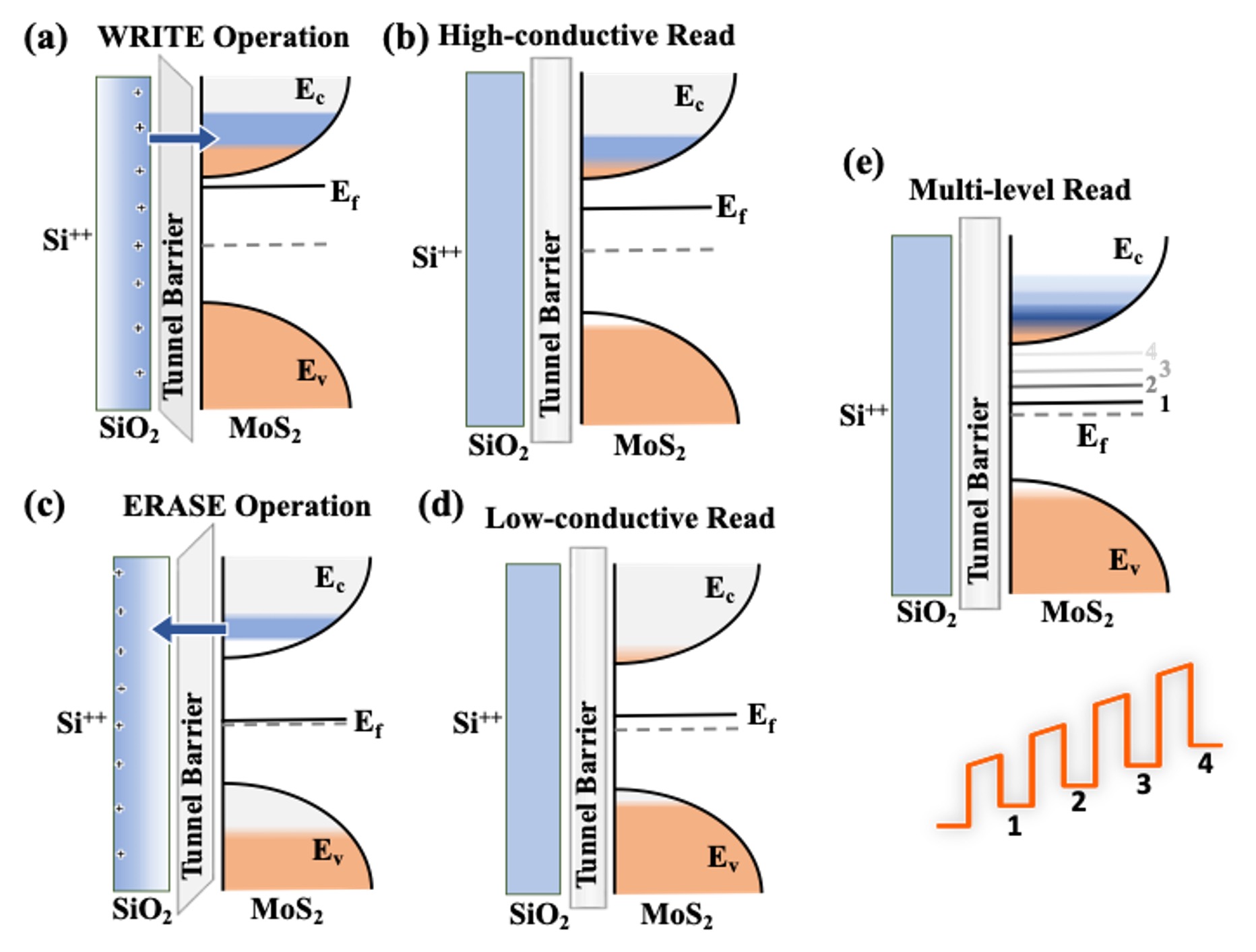}
	\caption{The proposed mechanism with suggested band diagrams of MoS$_2$/SiO$_2$/Si to explain the single and multi-level memory effects in MoS$_2$ FETs. (a) WRITE operation at V$_g$ $>$ 0. (b) High conductive READ state at V$_g$ = 0. (c) ERASE operation at V$_g$ $<$ 0 (d) Low conductive READ state at V$_g$ = 0. The orange filled color represents the carrier accumulation during the transistor switch-on, and the blue filled color represents the carrier tunneling due to the accumulation of active ions at the MoS$_2$/SiO$_2$ interface. Arrow represents the tunneling direction. (e) Multi-level READ state at V$_g$ = 0V. Different color bars represent multiple charge carrier injections at every implemented programmed pulse cycle and the successive elevation of Fermi levels.}
	\label{Fig_4ex2}
\end{figure}

To explain the non-volatile memory states in MoS$_2$ FETs originated by the thermally-driven accumulation of ions at the MoS$_2$/SiO$_2$ interface, we propose a hypothesized mechanism involving WRITE, READ, and ERASE schemes. Figure \ref{Fig_4ex2}(a-e) demonstrates the energy band diagrams for different gate voltage polarities to acknowledge basic n-type transistor functions and extrapolation of programmable conductance states for both single and multi-level memory capabilities. As shown in Fig.\ref{Fig_4ex2}(a), when a high positive voltage (V$_g$ $>$ 0) is applied to the gate terminal of the MoS$_2$ FETs, the Fermi-level shifts up, and accumulation of carriers (shown in orange) in the conduction band of MoS$_2$ essentially switch on the transistor offering a n-type FET behavior. Additionally, as the device is characterized at elevated temperature, the thermally induced ions drift and migrate towards MoS$_2$/SiO$_2$ interface due to high positive gate bias. Many accumulated ions persuade high carrier densities at the interface during positive gate stress. As a result, lowered tunneling barrier favors injecting additional charge carriers (shown in blue) into the conduction band of MoS$_2$. 

It should be noted that this injection process continues till the positive gate voltage is maintained, which constantly populates the conduction electrons in the MoS$_2$ channel. This behavior is consistent with the observed results in Fig.\ref{Fig_4.3}(d), where the drain current progressively increases during the pulsed gate voltage maintained at 30 V. This is denoted by the WRITE process, where conductance states can be achieved by allowing continuous migration and accumulation of ions near the SiO$_2$/MoS$_2$ interface, increasing the carrier densities in the MoS$_2$ channel over a progressive period by applying a positive gate voltage pulse. Figure \ref{Fig_4ex2}(b) demonstrates the band diagram for the high conductivity read state after immediate withdrawal of the gate field (V$_g$ = 0V). The Fermi level is lowered as the transistor goes to the off state at zero gate bias. However, the injected carriers mediated by the ions have a low probability of tunneling back through the barrier. As a result, despite the transistor being in the off state at zero gate bias, a finite density of carriers (shown in blue) is retained in the conduction channel of the FET, giving rise to a high conductivity state after the writing process. When a high negative bias is applied to the gate terminal (V$_g$ < 0), the transistor remains in the off state, and the Fermi level is further lowered. However, as shown in Fig.\ref{Fig_4ex2}(c), the accumulated ions diffuse and migrate away from the MoS$_2$/SiO$_2$ interface. This is known as ERASE process, where the additional injected carriers are removed from the conduction channel at a high negative gate field, regressing the device to the original neutral state. The read process is further performed at V$_g$ = 0 V, giving rise to a low conductivity state, where the carriers mediated by ions don’t contribute to the resultant drain current, as shown in Fig.\ref{Fig_4ex2}(d). Apart from single memory operation, which consists of a pair of (high and low) conductive read states, the proposed mechanism could adequately explain the multi-level memory effects arising in our MoS$_2$ FETs. As illustrated in Fig.\ref{Fig_4ex2}(e), successive WRITE operations lower the barrier allowing the charge carriers to tunnel into the conduction band of MoS$_2$, independently increasing the carrier densities and lifting the Fermi level at each pulse cycle. Continuing this process generates higher conductance states where each state corresponds to pulse cycle biasing history. This way, we could observe multi-level memory effects in our MoS$_2$ FETs. 

\section{Multi-level memory operations}
\begin{figure}
	\centering
	\includegraphics[width=0.9\textwidth]{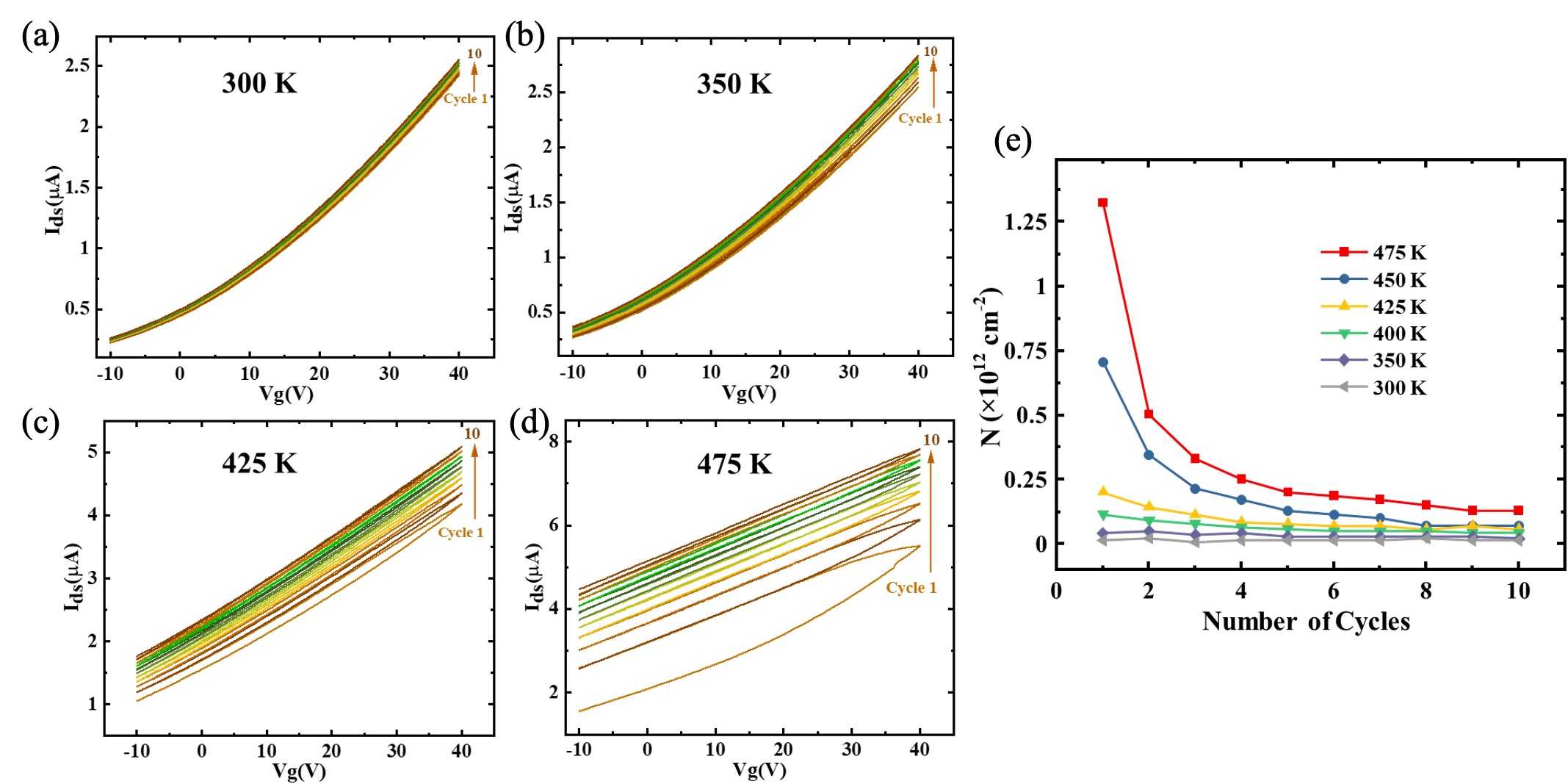}
	\caption{Dual sweep transfer characteristics with varying hysteresis widths for 10 successive cycles at (a) 300 K (b) 350 K (c) 425 K (d) 475 K. In each cycle, the gate voltage is swept from -10 V to 40 V and back to -10 V. With repeating cycles, the hysteresis curves initially provide overlapping current loops at 300 K. However, at elevated temperatures, the current loops shift upwards with each successive cycle, particularly at 475 K. This illustrates multiple conductance states can be achieved corresponding to several dual sweep cycles, which spurred the attainment of multi-level memory generation of our MoS$_2$ FETs. (e) Cation-mediated accumulated charge carrier densities cause anti-clockwise hysteresis loops at varying temperatures.}
	\label{Fig_4ex3}
\end{figure}

The development of a novel thermally-driven and ion-mediated non-volatile multi-level storage capacity can further be realized owing to the anti-clockwise hysteresis nature of our MoS$_2$ FETs at high temperatures. Figure \ref{Fig_4ex3} (see supporting information) shows dual sweep transfer characteristics with varying hysteresis widths for 10 successive cycles at four different temperatures. In each cycle, the gate voltage is swept from -10 to 40 V and back to -10 V. With repeating cycles, the hysteresis curves initially provide overlapping current loops at 300 K. However, the current loops shift upwards at elevated temperatures with each successive cycle, particularly at 475 K. This illustrates multiple conductance states can be achieved corresponding to several dual sweep cycles, which spurred the attainment of multi-level memory generation of our MoS$_2$ FETs. 

\begin{figure}
	\centering
	\includegraphics[width=0.9\textwidth]{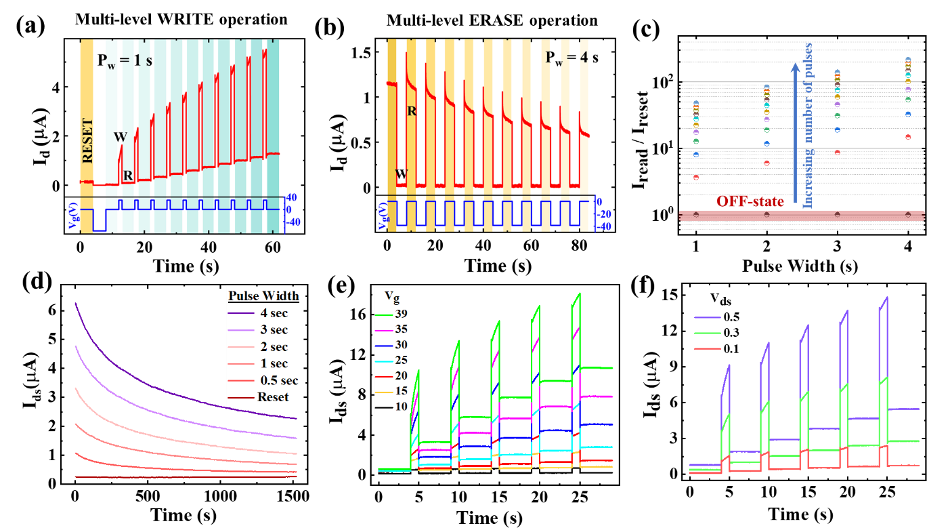}
	\caption{Dynamic responses of multi-level ionotronic memory (a) Multi-level programmed state operations at 450 K driven by gate pulse trains with pulse width 1 s. A Positive gate pulse train (10 cycles) with an amplitude of 30 V is provided for multiple writing processes, as shown in the lower panel (blue waveforms). The upper panel (red waveforms) shows the fleeting drain currents during RESET, WRITE, and READ programs. The green gradient bars represent the individual READ processes after periodic WRITE operations, which assimilate to increased distinct conductance states with several pulse cycles. (b) Multi-level ERASE operation with a programmable pulse train of pulse width 4 s where the ERASE and READ processes are performed at -40 V and 0 V, respectively (see the voltage waveform in the lower panel). (c) READ-to-RESET ratios for 10 successive pulses of widths 1, 2, 3, and 4 s. (d) The retention curves of the observed memory behavior after 3 successive WRITE operations for 1500 s with various pulse width conditions. Multi-level programmed state operations at 450 K driven by gate pulse trains (e) with gate pulse amplitude varied from 10 to 40 V (f) with drain-source voltage for 0.1, 0.2, and 0.3 V.}
	\label{Fig_4.4}
\end{figure}

Figure \ref{Fig_4.4}(a) shows multi-level programmed state operations driven by gate pulse trains having pulse width 1 s. Based on our single-state memory function at 450 K, a positive gate pulse train (10 cycles) with an amplitude of 30 V is provided for multiple writing processes, as shown in the lower panel (blue waveforms) of Fig. \ref{Fig_4.4}(a). The reading of distinct conductance states is performed at V$_g$ = 0 V after each writing process establishing a multi-level storage unit in the case of ML MoS$_2$ FET. Note that a complete RESET of the device is necessary (with V$_g$ = -70 V) before the initial writing process to recover from any higher conductance state (blue bar). The drain-source voltage is kept fixed at 300 mV at every step of the programmable memory operations. The upper panel (red waveforms) shows the fleeting drain currents during RESET, WRITE, and READ programs. The green gradient bars (a guide to eyes) represent the individual READ processes after periodic WRITE operations, which assimilate increased distinct conductance states with several pulse cycles. The READ current rises progressively with increasing the pulse number, a phenomenon that represents the continual accumulation of carriers in MoS$_2$ as prolonging the gate stress on the memory. 

To further explore the pulse-controlled memory behaviour, the multi-level memory cycles are expanded to different programming pulsed width measurements, attributed to distinct conductance states based on the number of cycles. It is worth noticing that using the programmed memory operations with longer pulse widths yields higher conductance states, which reflects that n-bit memory can be achieved by varying the pulse time. Additionally, we witness an interesting result on data erasing, i.e., the multi-level storage information can be erased gradually by applying low negative voltages accompanying the complete RESET at a high negative bias (-70 V). Figure \ref{Fig_4.4}(b) represents a multi-level ERASE operation with programmable erase pulse train of pulse width 4 s where the ERASE and READ processes are performed at -40 and 0 V, respectively (see the voltage waveform in the lower panel). Similar to the multi-level WRITE operation, a progressive reduction of the READ current with successive ERASE operations indicates continual decumulation of charge carriers in the conduction channel. This means any desired lower conductance state can be achieved from a higher conductance state by performing ERASE operations with n-cycle pulse trains. The multi-level WRITE and ERASE operations in our device show exceptional pulse control of charge injection and release, providing a complete yet simplest way to achieve large-scale data storage capabilities in MoS$_2$ FETs. The increment and decrement of current updates with the number of pulses provide a potential platform to achieve high temperature in-memory neuromorphic computing for online learning, discussed subsequently.

The thermally-driven multi-level memory functions discussed above can generate large set of data storage possibilities by controlling the pulse trains of different pulse widths with n-successive cycles. The READ-to-RESET ratio, a desirable feature of a memory device, is calculated for all the pulse-controlled READ states shown in Figure \ref{Fig_4.4}(c). For each pulse width, the ratio values are plotted in logarithmic scale for 10 successive pulses represented by colored half-filled circles spanned over a large range from 1 to 220. The ratio value 1 is represented by the OFF-state in all pulse widths. The ON/OFF ratio for the first pulse varies from 3 to 16 as the pulse width varies from 1s to 4 s. Interestingly, the pulse width with longer time has higher and distinct READ window values showing strong dependence on various pulse conditions for each corresponding cycle. An attempt has been made to realize the ON/OFF ratios in thermally-aided mem-transistors in 2D materials; for example, Goyal et. al. reported READ windows of about 1.9 and 7.4 in few-layered MoS$_2$ and ReS$_2$ memory devices, respectively, for single-step memory operations.\cite{Goyal2020Feb} Our results show unambiguously higher READ values in the case of ML MoS$_2$ FETs. More importantly, our studies show multi-level memory states in our device can take this ratio up to $\sim$ 220 and beyond by using longer pulse trains with varying pulse widths. It is also worth noting that combinational pulse waveforms can get desired READ window and store information in multiple storage cells. The retention curves of the observed memory behavior are obtained after 3 successive WRITE operations for various pulse width conditions, shown in Fig.\ref{Fig_4.4}(d). The retention behavior shows that the data levels of different pulse conditions are well-discernable after 1500 s. It may also be noted that our device shows much better retention capabilities with increasing pulse widths. Such memory devices can be used in electronic components that require low-scale information storage functionalities such as cache memory, disposable electronic tags, buffer memories, etc.

Apart from various pulse conditions, i.e., several pulses and pulse width, multi-level data storage can be obtained by changing the gate voltage amplitudes. In previous reports using floating gate memory configurations, optical response, and plasma treatment, the gate voltage-dependent multi-bit generation is achieved in 2D materials and heterostructures.\cite{Lee2016Nov, Kim2019Jul, Chen2014Apr, Gwon2021Oct, Wen2021Nov} Here, we also find a strong dependence on the drain current levels as we increase the V$_g$ from 10V to 40V, as shown in Fig.\ref{Fig_4.4}(e). The programmable memory operations are obtained for 5 successive pulses, which represent a similar progressive increase in READ states, i.e., the higher gate voltage amplitude corresponds to larger READ windows. This indicates gate voltage amplitude is crucial for obtaining stable and distinct memory states. We are also keen to observe the effects of drain-source voltage, i.e., V$_{ds}$, on the memory operations in the MoS$_2$ FETs, shown in Fig.\ref{Fig_4.4}(f). All these above experiments strongly vow the versatile nature of our memory device, which can be tuned with different controllable parameters such as the number of pulses, pulse-width, pulse-amplitude, and drain-source voltage.
\section{Chapter summary}
In summary, we have demonstrated a novel functional three terminal device with salt-assisted CVD-grown ML MoS$_2$ as channel material which serves as a high geared intrinsic transistor at room temperature and becomes a multi-level memory cell at relatively high temperature. The memory behavior is attributed to the hysteresis collapse and switch from clockwise to anticlockwise, which further expands with increasing temperature. The reverse hysteresis is ascribed to the migration of active ions and accumulation at the MoS$_2$/SiO$_2$ interface due to electrostatic gating at a higher temperature of around 450 K. The proposed thermally-driven programmable memory operations, i.e., READ, WRITE, ERASE, are found to be modulated by gate voltage biasing history. The step-like READ-RESET ratio and retention curves are obtained for various pulse conditions, which stipulates vigorous data storage capabilities in thermally-driven memory cells. Furthermore, the multi-level memory states are investigated with varying pulsed width, gate voltage amplitudes, and drain-source voltages. We corroborate these results with the relative energy band diagrams to explain the single and multi-level memory effects in MoS$_2$ FETs. Moreover, our device demonstrates excellent linearity and symmetry upon electrical potentiation and depression for high-temperature synaptic applications. As a result, a high classification accuracy of 95$\%$ is achieved during training and testing of the Modified National Institute of Standards and Technology (MNIST) datasets in artificial neural network (ANN) simulation. Our work serves as a precursor to novel pathways in the realm of 2D semiconductors, aimed towards achieving robust high temperature in-memory computing applications with enhanced memory capabilities for artificial cognitive development after the human brain.

\blankpage 
\chapter{Ionotronic monolayer WS$_2$ mem-transistors for neuromorphic computing applications}
\label{C5} 
\section{Introduction}
The amount of digital data generated by humans, which can be quantified in terms of the number of bits produced yearly, is continuously expanding and is projected to exceed 100 zettabytes by 2023. To effectively manage the massive amount of data and global memory requirements, the upcoming nonvolatile memory (NVM) technologies must furnish multi-bit ultra-high-density storage capacity with a significant extinction ratio, endurance, retention, and low energy consumption.\cite{Hou2019Sep} Recently, there has been an intensive exploration of  NVMs based on 2D materials and their heterostructures which utilize revolutionary device architecture and unusual mechanisms to overcome the limitations of conventional flash memory based on silicon..\cite{Liu2020Jul} Opto-electronically controlled floating gate (FG) and charge trap based devices can produce high operation speed, and low power memory characteristics.\cite{Lee2016Nov, Wen2021Nov, Lee2017Mar, Wang2018Apr, Yang2019Jan, Tran2019Feb, Huang2019Sep, Gao2021Nov, Liu2021Jun, Zhang2021Jul} Nevertheless, it is difficult to develop high-capacity multi-bit generation that can work at elevated temperatures. Recent report by Li et. al.\cite{Li2022Aug} proposes MoS$_2$/hBN/MoS$_2$/GDYO/WSe$_2$ heterostructure as an active layer to generate 8 memory states18 whereas, Lai et. al.\cite{Lai2022May} uses top FGM with stacked MoS$_2$/hBN/2D-RPP to produce 22 distinct memory states. Furthermore, the production of such memory devices necessitates cumbersome state-of-the-art methodologies, which require accurate deposition of various semiconductor layers, and multiple lithography processes, culminating in intricate mem-transistors with numerous floating metals and oxide layers. As the yearning for energy conservation is more prevalent than ever, advancing the scale-down of circuits beyond Moore's law will mandate the exploration of 2D materials based architectures.\cite{Zhu2023Jun} 

On the contrary, the electrolyte-gated transistors show analogue switching performance with superior storage capacity benefitted from their ion-gating mechanism over electrostatic charge trap phenomena, making them viable candidates for brain-inspired computation and logic-in-memory applications.\cite{Yang2017Jul, Jiang2017Aug, Yang2018Oct, Zhu2018May, Yao2021Jun} Post stimulation, the diffusive dynamics of mobile ions (such as Li$^+$, Na$^+$) possess linear and symmetric weight updates of the discrete multilevel states, making them more effective than FGM-based arrays in processing artificial neural networks (ANNs).\cite{Park2022Nov, Li2022Jan, Wang2021Dec} Despite the recent development of synaptic ion-gated transistors, multi-bit memory based on ion gating mechanisms is still lacking. With the increasing spatial density of FET arrays on a solitary wafer, Integrated Circuits (ICs) of superior performance can attain an operating temperature as high as 450 K.\cite{Schroder2003Jul} Therefore, understanding and leveraging the properties of 2D materials-based devices is crucial for their high-temperature potential applications in harsh environments such as aerospace industries, military, automotives, sensors, well logging, oil refinery etc.\cite{Watson2015Dec} However, no consensus has been reported to address stable multilevel memory switching at elevated temperatures. Unfortunately, the use of liquid electrolytes severely demarcates the integration density of devices and limits their applicability to high-temperature environments. As a result, it is imperative to introduce  multi-level memories with eccentric synaptic functionalities that can enable simpler device architectures, thermally-stable reliable switching, and enhanced throughput. 

This chapter focuses on development of monolayer WS$_2$-based novel functional mem-transistor devices that address nonvolatility and synaptic operations at high temperatures. The ionotronic memory devices based on WS$_2$ exhibit reverse hysteresis with memory windows larger than 25 V, and extinction ratio greater than 10$^6$. The mem-transistors show stable retention and endurance greater than 100 sweep cycles and 400 pulse cycles in addition to 6-bit (64 distinct nonvolatile storage levels) pulse-programmable memory features ranging over six orders of current magnitudes (10$^{-12}$ to 10$^{-6}$ A). The origin of the multi-bit states is attributed to the carrier dynamics under electrostatic doping fluctuations induced by mobile ions, which is illustrated by employing a fingerprint mechanism including band-bending pictures. The credibility of all the storage states is confirmed by obtaining reliable signal-to-noise ratios. We also demonstrate key neuromorphic behaviours, such as synaptic plasticity, near linear potentiation, and depression, rendering it suitable for successful implementation in high-temperature neuromorphic computing. Furthermore, artificial neural network simulations based on the conductance weight update characteristics of the proposed ionotronic mem-transistors are performed to explore the potency for accurate image recognition. Our findings showcase a new class of thermally aided memories based on 2D semiconductors unlocking promising avenues for high-temperature memory applications in demanding electronics and forthcoming neuromorphic computing technologies.
\section{Fabrication of ion-gated WS$_2$ mem-transistors}
\begin{figure}
	\centering
	\includegraphics[width=0.9\textwidth]{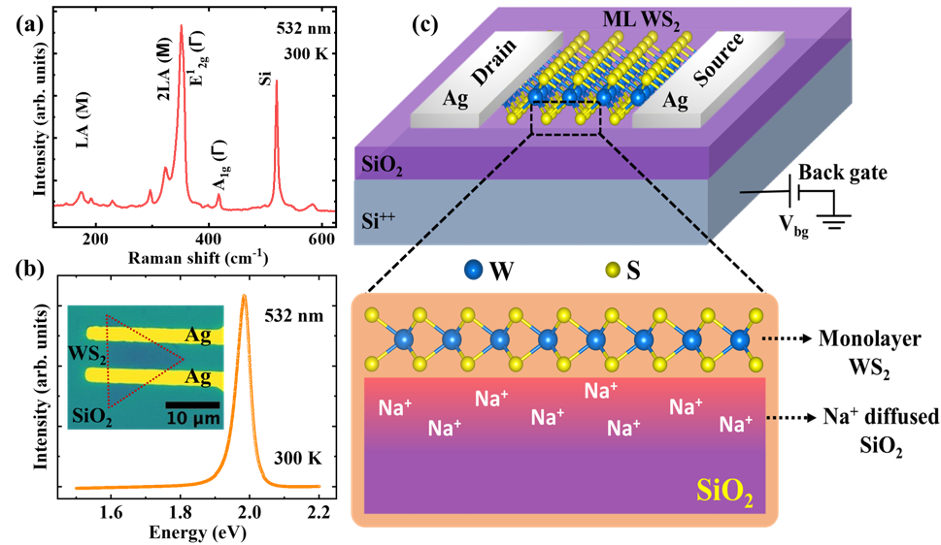}
	\caption{Structural characterization, and device architecture of ion-gated WS$_2$ transistors. (a) Room Temperature Raman spectrum of as-grown ML WS$_2$ supported by SiO$_2$ (285 nm)/Si substrate. (b) The photoluminescence spectrum confirms the monolayer nature of WS$_2$. Inset shows the optical micrograph (top view) of a WS$_2$ transistor with silver (Ag) metal as drain/source electrodes. (c) Schematic representation of the ML WS$_2$-based device model with necessary electrical contacts. The magnified cross-sectional view at the WS$_2$/SiO$_2$ interface illustrates the diffusion of Na$^+$ ions inside SiO$_2$.}
	\label{Fig_5.2}
\end{figure}

The monolayer (ML) WS$_2$ is synthesized via a salt-assisted chemical vapour deposition (CVD) technique, wherein the synthesis parameters are fine-tuned to achieve large-scale triangular domain growth with high crystallinity, as described in the previous section \ref{S 2.7}. The as-grown samples are characterized with Raman spectroscopy at room temperature, utilizing a $532$ nm excitation wavelength. The WS$_2$ exhibits two prominent Raman active modes, namely E$^1_{2g}$ and A$_{1g}$, as illustrated in Fig. \ref{Fig_5.2} (a), corresponding to the in-plane vibration of W and S atoms and the out-of-plane vibrations of the S atoms, respectively. The frequency difference ($\Delta$ $\omega$ = 61 cm$^{-1}$) between them indicates the ML nature of the as-grown WS$_2$ samples. Figure \ref{Fig_5.2} (b) displays the photoluminescence (PL) spectrum with a very sharp and intense luminescence at 1.96 eV which further confirms the high optical grade quality of salt-assisted CVD-grown ML WS$_2$.\cite{McCreary2016Oct} The inset shows the optical micrograph (top view) of a WS$_2$ transistor with silver (Ag) metal as drain/source electrodes. Figure \ref{Fig_5.2} (c) depicts the schematics of the device architecture with necessary electrical contacts where the back gate is facilitated by heavily doped silicon (Si$^{++}$). The WS$_2$/SiO$_2$ interface is portrayed in a separate magnified cross-sectional cartoon where Na$^+$ ions are shown to be diffused inside the SiO$_2$ near the interface region. Recently, Kaushik et. al.\cite{Kaushik2020Oct} reported the adsorption of lighter impurities such as Na$^+$ ions, on the SiO$_2$ surface during NaCl-aided CVD growth. These Na$^+$ ions get diffused inside the SiO$_2$ upon high-temperature annealing which can be employed to realize a solid-state ion-gated medium in our case, similar to electrolyte/ionic liquid-gated transistors.\cite{Jiang2017Aug, Yang2018Oct}
\section{Thermally-driven hysteresis evolution}
\begin{figure}
	\centering
	\includegraphics[width=0.9\textwidth]{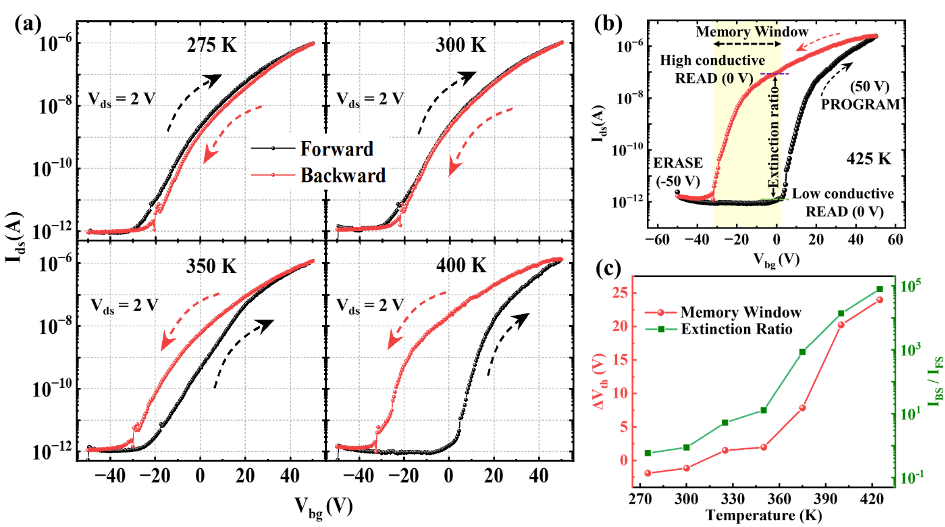}
	\caption{Temperature-dependent hysteresis evolution. (a) The dual sweep transfer characteristics at temperatures ranging from 275 to 400 K. The black and red arrows represent forward and backward sweep direction, respectively and nature of the hysteresis loop. (b) Dual sweep reverse hysteretic transfer curves with memory window of 25 V and extintiction ratio of $\sim$10$^5$ is demonstrated at 425 K for gate voltage sweep range of $\pm$50 V. The PROGRAM, ERASE, and READ operation regime are also demonstrated. \textit{(c)} The relation between memory window ($\Delta V_{th}$) and current switching ratios (IBS/IFS) with temperature at zero gate voltages during dual sweep measurements.}
	\label{Fig_5.3}
\end{figure}

The temperature-dependent dual sweep transfer characteristics are conducted under high vacuum ($10^{-6}$ mbar) after annealing the device at 200 $^{\circ}$C for 36 hours. Annealing is found to be crucial to reduce contact resistances without degrading the device performance, as reported earlier. Figure \ref{Fig_5.3}(a-b) depicts the evolution of hysteresis curves for temperatures ranging from 275 to 425 K during a gate voltage sweep range of $\pm$50 V. The black and red arrows represent forward and backward sweep directions and the threshold voltage differences between them indicate the memory window ($\Delta V_th$). Initially at 275 K, the nature of the hysteresis is found to be clockwise (CW), while at room temperature, the forward and backward sweeps overlap with each other collapsing the hysteresis completely. Interestingly, upon increasing the temperature to 350 K, an anti-clockwise (ACW) hysteresis loop is evolved which inflates further with elevated temperatures. It is worth to mention that, at 425 K the memory window is increased to as high as 25 V with extinction ratio (IBS/IFS) rises to the orders of 10$^5$ at zero gate bias. Moreover, in Fig.\ref{Fig_5.3}(b), we demonstrate the potential application of thermally-assisted reverse hysteresis in nonvolatile memory operations. The memory characteristics can be obtained by utilizing specific regions of the transfer curves to define the PROGRAM, ERASE, and READ operations. During the forward sweep, the ERASE operation can be executed by applying a gate voltage of -50 V, followed by a low conductive READ operation at V$_{bg}$ = 0 V. Similarly, by applying a high positive gate voltage of 50 V, we can PROGRAM our memory device to achieve a higher conductance state, which can be subsequently READ at V$_{bg}$ = 0 V during backward sweep. Recently, temperature-dependent hysteresis crossover/inversion has been reported in ML and few-layer MoS$_2$.\cite{He2016Oct, Kaushik2017Oct} However, the nature and origin of the hysteresis loops and their proposed mechanisms are very different from the results we observed in our case. For example, in the article by He et. al.,\cite{He2016Oct} the hysteresis loop contains two different regions i.e. anti-clockwise loop below V$_{bg}$ = 0 V and clockwise loop above V$_{bg}$ = 0 V, which also limits the extinction ratio to $\sim$10$^2$. Also, the origin of such hysteresis is attributed to carrier injection from Si into SiO$_2$ with minimal role of channel MoS$_2$. However, in our case, the device attains a single large anticlockwise hysteresis with extinction ratio of $\sim$10$^6$ and the memory is governed by the dielectric/channel interfacial physics. Figure \ref{Fig_5.3}(c) represents the progression of memory windows and extinction ratios as a function of temperature. Both the parameters show systematic and significant enhancements when compared with their respective room temperature values. Here we want to emphasize that our recent findings on CVD (without NaCl) grown monolayer MoS$_2$ show a high-temperature clockwise hysteresis behaviour owing to the accelerated charge trapping/detrapping, however, the anticlockwise hysteresis is only observed in NaCl-assisted CVD grown MoS$_2$.\cite{Mallik2023Aug} This further confirms the Na$^+$ driven ion-gating induced large reverse hysteresis in our NaCl-assisted CVD grown WS$_2$ based transistors.
\section{Operation mechanism of ionotronic memory}
\begin{figure} [hbt!]
	\centering
	\includegraphics[width=1.0\textwidth]{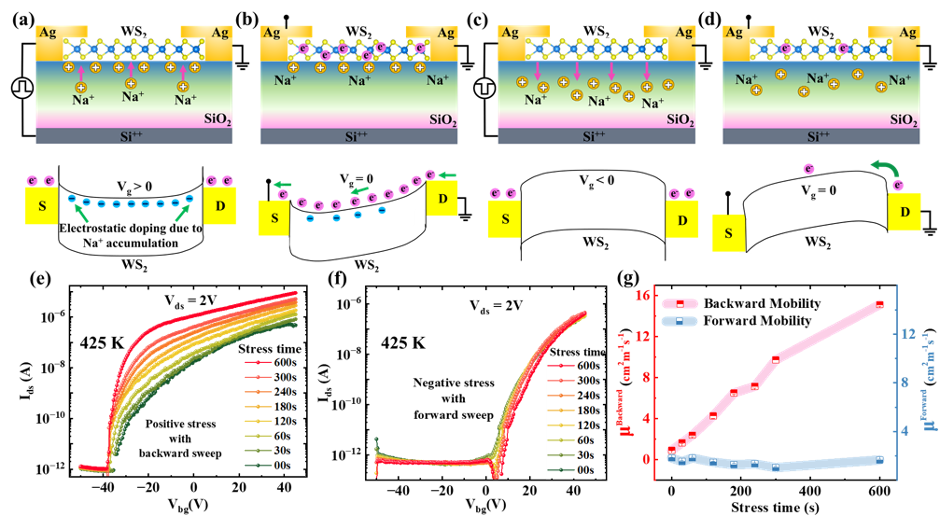}
	\caption{Operation mechanism of ionotronic memory. The cross-sectional schematic view of the WS$_2$ mem-transistor (upper panel), Energy band diagram of source-channel-drain (Lower panel) for (a) PROGRAM (b) High Conductive Read (c) ERASE, and (d) Low Conductive Read operations. For Read operations the bias voltage V$_{ds}$ is fixed at a finite voltage (in our case, 2 V) to measure the drain current values after PROGRAM/ERASE. The PROGRAM operation is facilitated by the Na$^+$ accumulation at the WS$_2$/SiO$_2$ interface, whereas in the ERASE operation, the depletion of the ion layer reduces the effective electrostatic doping resetting the device to the original state. The green arrows in the energy band diagrams indicate the direction of the flow of currents through the channel. Effect of (e) Positive (+50 V) and (f) Negative (-50 V) stress under different stress times of ML WS$_2$ mem-transistor. (g) Mobilities extracted from forward and backward transfer curves vs. stress time, at V$_{ds}$=2 V.}
	\label{Fig_5.4}
\end{figure}

In conventional WS$_2$ transistors, a relatively high surface-to-volume ratio necessitates an inescapable exposure of the atomically thin channel to the oxide trap/defect states, which induce clockwise hysteresis and high threshold voltage instabilities, leading to reliability issues.\cite{Lan2020Dec} Nonetheless, the substantial hysteresis generated during the dual sweep and pulsed gate operations in most 2D transistor devices has been meticulously utilized for nonvolatile (which encompass 2D flash memory, magnetic random access memory, resistive random access memory) and volatile (such as dynamic random access memory, semi-floating gate transistor) memory applications.\cite{Hou2019Sep} However, in our case, temperature-modulated hysteresis collapse and switch from clockwise to anti-clockwise hysteresis indicates ion-gating induced charge transfer and charge storage.\cite{Yao2021Jun} The increased kinetic energy of Na$^+$ inside SiO$_2$ at high temperatures generates an efficient ion gating over conventional dielectric gating for transistor applications.\cite{Wang2021Dec} To shed light on the effect of Na$^+$ ions on hysteresis, we propose a fingerprint mechanism as depicted in Fig.\ref{Fig_5.4}(a-d), that adequately illustrates the evolution of hysteresis with temperature, providing the possibilities to attain thermally stable nonvolatile memory states in ML WS$_2$. The upper and lower panel shows the cross-sectional schematic view of the WS$_2$/Na$^+$ diffused SiO$_2$ interface and energy band diagram of source-channel-drain, respectively during various programming operations. Applying a positive gate-source voltage induces a drift of mobile Na$^+$ ions, causing them to move towards the WS$_2$ channel. This accumulation beneath the interface, illustrated in Figure \ref{Fig_5.4}(a), results in the buildup of spatial charges near the interface. This accumulation has the effect of further electrostatically doping the WS$_2$ channel, an operation referred to as ‘PROGRAM’. The added n-doping of WS$_2$ reduces its energy band below the fermi level of Ag, creating an easily accessible pathway for conduction electrons to tunnel through.

Figure \ref{Fig_5.4}(b) illustrates the schematic for the ‘High Conductive Read’ operation with a bias voltage (V$_{ds}$) of 2 V at V$_{bg}$ = 0 V. Following the ‘PROGRAM’ operation, the collected Na$^+$ ions progressively enhance the drain current, reaching around $\sim$10$^{-6}$ A even in the absence of gate voltage during the backward sweep. The corresponding band diagram highlights the increased carrier concentration within the WS$_2$ channel under the conditions of the ‘High Conductive Read’ state. The Na$^+$ ions accumulated at the interface exhibit high retention due to their low probability of returning to their original state, even after removing the external electric field, resulting in band lowering and hence high conductance of the WS$_2$ channel. Thus it can only be depleted by employing a highly negative V$_{bg}$, an ‘ERASE’ operation. In such cases, the Na$^+$ ions migrate away from the interface and towards the opposite direction, facilitated by the electrostatic polarization effect. The ‘ERASE’ operation increases the barrier width at the Schottky junction, as displayed in Fig.\ref{Fig_5.4}(c). A further ‘Read’ operation can furnish low current values up to $\sim$10$^{-12}$ A, akin to the case of an undoped WS$_2$ channel. The schematic representation of Fig.\ref{Fig_5.4}(d) illustrates low carrier channel densities with corresponding band bending in the ‘Low Conductive Read’ operation. Hence, by administering a positive or negative gate voltage of approximately $\pm$50 V, the doping concentration of the WS$_2$ channel can be tuned, resulting in a subsequent modification of its resistivity.

Considering the intricate dynamics of Na$^+$ ions during gating, it is imperative to scrutinize the impact of gate bias stress in our ML WS$_2$ mem-transistors. Specifically, to induce positive (negative) gate bias stress, a uniform gate voltage of 50 V (-50 V) was applied for a defined duration before measuring the transfer characteristics. As demonstrated in Fig.\ref{Fig_5.4}(e), the transfer curve gradually shifts upwards with prolonged positive gate bias stress durations. The maximum output current (I$_{ds}^{max}$) augments linearly with increasing stress durations. Interestingly, the transfer curves remain unaltered for mounting negative gate bias stress durations as shown in Fig.\ref{Fig_5.4}(f). Considering the fact that the Na$^+$ are mobile in few nm range inside the diffused SiO$_2$ region, the ions can’t be pushed away to the gate/dielectric interface, unlike the case of single ion conductor.\cite{Xu2019Oct} Therefore, we expect the formation of a weaker negatively charged depletion layer at the channel/dielectric interface on the application of negative gate voltage which will have a negligible impact on the device performance. The values of mobilities obtained from multiple transfer curves with various positive and negative stress durations are determined and presented in a graph against the duration of stress in Fig.\ref{Fig_5.4}(g). It is observed that the positively stressed mobility values vary linearly with increasing stress durations, unlike the case for negatively stressed mobility values. We have also estimated the carrier doping densities of $\sim$10$^{12}$ cm$^{-2}$ for different stress time durations. 
\section{Impact of voltage stress on hysteresis}
\begin{figure}
	\centering
	\includegraphics[width=0.9\textwidth]{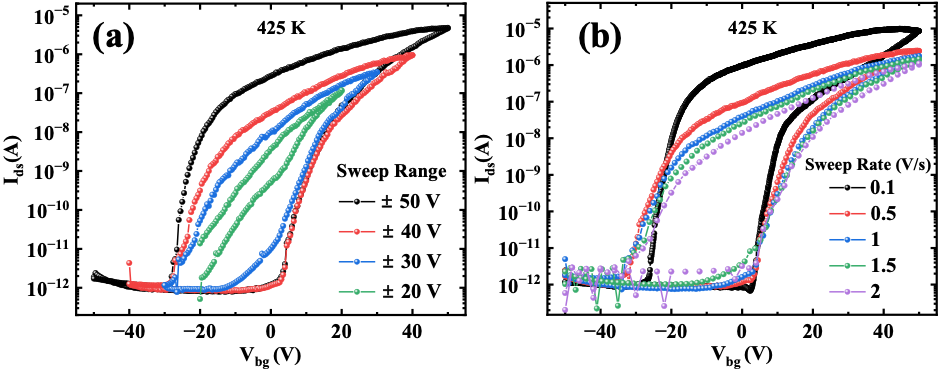}
	\caption{Dual sweep transfer characteristics of various gate voltage sweep conditions i.e. (a) different ranges from $\pm$20 V to $\pm$50 V at constant sweep rate of 0.5 V/s (b) different sweep rate from 0.1 V/s to 2 V/s at a fixed range between $\pm$50 V.}
	\label{Fig_5.13}
\end{figure}

It may also be noted that the doping concentrations in our WS$_2$ mem-transistors increase with increasing positive stress times. The existence of hysteresis in the transistor current suggests a dynamic characteristic, which we investigate in Fig.\ref{Fig_5.13} by measuring transfer curves with varying sweep rates and ranges. It may be deduced that since this is an analog memory, changing the voltage sweep range and step size will have an impact on the extinction ratio and hysteresis window because the effective program time will change drastically. We observe an interesting trend by varying the sweep rate from 2 V/s to 0.1 V/s and the range from $\pm$20 V to $\pm$50 V. As the sweep range (rate) is increased (decreased), the current jump observed in the forward sweep becomes more significant. However, a considerable increase in the overall current level is observed toward a lower sweep rate in the backward sweep case, indicating the existence of distinct time scales for different segments of the hysteresis curve. 

\begin{figure}
	\centering
	\includegraphics[width=1\textwidth]{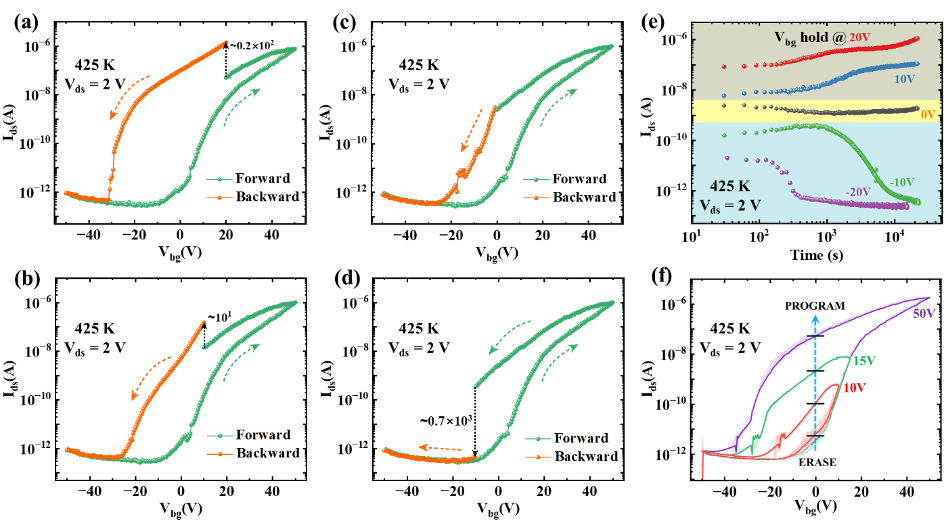}
	\caption{The effect of hold voltage on the hysteretic behavior is demonstrated. Initially, the gate voltage is swept from the negative to positive voltage range, and during the backward sweep, it is halted at a specific intermediate voltage called V$_{bg}$ hold: (a) 20 V, (b) 10 V, (c) 0 V, and (d) -10 V. Then, the current progression is recorded over a period exceeding 20,000 s. After the hold, the sweep is resumed, and the gate voltage returns to the negative range. The hysteresis curve is indicated in green color before the hold and is represented by orange after the delay. (e) The I$_{ds}$ vs. time measured during the hold phase of the experiments demonstrated in (a-d). (f) Dual sweep transfer curves lead to distinct PROGRAM states by varying V$_{bg}$ amplitudes.}
	\label{Fig_5.5}
\end{figure}

To further probe the dynamics, we terminate the backward sweep at an intermediate gate voltage (V$_{bg}^{hold}$) and closely observe the time-dependent fluctuations of I$_{ds}$ over a specific time interval prior to recommencing the sweep as shown in Fig.\ref{Fig_5.5}(a-d) when we hold at 20 V, the drain current increases by approximately 10$^{2}$, whereas holding at 10 V for 20,000 seconds can elevate the current level to the order of $\sim$10. In contrast, holding at negative V$_{bg}^{hold}$ can decrease the current level. Furthermore, Figure \ref{Fig_5.5}(e) displays the temporal evolution of the current achieved at distinct V$_{bg}^{hold}$ levels. Upon halting at a positive V$_{bg}^{hold}$, a gradual increase in the drain current is witnessed over a prolonged time period. However, holding at V$_{bg}^{hold}$= 0 V, we observe no significant alterations in drain current values after the holding period, indicating zero gate bias as an ideal condition for READ voltage. For the negative V$_{bg}^{hold}$, the current exhibits a remarkably rapid decrease in a time frame that is several orders of magnitude shorter. These findings are in sharp contrast to the transient analysis reported earlier, where no current rises were observed for any V$_{bg}^{hold}$ cases.\cite{He2016Oct} Additionally, we showcase an alternative trial by altering the V$_{bg}$ sweep amplitudes to accomplish discrete PROGRAM states. As depicted in Figure \ref{Fig_5.5}(f), a rise in the positive V$_{bg}$ from 10 V to 50 V leads to an increase in the hysteresis window and extinction ratios, with the expectation of achieving diverse PROGRAM states. While conducting a dual sweep, we execute retention measurements at V$_{bg}$ = 0 V to observe the loss of carriers over time. However, the satisfactory retention behaviour of over 10$^4$ for various PROGRAM states exemplifies the exceptional dynamic behaviour of charge injection and release of our ion-gated WS$_2$ mem-transistor.\cite{Mallik2023Sep}
\section{Robustness of nonvolatile memory in ion-gated transistors}
\subsection{Device to device variability}
\begin{figure} [hbt!]
	\centering
	\includegraphics[width=1\textwidth]{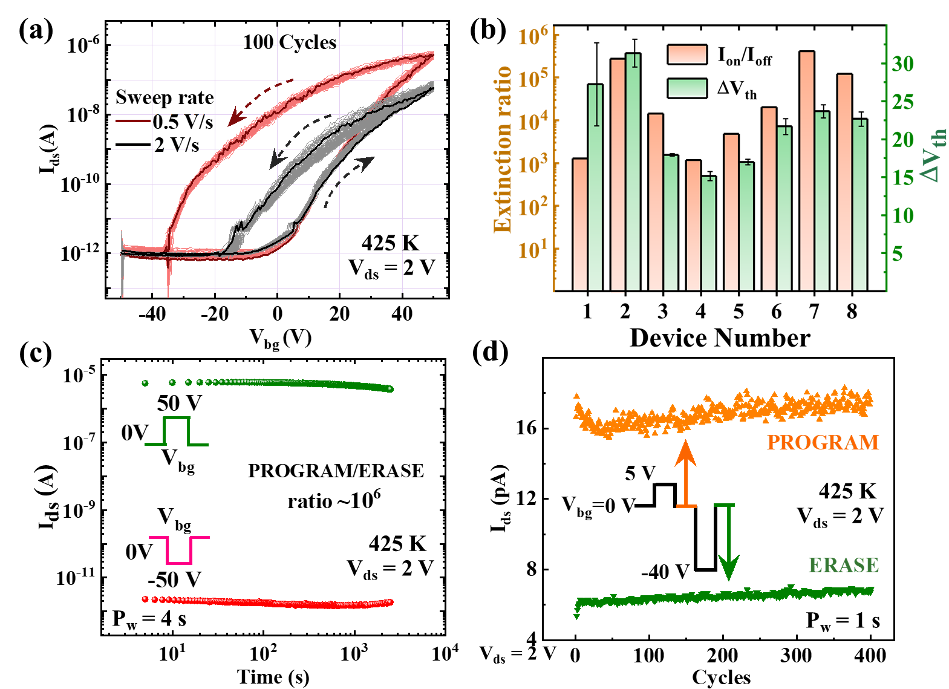}
	\caption{Robustness of nonvolatile memory in ion-gated transistors: (a) The endurance tests of 100 dual sweep hysteresis cycles for varied sweep rates. The gate voltages are swept from -50 V to +50 V and back to again -50 V. Transfer curves (black) of faster sweep rates introduce a stable extinction ratio of 10$^2$, whereas the Transfer curves (black) of slower sweep rates provide a stable I$_{on}$/I$_{off}$ ratio of 10$^2$. (b) Extracted parameters, i.e., I$_{on}$/I$_{off}$ ratios (orange) and memory windows ($\Delta$V) (green) of transfer curves from 8 different devices illustrating device-to-device variations. (c) Retention characteristics of the device after PROGRAM (pulse of +50 V, 4 s) and ERASE (pulse of -50 V, 4 s) operations for 2000 s. (d) The ionotronic memory device over 400 cycles of operations showing endurance characteristics of the programmable waveforms at (0 V, 1 s) READ pulse (orange) succeeded by a (5 V, 1 s) PROGRAM pulse. After that, an ERASE pulse of (–40 V, 1 s) is supplied, and another READ pulse (green) for 1 s is followed. The orange (green) triangles represent High (low) conductive READ currents in each cycle.}
	\label{Fig_5.6}
\end{figure}

Transfer curves have been obtained for eight different devices to establish stability amI$_{ds}$t device-to-device variations. The extracted parameters, encompassing extinction ratios and memory windows ($\Delta V$) of different devices, are depicted in Figure \ref{Fig_5.6}(b). The large extinction ratios observed in these ionotronic memory devices concur with the previous reports based on the floating gate and electrolyte-gated mem-transistors.
\subsection{Endurance}
\begin{figure}
	\centering
	\includegraphics[width=0.9\textwidth]{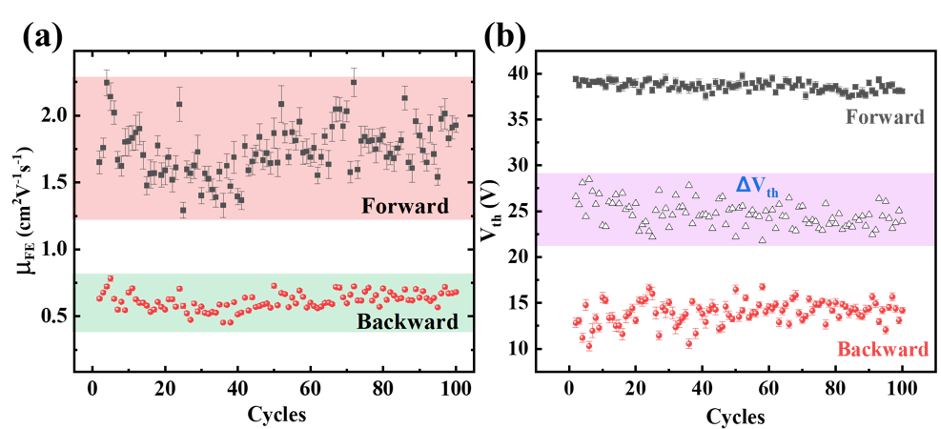}
	\caption{(a) Mobility during the forward sweep and backward sweep of 100 repeated cycles forming a range indicated by the red and green band, respectively, (b) Threshold voltage (V$_{th}$) of the transfer characteristic during forward and backward sweep is indicated by black square and red circle respectively. The white triangles in the pink band represent the difference between both V$_{th}$, also referred to as the memory window.}
	\label{Fig_5.15}
\end{figure}

The WS$_2$-based mem-transistors are subjected to endurance tests, wherein more than 100 hysteresis cycles with varying sweep rates are performed. As shown in Figure \ref{Fig_5.6}(a), the tests reveal that our memory device sustains high on/off ratios without significant decay, demonstrating its superior memory capacity and reliability. In addition, we have extracted the forward and backward mobilities and the threshold voltage shifts ($\Delta V_{th}$), of the WS$_2$ channel from the transfer curves acquired at a sweep rate of 0.5 V/s, which have been briefly summarized in Fig.\ref{Fig_5.15}. The mobilities and $\Delta V$ values of 100 hysteresis cycles demonstrate the reliability of these memory devices, with minimal fluctuation bands
\subsection{Retention}
An additional performance index of a durable memory device is its retention behavior over an extended duration. In this particular case, the retention characteristics following PROGRAM (a pulse of +50 V, 4 s) and ERASE (a pulse of -50 V, 4 s) operations have been illustrated in Figure \ref{Fig_5.6}(c). Our device exhibits a stable extinction ratio exceeding 10$^6$ after 2000 s at elevated temperatures, thereby opening up opportunities for practical device applications in innovative nonvolatile memories. Building memory devices that can operate at high temperatures offers numerous challenges, including the shortened data retention time resulting from increased thermal screening.\cite{Li2020Nov} Additionally, an augmented carrier trapping and de-trapping leads to significant off-state currents, which results in lower on/off ratios.\cite{Watson2015Dec} To our knowledge, no report on such high-temperature retention characteristics with a larger extinction ratio is available. Furthermore, we investigate the programmable cyclic endurance of the device by applying alternating positive and negative pulses. The applied periodic waveform includes a PROGRAM (5 V, 1 s) pulse, followed by a READ (0 V, 1 s) pulse, and then an ERASE (–40 V, 1 s) pulse followed by another similar READ pulse, as shown in Figure \ref{Fig_5.6}(d). The high/low conductive READ currents remain nearly invariant even after subjecting the devices to 400 periodic cycles, indicating the exceptional endurance properties of these memory devices. Here, the novel mechanism involved in the 2D WS$_2$ nonvolatile memory devices opens up broader possibilities to use 2D semiconductors for robust data storage applications.
\section{Dynamic response and paradigm of multibit storage}
\begin{figure} [hbt!]
	\centering
	\includegraphics[width=1.0\textwidth]{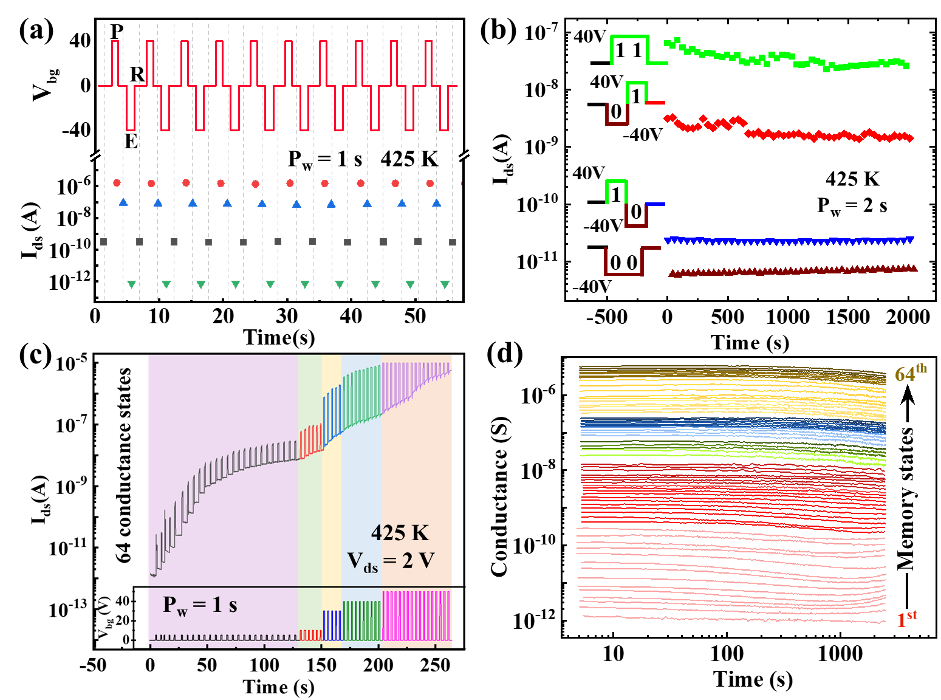}
	\caption{The dynamic response of ML WS$_2$ mem-transistors based multibit storage capabilities: (a) Switching between PROGRAM (P, high current, device ON, red circles) and ERASE (E, low current, device OFF, green triangles) states (lower panel) induced by the application of alternating V$_{bg}$ pulses ($\pm$40 V,1 s), upper panel. The READ (R) operations are performed by using V$_{bg}$ = 0 V. The high (low) conductive READs are blue triangles (black squares). (b) V$_{bg}$ pulses used for programming the mem-transistor into 4 distinguishable storage states (‘00’, ‘01’, ‘10’, ‘11’), i.e., a 2-bit memory device. Retention characteristics of 4 states are also shown for 2000 s at V$_{bg}$= 0 V and V$_{ds}$= 1 V.  (c) Transient drain current memory operations by using a combinational gate pulse of various amplitudes, i.e., 5 V, 10 V, 30 V, 40 V, and 50 V, displaying over 64 distinct current levels. (d) The retention of 64 conductance states for more than 1000 s, taken at V$_{bg}$ = 0 V, V$_{ds}$ = 2 V.}
	\label{Fig_5.7}
\end{figure}
\subsection{Single memory cell}
Multibit memories, featuring the capacity to stockpile over a solitary bit of data within a single device, have been widely acknowledged as an effective mechanism to amplify the data retention ability in forthcoming miniaturized electronics. To examine the dynamic response of our memory device, we alter the V$_{bg}$ pulses in a programmable operation, fluctuating between $\pm$40 V for a duration of 1 s, as depicted in Fig.\ref{Fig_5.7}(a). The switching process between READ states subsequent to each PROGRAM/ERASE cycle is accomplished via the attrition of Na$^+$ ions in the vicinity of the WS$_2$/SiO$_2$ interface. At the same time, the drain is grounded, and the source is biased at 1 V. The device is first set to the OFF state ($\sim$10$^{-10}$ A) by implementing a high negative gate voltage (V$_{bg}$= - 40 V) pulse. To activate the device, a positive voltage pulse of 40V for 1 s causes attrition of charges in the WS$_2$ channel, signified by the $\sim$10$^{-6}$ A current surge. Upon resetting the gate voltage to 0 V, the device persists in a high conductive READ level with a steady OFF current of $\sim$10$^{-8}$ A. The imposition of a similar negative pulse of -40V for 1 s restores the initial OFF state ($\sim$10$^{-10}$ A), also called a low conductive READ state. 
\subsection{2-bit storage}
To attain a storage capacity of n-bits, a mem-transistor must comprise no less than 2$^n$ distinct storage levels.\cite{Chen2014Apr} These devices can be exploited for a 2-bit storage potential that mandates 4 data levels. Figure \ref{Fig_5.7}(b) depicts four V$_{bg}$ operations implemented to configure our mem-transistor into 4 data states (‘00’, ‘10’, ‘01’, ‘11’). The V$_{bg}$ pulses denoted by ‘00’ and ‘11’ correspond to the pulses employed to establish the device into binary ‘ERASE’ and ‘PROGRAM’ states, respectively. However, the V$_{bg}$ pulses for collecting the ‘01’ and ‘10’ states are combinational, consisting of two successive pulses of  (-40 V, 1 s) and (40 V, 1 s) and vice versa. These operations generate two additional intermediate PROGRAM states with stable READ currents traversing four orders of current magnitudes across the spectrum. The conservation of the corresponding READ states is additionally probed for 2000 s at a steady drain-source bias of 2 V, as depicted in Fig.\ref{Fig_5.7}(b). 
\subsection{6-bit storage}
By scrutinizing their time-dependent behavior, it can be inferred that ion-gated transistors based on 2D materials can cater to the pressing need for high-temperature multi-level storage devices. The extensive extinction ratios and retention capabilities of these ionotronic WS$_2$ mem-transistors signify the potential of multibit memory with exceptional storage capacities. Although floating-gate multibit memories based on 2D materials, such as MoS$_2$, manifest high switching ratios exceeding 10$^6$, the ungovernable laws of the charge trapping mechanism lead to a restricted storage capability with less than 16 storage levels (4 bits).\cite{Zhang2021Jul, Li2022Aug} Nevertheless, the electrolyte-gated transistors exhibit enhanced conductance modulation due to their ion-gating mechanism compared to the electrostatic charge trap phenomena, rendering them suitable contenders for high storage densities.\cite{Bisri2017Jul, Lee2020Oct} We execute several voltage operations to demonstrate the multibit storage potential of our solid-state ion-gated WS$_2$ mem-transistors. Initially, the device is subjected to repetitive stimulation by a gate pulse of (5 V, 1 s) duration. The phenomenon observed is the gradual increase in storage current with increasing the number of pulses indicating the persistent accumulation of charge carriers in WS$_2$. Furthermore, we utilize combinational gate pulses of various amplitudes, including 5 V, 10 V, 30 V, 40 V, and 50 V, as depicted in Fig. c (c). This approach yielded 65 distinct storage currents, spanning over six orders of magnitude of current levels, thereby revealing the potential for 6-bit data storage capabilities in our ion-gated WS$_2$ mem-transistors compared to existing literature.\cite{Yang2019Jan, Tran2019Feb, Huang2019Sep, Gao2021Nov, Liu2021Jun, Zhang2021Jul} Our results, combined with recent studies on multibit memory that use two- and three-terminal device configurations on 2D materials and heterostructures, have been consolidated in Table \ref{table 5.1}. The table highlights various memory parameters such as retention, extinction ratios, and thermal stabilities.
\subsection{Signal to noise ratio}
\begin{figure}
	\centering
	\includegraphics[width=0.95\textwidth]{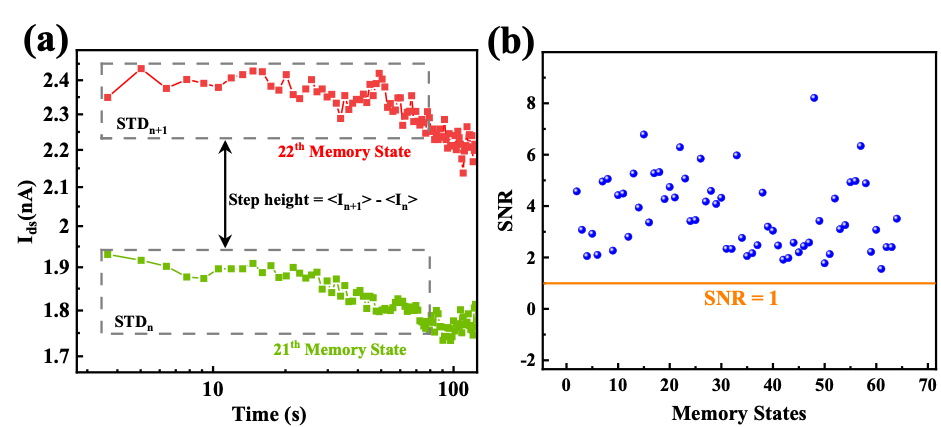}
	\caption{(a) Calculated standard deviation between storage currents states of 28$^{th}$ and 29$^{th}$.(b) Signal-to-noise ratio for all the storage states.}
	\label{Fig_5.16}
\end{figure}
Notably, our study presents, for the first time, the capability of achieving such high data storage states ($>$ 6 bits) with intriguing memory characteristics at high temperatures (425 K) using NaCl-assisted CVD-grown monolayer WS$_2$. Figure \ref{Fig_5.7}(d) depicts the stable nonvolatile retention characteristics of 64 storage states, even after 1000 s of programming using combinational gate pulses. We explicitly observe 64 levels spread across six orders of current, with discernible gaps between consecutive levels. To further confirm the credibility of all the storage states, we calculated the signal-to-noise ratio (SNR) as illustrated in Fig.\ref{Fig_5.16}. All the computed ratios surpass the critical limit of SNR = 1, anticipating the validity of the 6-bit ionotronic memory in the present study. The device's unwavering dependability and thermal stability can be attributed to the highly effective conductance modulation through ion-gating. Such reliable multi-bit data storage is highly desirable in high performance neuromorphic computing applications.\cite{Li2020Nov}
\begin{table}[hbt!]
\caption{Comparison list of memory parameters of various multibit memory device configurations using 2D materials and heterostructures in recent years.}
\label{table 5.1}
\centering{}
\begin{threeparttable}
\renewcommand{\tabcolsep}{0.07cm}
\renewcommand{\arraystretch}{2.0}
\scalebox{0.75}{
\begin{tabular}{|l|l|l|l|l|l|l|}
\hline
\hspace{0.4cm} Device structure &
  \hspace{1cm}Active material &
  \begin{tabular}[x]{@{}c@{}} Extinction \\ ratio
  \end{tabular} &
  Retention (s) &
  \begin{tabular}[x]{@{}c@{}} Memory states \\ ($\#$ of bit) \end{tabular} &
  \begin{tabular}[x]{@{}c@{}} High temp. \\ application \end{tabular} &
  References \\ \hline
Ionotronic transistor & CVD-grown WS$_2$ & \hspace{0.6cm}10$^6$ & \hspace{0.6cm}$>$10$^3$ & \hspace{0.5cm}64 (6-bit) & \hspace{0.5cm}150 $^{\circ}$C & This work \\ \hline
\hspace{1cm} FGM & \begin{tabular}[c]{@{}l@{}}MoS$_2$/hBN/MoS/GDYO/\\ WSe$_2$\end{tabular} & \hspace{0.6cm}10$^7$ & \hspace{0.6cm}$>$10$^4$ & \hspace{0.5cm}8 (3-bit) & \hspace{0.8cm}--- & \hspace{0.6cm}\cite{Li2022Aug} \\ \hline
\hspace{1cm} FGM          & MoS$_2$/hBN/2D-RPP          & \hspace{0.6cm}10$^4$ & \hspace{0.6cm}$>$10$^3$ & \hspace{0.5cm}22 (4-bit) & \hspace{0.8cm}---     & \hspace{0.6cm}\cite{Lai2022May} \\ \hline
\hspace{1cm} FGM          & MoS$_2$/hBN/Gr              & \hspace{0.6cm}10$^2$ & \hspace{0.6cm}$>$10$^4$ & \hspace{0.5cm}4 (2-bit)  & \hspace{0.8cm}---     & \hspace{0.6cm}\cite{Gwon2021Oct} \\ \hline
\hspace{1cm} CTM          & MoS$_2$/hBN/GDY/Gr          & \hspace{0.6cm}10$^6$ & \hspace{0.6cm}$>$10$^4$ & \hspace{0.5cm}6 (2-bit)  & \hspace{0.8cm}---     & \hspace{0.6cm}\cite{Zhang2021Jul} \\ \hline
\hspace{1cm} CTM          & MoS$_2$/GDY                 & \hspace{0.6cm}10$^8$ & \hspace{0.6cm}$>$10$^4$ & \hspace{0.5cm}10 (3-bit) & \hspace{0.8cm}---     & \hspace{0.6cm}\cite{Wen2021Nov} \\ \hline
\hspace{1cm} OM           & MoS$_2$/BP/MoS$_2$             & \hspace{0.6cm}10$^7$ & \hspace{0.6cm}$>$10$^4$ & \hspace{0.5cm}11 (3-bit) & \hspace{0.8cm}---     & \hspace{0.6cm}\cite{Liu2021Jun} \\ \hline
\hspace{1cm} FGM          & SnS$_2$/hBN/Gr             & \hspace{0.6cm}10$^6$ & \hspace{0.6cm}$>$10$^3$ & \hspace{0.5cm}50 (5-bit) & \hspace{0.8cm}---     & \hspace{0.6cm}\cite{Gao2021Nov} \\ \hline
\hspace{0.6cm} Tribo-tronic & MoS$_2$/hBN/MoS$_2$            & \hspace{0.6cm}10$^5$ & \hspace{0.6cm}$>$10$^3$ & \hspace{0.5cm}14 (3-bit) & \hspace{0.8cm}---     & \hspace{0.6cm}\cite{Jia2021May} \\ \hline
\hspace{1cm} FGM          & MoS$_2$/hBN/Gr              & \hspace{0.6cm}10$^6$ & \hspace{0.6cm}$>$10$^4$ & \hspace{0.5cm}13 (3-bit) & \hspace{0.8cm}---     & \hspace{0.6cm}\cite{Huang2019Sep} \\ \hline
\hspace{1cm} CTM          & MoS$_2$/SWCNT network       & \hspace{0.6cm}10$^6$ & \hspace{0.6cm}$>$10$^3$ & \hspace{0.5cm}10 (3-bit) & \hspace{0.8cm}---     & \hspace{0.6cm}\cite{Yang2019Jan} \\ \hline
\hspace{0.6cm} OM CTM       & MoS$_2$/hBN/Gr              & \hspace{0.6cm}10$^6$ & \hspace{0.6cm}$>$10$^4$ & \hspace{0.5cm}18 (4-bit) & \hspace{0.8cm}---     & \hspace{0.6cm}\cite{Tran2019Feb} \\ \hline
\hspace{1cm} CTM          & MoS$_2$/PbS                 & \hspace{0.6cm}2      & \hspace{0.6cm}$>$10$^4$ & \hspace{0.5cm}4 (2-bit)  & \hspace{0.8cm}---     & \hspace{0.6cm}\cite{Wang2018Apr} \\ \hline
\hspace{0.6cm} Memristor    & Gr/MoS$_{2–x}O_x$/Gr           & \hspace{0.6cm}10     & \hspace{0.6cm}$>$10$^4$ & \hspace{1cm}---          & \hspace{0.5cm}340 $^{\circ}$C & \hspace{0.6cm}\cite{Wang2018Feb} \\ \hline
\hspace{1cm} OM           & MoS$_2$/plasma-treated SiO2 & \hspace{0.6cm}4700   & \hspace{0.6cm}$>$10$^4$ & \hspace{0.5cm}8 (3-bit)  & \hspace{0.8cm}---     & \hspace{0.6cm}\cite{Lee2017Mar} \\ \hline
\hspace{1cm} OM           & MoS$_2$/cPVP/AuNPs          & \hspace{0.6cm}10$^7$ & \hspace{0.6cm}$>$10$^4$ & \hspace{0.5cm}8 (3-bit)  & \hspace{0.8cm}---     & \hspace{0.6cm}\cite{Lee2016Nov} \\ \hline
\hspace{0.6 cm} Transistor   & Plasma-treated MoS$_2$      & \hspace{0.6cm}10$^3$ & \hspace{0.6cm}$>$10$^4$ & \hspace{0.5cm}8 (3-bit)  & \hspace{0.8cm}---     & \hspace{0.6cm}\cite{Chen2014Apr} \\ \hline
\end{tabular}}
\begin{tablenotes}
\setlength\labelsep{0pt}
\item[] 
FGM floating gate memory, CTM charge trap-based memory, OM optoelectronic \\ memory, GDYO graphdiyne oxide, 2D-RPP two-dimensional Ruddlesden–Popper \\ perovskite, Gr graphene, GDY graphdiyne, SWCNT single-walled carbon nanotube, \\ cPVP crosslinked poly(4-vinylphenol), AuNP metallicgold nanoparticle.
\end{tablenotes}
\end{threeparttable}
\end{table}
\section{Neuromorphic adaptation with 2D materials at high temperature}
Non-volatile memory (NVM) devices present a potential application as an in-memory computing element,\cite{Yang2021May} which can be accomplished via the multi-level conductance response of an NVM, providing the capability to store analog synaptic weights of an Artificial Neural Network (ANN) on-chip.\cite{Islam2019Jan} ANNs draw inspiration from the neural connectivity of the human brain, although they do not align with any particular biological learning paradigm. Notably, in-memory computing is capable of executing matrix-vector multiplication (MVM) operations, which are the primary computations utilized in the domain of Artificial Intelligence (AI).\cite{Wang2021Jun} According to this approach, a crossbar array of NVM devices can undertake the MVM operation by encoding the input vector as an analog voltage and the weight matrix as analog conductance values stored in the memory devices. It may be worth mentioning that the main focus of most of the neuromorphic computing literature is based on conductance weight updates at room temperature for online learning.\cite{Li2020NovN, Wang2019Jun, Jiang2017Aug} As the field of neuromorphic computing is growing rapidly, it is crucial to shed light on its high-temperature applications. However, a few attempts have been made to understand high-temperature neuromorphic learning aspects in integrated synaptic devices.\cite{Seo2022Oct, Wang2021Dec} In this regard, by tuning the presynaptic voltage pulses, we demonstrate near linear synaptic weight updates for high temperature brain-inspired online learning behavior in 2D material-based mem-transistor, and the results are discussed in the following section.
\begin{figure}
	\centering
	\includegraphics[width=1.0\textwidth]{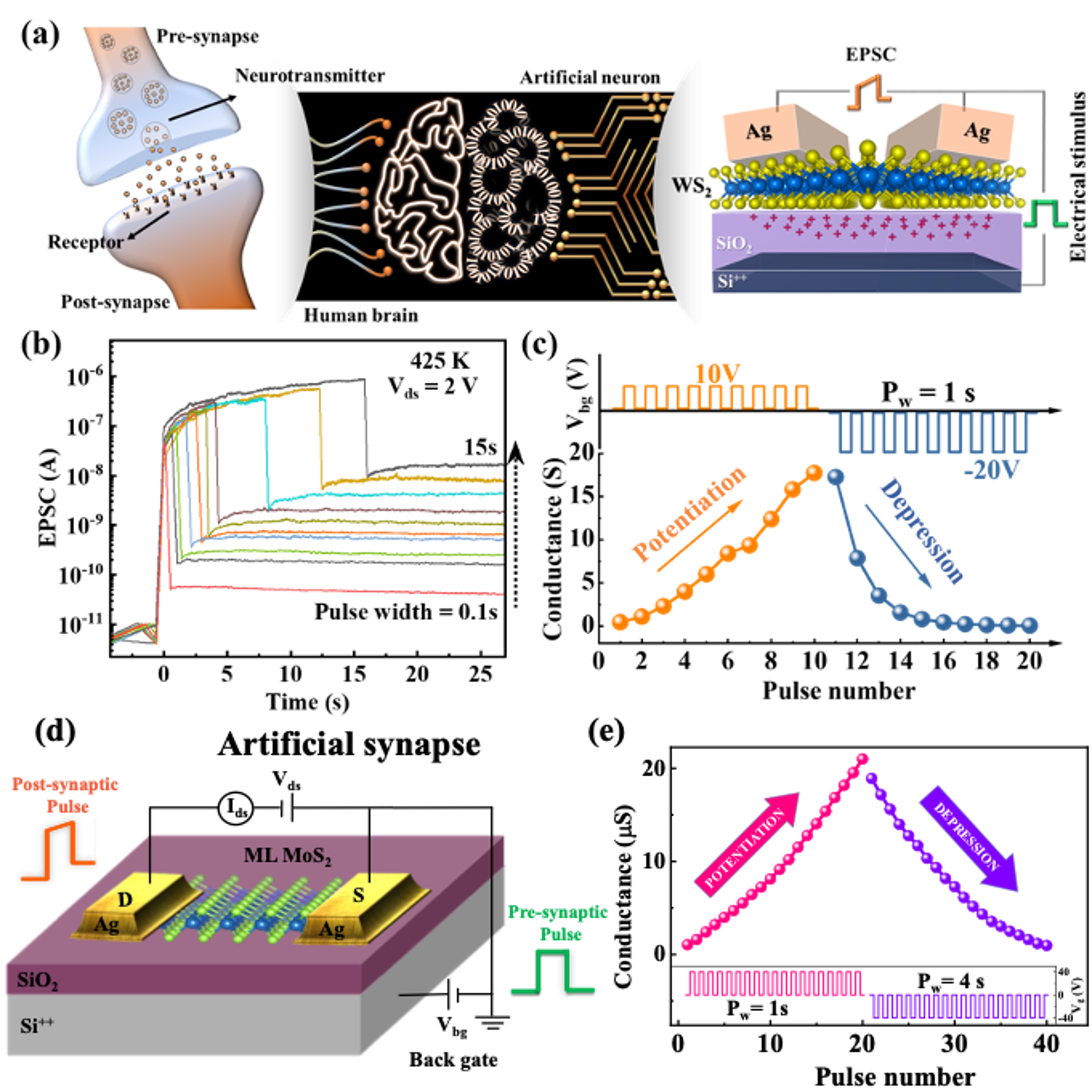}
	\caption{Synaptic characteristics of ionotronic WS$_2$ mem-transistors (a) Schematic illustration of human and artificial brain synaptic elements. (left: biological synapse operation by pre and post-synaptic neurons and right: artificial synapse operation by using the proposed ion-gated WS$_2$-based memtransistors) (b) Excitatory postsynaptic current (EPSC) stimulated by gate voltages with an amplitude of 40 V and different duration times (0.1 s to 15 s) showing long-term plasticity (c) Emulating long-term potentiation (LTP) and depression (LTD) with consecutive identical pulses of (10 V, 1 s), and (-20 V, 1 s), respectively.(d) Schematic illustration of our proposed MoS$_2$ based artificial synapse in accordance with biological response. (e) Long-term plasticity characteristics of the proposed device emulating long term potentiation (LTP) and depression (LTD) with 20 consecutive identical pulses of (40 V, 1 s), and (-40 V, 4 s), respectively.}
	\label{fig ex2}
\end{figure}
\subsection{Excitatory post synaptic current (EPSC)}
The robust multi-bit storage of our proposed WS$_2$ memtransistors, provides an upper hand in achieving reliable neuromorphic computing at high temperatures, as most previous literatures on ion-gated synaptic transistors are based on room-temperature neuromorphic performances.\cite{Jiang2017Aug, Li2022Jan} The proposed WS$_2$ mem-transistor operating at 425 K can be treated as an artificial synapse, and the structural analogy with a biological synapse is illustrated in Fig.\ref{fig ex2}(a). In biological synapses, the information transmits between presynaptic and postsynaptic neurons by sending neurotransmitters across the synapse to dock with receptors on the postsynaptic neuron. If enough neurotransmitters dock with the receptors, a signal is sent down that neuron to generate excitatory post-synaptic current (EPSC). Similarly, in the case of our artificial synaptic device, the gate acts as the presynapse to input modulatory signals through Na$^+$ diffused SiO$_2$, while the drain electrode serves as the postsynapse to measure the EPSC from the WS$_2$ channel. To probe the high-temperature neuromorphic characteristics of WS$_2$ mem-transistors, sequential voltage pulses of equivalent magnitudes (40 V) and diverse time scales (from 0.1 to 15 s) are exerted on the gate electrode as an external action potential at 425 K. The corresponding excitatory postsynaptic current (EPSC) is measured at a fixed V$_{ds}$ of 2 V. As shown in Fig.\ref{fig ex2}(b), positive voltages applied to the gate terminal significantly increase synaptic weights due to the accumulation of Na$^+$ ions at the SiO$_2$/WS$_2$ interface.  The EPSCs exhibit an incremental peak value as the duration of the voltage pulses increases in a manner similar to that of biological excitatory synapses. Moreover, a linear correlation is observed between the duration of the pulse width and the EPSC response. Furthermore, we observe negligible decay in post-stimulated read currents attributed to the robust non-volatile EPSC response at 425 K from these ion-gated WS$_2$ mem-transistors. 
\subsection{Potentiation and depression}
Long-term potentiation (LTP) and depression (LTD), essential for implementations of ANNs in neuromorphic learning, are also emulated at high temperatures by tuning the conductance states of our ionotronic memory devices with repetitive presynaptic stimulation pulses. As depicted in Fig.\ref{fig ex2}(c), the potentiation triggered using successive voltage pulses of 10 V exhibits near linear synaptic weight changes even at elevated temperatures, which entails the robust ion-gating mechanism in our case. Similarly, a depression characteristic is emulated by employing a series of voltage pulses of -20 V, illustrating an inhibitory post-synaptic current behaviour. In this process, we anticipate decumulations of Na$^+$ back to the original state under negative gate bias. The proposed ionotronic device based on ML MoS$_2$ can function as an artificial synaptic transistor, emulating the biological synapse, as presented in Figure \ref{fig ex2}(d). MoS$_2$ plays the role of the synapse, where the back-gate voltage is considered as the presynaptic stimulation, and the drain/source electrode acts as the postsynaptic terminal responsible for accumulating EPSC. Figure \ref{fig ex2}(e) depicts the change in channel conductance with the number of repetitive presynaptic stimulation pulses. To imitate long-term synaptic plasticity in our device, identical potentiation (consisting of voltage pulses with amplitude of 40 V and width of 1 s) and depression (consisting of voltage pulses with amplitude of -40 V and width of 4 s) pulses are consecutively implemented to the gate terminal for 20 times as a presynaptic spike. To read the conductance states, V$_{ds}$ is fixed at 0.1 V during these operations. 
\section{Image recognition of 2D memtransistor based synaptic devices}
\begin{figure}
	\centering
	\includegraphics[width=0.88\textwidth]{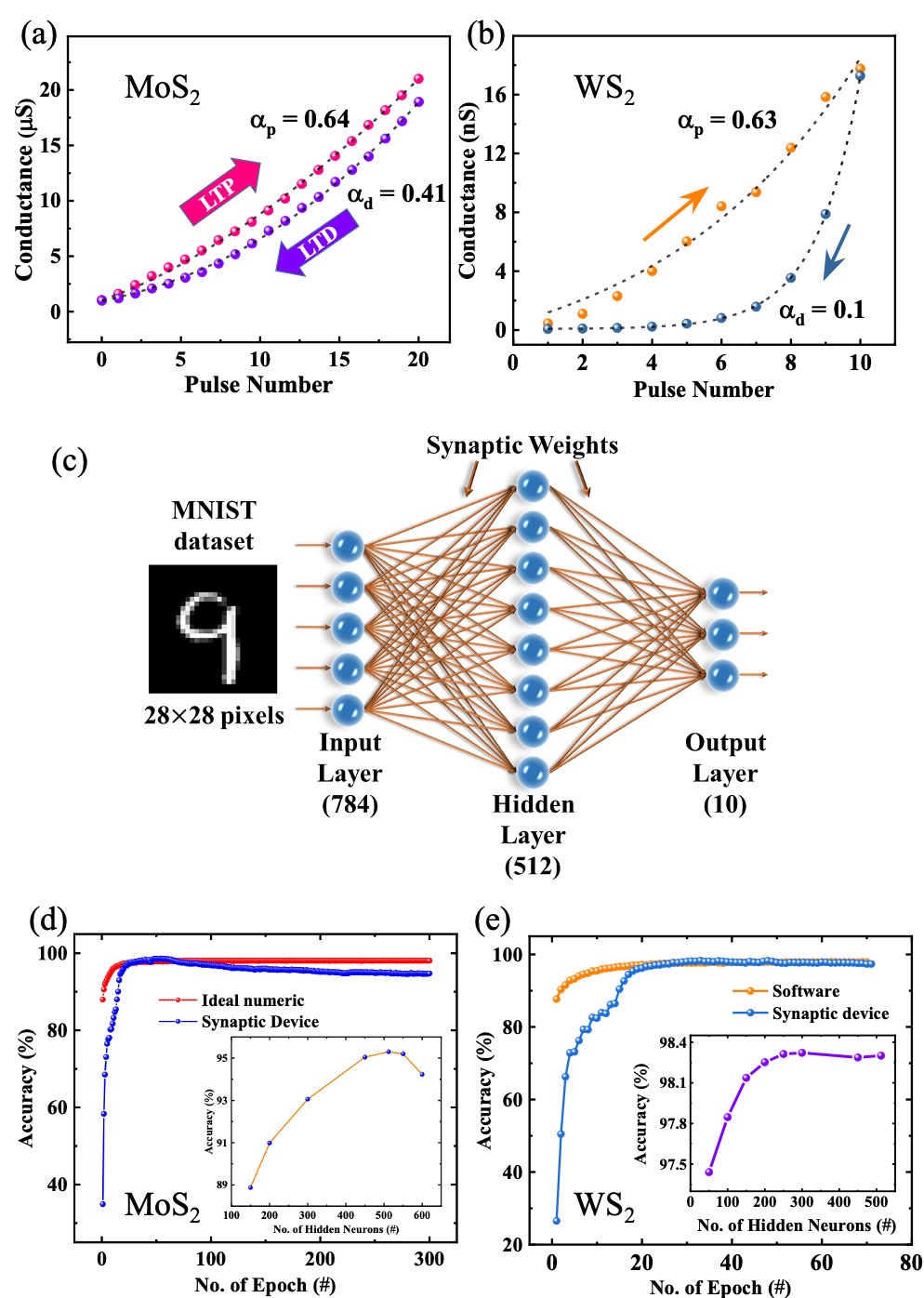}
	\caption{Hand-written digital image recognition of WS$_2$ memory devices using ANN. The nonlinearity factors ($\alpha_p$ and $\alpha_d$) are extracted from LTP and LTD characteristics of (a) MoS$_2$, and (b) WS$_2$ based synaptic transistor. (c) The schematic of a three-layer ANN structure consisting of the input layer (784 nodes), hidden layer (512 nodes), and output layer (10 nodes) for recognition of MNIST images.  The recognition accuracy of our synaptic device with respect to training epoch in case of (d) MoS$_2$, and (e) WS$_2$. Software limit is shown for comparison. Inset shows the relation between accuracy with increasing nodes in hidden layers.}
	\label{fig ex3}
\end{figure}

As shown in Fig.\ref{fig ex2}, our proposed 2D synaptic transistors show linear and symmetrical conductance updates. The dynamics of potentiation and depression, specifically the dynamic range, linearity, and asymmetry, play a pivotal role in ensuring the precision of learning and recognition simulation.\cite{Park2023Mar} The outstanding linearity in the case of MoS$_2$ mem-transistors could provide a potential opportunity to achieve high classification accuracy in ANN training. We have attained a dynamic range of conductance ratio [$G_{\rm Max}$/$G_{\rm Min}$] equal to $\sim$20, which may be deemed as the on/off ratio value for accomplishing superior performance in ANN tasks. We ascribe the enhanced linearity of the conductance updates to the ion-driven charge transfer, storage, and release. Additionally, we employ a synaptic model to replicate the LTP and LTD features.\cite{Hao2021Feb} Fig.\ref{fig ex3}(a-b) shows the nonlinearity factors of LTP and LTD characteristics of MoS$_2$, and WS$_2$ memtransistors which are extracted using the following equation.
\begin{equation}
    G =\left\{
    \begin{array}{l l}
    ((G^\alpha_{\rm Max} - G^\alpha_{\rm Min})\times\omega + G^\alpha_{\rm Min})^{1/\alpha}, & \quad \mbox{if $\alpha \neq 0$}\\\\
    G^\alpha_{\rm Min} \times (G_{\rm Max}/G_{\rm Min})^\omega, & \quad \mbox{if $\alpha=0$} 
    \end{array}
    \right.
\end{equation}
where $G_{\rm Max}$ and $G_{\rm Min}$ represent maximum and minimum conductances, $\omega$ is an internal variable and $\alpha_p$, $\alpha_d$ are nonlinearity coefficients of potentiation and depression, respectively. As shown in Fig.\ref{fig ex3}(a), the values of $\alpha_p$, and $\alpha_d$ are 0.64 and 0.41, respectively for MoS$_2$ memtransistors. In case of WS$_2$, the nonlinearity coefficients obtained from the fitted LTP and LTD curves are 0.63 and 0.1, respectively. It is important to note that the value of $\alpha$ = 1 for ideal device cases. These values are key metrics for evaluating the high classification accuracy rate in neural network simulations.\cite{Jena2023Jan}
\subsection{Artificial neural network simulation}
Based on excellent memory capabilities with key synaptic performances, we explore the potency of our 2D material-based memory devices for accurate image recognition by conducting ANN simulations using the open-source Pytorch package.\cite{Paszke2019}  In this approach, a three-layer neural network is constructed to train and test the hand-written Modified National Institute of Standards and Technology (MNIST) data sets,\cite{Deng2012Oct} as shown in Fig.\ref{fig ex3}(c). The training images are divided into batches of 32, with each batch image linearized to a 784$\times$1 input matrix of pixel intensities normalized to [0, 1]. The input, hidden, and output layers are fully connected through 784$\times$512$\times$10 synaptic weights. The 784 input neurons correspond to the 28$\times$28 pixels input MNIST image, whereas the 10 output neurons correspond to the digit ranging from “0” to “9”. Moreover, The multi-bit memory states of our ionotronic device are treated as synaptic weights during the training of 60,000 data sets. The backpropagation algorithm updates the synaptic weights in our ANN module with cross-entropy loss as the cost function at zero bias condition. A nonlinear rectified linear unit (ReLU) activation function is used during information propagation from one layer to another. Each training session is repeated 5 times, and the mean accuracy is plotted against the number of epochs. As shown in Fig.\ref{fig ex3}(d), the MoS$_2$ synaptic device achieves a maximum training accuracy of $\sim$98$\%$ and eventually drops to a stable limit of $\sim$95$\%$ for the pattern recognition task at the end of the training cycles, which strongly suggests the neuromorphic adaptation of our artificial synaptic transistors. The inset displays the relationship between the number of hidden nodes and accuracy at 200 epochs. As shown in the graph, an increase in hidden nodes leads to a higher recognition rate, with a maximum rate achieved at 512 nodes. Similarly, as displayed in Fig.\ref{fig ex3}(e), the validation accuracy reached over 90 after training for 20 epochs establishing an excellent recognition rate using our proposed WS$_2$ memory devices. Moreover, the maximum accuracy approaches the computational limit of $\sim$98, which delineates excellent neuromorphic adaptation with superior near-ideal synaptic characteristics in our case, which are also consistent with previous reports.\cite{Jena2023Jan} The inset shows the improvement of test accuracy with an increasing number of neurons in hidden layers at a specific epoch. These findings indicate the potential of ion-driven MoS$_2$-based transistors for building multilevel in-memory synapse arrays, facilitating complex data processing tasks. 
\subsection{Confusion matrix}
\begin{figure} [hbt!]
	\centering
	\includegraphics[width=0.6\textwidth]{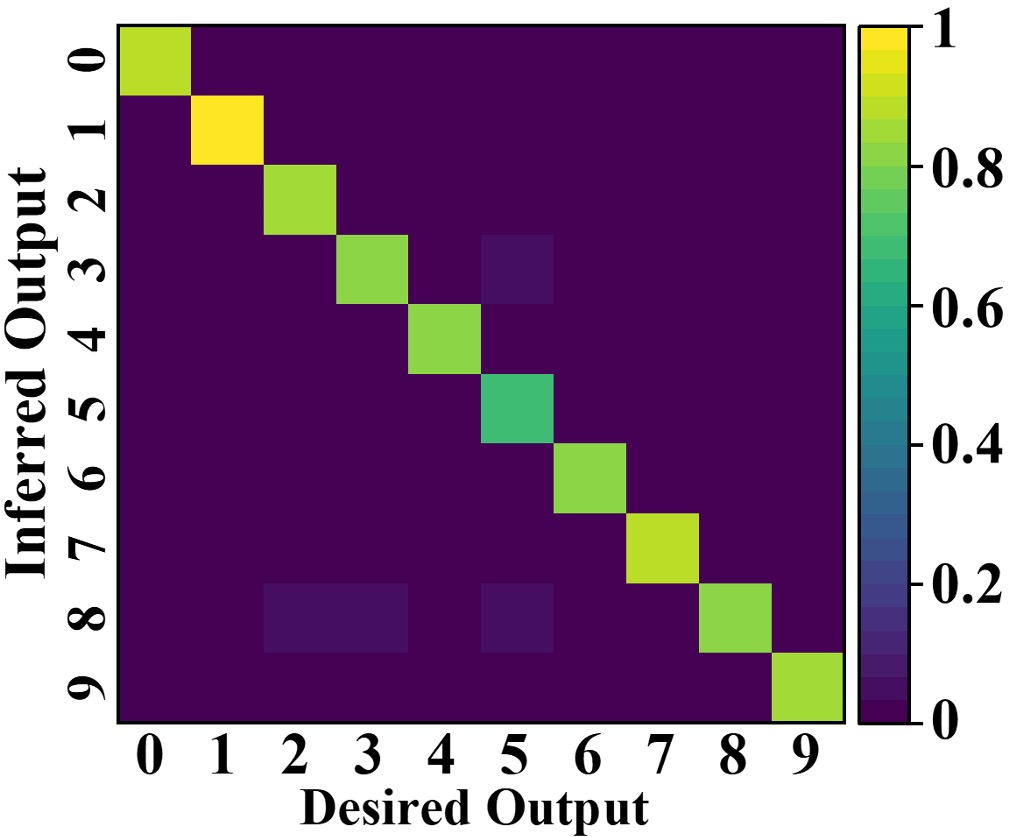}
	\caption{Confusion matrix showing the digit classification results for hand-written MNIST training datasets. The diagonal form represents most of the digits are correctly classified.}
	\label{fig ex4}
\end{figure}

To further support the high accuracy obtained in our pattern recognition task, we have computed the confusion matrix in case of WS$_2$ mem-transistors, shown in Fig.\ref{fig ex4}. The resulting confusion matrix can be used to evaluate the performance of our ANN model in terms of output class separability. The diagonal elements denote the normalised ANN model predictions that match the true labels in the test data. The observed high diagonal and low off-diagonal values clearly indicate the class separation capability of our device-based ANN implementation. The proposed ion-gated 2D material mem-transistors in this work offer potential avenues for employing other 2D materials in high-temperature multifunctional nonvolatile memories with extensive storage capacities and future neuromorphic computing applications.
\section{Chapter summary}
In this chapter, we demonstrate a novel mem-transistor device based on monolayer WS$_2$ that addresses more than 64 storage states (6-bit memory) and synaptic operations, emulating biomimetic plasticity. The direct salt-assisted synthesis of monolayer WS$_2$ on SiO$_2$/Si$^{++}$ substrate using a cost-effective CVD method has the potential for scalability. In addition, it is advantageous to directly fabricate three-terminal 2D memory devices with robust ion-gating mechanisms and device functionalities. The Na$^+$ diffused SiO$_2$ can be treated as an ionic gating medium instead of dielectric gating at high temperatures due to the thermal activation of mobile Na$^+$ ions. Using band-bending characteristics, we employ a fingerprint mechanism to address the carrier dynamics under electrostatic doping fluctuations induced by local ion movements. The WS$_2$-based ionotronic memory is equipped with endurance tests ($>$ 400 pulse cycles), 10$^5$ extinction ratio with stable retention, low device-to-device variations, and 64 distinct nonvolatile storage levels with reliable signal-to-noise ratios using combinational pulses. We also demonstrate the high-temperature neuromorphic adaptation of the proposed ionotronic 2D material-based memory devices such as the excitatory postsynaptic current with varying time scales, near linear potentiation, and depression for successful implementation in ANNs. We anticipate that the simple and cost-effective synthesis and fabrication methods presented in this work would substantially enhance the distinctive attributes of emerging layered 2D materials, thus representing a remarkable potential for their advancement as scalable memory solutions. The associated bio-inspired synaptic capability delineates a viable path to constructing next-generation 2D mem-transistors towards advanced computing platforms at high temperatures.
\chapter{Two dimensional dilute magnetic semiconductors for spintronics applications: A first principle study}
\label{C6} 
\section{Introduction}
Renewed interests in dilute magnetic semiconductors (DMS)\cite{Sato2010May} based on (non)-magnetic ion doping in 2D materials such as graphene,\cite{Krasheninnikov2009Mar, Santos2008Nov} phosphorene,\cite{Yu2016Dec} h-BN,\cite{Ataca2010Oct} and transition metal dichalcogenides (TMDCs)\cite{Cheng2013Mar, Mishra2013Oct, Karthikeyan2019Jul, Zhang2020Dec, Yun2020May} have been the focus of extensive research for the past few years to meet several applications in nanoelectronics and spintronics.\cite{Liu2020Dec} The criteria for strong ferromagnetism and large magnetic anisotropy in 2D DMS demands investigations of magnetic impurities along with various doping strategies. So far, a variety of methods have been performed to induce strong and stable magnetism in non-magnetic semiconductors such as defect configurations,\cite{Cai2015Feb, Mathew2012Sep, Zhang2013Dec} adatom adsorption,\cite{Ataca2011Jul} hydrogenation and strain engineering, \cite{Shi2013Nov} transition metal (TM) doping,\cite{Huang2013Jul, Yang2016Apr, Xiang2015Jun, Huang2017Mar, Kang2020Dec, Fu2020Apr} etc. However, the doping of TM elements such as Mn, Co, Fe, etc. is proved to be the most efficient and facile one when it comes to the realization of DMS.\cite{Karthikeyan2019Jul} Additionally, in such a method, the tuning of magnetic properties is less challenging and can be controlled by factors such as doping concentrations, configurations, etc. More recently, The experimental observations of room temperature DMS in doped MoS$_2$ and WS$_2$ systems have been achieved with high structural, chemical stability, and minimal dopant aggregation providing the routes for magneto-electric and magneto-optics applications.\cite{Zhou2020Mar, Li2020Apr, Xia2016Mar}

With the development of advanced synthesis techniques, the recent experimental and theoretical studies of TMDCs van der Waals (vdW) hetero-structures have now been incorporated with modern electronics, enhanced FETs, sensors, and Li ion batteries.\cite{Huo2014Nov, Wang2016Mar, Li2020AprM} The type-II band alignment, strong interlayer coupling, and valley pseudospin offer an unprecedented platform for ultrafast charge transfer induced large valley Zeeman splitting and magnetic tuning of polarization in MoS$_2$/WS$_2$ hetero-BLs.\cite{Zhang2016Mar, Zhang2019Sep, Sahoo2021Oct} Even though these vdW heterostructures have been associated with superior electrical and optical properties than individual TMDC components, the magnetic behavior in doped hetero BLs remains an open question that needs to be addressed to expand its fundamental and technical applications in spintronics. Furthermore, the factors influencing the magnetic exchange interactions between the dopants such as doping elements, concentrations, and configurations are yet to be understood. This is natural to wonder whether the vdW hetero-BLs have the potential to be used as a 2D DMS. In this chapter, with the implementation of extensive first-principles calculations, we demonstrate the MS$_2$ (M = Mo, W) monolayers, as well as their van der Waals (vdW) hetero-bilayers as promising candidates for the successful realization of 2D DMS with the incorporation of Mn and Co dopants. Under various pairwise doping configurations at different host atom sites, we report the electronic properties modifications induced change in magnetic exchange interactions. The magnetic coupling among the dopant pairs can be tuned between ferromagnetic (FM) and anti-ferromagnetic (AFM) orderings via suitable doping adjustments. The developed interlayer exchange coupling between the vdW layers leads to strong and long-ranged FM interactions which unleash robust magnetic moments with stable doping configurations. Our findings address the novel magnetic behavior of the layered vdW heterostructures and may further guide future experimental efforts for the possible applications in modern electronics and nanoscale magnetic storage devices. 

Strain-mediated magnetism in 2D materials and dilute magnetic semiconductors hold multi-functional applications for future nano-electronics. Furthermore, lattice strain can be adapted to modulate the physical properties of TMDCs. We studied a possible emergence of ferromagnetism in Co-doped WS$_2$ ML under strain engineering by using first-principles DFT calculations and micromagnetic simulation, which have tremendous applications in TMDC-based straintronics. Co-doping marks a significant change in magnetic properties with an impressive magnetic moment of 2.58 $\mu _B$. The magnetic exchange interaction is found to be double exchange coupling between Co and W and strong p-d hybridization between Co and nearest S, which is further verified from spin density distribution. We find that the resultant impurity bands of the Co-doped WS$_2$ play the role of seed to drive novel electronic and magnetic properties under applied strain. Among several biaxial strains, the magnetic moment is found to be a maximum of 3.25 $\mu _B$ at 2 $\%$ tensile and 2.69 $\mu _B$ at -2 $\%$ compressive strain due to strong double exchange coupling and p-d hybridization among foreign Co and W/S. Further magnetic moments at higher applied strain are decreased due to reduced spin polarization. In addition, the competition between exchange splitting and crystal field splitting of Co d-orbital plays a significant role in determining these values of magnetic moments under the application of strain. From the micromagnetic simulation, we found that the Co-doped WS$_2$ monolayer shows slanted ferromagnetic hysteresis with a low coercive field. The effect of higher strain suppresses the ferromagnetic nature, which has a good agreement with the results obtained from DFT calculations. Our findings indicate that induced magnetism in WS$_2$ monolayer under Co-doping promotes the application of 2D TMDCs for the nano-scale spintronics, and especially, the strain-mediated magnetism can be a promising candidate for future straintronics applications.
\section{First principle density functional theory calculation}
The spin-polarized first-principles calculations are carried out by the density functional theory employed in the Vienna Ab initio Simulation Package (VASP).\cite{Kresse1996Oct} The electron–ion interaction is described through the projector augmented wave method (PAW) whereas the core and valence electron is approximated by Perdew-Burke-Ernzerhof (PBE) pseudopotential within the frozen core approximation.\cite{Blochl1994Dec, Kresse1999Jan} The exchange–correlation functional is employed by generalized gradient approximation (GGA) functional.\cite{Perdew1996Oct} To include the interlayer van der Waals interactions, we follow the DFT-D2 method developed by Grimme. To expand the wavefunctions as plane-wave basis sets, the kinetic energy cut-off is fixed to 450 eV. The Brillouin zone integrations are performed by the Gamma-centered Monkhorst pack k-mesh of 4$\times$4$\times$1 and 10$\times$10$\times$1 for self-consistent and projected density calculations. Doping calculations are proceeded with the 5$\times$5$\times$1 supercells of MLs and commensurate hetero-BL systems with lattice-constant of 15.82 $\AA$. Ionic relaxations of the systems are carried out by the conjugate-gradient scheme until the Hellmann-Feynman forces and energy differences are achieved up to 10$^{-3}$ eV/$\AA$ and 10$^{-6}$ eV respectively. A sufficient vacuum of 22 $\AA$ and 35 $\AA$ are imposed in the z-direction for ML and BL systems respectively, to avoid the fictitious interactions between the periodic images. In case of hetero-BL systems, the interlayer distance between the top (W-atom) and bottom (Mo-atom) layers is optimized to 6.5 $\AA$ as per the energy minimization. The charge densities are plotted using the visualization software VESTA\cite{Momma2011Dec} and the postprocessing of the VASP calculated data is carried out by VASPKIT.\cite{Wang2021Oct} The correction of a Hubbard term U (GGA + U) is necessary to describe the strongly correlated 3d orbitals of defect impurities TM element.\cite{Sahu2021Feb} However, the magnetic properties of the half-metallic doped TMDC systems are independent of U values. Our primary focus is to study the magnetic moments of these systems and the effect of the magnetic orderings with different doping configurations. However, the value of the magnetic moments remains little affected with GGA + U or hybrid method. Henceforth, to reduce the computational cost and time, we adopted the GGA method in our calculations.
\section{TM dopants in monolayer MS$_2$ (M = Mo, W)}
\subsection{Doping configuration}
\begin{figure}[hbt!]
	\centering
	\includegraphics[width=0.9\textwidth]{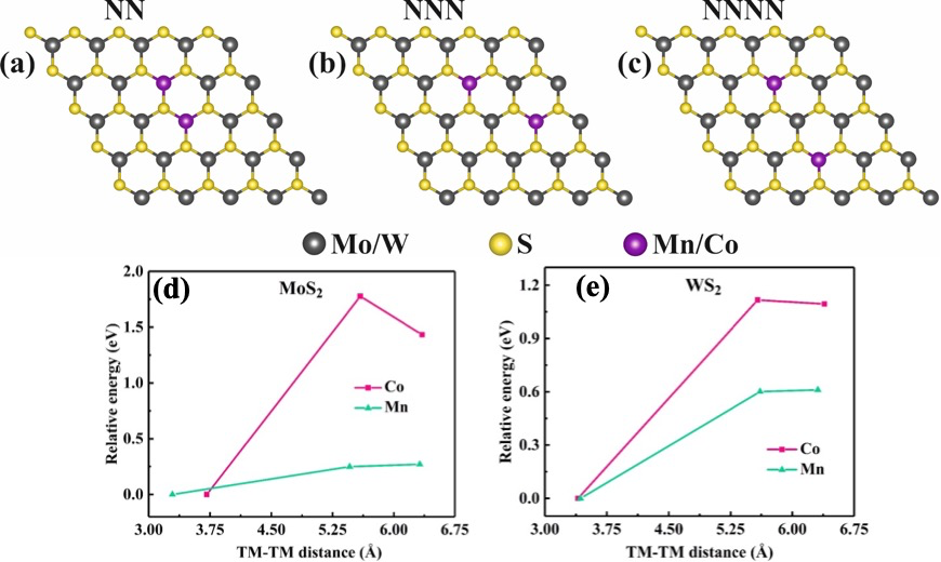}
	\caption{Top views of the 5$\times$5$\times$1 MS$_2$ supercells with TM dopants having doping concentration 8 $\%$ (a) first nearest neighbor (NN) configuration (b) second nearest neighbor (NNN) configuration (c) third nearest neighbor (NNNN) configuration. The grey, yellow and pink color atoms represent the host TM atoms, sulfur, and dopants respectively. Relative energy curves for Mn (green) and Co (pink) dopant pairs for three different configurations in (d) MoS$_2$ ML and (e) WS$_2$ ML.}
	\label{Pic_1.1}
\end{figure}

We begin our studies by analyzing the geometrical, structural, electronic and magnetic properties of TM doped ML MS$_2$ with varying doping configurations. A pair of host Mo and W atoms have been replaced with two TM elements in a 5$\times$5$\times$1 supercell of MoS$_2$ and WS$_2$ MLs respectively, representing a doping concentration of about 8 $\%$, which is sufficient to study the DMS behaviors in view of previously reported results.\cite{Cheng2013Mar, Pan2020Oct} Depending upon the separation, the two TM atoms are substituted at different M-sites generating 3 configurations as shown in Fig.\ref{Pic_1.1}(a) first nearest neighbor (NN)  (b) second nearest neighbor (NNN), and (c) third nearest neighbor (NNNN). The NN and NNNN configurations represent the TM doping in the zigzag chain whereas, in the NNN configuration, the TM pairs are placed in an armchair chain. The structural stabilities of the different doping configurations are determined by considering the relative energies of the relaxed structures of MS$_2$ MLs with respect to the TM-TM separation as performed in previous theoretical studies,\cite{Smiri2021May} which is depicted in Fig.\ref{Pic_1.1}(d-e). In the case of Mn doping, it is observed that the relative energies of both MoS$_2$ and WS$_2$ MLs have monotonous dependence with Mn-Mn distance whereas, in the case of Co doping, the NNNN configuration is seen to have slightly lower energy than NNN. This could be ascribed due to the effect of substitution in the zigzag chain over the armchair chain.\cite{Cong2015Mar} However, in all cases, the NN configurations are quite stable with the lowest energies which explain the preference for cluster formation of the dopants.\cite{Smiri2019Aug} The optimized TM-TM distances (d$_{TM1-TM2}$) for Mn and Co doping configurations in MS$_2$ MLs are listed in Table \ref{table 6.1}. It is noted that irrespective of the types of dopants, in the case of WS$_2$ MLs the (d$_{TM1-TM2}$) don’t vary substantially for respective doping configurations which are unlike in the case of MoS$_2$. This behavior is attributed to the large Jahn teller effect in MoS$_2$ MLs upon TM substitution than WS$_2$.

\begin{table}[]
\centering
\caption{The relaxed dopant pair separation d$_{TM1-TM2}$, onsite magnetic moments $\mu_{TM1}$, $\mu_{TM1}$ and total magnetic moment $\mu_{tot}$ of Mn, Co-soped MoS$_2$ (left panel) and WS$_2$ (right panel) MLs for three different doping configurations.}
\label{table 6.1}
\begin{tabular}{|llllll|llll|}
\hline
\multicolumn{6}{|l|}{Doping at Mo sites (ML-MoS$_2$)} &
  \multicolumn{4}{l|}{Doping at W sites (ML-WS$_2$)} \\ \hline
\multicolumn{2}{|l|}{Configurations} &
  \multicolumn{1}{l|}{\begin{tabular}[c]{@{}l@{}}d$_{TM1-TM2}$ \\ $(\AA)$\end{tabular}} &
  \multicolumn{1}{l|}{$\mu_{TM1}$} &
  \multicolumn{1}{l|}{$\mu_{TM2}$} &
  $\mu_{tot}$ &
  \multicolumn{1}{l|}{\begin{tabular}[c]{@{}l@{}}d$_{TM1-TM2}$\\ $(\AA)$\end{tabular}} &
  \multicolumn{1}{l|}{$\mu_{TM1}$} &
  \multicolumn{1}{l|}{$\mu_{TM2}$} &
  $\mu_{tot}$ \\ \hline
\multicolumn{1}{|l|}{\multirow{3}{*}{Mn}} &
  \multicolumn{1}{l|}{NN} &
  \multicolumn{1}{l|}{3.29} &
  \multicolumn{1}{l|}{-0.12} &
  \multicolumn{1}{l|}{-0.12} &
  0.00 &
  \multicolumn{1}{l|}{3.43} &
  \multicolumn{1}{l|}{0.00} &
  \multicolumn{1}{l|}{0.00} &
  0.00 \\ \cline{2-10} 
\multicolumn{1}{|l|}{} &
  \multicolumn{1}{l|}{NNN} &
  \multicolumn{1}{l|}{5.46} &
  \multicolumn{1}{l|}{0.82} &
  \multicolumn{1}{l|}{-0.91} &
  -0.01 &
  \multicolumn{1}{l|}{5.61} &
  \multicolumn{1}{l|}{1.22} &
  \multicolumn{1}{l|}{0.98} &
  2.05 \\ \cline{2-10} 
\multicolumn{1}{|l|}{} &
  \multicolumn{1}{l|}{NNNN} &
  \multicolumn{1}{l|}{6.32} &
  \multicolumn{1}{l|}{1.10} &
  \multicolumn{1}{l|}{1.11} &
  2.06 &
  \multicolumn{1}{l|}{6.31} &
  \multicolumn{1}{l|}{1.06} &
  \multicolumn{1}{l|}{1.07} &
  2.03 \\ \hline
\multicolumn{1}{|l|}{Co} &
  \multicolumn{1}{l|}{NN} &
  \multicolumn{1}{l|}{3.71} &
  \multicolumn{1}{l|}{0.27} &
  \multicolumn{1}{l|}{0.28} &
  1.81 &
  \multicolumn{1}{l|}{3.40} &
  \multicolumn{1}{l|}{1.07} &
  \multicolumn{1}{l|}{1.05} &
  3.51 \\ \hline
\multicolumn{1}{|l|}{} &
  \multicolumn{1}{l|}{NNN} &
  \multicolumn{1}{l|}{5.59} &
  \multicolumn{1}{l|}{1.91} &
  \multicolumn{1}{l|}{1.95} &
  5.28 &
  \multicolumn{1}{l|}{5.58} &
  \multicolumn{1}{l|}{1.95} &
  \multicolumn{1}{l|}{1.90} &
  5.30 \\ \hline
\multicolumn{1}{|l|}{} &
  \multicolumn{1}{l|}{NNNN} &
  \multicolumn{1}{l|}{6.35} &
  \multicolumn{1}{l|}{1.23} &
  \multicolumn{1}{l|}{1.22} &
  5.26 &
  \multicolumn{1}{l|}{6.39} &
  \multicolumn{1}{l|}{1.68} &
  \multicolumn{1}{l|}{1.67} &
  5.29 \\ \hline
\end{tabular}
\end{table}
\subsection{Density of states}
Moving on to the electronic properties, we investigate the spin-polarized density of states (DOS) for different atomic configurations in doped MoS$_2$ as plotted in Fig. \ref{fig:6.2-1}. Both the total and partial DOS (containing Mn-d, Mo-d, and S-p orbitals) have been shown in order to collate the contributions to the spin states arising in the bandgap region. For fair comparison, the similar PDOS calculations have been performed for ML MoS$_2$, WS$_2$, and their hetero-BLs without TM doping.\cite{KumarMallik2022Oct}
\begin{figure*}
	\centering
	\includegraphics[width=0.9\textwidth]{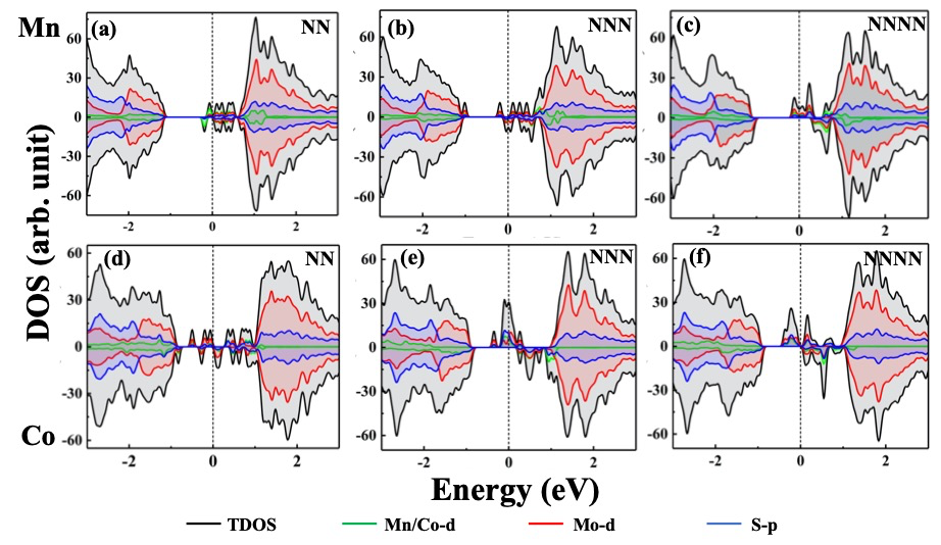}
	\caption{Total and partial density of states of the doped MoS$_2$ ML with three doping configurations; for Mn dopants (a) Mn-NN (b) Mn-NNN (c) Mn-NNNN and for Co dopants (d) Co-NN (e) Co-NNN (f) Co-NNNN. The dotted black line indicates the Fermi level which is set to energy zero.}
	\label{Pic_1.2}
\end{figure*}
It is evident that no additional spin states are observed in bandgap regions of the un-doped systems indicating their non-magnetic nature. However, in case of doped MLs, the localized spin states arise in the bandgap as depicted in Fig.\ref{Pic_1.2}. It is apparent that the spin polarization in the case of Mn:MoS$_2$ and Co:MoS$_2$ are observed to be unalike and also varies with dopant pair separations. The Co pair substitutions drastically thin down the bandgap of the MLs whereas, in the case of Mn pair substitution, a comparatively large bandgap nature is preserved. These dopant pairs are also responsible to induce spin states in their nearest neighbor S and followed by Mo atoms. One can notice highly localized spin states around the Fermi level in NN configurations for both types of dopants from Fig.\ref{Pic_1.2}(a), and (d). However, partially filled broader states can be observed for the rest of the configurations, and the Fermi levels are seen to pass through these broad bands illustrating the semi-metallic behavior of the doped system. Interestingly, in Mn:WS$_2$ system with NN configuration,\cite{KumarMallik2022Oct} though the localized spin states arise in the bandgap, the zero spin polarization indicates the complete overlapping of the majority and minority spin states leaving the system as non-magnetic. However, other configurations of doped WS$_2$ MLs favor strong magnetic orderings with fully or partially occupied majority bands of hybridized orbitals depending on the dopant pair separations.
\begin{figure}
	\centering
	\includegraphics[width=0.88\textwidth]{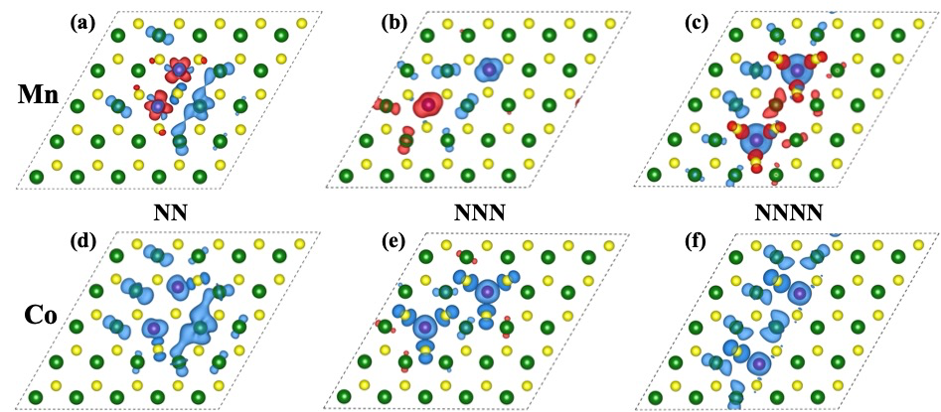}
	\caption{Top views of the isosurface plots showing the spin density distribution around the impurity pairs of the doped MoS$_2$ ML with three doping configurations; for Mn dopants (a) Mn-NN (b) Mn-NNN (c) Mn-NNNN and for Co dopants (d) Co-NN (e) Co-NNN (f) Co-NNNN. The blue and red isosurfaces represent positive and negative spin density, respectively. The value of the iso-surface is taken at 0.002 e/$\AA$$^3$. }
	\label{Pic_1.3}
\end{figure}

\subsection{Spin density distribution}
In order to further analyze the magnetic ordering and exchange couplings between the dopant pairs in doped ML systems, the iso-surface spin density ($\rho\uparrow$ - $\rho\downarrow$) distributions are investigated for different cases. According to Fang et al.\cite{Fang2017Dec} Mn dopants lead to ferromagnetic (FM) coupling in ML MoS$_2$ and magnetic coupling between them can be mediated by nearest neighbor S. However, considering the dopant configurations, the coupling between the Mn pairs is observed to be FM initially with NN configuration, followed by antiferromagnetic (AFM) in NNN and again FM in NNNN configurations as shown in Fig.\ref{Pic_1.3}(a-c). Despite being FM nature of the Mn pairs, the large AFM contribution from the sub-neighbor Mo atoms cancels the effective magnetic moments arising from the dopant pairs in NN configuration. Likewise, due to the AFM alignment of the dopant pairs in NNN configuration and similar spin density arrangements of the sub-neighbor Mo atoms, we expect zero net magnetic moment. However, in the NNNN configuration, the local spin density at the two Mn pairs is perceived to be symmetrical with FM coupling adding up the magnetic moments. The local magnetic moments of the dopant pairs and total magnetic moments are listed in Table \ref{table 6.1}. It can be inferred that the NNNN configuration of Mn:MoS$_2$ gives rise to the maximum net magnetic moment, contributed mainly from the dopant pairs. The AFM alignment of the foreign Mn pairs with the neighboring S atoms is governed by a successive spin polarization (SSP) based magnetic interaction as discussed in Andriotis et al.\cite{Andriotis2014Sep}.

The spin polarization of the S atoms are dictated by the neighboring Mn pairs as evident from Fig.\ref{Pic_1.3}(c). The induced spin polarization of the commonly shared Mo atom is found to be opposite to Mn atoms as a result of d-d electron transfer between Mo and two dopants confirming the SSP FM interaction between the impurity pairs. In contrast, for other neighboring Mo atoms, the induced spin is dictated by the polarised S atoms due to p-d hybridization. However, for all the doping configurations in the case of Co:MoS$_2$, the high parallel spin polarization of the host Mo and S atoms with foreign dopants is observed in Fig.\ref{Pic_1.3}(d-f). The long-ranged FM ordering can also be seen in the case of Co:MoS$_2$ as the distance between the dopant pairs increases. The spin density distributions of Mn and Co-doped WS$_2$ illustrate similar kinds of above-studied FM interactions between dopant pairs for NNNN configurations.\cite{KumarMallik2022Oct} The coupling between the impurities and S atoms is mostly AFM in the case of Mn doping and FM in Co doping. From Table \ref{table 6.1}, it is noted that the NN configurations in the case of both Mn and Co doping give rise to low net magnetic moments as compared to other configurations. This indicates despite the stability of the dopant ions in the host system, the magnetic exchange interaction depends on other prime factors i.e. the relative valency of the magnetic impurity with the atom it substitutes, the difference between the strength of the crystal field splitting with respect to that of the exchange splitting of the d-orbital of TM impurities.

In the previous section, we studied that the MoS$_2$ and WS$_2$ MLs upon TM doping show great potential as a 2D DMS, and the magnetic interactions between the foreign and host atoms can be modulated by varying the doping elements, and configurations. The embodiment of the vdW hetero-BLs in modern electronics with advanced synthesis techniques and improved device performances proves its multi-functionality characteristics in leading research areas. The MoS$_2$/WS$_2$ hetero-BL is becoming a new prototype vdW heterostructures owing to its large bandgap modulation, strong interlayer coupling, and valley pseudospin splitting.\cite{Zhang2019Sep} Such physical properties offer a remarkable platform to study the magnetic coupling behavior in TM doped MoS$_2$/WS$_2$ hetero-BL.
\section{TM dopants in MS$_2$ hetero-BLs (M = Mo, W)}
\begin{figure}
	\centering
	\includegraphics[width=0.9\textwidth]{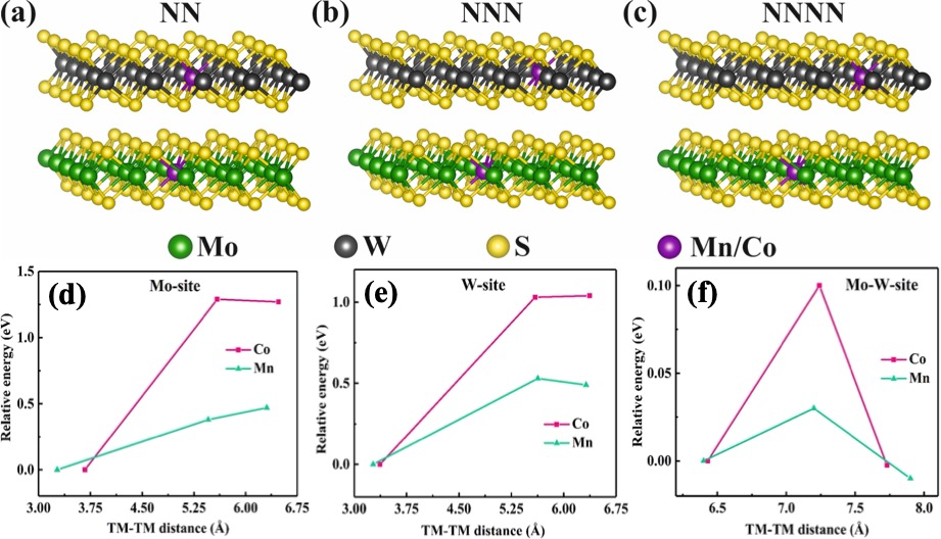}
	\caption{Side views of 5x5x1 supercell of MoS$_2$/WS$_2$ hetero-BLs with TM dopants at both Mo and W sites for (a) first nearest neighbor (NN) (b) second nearest neighbor (NNN), and (c) third nearest neighbor (NNNN) configurations. The green/grey, yellow, and pink color atoms represent the host TM atoms, sulfur, and dopants respectively. Relative energy curves for Mn (green) and Co (pink) dopant pairs for aforementioned three configurations with dopant pairs occupying (d) Mo-sites (e) W-sites, and (f) both Mo-W-sites in MoS$_2$/WS$_2$ hetero-BL.}
	\label{Pic_1.4}
\end{figure}

In this section, we present the ab initio results to study the structural, electronic, and magnetic properties of the Mn and Co-doped MoS$_2$/WS$_2$ hetero-BLs with varying doping configurations. Observing the negligible lattice mismatch ($\sim$0.01 $\%$) between the MoS$_2$ and WS$_2$ unit cells, we construct the MoS$_2$/WS$_2$ hetero-BL by considering the relaxed lattice constants of the individual MLs. To analogize with the doped ML systems, a similar doping environment is created in hetero-BLs to investigate the DMS behavior. In Case-I scenario, a pair of TM dopants are substituted in Mo(W)-sites upon removing two Mo(W)-atoms leaving the W(Mo)S$_2$ layer unperturbed in MoS$_2$/WS$_2$ hetero-BL. The Case-II and the most interesting part of the present study corresponds to the doping of TM elements at both layers which entails single substitution at upper W-site and lower Mo-site as depicted in Fig.\ref{Pic_1.4}(a-c). In both cases, similar doping arrangements are considered with NN, NNN, and NNNN configurations. However, in Case-II, the distances between the impurity pairs in all configurations are relatively larger than the respective configurations of the Case-I. Similar to ML case, here also we consider the structural stability analysis based on the relative energies of various doping configurations with dopant pair separations after the geometric relaxation, which is shown in Fig.\ref{Pic_1.4}(d-f). The doping of Mn and Co pairs at Mo and W sites in the MoS$_2$/WS$_2$ hetero-BLs are most stable at NN configurations preferring the cluster formation of the dopant pairs which is similar to the case of doping in individual MLs. Surprisingly, in Case-II (when doped at Mo-W-sites), as depicted in Fig.\ref{fig:6.4-1}(f), the relative energy values manifest a sharp contradicting trend with the increase in dopant pair separation effectuating the NNNN configurations as the most stable ones. This indicates that the cluster formation of the magnetic impurities can be eschewed by the doping of TM elements at both Mo-W-sites evincing the usual DMS behavior. The NNN configurations are found to be the least stable ones for both the foreign dopants. However, on comparing with the Case-I, the difference between the relative energies in the Case-II provides much smaller values than the former ones implying the near stable configurations. 
\begin{table}
\centering
\caption{The relaxed dopant pair separation d$_{TM1-TM2}$, onsite magnetic moments $\mu_{TM1}$, $\mu_{TM1}$ and total magnetic moment $\mu_{tot}$ of MoS$_2$/WS$_2$ hetero BL doping at Mo and W sites, separately for three different doping configurations.}
\label{table 6.2}
\begin{tabular}{|llllll|llll|}
\hline
\multicolumn{6}{|l|}{Doping at Mo sites (MoS$_2$/WS$_2$)} &
  \multicolumn{4}{l|}{Doping at W sites (MoS$_2$/WS$_2$)} \\ \hline
\multicolumn{2}{|l|}{Configurations} &
  \multicolumn{1}{l|}{\begin{tabular}[c]{@{}l@{}}d$_{TM1-TM2}$ \\ $(\AA)$\end{tabular}} &
  \multicolumn{1}{l|}{$\mu_{TM1}$} &
  \multicolumn{1}{l|}{$\mu_{TM2}$} &
  $\mu_{tot}$ &
  \multicolumn{1}{l|}{\begin{tabular}[c]{@{}l@{}}d$_{TM1-TM2}$\\ $(\AA)$\end{tabular}} &
  \multicolumn{1}{l|}{$\mu_{TM1}$} &
  \multicolumn{1}{l|}{$\mu_{TM2}$} &
  $\mu_{tot}$ \\ \hline
\multicolumn{1}{|l|}{\multirow{3}{*}{Mn}} &
  \multicolumn{1}{l|}{NN} &
  \multicolumn{1}{l|}{3.27} &
  \multicolumn{1}{l|}{1.36} &
  \multicolumn{1}{l|}{1.36} &
  2.11 &
  \multicolumn{1}{l|}{3.27} &
  \multicolumn{1}{l|}{1.22} &
  \multicolumn{1}{l|}{1.22} &
  2.06 \\ \cline{2-10} 
\multicolumn{1}{|l|}{} &
  \multicolumn{1}{l|}{NNN} &
  \multicolumn{1}{l|}{5.46} &
  \multicolumn{1}{l|}{0.87} &
  \multicolumn{1}{l|}{1.18} &
  2.04 &
  \multicolumn{1}{l|}{5.63} &
  \multicolumn{1}{l|}{-0.59} &
  \multicolumn{1}{l|}{0.71} &
  0.01 \\ \cline{2-10} 
\multicolumn{1}{|l|}{} &
  \multicolumn{1}{l|}{NNNN} &
  \multicolumn{1}{l|}{6.31} &
  \multicolumn{1}{l|}{1.07} &
  \multicolumn{1}{l|}{1.07} &
  2.06 &
  \multicolumn{1}{l|}{6.32} &
  \multicolumn{1}{l|}{-1.05} &
  \multicolumn{1}{l|}{-1.04} &
  -2.03 \\ \hline
\multicolumn{1}{|l|}{Co} &
  \multicolumn{1}{l|}{NN} &
  \multicolumn{1}{l|}{3.67} &
  \multicolumn{1}{l|}{0.28} &
  \multicolumn{1}{l|}{0.29} &
  1.82 &
  \multicolumn{1}{l|}{3.37} &
  \multicolumn{1}{l|}{1.06} &
  \multicolumn{1}{l|}{1.05} &
  3.52 \\ \hline
\multicolumn{1}{|l|}{} &
  \multicolumn{1}{l|}{NNN} &
  \multicolumn{1}{l|}{5.59} &
  \multicolumn{1}{l|}{0.78} &
  \multicolumn{1}{l|}{1.46} &
  5.36 &
  \multicolumn{1}{l|}{5.59} &
  \multicolumn{1}{l|}{1.94} &
  \multicolumn{1}{l|}{1.89} &
  5.30 \\ \hline
\multicolumn{1}{|l|}{} &
  \multicolumn{1}{l|}{NNNN} &
  \multicolumn{1}{l|}{6.48} &
  \multicolumn{1}{l|}{1.15} &
  \multicolumn{1}{l|}{1.16} &
  5.38 &
  \multicolumn{1}{l|}{6.37} &
  \multicolumn{1}{l|}{1.72} &
  \multicolumn{1}{l|}{1.72} &
  5.32 \\ \hline
\end{tabular}
\end{table}
The calculated total and onsite magnetic moments (MM) along with the optimized separation between the dopant pairs for Mo-site and W-site configurations are represented in Table \ref{table 6.2}. In the case of Mn doping at Mo-sites, the total MM is found to be higher for NN than the other two configurations and the local MMs at Mn sites contribute equally. This behavior is highly unlike with the previously discussed doped ML MoS$_2$ system, where no net magnetic ordering is observed due to quenching of MMs from the AFM alignments though the separation between the dopant pairs is nearly equal in both doped ML and hetero BL. The finite MM values in Mn substituted at Mo-sites in hetero-BLs provide necessary ferromagnetic ordering as evident from Table \ref{table 6.2}. Similarly, The Mn doping at W-sites produces a net MM of 2.06 $\mu _B$ whereas the doped WS$_2$ ML system renders its nonmagnetic nature upon Mn doping under NN configuration. This indicates the dependence of magnetic exchange coupling on the vdW stacked layers and the possible tuning of magnetic properties by suitably choosing doping sites in 2D vdW heterostructures. In the case of Co-doping, the MM values are monotonically increasing with the increase of the dopant separation for respective Mo-site and W-site configurations. Though the onsite small MMs, the net increase in MMs is observed owing to the significant contributions from the neighboring Mo/W and S atoms. In Table \ref{table 6.3}, the magnetic parameters are presented for the TM substitutions at both Mo-W-sites (Case-II). The total MM in the case of Mn doping is doused for the NN configuration leaving a nonmagnetic signature whereas it increases with Mn separation and becomes maximum at the stable NNNN configuration. However, the doping of Co at both Mo-W-sites makes the system magnetically unaltered with the dopant pair separation providing robust magnetic moments for the realization of 2D DMS.
\begin{table}
\centering
\caption{The relaxed dopant pair separation d$_{TM1-TM2}$, onsite magnetic moments $\mu_{TM1}$, $\mu_{TM1}$ and total magnetic moment $\mu_{tot}$ of Mn, Co-doped MoS$_2$/WS$_2$ hetero BL at Mo-W-sites for three different doping configurations.}
\label{table 6.3}
\begin{tabular}{|llllll|}
\hline
\multicolumn{6}{|l|}{Doping at Mo and W sites} \\ \hline
\multicolumn{2}{|l|}{Configurations} &
  \multicolumn{1}{l|}{d$_{TM1-TM2} (\AA)$} &
  \multicolumn{1}{l|}{$\mu_{TM1}$} &
  \multicolumn{1}{l|}{$\mu_{TM2}$} &
  $\mu_{tot}$ \\ \hline
\multicolumn{1}{|l|}{\multirow{3}{*}{Mn}} &
  \multicolumn{1}{l|}{NN} &
  \multicolumn{1}{l|}{6.40} &
  \multicolumn{1}{l|}{-0.91} &
  \multicolumn{1}{l|}{0.94} &
  -0.01 \\ \cline{2-6} 
\multicolumn{1}{|l|}{}   & \multicolumn{1}{l|}{NNN}  & \multicolumn{1}{l|}{7.20} & \multicolumn{1}{l|}{1.14} & \multicolumn{1}{l|}{0.73} & 1.20 \\ \cline{2-6} 
\multicolumn{1}{|l|}{}   & \multicolumn{1}{l|}{NNNN} & \multicolumn{1}{l|}{7.90} & \multicolumn{1}{l|}{1.01} & \multicolumn{1}{l|}{0.92} & 2.01 \\ \hline
\multicolumn{1}{|l|}{Co} & \multicolumn{1}{l|}{NN}   & \multicolumn{1}{l|}{6.43} & \multicolumn{1}{l|}{0.93} & \multicolumn{1}{l|}{1.87} & 5.35 \\ \hline
\multicolumn{1}{|l|}{}   & \multicolumn{1}{l|}{NNN}  & \multicolumn{1}{l|}{7.24} & \multicolumn{1}{l|}{1.44} & \multicolumn{1}{l|}{1.77} & 5.31 \\ \hline
\multicolumn{1}{|l|}{}   & \multicolumn{1}{l|}{NNNN} & \multicolumn{1}{l|}{7.73} & \multicolumn{1}{l|}{1.12} & \multicolumn{1}{l|}{1.83} & 5.26 \\ \hline
\end{tabular}
\end{table}

\begin{figure}
	\centering
	\includegraphics[width=0.9\textwidth]{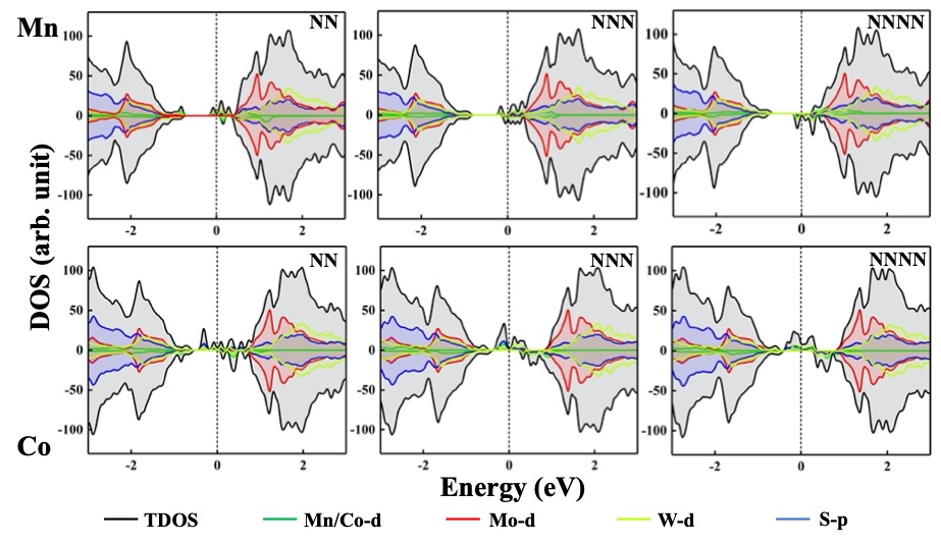}
	\caption{Total and partial density of states of the MoS$_2$/WS$_2$ hetero-BL doped at W-sites with three doping configurations; for Mn dopants (a) Mn-NN (b) Mn-NNN (c) Mn-NNNN and for Co dopants (d) Co-NN (e) Co-NNN (f) Co-NNNN. The dotted black line indicates the Fermi level which is set to energy zero.}
	\label{Pic_1.5}
\end{figure}
\subsection{Case-I: MoS$_2$/WS$_2$ hetero-BL doped in same layers}
\label{rmf_edf}
To further gain insights into the origin of magnetism and understand the dependence of magnetic behaviors with electronic structure modifications for various doping configurations in MoS$_2$/WS$_2$ hetero BLs, we perform the TDOS and PDOS calculations with high K-point grids. The spin-polarized DOS for Mn and Co doping at W-sites are represented in Fig.\ref{Pic_1.5}. The originated impurity states around the Fermi level and the asymmetry between the spin-up and spin-down states establish the net-driven magnetic moments in each case. An individual Mn atom has 4s$^2$3d$^5$ configuration contributing one additional valence electron when substituted at Mo/W sites whereas in the case of Co (with electronic configuration 4s$^2$3d$^7$) substitution three excess valence electrons govern the magnetic interactions among foreign and host atoms. In the case of Mn doping, the impurity states are observed near to the conduction band indicating the n-type doping with a finite gap from the valence band maximum. Nevertheless, the spin states corresponding to the Co doping are highly dispersed and extend from valence band edge to conduction band edge which indicates the availability of conduction charge carriers mediating the exchange interactions. The magnetic coupling behavior in a TM doped TMDCs systems can also be tuned by the introduction of cation and anion vacancies.\cite{Wang2016Apr} In such cases, the trapped electrons in the vacancies couple with the magnetic moments of the TM impurity ions form bound magnetic polarons (BMP) within the Bohr radius of the impurity site. The recently studied experimental results confirms the strong ferromagnetism induced by BMP driven by S-vacancies in MoS$_2$ nanosheet.\cite{Sanikop2020Jan} However, in our DFT calculations, the magnetic properties of TM-doped hetero-BL system have been established by considering the S-rich environment due to their lower formation energies. Henceforth, The contribution from BMP and S-deficiency-induced spin delocalization has weak effect on ferromagnetism. From the PDOS results, it is evident that in the case of doping at Mo-sites, the net magnetic contributions come from the majority (occupied) and minority (unoccupied) spin states of the foreign Mn/Co-d, host Mo-d, and neighboring S-p valence orbitals.\cite{KumarMallik2022Oct} The W-d orbital doesn’t show any spin polarization and hence no contribution to the net magnetic moments, which is similar to Mo-d valence states in the case of doping at W-sites as depicted in Fig.\ref{Pic_1.5}.
\begin{figure}
	\centering
	\includegraphics[width=0.9\textwidth]{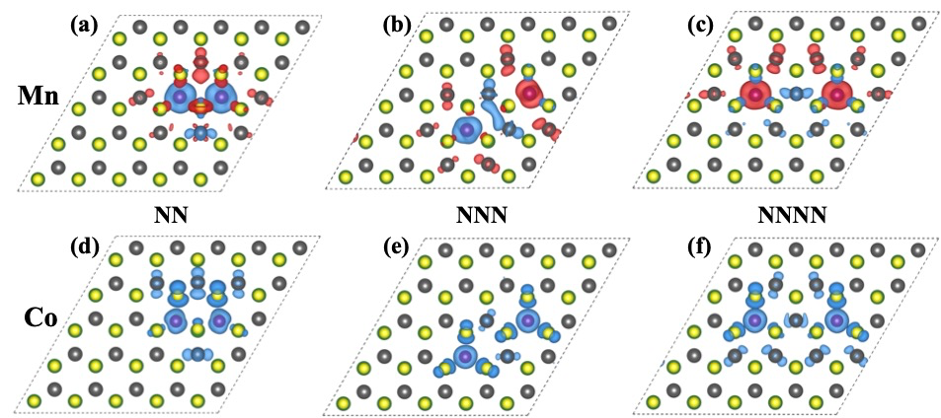}
	\caption{Top views of the isosurface plots showing the spin density distribution around the impurity pairs of the MoS$_2$/WS$_2$ hetero-BL doped at W-sites with three doping configurations; for Mn dopants (a) Mn-NN (b) Mn-NNN (c) Mn-NNNN and for Co dopants (d) Co-NN (e) Co-NNN (f) Co-NNNN. The blue and red isosurfaces represent positive and negative spin density, respectively. The value of the iso-surface is taken at 0.002 e/$\AA$$^3$.}
	\label{Pic_1.6}
\end{figure}
\begin{figure*}
	\centering
	\includegraphics[width=0.9\textwidth]{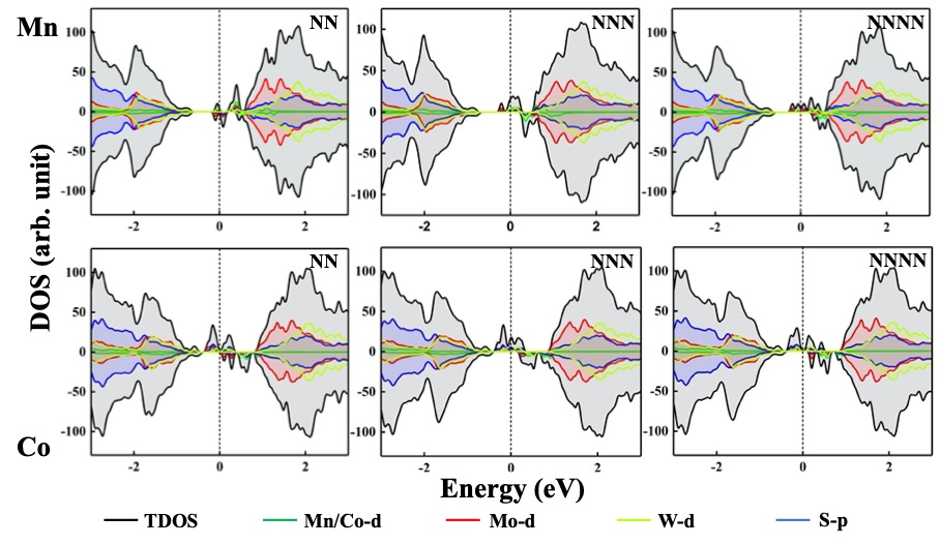}
	\caption{Total and partial density of states of the MoS$_2$/WS$_2$ hetero-BL doped at Mo-W-sites with three doping configurations; for Mn dopants (a) Mn-NN (b) Mn-NNN (c) Mn-NNNN and for Co dopants (d) Co-NN (e) Co-NNN (f) Co-NNNN. The dotted black line indicates the Fermi level which is set to energy zero. }
	\label{Pic_1.7}
\end{figure*}
The iso-surface spin density distribution is performed to stipulate the magnetic orderings between the dopant pairs and the coupling between the foreign and host atoms of the MoS$_2$/WS$_2$ hetero-BL doped at various metal-sites. More importantly, the switching from FM to AFM and again FM coupling is observed with an increase in dopant separations for Mn doping at W-sites as shown in Fig.\ref{Pic_1.6}. The AFM coupling in the NNN configuration results in quenching the net magnetic moment for this case. The commonly shared W atoms could be responsible for the dictated spin polarization of respective Mn dopants. However, in Co-doping, similar long-ranged FM coupling is observed between the dopant pairs and with its sub-neighboring host atoms when compared to ML systems. Unlike the case of Mn-doped ML MoS$_2$, the spin arrangements depict FM interaction between Mn pairs when doped at Mo-sites in MoS$_2$/WS$_2$ hetero BL.\cite{KumarMallik2022Oct} The AFM coupling is observed between the Mn dopants and neighboring S/Mo atoms for NN configuration while it switches to FM coupling with NNN doping arrangements. However, the spin alignments are noticed to be unaltered with induced parallel spins at each configuration of Co-doping giving rise to robust ferromagnetism.
\subsection{Case-II: MoS$_2$/WS$_2$ hetero-BL doped in different layers}
\label{rmf_edf}
On the comparison to the doping at the same layers (Case-I), the doping at different layers i.e. at Mo-W-sites induces partially/fully occupied hybridized orbitals of both Mo and W states contributing to the net magnetization stipulating the interlayer magnetic interaction between two vdW layers as evident from Fig.\ref{Pic_1.7}. The interlayer magnetic interactions among the doped impurities at Mo-W-sites for different doping configurations arise via nonmagnetic conduction charge carriers from the bottom and top layer of S-atoms of the WS$_2$ and MoS$_2$ respectively. The Co atoms are strongly coupled with their neighbouring S atoms around doping sites due to the hybridization of out-of-plane Co 3d and S 3p orbitals creating considerable unbalanced spin populations in spin-split impurity bands near Fermi level. The broad and highly dispersed S-p states provide major contributions to the impurity states in Co-doped Mo-W-site doping configurations which ensure the coupling between the magnetic dopants. These modifications in electronic properties with the various doping configurations of the TM elements into the MoS$_2$/WS$_2$ hetero BLs hold the platform to tailor the nature of the magnetic behavior.
\begin{figure*}
	\centering
	\includegraphics[width=0.9\textwidth]{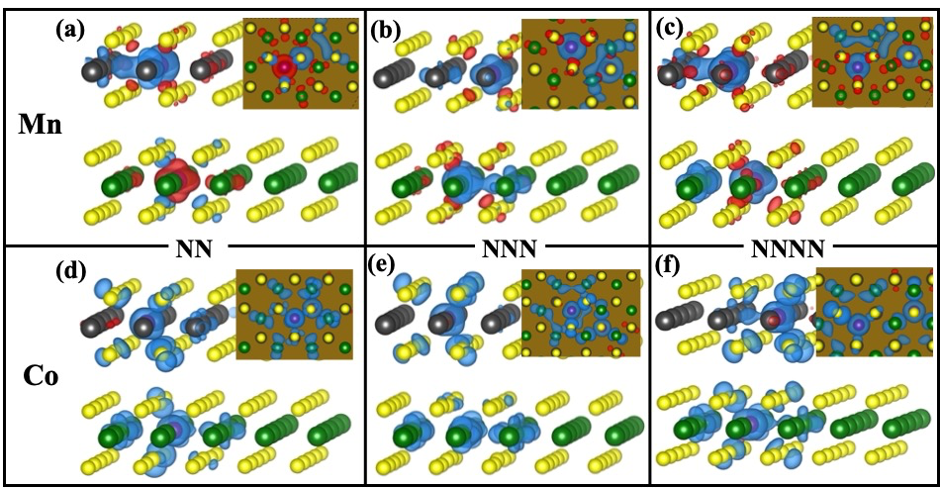}
	\caption{Top (inset) and side views of the isosurface plots showing the spin density distribution around the impurity pairs of the MoS$_2$/WS$_2$ hetero-BL doped with Mn and Co dopants at Mo-W-sites with six doping configurations (a) Mn-NN (b) Mn-NNN (c) Mn-NNNN (d) Co-NN (e) Co-NNN (f) Co-NNNN. The blue and red isosurfaces represent positive and negative spin density, respectively. The value of the iso-surface is taken at 0.002 e/$\AA$$^3$.}
	\label{Pic_1.8}
\end{figure*}
The interlayer magnetic coupling behavior in the case of doping at Mo-W-sites is found to be mediated by nonmagnetic S anions. Figure \ref{Pic_1.8}(a) depicts the Mn atom doped in MoS$_2$ layer anti-ferromagnetically coupled to the Mn atom doped in WS$_2$ layer and the induced spin polarization in the neighboring S atoms are opposite for NN configuration. However, the other two Mn-doped configurations retain their FM nature and the neighboring S atoms are anti-ferromagnetically coupled to the Mn dopants. Moreover, from Fig.\ref{Pic_1.8}(d-f), it can be inferred that the induced spin polarization on the S atoms by the Co pairs is large when compared with the spin polarization on the S atoms induced by the Mn pairs. It reveals a stronger interlayer magnetic exchange coupling between the vdW layers in Co-doped scenario driven by delocalized S-p orbitals. Our results reflect the possible emanations of magnetic orderings in novel vdW non-magnetic heterostructures and tailoring the magnetic properties with TM doping engineering.
\section{Strain engineering in Co-doped in monolayer WS$_2$}
\begin{figure*}
	\centering
	\includegraphics[width=0.9\textwidth]{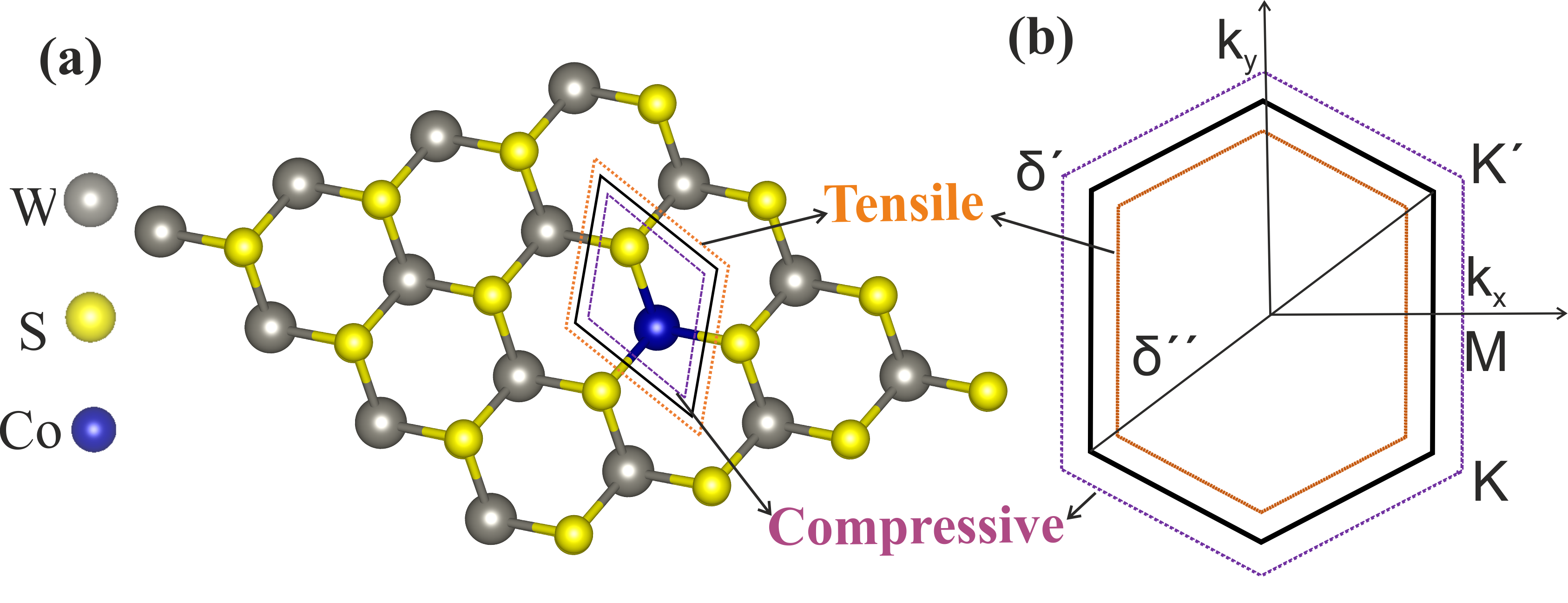}
	\caption{Schematic structure showing (a) top view of 4$\times$4$\times$1 supercell Co-doped WS$_2$ monolayer. The black line represents an unstrained unit cell. The orange and violet dotted lines represent unit cells for the isotropic tensile and compressive strain, respectively. (b) The Brillouin Zone of $(Co, W)S_2$ ML under applied biaxial compressive (dotted orange line) and tensile strain (dotted violet line). $\delta^{\prime}$ and $\delta^{\prime\prime}$ are strain variation in percentage at compressive and tensile strain, respectively.}
	\label{Pic_2.1}
\end{figure*}

To study the effects of Co-doping for the magnetic properties of WS$_2$ monolayer (ML), we replace one host W-atom by foreign Co-atom in 4$\times$4$\times$1 [$(Co, W)S_2$] and extended 5$\times$5$\times$1 super-cell with doping concentrations 6.25 $\%$ and 4.16 $\%$ respectively. The graphical representation of $(Co, W)S_2$ ML with its hexagonal unit cell (black solid line) is shown in Fig. \ref{Pic_2.1}(a), where the optimized lattice constant is found to be 3.16 $\AA$. The structural and magnetic parameters extracted from the calculated results are listed in Table \ref{tab:table1}.
The obtained results show that the bond lengths of W-S shrink around the doping sites locally after Co-doping when compared to pristine WS$_2$. However, the Co-S bond length remains the same as the W-S bond length in pristine WS$_2$ at various doping concentrations. Such behavior may be attributed to various factors such as electron affinities, ionic radii differences between the Co and W ions after the geometrical relaxation and the changed electrostatic forces that arise due to the occurrence of multivalent Co ions at the host W site\cite{doi:10.1063/1.4891495}. It is apparent to notice that WS$_2$ exhibits effective magnetic ordering with net magnetic moments 2.62 $\mu_B$ and 2.58 $\mu_B$ at different Co-doping concentrations of 4.16 $\%$ and 6.25 $\%$, respectively. The obtained results offer similar/higher magnetic moments than the previously studied transition metal-doped WS$_2$ ML system \cite{ZHAO201645}. The observation of such high magnetic moments in Co-doped WS$_2$ ML suggests its potential candidature for 2D DMSs. In order to investigate the effects of external perturbation such as strain on the magnetic properties, both compressive (-5 $\%$ to 0 $\%$) and tensile (0 $\%$ to 5 $\%$) isotropic biaxial strain (along x- and y-axis simultaneously) have been applied to $(Co, W)S_2$ ML. 
\begin{figure*}
	\centering
	\includegraphics[width=0.9\textwidth]{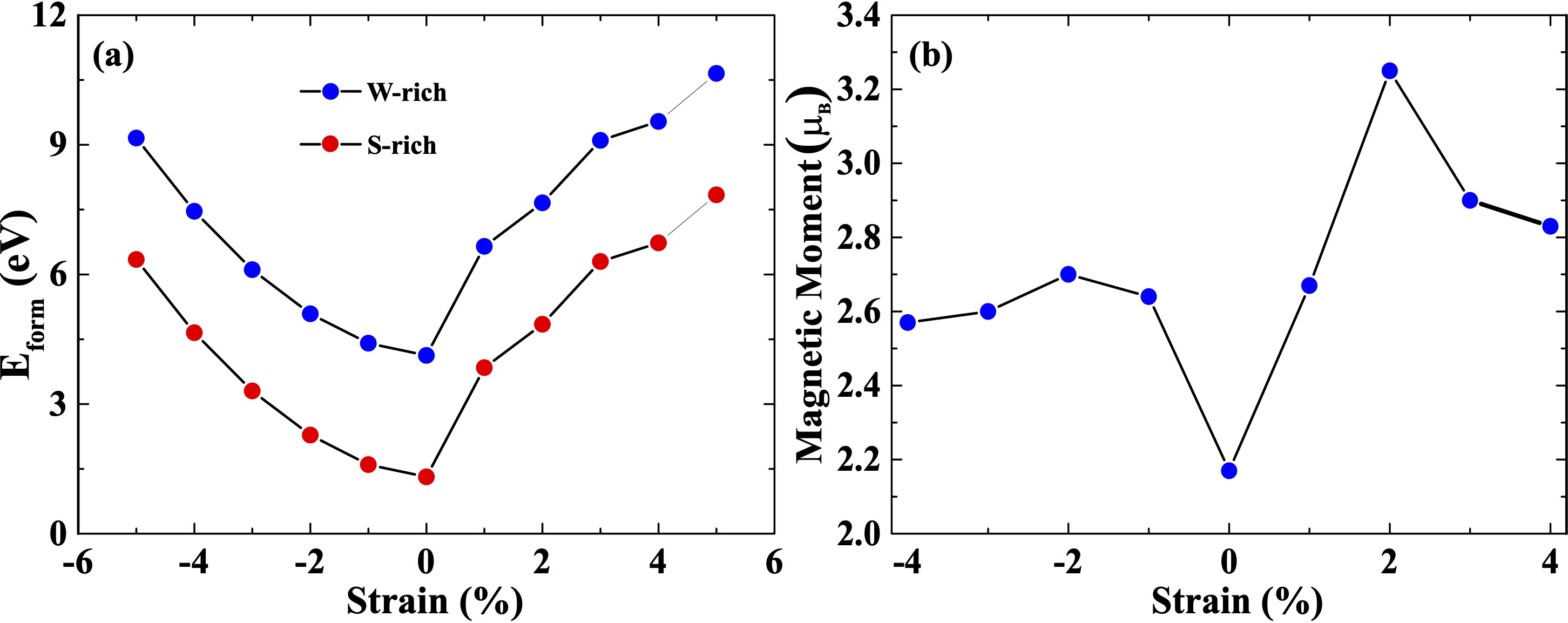}
	\caption{(a) The formation energy ($E_{form}$) of W-rich and S-rich calculated and (b) calculated magnetic moment for $(Co, W)S_2$ ML under biaxial strain, ranging from compressive -5\% to tensile +5\% strains.}
	\label{Pic_2.2}
\end{figure*}
\begin{table}
\centering
 \caption{\label{tab:table1} The calculated structural parameters, magnetic moments, the total energies for undoped and Co-doped WS$_2$ ML at various Co-doping concentration.}
\begin{tabular}{lllllll}
\hline \hline
Supercell  &$d_{W-S}$  &$d_{Co-S}$  &$M_{total}$  &$M_{Co}$   &$M_S$       &$M_W$     \\
           &(\AA)     &(\AA)       &$(\mu_B)$    &$(\mu_B)$  &$(\mu_B)$   &$(\mu_B)$ \\
\hline
WS$_2$                 &2.41        &-       &-       &-      &-           &-                 \\
(W, Co:6.25 $\%$)S$_2$    &2.39        &2.41    &2.58    &1.67   &0.11        &0.03              \\
(W, Co:4.16 $\%$)S$_2$    &2.39        &2.41    &2.62    &1.83   &0.10        &0.01              \\
\hline \hline
\end{tabular}
\end{table}
From Fig.\ref{Pic_2.1}(a), strained unit cells (dotted orange and violet cells for tensile and compressive strain, respectively) have been shown schematically along with the unstrained unit cell. It has been previously observed that the biaxial strain attributes isotropic changes to the magnitude of the lattice vectors of various TMDCs ML\cite{doi:10.1021/acsanm.9b01164}. The relationship between the reciprocal lattice in the Brillouin zone and the primitive vectors in real space is expressed as:   
\begin{equation}
\centering
 a_i.b_i= 2\pi \delta_{i,j} =\begin{cases}
2\pi & i=j\\
0    & i\neq j (i, j=1, 2, 3.....)
\end{cases}  
\end{equation}
where $a_i$ and $b_i$ are the primitive vectors and reciprocal lattice, respectively. Accordingly, under an isotropic/uniform biaxial tensile (compressive) strain, the lattice constant will be increased (decreased), and the reciprocal lattice will be shrunk (enlarged), which in turn reduces (expands) the Brillouin zone as shown in Fig.\ref{Pic_2.1}(b) \cite{doi:10.1021/acsanm.9b01164}. During our calculations, the periodicity is well preserved under both biaxial strains. 
\section{Formation energy of the strained WS$_2$}
The structural stability of $(Co, W)S_2$ ML under isotropic biaxial strain can be estimated by formation energies, which can be expressed as:
\begin{equation}
\centering  
E_{\rm FE}= E_{\rm doped}-E_{\rm pure}+n(\mu_W-\mu_{\rm Co}) \, ,
\end{equation}
where $E_{\rm pure}$ and E$_{\rm doped}$ represent the total energy of the pristine WS$_2$ and Co-doped WS$_2$ ML, respectively. $\mu_W$ and $\mu_{\rm Co}$ are the chemical potential for host W and foreign Co-atom, respectively. \textit{n} is the number of dopants in the studied supercell. The formation energies at various biaxial strains at W- and S-rich conditions are listed in Table \ref{tab:table2}. The formation energies of the strained systems are seen to increase monotonically with the increase in both compressive and tensile strains, as displayed in Fig.\ref{Pic_2.2}(a). Previous experimental results evident that the S-rich condition is more suitable for the practical growth of pristine WS$_2$ monolayer than the W-rich condition \cite{LI20162364}. For our system, it can be noticed that $E_{form}$ for S-rich is lower when compared to W-rich conditions at each studied strain system, as supported by the experimental condition. From Fig.\ref{Pic_2.2}(b), it can be worth noting that the magnetic moments are further modified under various applied strains, which explore possible directions for TMDCs based spintronic and straintronic application \cite{Ahn2020, doi:10.1063/1.5042598}. Among all the applied strains, the maximum magnetic moment of 3.25 $\mu_B$ and 2.69 $\mu_B$ are achieved for +2 $\%$  tensile and -2 $\%$ compressive strain, respectively. The obtained results ensure a higher magnetic moment at low strain as compared to other TM-doped TMDCs \cite{Luo2017, doi:10.1080/00150193.2018.1497417, MIAO2020109459}. According to Luo \textit{et. al.}, the enhancement in the magnetic moments was only observed under compressive strain in Al-doped WS$_2$ ML \cite{doi:10.1080/00150193.2018.1497417}; however, the present study reveals the improvement of the magnetic moment under both compressive and tensile strain by Co-doping. The enhanced magnetic properties under strained conditions may be attributed to the overlapping of spin-polarized electronic band alignments originating from both magnetic cations and nearest neighbour anions.  
\begin{figure*}
	\centering
	\includegraphics[width=0.9\textwidth]{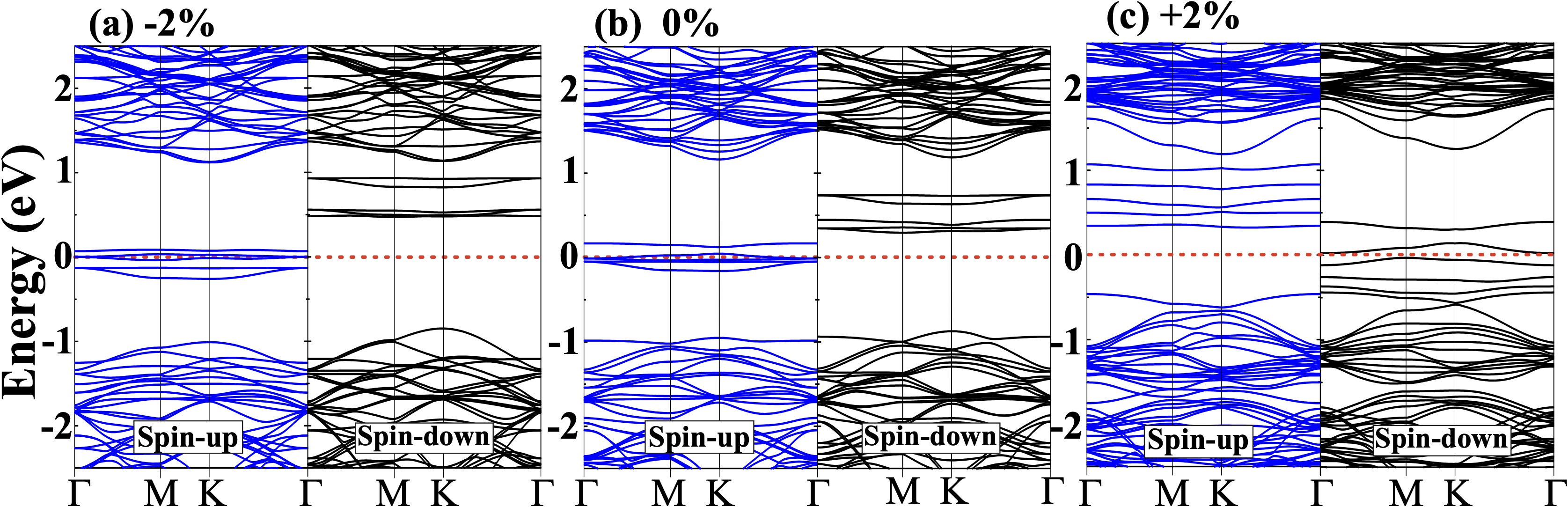}
	\caption{Spin-polarized band structure of (a) -2\% compressive, (b) 0\% unstrained and (c) +2\% tensile strain for single Co-atom doped $(Co, W)S_2$ monolayer. The blue and black line represents the spin-up and spin-down channels, respectively. The horizontal red dotted line indicates that the Fermi level ($E_f$) is set at zero.}
	\label{Pic_2.3}
\end{figure*}
\begin{table}
\centering
\caption{\label{tab:table2} The calculated bond lengths \textit{Co-S} and \textit{W-S}, bond angle \textit{S-Co-S} and formation energy ($E_{\rm form}$) of $(Co, W)S_2$ ML at different compressive and tensile strains.}
\begin{tabular}{lllllll}
\hline\hline
Strain  &$d_{Co-S}$ &$d_{W-S}$  &$\theta_{S-Co-S}$ &$M_T$ & \multicolumn{2}{c}{$E_{\rm form}$ (eV)} \\
 ($\%$)           &(\AA)        &(\AA)      &(\AA)  &$(\mu_B)$  &W-rich &S-rich \\
            
\hline
-5    &2.351   &2.387  &77.128   &2.55      &9.15   &6.34\\

-4    &2.358   &2.390  &78.148   &2.56      &7.46   &4.65\\

-3    &2.376   &2.381  &79.911   &2.60      &6.11   &3.30\\

-2    &2.381   &2.384  &80.914   &2.69      &5.09   &2.28\\

-1    &2.397   &2.389  &81.821   &2.64      &4.40   &1.59\\

0     &2.409   &2.394  &82.709   &2.58      &4.12   &1.13\\

1     &2.431   &2.396  &84.521   &2.67      &6.65   &3.84\\

2     &2.447   &2.397  &86.071   &3.25      &7.65   &4.84\\

3     &2.451   &2.409  &85.285   &2.89      &9.10   &6.29\\

4     &2.468   &2.415  &86.122   &2.82      &9.54   &6.73\\

5     &2.472   &2.421  &87.521   &2.58      &10.65  &7.84\\
\hline \hline
\end{tabular}
\end{table}
\section{Band structure and density of states}
The pristine WS$_2$ ML exhibits a direct bandgap of E$_g$ = 1.87 eV in our calculation\cite{Jena2022Feb} at the K-point having very close approximation with both experimental (E$_g$ = 1.88 eV) \cite{PhysRevB.64.205416} and other theoretical (E$_g$ = 1.68eV ) results \cite{PhysRevB.86.115409}. Figure \ref{Pic_2.3} elucidates the spin-polarized band structures of unstrained and strained $(Co, W)S_2$ ML at their equilibrium lattice constant. However, it can be noticed from Fig.\ref{Pic_2.3}(b) that some impurity states appear within the bandgap when a single W-atom is replaced by Co in both spin-up and spin-down channels. These impurity states are mainly contributed by Co-atom, while the effects from the nearest neighboring W and S atoms can be neglected, which has been explained in other TM-doped TMDCs \cite{ZHAO201645}. The asymmetric behavior of spin-up and spin-down components of Co-doped WS$_2$ ML indicates WS$_2$ behaves magnetically active after Co-doping. Moreover, the doped ML systems show half-metallic characteristics due to the suppression of bandgap. The spin-up component in the doped band structure shows dispersion behavior due to the crossing of band lines across the Fermi level, as shown in Fig.\ref{Pic_2.3}(b). Similarly, the spin-down component of the doped system has a band gap of 1.168 eV with a half-metallic spacing of 0.294 eV for $(Co, W)S_2$ ML. 

Furthermore, the tuning of bandgap at various isotropic biaxial strains brings out effective variations in the magnetic properties of the doped system. The increase in the bond lengths and bond angles under various isotropic tensile strains are listed in Table \ref{tab:table2}. This indicates the decrease in the bond energy between TM elements and S atoms leads to an increase in the number of spin-polarized electrons at the vicinity of the Fermi level. As a result of which, the maximum obtained magnetic moment (3.25 $\mu_B$) is observed at +2 $\%$ tensile strain. Figure \ref{Pic_2.3}(a) and Fig.\ref{Pic_2.3}(c) shows the spin-polarized band structure under the application of -2 $\%$ compressive and +2 $\%$ tensile strain, respectively. Moreover, one can infer that in the case of -2 $\%$, spin-up impurity states are localized near the Fermi level occupying Co-d states, whereas, for +2 $\%$, the spin-down Co-d states are filled to produce the magnetism. This behavior can be correlated to the opposite nature of the magnetic ordering in compressive and tensile strains and may show good performance in transferring one particular spin-oriented electron such as in spin-filtering devices. The highly dispersive impurity d-band lines around the fermi level in case of +2 $\%$ tensile strain renders more electrons on the edge of Co atom spin-polarised when compared to the less dispersive d-band lines in case of -2 $\%$ compressive strain. Interestingly, the bandgap is minimum at strain $\pm$ 2 $\%$, which ensures the bandgap tunability plays a significant role in TMDC-based DMSs. The variation of half-metallic gaps with two different U values (U = 2.5, 3) under unstrained and strained (both compressive and tensile) conditions have also been incorporated.\cite{Jena2022Feb} However, the bandgap is further increased at higher applied strains resulting in the decrease of the magnetic moments in both strain directions.\cite{Jena2022Feb}
\begin{figure}[hbt!]
	\centering
	\includegraphics[width=0.9\textwidth]{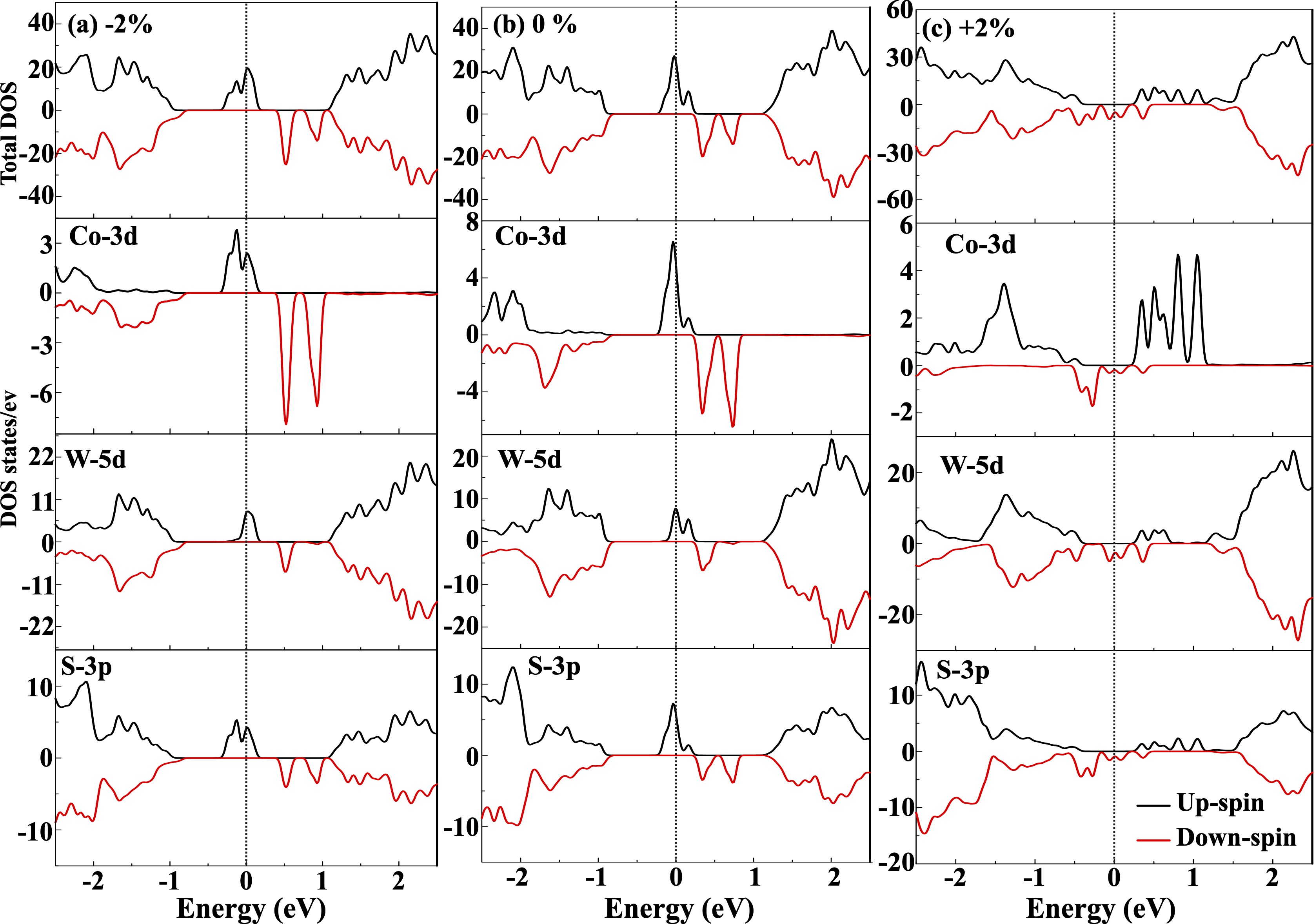}
	\caption{The total density of states (TDOS) and partial density of states (PDOS) of (a) -2 $\%$ compressive, (b) 0 $\%$ unstrained and (c) +2 $\%$ tensile strain for 4$\times$4$\times$1 supercell $(Co, W)S_2$ monolayer. The horizontal line indicates that the Fermi level (E$_f$) is set at zero.}
	\label{Pic_2.4}
\end{figure}

To further elucidate the electronic and magnetic properties in strained $(Co, W)S_2$ ML, the total density of states (TDOS) and partial density of states (PDOS) are plotted in Fig.\ref{Pic_2.4}. The pristine WS$_2$ behaves as a non-magnetic semiconductor, that can be inferred from the symmetric nature of spin-up and spin-down states,\cite{Jena2022Feb} which is well consistent with previously studied results \cite{PhysRevB.86.115409}. Unlike pristine WS$_2$, in $(Co, W)S_2$ ML, the splitting of spin states is observed, giving rise to net magnetic moment (2.58 $\mu_B$), as depicted in Fig.\ref{Pic_2.4}(b). The origin of magnetization in $(Co, W)S_2$ ML is mainly contributed from the additional three unpaired electrons of Co - 3d$^7$4s$^2$ than W - 5d$^4$4s$^2$ with 1.67 $\mu_B$ per Co-atom giving rise to n-type doping. However, the contribution from the nearest neighbor W (0.03 $\mu_B$ per W atom) and S (0.11 $\mu_B$ per S atom) atoms to the net magnetic moment is less than the contribution arising from Co. Similarly, the contribution from the Co-atom and the nearest neighbour W, S atoms at various studied compressive and tensile strains are listed (see SI: Table S1). The TDOS and PDOS for extended 5$\times$5$\times$1 supercell show no significant modification in magnetization after Co-doping \cite{Jena2022Feb}. To understand the magnetic exchange behavior in $(Co, W)S_2$ ML, the interaction between foreign Co-\textit{d} with neighbouring W-\textit{d} and S-\textit{p} needs to be considered. The values of magnetic moments depend on the hybridization among the Co-\textit{d}, W-\textit{d}, and S-\textit{p}. The occupied states near the Fermi level mainly arise from Co-3d and W-5d orbitals in the majority spin channel, as evident from Fig.\ref{Pic_2.4}(b). As the Fermi level lies within the partially occupied majority band of the impurity states, expecting a double exchange coupling between Co and W \cite{Dietl1019, RevModPhys.82.1633}. Similarly, from Fig.\ref{Pic_2.4}(a), it can be inferred that the broadening of the majority bands along with the Fermi level passing through them indicates a stronger double exchange mechanism under compressive strain \cite{MIAO2020109459}. However, in tensile strain, the broadening of the minority charge carriers can be noticed across the Fermi level forming highly delocalized minority spin channels which allows maximum magnetic moment under strained $(Co, W)S_2$ ML. The interaction of Co-\textit{d} with the nearest adjacent S-\textit{p} can be explained in terms of \textit{p-d} hybridization mechanism. The Co atoms are strongly coupled with their neighboring S atoms around the doping sites due to the hybridization of out-of-plane Co-\textit{3d} and S-\textit{3p} orbitals creating considerable unbalanced spin populations in spin-split impurity bands near the Fermi level. It is worth noting from Table \ref{tab:table2} that the Co-S bond length is lower than the W-S bond under compressive strain, which is found to be reversed in the case of applied tensile strain. This change in bond lengths leads to different hybridization mechanisms for varied strains. From the above analysis, it can be believed that the competition between the Co-S and W-S bond lengths may be considered as the prime factor in modulating the magnetic properties under strain engineering. 
\begin{figure}
	\centering
	\includegraphics[width=0.8\textwidth]{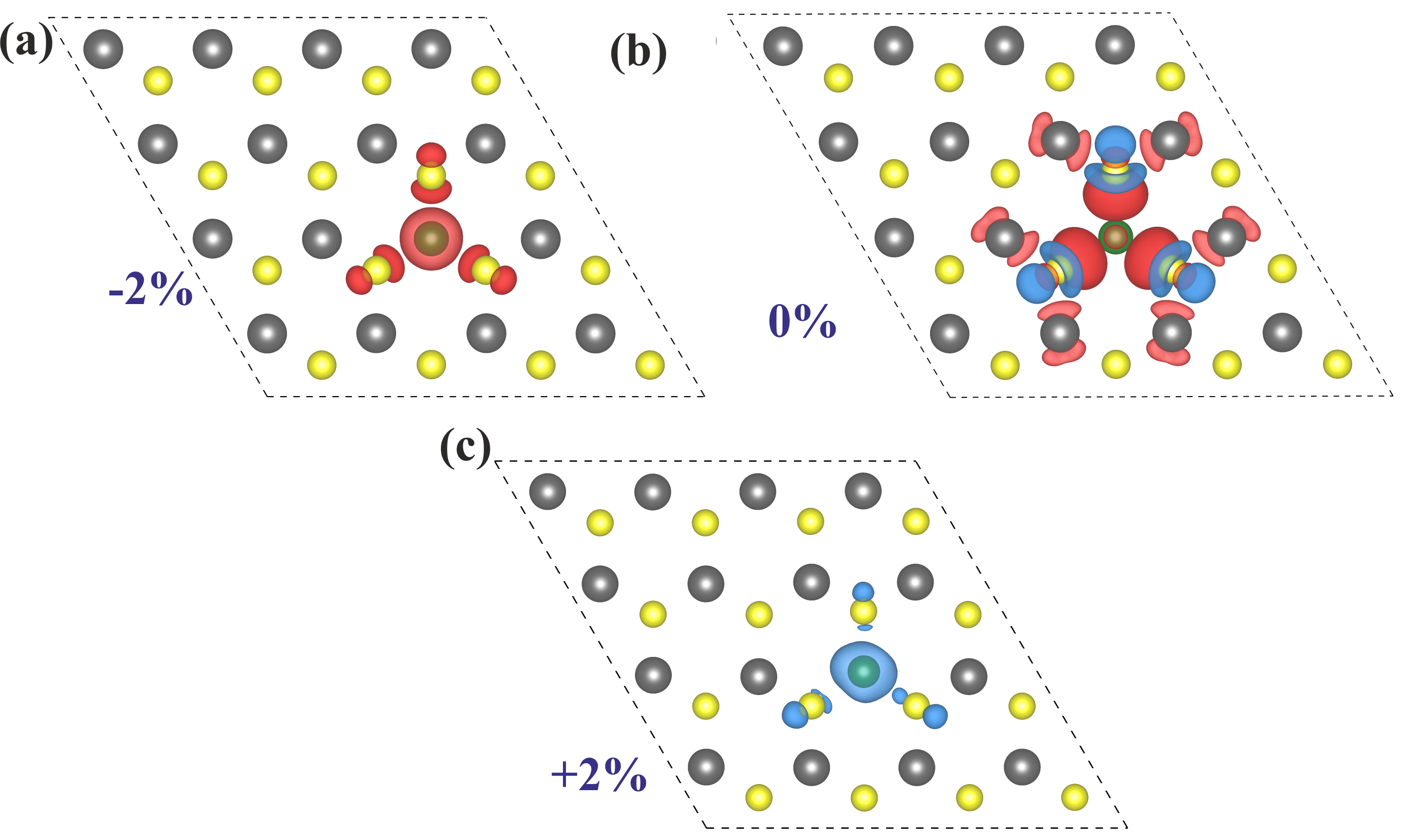}
	\caption{Spin density for a single Co-atom doped $(Co, W)S_2)$ ML at (a) -2 $\%$ compressive, (b) unstrained and (c) +2 $\%$. Red and blue isosurfaces represent positive and negative spin densities ($\pm$ 0.008 e$\AA^3$), respectively.}
	\label{Pic_2.5}
\end{figure}
\section{Spin density distribution}
Next, we consider the spin density distribution to support our understanding of the exchange coupling from the PDOS near the Fermi level, shown in Fig.\ref{Pic_2.5}. Figure \ref{Pic_2.5}(b) elucidates the spin polarization between the Co atom and its nearest W/S in unstrained $(Co, W)S_2$ ML. The coupling between the foreign Co-atom and host W-atom results in parallel spin alignment indicating a double exchange interaction as evident from PDOS in Fig.\ref{Pic_2.5}(b). However, the interaction between Co and nearest neighbor S results in \textit{p-d} hybridization from the out-of-plane orbitals as depicted in Fig.\ref{Pic_2.5}(b). The magnetic coupling behavior in a TM-doped TMDCs system can also be tuned by the introduction of cation and anion vacancies.\cite{Wang2016Apr} In such cases, the trapped electrons in the vacancies coupled with the magnetic moments of the TM impurity ions form bound magnetic polarons (BMP) within the Bohr radius of the impurity site. The recently studied experimental results confirm the strong ferromagnetism induced by BMP driven by S-vacancies in MoS$_2$ nanosheet.\cite{Sanikop2020Jan} However, in our DFT calculations, the magnetic properties of the Co-doped WS$_2$ monolayer have been established by considering the S-rich environment due to their lower formation energies. Additionally, The contribution from BMP and S-deficiency-induced spin delocalization in Co-doped WS$_2$ monolayer under both unstrained and strained conditions has weak effects on the ferromagnetism. From the spin density distribution of $(Co, W)S_2$ ML, it can be perceived that Co-doping induces a long-range magnetic interaction with the nearest neighboring W/S atoms. Moreover, the spin distribution is more localized around its magnetic centers for strained systems when compared to the spin distribution around the magnetic centers of unstrained systems. In the case of unstrained $(Co, W)S_2$ ML, a continuous network of the magnetically coupled TM impurities over local clusters establishes a long-range magnetic interaction leading to a lower percolation threshold. However, under the application of compressive and tensile strains, the local spin clusters around the impurity atoms establish a short-range magnetic interaction, which limits to the nearest neighbour S atoms. In such a scenario, a higher doping concentration is required to reach the percolation threshold. \cite{LAMBRECHT201643} Additionally, it can be concluded that the induced spin density at the dopant site is maximum under +2 $\%$ tensile strain, which reflects the ultimate magnetic moment in this case. For the extended 5$\times$5$\times$1 Co-doped WS$_2$ supercell, the magnetic coupling between dopant and host atoms shows similar behavior.\cite{Jena2022Feb} However, with the increase in either compressive or tensile strain, the magnetic moment decreases due to lower spin polarization.
\begin{figure}
	\centering
	\includegraphics[width=0.8\textwidth]{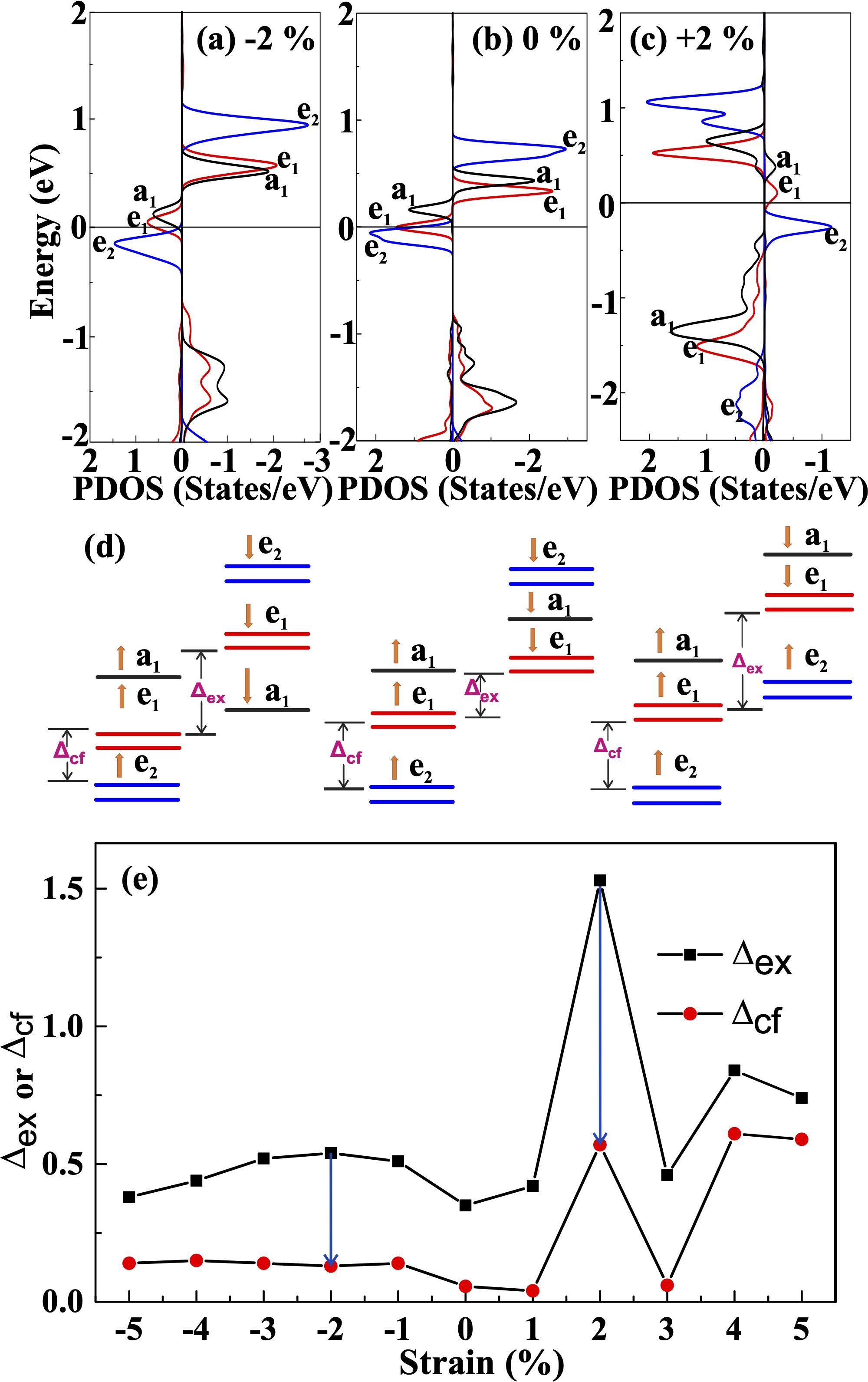}
	\caption{Orbital decomposed partial density of states (PDOS) of Co-doped WS$_2$ ML at (a) -2 $\%$, (b) 0 $\%$, and (c) +2 $\%$. (d) d-orbital splitting of Co-dopant at various applied strain. $\Delta_{ex}$ and $\Delta_{cf}$ represent the intra-atomic Hund's exchange splitting and crystal field splitting, respectively. (e) The exchange splitting ($\Delta_{ex}$) and crystal-filled splitting ($\Delta_{cf}$) at various applied compressive and tensile strains. The vertical arrow indicates the separation between $\Delta_{ex}$ and $\Delta_{cf}$. }
	\label{Pic_2.6}
\end{figure}
\section{Crystal field splitting and Exchange field splitting }
The origin of FM behavior in WS$_2$ ML after doping and strain engineering can be further explained based on orbital decomposed PDOS analysis of the Co-atom, as shown in Fig.\ref{Pic_2.6}. According to ligand field theory, the 3d states of Co atom can be split into single [a$_1$ (d$_{Z^2}$)] and two two-fold degenerate [e$_1$(d$_{xy, x^2-y^2}$), e$_2$ (d$_{xz,yz}$] states. Intra-atomic Hund's exchange splitting ($\Delta_{ex}$) is determined by the energy difference of $e_1$ orbital between the spin-up and spin-down states, whereas the energy difference between $e_1$ and $e_2$ orbitals is referred to as crystal field splitting ($\Delta_{cf}$) \cite{Cheng2013Mar, Kresse1996Oct}. The spin-splitting in Co-doped and strain-engineered WS$_2$ near the Fermi level mainly results from the exchange splitting and crystal field splitting. As suggested by Pan \textit{et. al.}, the FM behavior in TMDCs due to TM doping arises from the competition between the exchange splitting and crystal field splitting \cite{PAN2019125883}. Figure \ref{Pic_2.6}(d) represents the schematic for exchange and crystal field splitting under -2 $\%$ compressive and +2 $\%$ tensile strain compared with the unstrained condition. Moreover, the exchange splitting dominates over crystal field splitting in our studied system due to n-type Co-dopant. From Fig.\ref{Pic_2.6}(d), it can be observed that the difference between $\Delta_{cf}$ and $\Delta_{ex}$ increases under the application of strains, which in turn reflects the increased magnetic moments, as listed from Table \ref{tab:table2}. This $\Delta_{ex}$ and $\Delta_{cf}$ at each studied strain is plotted in Fig.\ref{Pic_2.6}(e). The detailed analysis confirms that the larger the separation between $\Delta_{cf}$ and $\Delta_{ex}$, the greater will be magnetic moments, as evident from Fig.\ref{Pic_2.6}(e) \cite{PAN2019125883}. 
\section{Micro magnetic simulation}
Our detailed observation from the DFT calculation reveals that the Co-doped WS$_2$ ML behaves as FM in nature. In order to better implement an application point of view, the magnetization reversal of WS$_2$ under Co-doping engineering needs to be understood. After substituting Co at W-site, there are two possibilities of formation of anisotropy, one is uniaxial, and another one is biaxial anisotropy. As suggested from the previous reports, the uniaxial anisotropy value is high due to the formation of a larger coercive field (H$_c$) with the easy axis measurement \cite{doi:10.1063/1.360884}. In contrast, the biaxial anisotropy strength is lower, resulting in low H$_c$ values. Experimental results show that the Co-doped WS$_2$ bulk, nanosheets, and ML exhibit hysteresis at very low H$_c$ (few hundreds of Oe) at room temperature \cite{Mao2013, AHMED201825, Kang2020Dec}. A similar FM signature is also reported in the Co-doped WSe$_2$ system \cite{AHMED201825}. However, the nature of the hysteresis is found to be slanted in all these cases. This intrigues us to understand the behavior of magnetization in our system. However, the FM behavior in Co-doped in WS$_2$ has not been thoroughly understood in terms of magnetization reversals. The quest for slanted hysteresis in all the reported cases is still elusive in the scientific community. Here we address the behavior of magnetization reversal and the magnetic effects of the system under strain. We approach micromagnetic modeling by using an open freeware object-oriented-micromagnetic-framework (OOMMF) package to perform qualitative analysis for ferromagnetism \cite{41616}. This is to understand the intrinsic magnetic properties of the FM systems, which have very limited reports available on TMDCs materials. For practical applications, we need a sample dimension of nm-in-range. For micromagnetic modeling, we used a sample dimension of 50$\times$50$\times$1 nm$^3$, and cell size is 1$\times$1$\times$0.5 nm$^3$ to compute the simulation. This micromagnetic simulation governs by Landau-Lifshitz-Gilbert (LLG) equation, can be written as \cite{WANG2001357}:   

\begin{equation}
\frac{d\textbf M}{dt}=-\frac{\gamma}{1+{\alpha}^2}\textbf M \times \textbf{H}_{eff}-\frac{\gamma\alpha}{(1+{\alpha}^2) M_s}  \textbf M \times (\textbf M \times\textbf{H}_{eff})
\end{equation} 

 \noindent $\frac{d\textbf M}{dt}$ provides the information of the \textbf {M} over time, the first term represents the precision of moments, while the second term is responsible for damping. $\gamma$ denotes the gyromagnetic ratio, and $\alpha$ stands for the damping factor, these values are kept constant throughout the process of simulation. $M_s$ represents saturation magnetization, and $\textbf{H}_{eff}$ is the effective field of demagnetization and external magnetic field. $M_s$ is correlated with the total energy and $\textbf H_{eff}$ of the system, $\textbf H_{eff}= -(E_{total}/\textbf M)/(\mu_0 M_{s}$), where  $E_{total}$ is the total energy density of the system. $E_{total}$ would be the sum of all micromagnetics energies, which can be written as:
\begin{equation}
\centering
E_{\rm total}= E_{\rm exch} + E_{\rm anis} + E_{\rm demag} + E_{\rm Zeeman} + E_{\rm me}
\end{equation} 
$E_{\rm exch}$ is the exchange energy, $E_{\rm anis}$ is the magnetocrystalline anisotropy energy, $E_{\rm demag}$ is the demagnetizing or stray field energy, $E_{\rm Zeeman}$ is due to an external field, and $E_{\rm me}$ is the magnetoelastic energy. 
\begin{figure}[hbt!]
	\centering
	\includegraphics[width=0.9\textwidth]{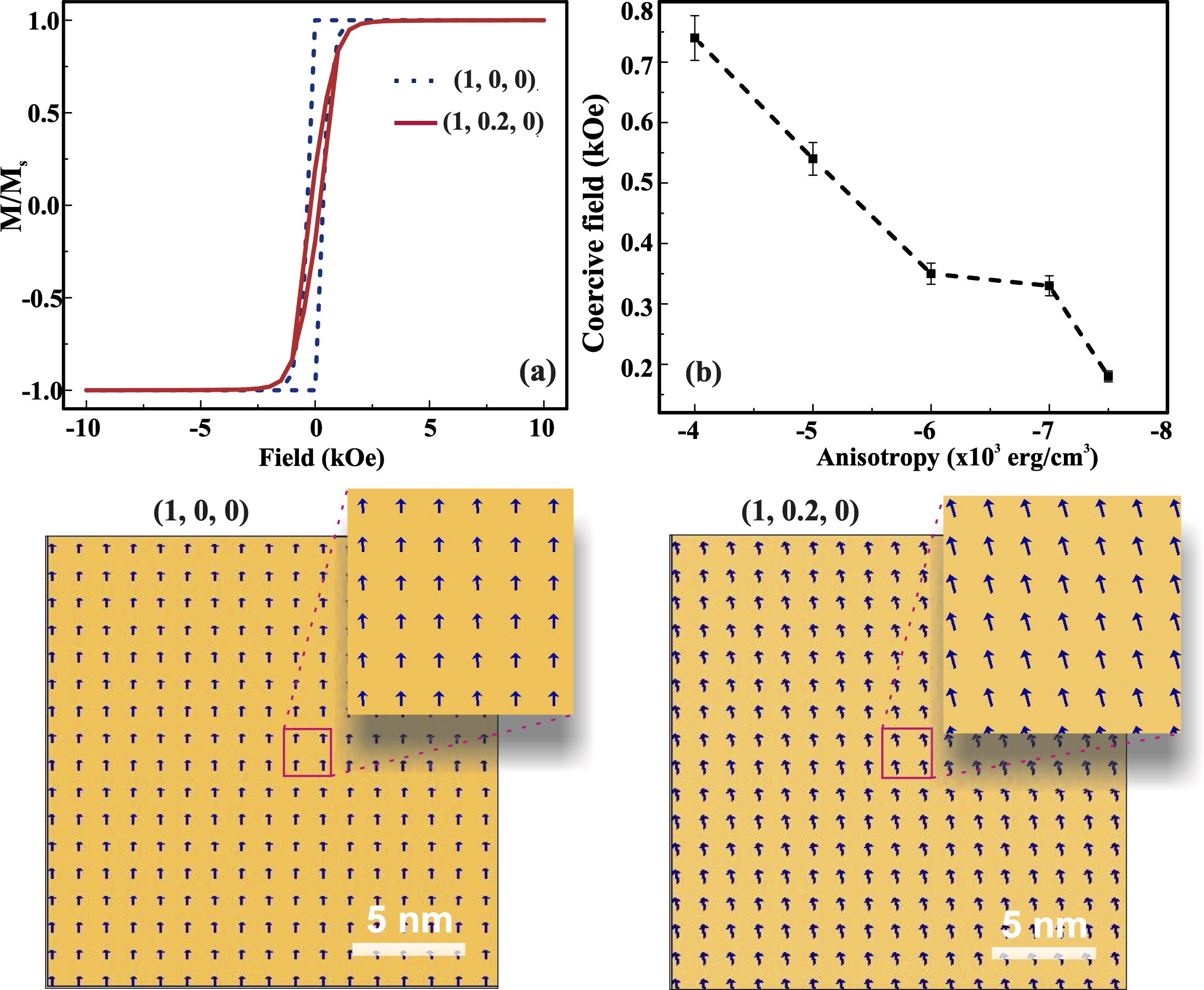}
	\caption{Simulated hysteresis and magnetic domains: (a) represents two loops obtained for (1, 0, 0), and (1, 0.2, 0) orientation. The ground state corresponding domains are marked bottom of it. (b) The variation of the coercive field with the anisotropy values represents the strain-induced ferromagnetism in Co-doped WS$_2$ monolayer.}
	\label{Pic_2.7}
\end{figure}

In the micromagnetic simulation, magnetic parameters are set the same throughout the cells. A stable state is achieved from a random energetic state. The simulation is performed by implementing finite difference methods. The parameters are optimized by looking at various Co-doped systems and the magnetic properties associated with atomic layers of Co films \cite{doi:10.1063/1.360884}. The optimized parameters like exchange length: 2.1 $\times$ 10$^{-6}$ erg/cm, saturation magnetization: 30 emu.cm$^{-3}$, anisotropy values: (4-6)$\times$10$^{-3}$ erg/cm$^3$ are considered for micromagnetics simulation. When Co is doped initially at the W-site, the magnetic moments are randomly dispersed in the cell matrix. On applying a sufficient magnetic field, the Co moments get oriented in the direction of the external field. The snapshots of the various simulated states for the orientation of the spin moments are taken while moving from a random state to a uniform ground state with a systematic increase in an applied magnetic field.\cite{Jena2022Feb} A single magnetic domain state is observed throughout the hysteresis. Figure \ref{Pic_2.7}(a) represents two kinds of hysteresis taken at two different easy axes. The hysteresis for (1, 0, 0) is quite square-in-nature, where the nucleation of domains occurred near the remanence. However, this signature of the loop contradicts the loop obtained for Co-doped in WSe$_2$, and WS$_2$ ML systems \cite{AHMED201825, Mao2013}. To achieve the experimental signature of the loop, which is slightly slanted and nucleation occurs at a distance from the remanence, we consider magnetization orientation tilted by 20 $\%$ off from the original orientation, i.e., (1, 0.2, 0) direction. As a result, the obtained hysteresis is slightly slanted from the previous orientation (1, 0, 0), which is nearly similar to the hysteresis obtained in the WS$_2$ system \cite{Mao2013}. From the previously studied experimental results, it has been observed that WSe$_2$ shows relatively strong ferromagnetism after doping with TM Co \cite{AHMED201825}. The experimental observation shows a tilting magnetization hysteresis curve at all studied temperatures ranging from very low 5 K to 300 K \cite{AHMED201825}. A similar FM response was reported in a wide range of temperatures including room temperature of WS$_2$ \cite{Mao2013}. Although the magnetic parameters of WS$_2$ system are tuned for various factors such as size-dependent (power and nanoribbons), thermal effect, doping, defect engineering, etc., the tilting behavior persists with all conditions \cite{DING2019740, doi:10.1063/1.4820470, AHMED201825, Mao2013, Kang2020Dec}. Hence, the behavior of tilting M-H behavior is unaffected under these conditions, which ensures the anisotropic energy of the system attributed to the tilting behavior in the magnetization hysteresis curve of TMDCs. The coercive field for (1, 0, 0) orientation is around 345 Oe, whereas the same has been reduced to 190 Oe for (1, 0.2, 0). This coercive field is quite agreeing with the result reported in the WS$_2$ system \cite{Mao2013}. Therefore, after Co doping into the system, it is worth mentioning that the effective magnetization orientation is slightly off from the easy anisotropic axis. 

From DFT calculations it is evident that the FM behavior can be tuned under the application of strain. Here, the micromagnetic simulation provides the effect of strain on the larger scale in Co-doped WS$_2$ ML. In this case, if strain is applied to the system, we expect an alteration in the anisotropic values. In the literature also, it is reported that strain can certainly control the anisotropy in thin films \cite{WOLLOCH2021167542, ZHOU2020167303}. Here, we vary anisotropy values to understand the effect of strain in the Co-doped WS$_2$ system. In this case, we tune the anisotropy values from -4 $\times$ 10$^{-3}$ to -7.5 $\times$ 10$^{-3}$ erg/cm$^3$ to observe the changes in the magnetic properties. The coercive field is gradually decreased with the increase in the anisotropy value, which is represented in Fig.\ref{Pic_2.7}(b). On the other way, FM properties are affected by enhancing anisotropy values. This behavior is quite analogous to the results obtained from DFT calculations, where the increase in the percentages in strain values leads to a decrease in FM nature. Our results reinvigorate the FM coupling behavior in Co-doped WS$_2$ ML under various strain conditions; a further understanding of magnetic reversal in this system may pave the way for next-generation spintronics and straintronics applications.
\section{Chapter summary}
In summary, We demonstrate a thorough investigation of the dopant stability, electronic, and magnetic properties in monolayer MoS$_2$ and WS$_2$ as well as MoS$_2$/WS$_2$ hetero-bilayer with the help of first principles calculation. By employing magnetic ions such as Mn, and Co at various host atom sites, we ratify a significant change in magnetic behavior inducing impressive magnetic moments with unwavering dilute doping concentration. The switching between the FM and AFM ordering among the dopant pairs is observed with the dopant pair separation leading to tunable magnetism under doping configurations. The interlayer magnetic exchange coupling is observed between the van der Waals layers with the doping of TMs at both Mo and W-sites and is found to have long ranged ferromagnetic ground state with Co doping. The electronic properties modification-induced change in magnetic exchange interaction is further validated by the spin density distributions. Our results meet the realization of the MoS$_2$/WS$_2$ hetero-BL as a 2D DMS and may further guide future experimental efforts for the precise control over doping of TM magnetic impurities.

Furthermore, we explore the strain-induced ferromagnetism in transition metal Co-doped WS$_2$ monolayer by using first-principles DFT calculations and micromagnetic simulation. Co-doping marks a significant change in magnetic properties with an impressive magnetic moment of 2.58 $\mu_B$. The magnetic exchange interaction is found to be double exchange coupling between Co and W and strong \textit{p-d} hybridization between Co and nearest S, which is further verified from spin density distribution. We find that the resultant impurity bands of the Co-doped WS$_2$ play the role of seed to drive novel electronic and magnetic properties under applied strain. Among several biaxial strains, the magnetic moment is found to be a maximum of 3.25 $\mu_B$ at 2 $\%$ tensile and 2.69 $\mu_B$ at -2 $\%$ compressive strain due to strong double exchange coupling and \textit{p-d} hybridization among foreign Co and W/S. Further magnetic moments at higher applied strain is decreased due to reduced spin polarization. In addition, the competition between exchange splitting and crystal field splitting of Co \textit{d}-orbital plays a significant role in determining these values of magnetic moments under the application of strain. From the micromagnetic simulation, it is confirmed that the Co-doped WS$_2$ monolayer shows slanted ferromagnetic hysteresis with a low coercive field. The effect of higher strain suppresses the ferromagnetic nature, which has a good agreement with the results obtained from DFT calculations. Our findings indicate that induced magnetism in WS$_2$ monolayer under Co-doping promotes the application of 2D TMDCs for the nano-scale spintronics, and especially, the strain-mediated magnetism can be a promising candidate for future straintronics applications.

\chapter{Summary and Outlook}
\label{C7}
The research work in this dissertation is composed of the CVD synthesis of atomically thin 2D materials such as monolayer TMDCs and their potential applications in various electronic devices. This chapter presents the summary of this thesis work and highlights the key advancements achieved in the devices such as transistors, memory, and neuromorphic computing applications. Furthermore, the results from the first-principle density functional theory analysis predict the realization of 2D dilute magnetic semiconductors using monolayer as well as heterostructure under various strain and doping conditions, which could be helpful in future spintronics applications. The scope for future work is also discussed at the end.
\section{Conclusions}

Monolayer Transition metal dichalcogenides (TMDCs, such as MoS$_2$ and WS$_2$ represent a cutting-edge frontier in semiconductor research due to their remarkable optoelectronic properties and potential for integration into existing silicon technology. In this thesis, we have synthesized various 2D TMDCs such as monolayer/multilayers of MoS$_2$, WS$_2$ using a modified CVD setup. We demonstrate an optimization and manipulation of NaCl-assisted CVD synthesis of layered MoS$_2$ under diverse growth conditions. The vibrational and optical properties of the CVD-grown samples are characterized by using Raman and PL spectroscopic techniques. We further discuss key growth parameters that regulate the growth dynamics and severely affect the size, shape, and domain orientation of the samples. Major new parameters include confinement in the geometry surrounding the metal precursors and their ratio with promoters such as NaCl. During the optimization to achieve high-quality large-scale samples, we have also encountered several growth challenges such as premature growth, defects, sulfur vacancies, and non-uniformity in synthesizing MoS$_2$ which has been discussed in this thesis. Understanding these parameters could be beneficial to the high-quality synthesis of 2D materials and their large-scale integration into modern devices. We also extend our study to understand the stacking of 2D materials by performing other important techniques such as low-frequency and polarization-dependent Raman characterizations on the CVD-grown MoS$_2$ samples. In the end, we also provide the salt-assisted synthesis route to obtain large-scale WS$_2$ monolayers.

We have fabricated field-effect transistors on monolayer MoS$_2$ by using the photolithography technique. We discuss crucial device parameters such as contact electrodes, contact resistances, and various device geometries i.e. two-probe method, transfer length method, etc. The optoelectronics response in optimally grown MoS$_2$ samples is presented to demonstrate tunable electronics properties under wavelength selective illuminations. Furthermore, to achieve high-performance hysteresis-free transistors we employ time-domain drain current measurements to minimize the fast transient charge-trapping effects in NaCl-assisted CVD synthesized high-quality monolayer MoS$_2$ FETs. We explore the interfacial properties between drain/source electrode (Ag) and monolayer MoS$_2$ by performing first-principle density functional theory calculations. The room temperature hysteresis studies using DC and pulsed I–V techniques are carried out under both ambient and vacuum conditions to inspect various trapping mechanisms to extract intrinsic FET parameters. An increase in the field-effect mobility up to 30 $cm^{2}V^{-1}s^{-1}$ is achieved in the short-pulse amplitude sweep measurement compared with that in DC measurements. Single-pulse transient measurements are carried out to get further insight into the fast-trapping phenomena. Our studies reinvigorate the origin of hysteresis in salt-aided monolayer MoS$_2$ samples; a further understanding of these significant trapping mechanisms is of prime importance for the next-generation device applications.

In another work, we demonstrate a facile one-step PMMA-assisted etchant-free technique to transfer monolayer MoS$_2$ device arrays with metal (Ag) contacts from one substrate to preferred target substrates. The delamination of device arrays from their growth substrates is facilitated by several optimizations of spin-coating parameters, baking time, temperature, etc. Furthermore, Raman and PL studies reveal the release of strain and interfacial coupling effects before and after transfer. Our direct device transfer (DDT) method yields improvement in the electrical transport outcomes of the transferred devices. In particular, the transistor performances such as ON/OFF ratios, carrier mobilities, Fermi-level pinning, nature of the hysteresis, and contact between Ag/MoS$_2$ are investigated. The temperature-dependent transport measurements are carried out in MoS$_2$ FETs to study the carrier dynamics and trapping mechanisms before and after the DDT method. Our transfer technique could be beneficial to efficiently transfer large array of 2D materials-based devices and sensors for flexible optoelectronic applications.   

This thesis also investigates the robust MoS$_2$/SiO$_2$/Si$^{++}$ three-terminal devices with very different functionality and a novel
mechanism that drives to a high-geared intrinsic transistor at room temperature and exhibits a multi-bit memory cell above room temperature. The proposed thermally-driven programmable
memory operations, i.e., READ, WRITE, ERASE, are
found to be modulated by gate voltage biasing history. The
step-like READ/RESET ratio and retention curves are
obtained for several pulse voltage conditions. The multilevel
states are investigated with varying pulsed width, gate voltage
amplitudes, and drain/source voltages. We also propose the
energy band diagrams that explain the single and multilevel
memory effects in MoS$_2$ FETs to validate the results. These findings indicate the potential of in-memory applications in ion-driven MoS$_2$-based transistors for building multilevel memory arrays, and facilitating complex data processing tasks.

Furthermore, we fabricate such kinds of mem-transistors using monolayer WS$_2$ and demonstrate multi-bit nonvolatile memory applications at high temperatures (~425 K). The WS$_2$ based mem-transistor manifests endurance properties greater than 100 sweep and 400 pulse cycles, high switching ratios, non-volatility with
stable retention and efficient conductance modulation. The
memory behavior is attributed to the interfacial electrostatic
doping effect of channel material through active cation migration
under the applied gate field. Furthermore, the three-terminal
device exhibits more than 64 distinct memory states, making it a
potential candidate for 6-bit memory operations. The multi-bit
memory operations in our device show exceptional pulse control
of charge injection and release, providing a complete yet
straightforward strategy to achieve large-scale data storage
capabilities in WS$_2$ mem-transistors. Moreover, key neuromorphic functionalities such as excitatory post-synaptic current (EPSC), longterm potentiation/depression (LTP/LTD) are demonstrated at elevated temperatures using various pulse schemes in 2D material based synaptic mem-transistors. specifically, the MoS$_2$ synaptic transistors show excellent linearity and symmetry upon electrical potentiation and depression. As a result, a high classification accuracy is achieved during the training and testing of the Modified National Institute of Standards and Technology (MNIST) datasets in artificial neural network (ANN) simulation. We utilize the ionotronic non-volatile conductance weights of the 2D synaptic transistors to establish excellent pattern recognition accuracy approaching software limits. Our results establish a different class of thermally-driven memories based on 2D semiconductors, representing a leap forward for developing high density memories and future neuromorphic computing technologies operating at high temperatures.

In the end, we focus on enriching the properties of MoS$_2$, WS$_2$ non-magnetic semiconductors by introducing spin degree of freedom by doping with magnetic ions. We employ rigorous first-principles density functional theory (DFT) calculations to demonstrate a new type of 2D dilute magnetic semiconductor (DMS) consisting of transition metal doped MoS$_2$/WS$_2$ hetero-bilayer(BL). The Mn and Co doping attribute the consequential change in magnetic properties giving rise to excellent magnetic moments with dilute doping concentration. To collate the results, we implement our calculations on the doped monolayer (ML) MoS$_2$ and WS$_2$ systems. We choose various doping arrangements at different host atom sites to achieve stable configurations under geometry modifications. We discuss the change in magnetic exchange interactions via electronic properties alterations with dopant pair separations. Both ferromagnetic (FM) and Anti-ferromagnetic (AFM) ordering between the dopant pairs is achievable with the tuning of dopant elements and positions. The developed unique interlayer magnetic coupling between the van der Waals layers is observed and discussed for the first time in a hetero-BL system. Furthermore, we have studied a possible emergence of FM in Co-doped WS$_2$ ML under strain engineering, which have potential applications in TMDC-based straintronics. We employed DFT calculations to understand the mechanism of FM behavior in Co-doped WS$_2$ ML at biaxial compressive and tensile strain. The exact behaviour of the FM nature is understood by using crystal-field and exchange-field splitting. The system is further studied by micromagnetic simulation to address the behavior of reversal magnetization and the magnetic effects under strain. The electronic and magnetic properties are also discussed, which will be significant for future spintronic applications. More importantly, this is the first attempt to understand ferromagnetism in TMDCs of low-dimension DFT calculations with nanoscale micromagnetic simulations. Finally, we discussed the synthesis of doped monolayers of MoS$_2$ and WS$_2$ using the CVD method. Magnetization measurements are conducted with a SQUID magnetometer revealing slanted magnetic hysteresis loops. Both low and room temperature which is an indication of weak ferromagnetic order at both 5 K and 300 K, confirms their status as dilute magnetic semiconductors. Nevertheless, additional characterizations are necessary to fully understand and harness the potential of these doped 2D materials for spintronic applications. 
\section{Future Prospective}
Taking the advantages of atomically thin nature, vdW crystal structure with dangling-bond-free surfaces, carrier confinement and tunable electrical and optical responses,  the 2D materials possess enormous potential in miniaturizing various components of modern electronics. Furthermore, they offer promise in advancing parallel computing through continual transistor size reduction, high-efficiency logic gates, and enhanced chip integration. While this thesis focuses on the development of large-scale 2D materials, particularly MoS$_2$ and WS$_2$, for diverse applications in transistors, memory, and neuromorphic computing, numerous challenges and unanswered queries underscore the need for further exploration in this promising field.

At the material level, achieving extensive coverage of 2D materials is imperative for integrated circuit design, while accurate doping is vital for complementary transistor technology. Despite the advancements in controlling the composition and layers of various TMDCs through modified CVD, the growth of high-quality, large-area TMDs remains still at an early stage, which necessitates thorough investigation by the scientific community. Precise control over factors like stoichiometry, thickness, defects, grain size, and surface roughness is essential in the preparation of multi-element, high-quality TMDCs and their heterostructures.

Emphasizing the enhancement of device performance and the integration of vdW materials into compact footprints across all 2D material architectures is crucial. Direct deposition of metal electrodes onto 2D materials disrupts the covalent bonds, atomic arrangements, introducing defects. For instance, a typical MoS$_2$ transistor exhibits a contact resistance of around 10 k$\Omega$ $\mu$m, significantly higher than that of a silicon transistor by more than 100 times.\cite{Kappera2014Dec} Mitigating this contact resistance can be achieved through several approaches, such as employing edge contacts, vdW electrodes, and techniques like strain and phase engineering. Furthermore, alternative technologies capable of boosting the driving current of 2D transistors could be explored, including various doping strategies and the hetero-integration of 2D and 3D stacked heterostructures.

Recent advancements in next-generation storage and computing paradigms have presented significant opportunities in 2D materials research. Expanding the dynamic and multilevel memory capacity can be achieved through the selection of appropriate 2D heterostructures with adjustable carrier doping. The simple architecture of our ionotronic devices facilitates linear synaptic weight updates, ensuring long-term retention suitable for hardware implementation of artificial neural networks. In the future, this technology holds promise for fabricating high-density crossbar arrays for neuromorphic computing and sensing applications at high-temperature. Additionally, optimizing synthesis techniques, device geometry, and dielectric thickness can contribute to reducing the device's power consumption.

\blankpage 
\addcontentsline{toc}{chapter}{REFERENCES}
\bibliographystyle{hcdas}
\bibliography{thesis}
\end{document}